\documentclass[seceqn]{elsart}   

\usepackage{amsmath,amssymb,latexsym,amssymb,tables,axodraw,psfig,mathrsfs}
\usepackage{macros,subfigure,psfrag,color,bbm,cite}
\usepackage[dvips]{epsfig}
\usepackage[dvips]{graphicx}

\input eqalign.sty

\def\@journal{Physics Reports}
\def\@date{April 2007}

\begin{document}

\begin{frontmatter}

\begin{flushright}
{\texttt{DESY 07-080}} \\
{\texttt{SFB/CPP-07-25}}
\end{flushright}

\title{Twisted mass lattice QCD
}

\author[DESY]{Andrea Shindler}\symbolfootnote[2]{Current address: 
    Theoretical Physics Division, Dept. of Mathematical Sciences,
    University of Liverpool, Liverpool L69 3BX, UK. \newline E-mail:{\tt andrea.shindler@liverpool.ac.uk}}
\address[DESY]{
NIC/Deutsches Elektronen-Synchrotron, DESY \\
Platanenallee 6, D-15738 Zeuthen, Germany
}

\maketitle
\begin{abstract}

I review the theoretical foundations, properties as well as the simulation results obtained so far
of a variant of the Wilson lattice QCD formulation: Wilson twisted mass lattice QCD. 
Emphasis is put on the discretization errors and on the effects of these discretization
errors on the phase structure for Wilson-like fermions in the chiral limit.
The possibility to use in lattice simulations 
different lattice actions for sea and valence quarks to ease the renormalization patterns of 
phenomenologically relevant local operators, is also discussed.

\end{abstract}

\\
\endkeyword

\end{frontmatter}
\cleardoublepage

\section{Introduction}
\label{sec:intro}

Quantum Chromodynamics (QCD) is today considered the fundamental theory of strong interactions.
The adimensional coupling of the theory $g_0$, once it is renormalized, will depend
on the energy scale of the considered physical process, and it gives a 
measure of the strength of the interaction at that energy scale.
The outstanding property of QCD, {\it asymptotic freedom}, tells us that 
the coupling decreases with increasing energies. This allows the usage of perturbation
theory to make phenomenological predictions for processes with large momentum transfered.
However the increase of the coupling with decreasing energies does not allow 
to use perturbative methods to compute physical quantities in the low energy 
region such as the mass spectrum or hadronic matrix elements.
A possible strategy in this case is to use a non-perturbative regularization
of the theory, introducing a discretized space-time (lattice)~\cite{Wilson:1974sk,Wilson:1975id}. 
This strategy has two main advantages: first it provides an ultraviolet regularization
and secondly it reduces the degrees of freedom of the theory 
to a numerable infinity.
Considering the theory in a finite volume lattice, it is possible to perform
numerical simulations of QCD through Monte Carlo methods.
The continuum QCD action has to be discretized in a sensible way, and a simple and
attractive lattice action is the one proposed by Wilson long time ago~\cite{Wilson:1974sk,Wilson:1975id}.

The relation between the coupling constant $g_0$ and the lattice spacing $a$ in physical
units is given, in lattice QCD, by the renormalization group equations,
and we are naturally interested in the continuum limit of the theory.
The simulations of lattice QCD are always performed for values of $g_0$ corresponding, 
in the renormalized theory, to a finite and non zero value of $a$.
This introduces in the results of the simulations, using the Wilson fermion action, 
errors of order $a$ and numerically these O($a$) errors could be of the order of $20-30\%$.
The O($a$) discretization errors could be eliminated, in principle, 
following the Symanzik's improvement program~\cite{Symanzik:1981hc,Symanzik:1983dc,Symanzik:1983gh},
where the O($a$) cutoff effects in on-shell quantities are
canceled by adding local O($a$) counterterms to the lattice action and to the
composite fields of 
interest~\cite{Luscher:1984xn,Sheikholeslami:1985ij,Wohlert:1987rf,Luscher:1996sc,Heatlie:1990kg}. 
A technical difficulty is that the improvement coefficients multiplying these counterterms
are not known a priori, and they should be all computed using Monte Carlo simulations.

A new intriguing possibility is the so called automatic O($a$) improvement~\cite{Frezzotti:2003ni}, 
where none of the improvement coefficients are needed in order to have O($a^2$) cutoff effects
in physical observables.
The basic idea is that the Wilson theory for fermions with a suitable infrared cutoff is in the massless limit
free from O($a$) errors.
We will see that to extend automatic O($a$) improvement
to a theory in infinite volume with a non zero mass term
we have to add the so called {\it twisted mass}, 
keeping the standard quark mass to be zero.
The twisted mass term, that in a way will act also as
a sharp infrared cutoff, can be obtained in continuum QCD 
via a non-anomalous chiral rotation from the standard mass term.
To be specific, if we consider QCD with a field $\chi$ describing a flavour doublet, 
the twisted mass term looks like
\be
i \mu_{\rm q} \chibar  \gamma_5 \tau^3 \chi
\label{eq:tm}
\ee
where $\tau^3$ is the Pauli matrix in flavour space and $\mu_{\rm q}$ is what is called twisted mass.

The twisted mass term in a lattice QCD action appears to my knowledge for the first time in 
ref.~\cite{Aoki:1984qi}, where it is given an ansatz for the phase structure of Wilson fermions
in the parameter space $m_0-g_0^2$, where $m_0$ is the bare quark mass.
Based on the analysis of the lattice Gross-Neveu model, 
and on the strong coupling expansion of Wilson lattice QCD, the author suggested that there are regions in 
the parameter space of $g_0^2$ and $m_0$ where the true vacuum has a non zero expectation
value of $i \chibar \gamma_5 \chi$ signalling the spontaneous breaking of parity symmetry.
It was then natural to propose, in order to pick up the real vacuum from numerical simulations, to add
an external field $i H\psibar \gamma_5 \psi$ to the original lagrangian and
to perform the limit $H \rightarrow 0^+$.
The $H$ is what now we would call twisted mass.
The twisted mass in this case is then just an external field used to probe
the structure of the vacuum of the theory and it has to be removed at the end of the computation.
We will come back in deatil to the chiral phase structure of the Wilson theory in sect.5.
Here I just would like to mention that in the same paper a new method to improve the scaling
behaviour of the chiral condensate was proposed based on the observation that the scaling violations 
of the condensate are odd under a change of sign of the coefficient of the Wilson term, and they can be 
easily averaged out. We will see in app.~\ref{app:E} that this is a possible starting point 
to understand automatic O($a$) improvement.

The twisted mass term breaks parity and flavour symmetry. It is a natural question whether this mass term
changes also the continuum action or just the discretization errors of the theory.
In this report we will show that with a Wilson fermion lattice action the twisted mass term generates
parity and flavour violating cutoff effects (in most cases of O($a^2$)) which go away performing the continuum limit.

The fact that the twisted mass actually induces only flavour 
and parity breaking cutoff effects, and it is actually equivalent to QCD, 
can be understood considering an old remark made by Gasser and Leutwyler.
In fact it was noticed many years ago~\cite{Gasser:1985gg} 
that the usage of what we would now call 
a twisted mass term is irrelevant in continuum QCD.
The fermionic part of the 2 flavours QCD 
Lagrangian is usually given in the form
\be
\mathcal{L}_{\rm QCD} = \chibar \big[ \gamma_\mu D_\mu + {\bf m} \big] \chi \,
\label{eq:ferm_cQCD}
\ee
where ${\bf m}$ is the mass matrix.
The quark masses in the standard model originate from the asymmetries of the electroweak vacuum.
Since the electroweak interactions do not preserve parity there is no reason a priori
for the quark mass term of QCD to be parity invariant, and can be generically written as
$\chibar ({\bf m} + i \gamma_5 {\boldsymbol{\mu}})\chi$. We assume here that {\boldmath{${\bf \mu}$}}
is a traceless matrix to avoid an unnecessary discussion of the QCD vacuum angle.
With a suitable non-anomalous chiral transformation of the quark fields the general mass term 
$\chibar ({\bf m} + i \gamma_5 {\boldsymbol{\mu}})\chi$ can always be brought to the standard form, where
the mass matrix is diagonal with real positive eigenvalues $m_{\rm u}$ and  $m_{\rm d}$.
The remaining part of the Lagrangian constrained by the requirement of renormalizability is 
left invariant by this chiral transformation.
In brief, with a change of variables in the functional integral, that leaves the measure invariant,
one can show that even if a general parity and flavour violating mass term is allowed, the request of having a 
renormalizable theory preserving gauge and Lorentz invariance, 
generates these ``accidental symmetries''.
This simple example shows that the specific form of the mass term in the 2 flavours
continuum QCD Lagrangian is actually irrelevant. 
The reason for this is the fact that the massless theory
is invariant under the chiral non-anomalous transformation that changes the form of the mass term.

This is just an example of a more general phenomenon.
Renormalizable theories that describe electro-weak and strong interactions, can
be considered as low energy effective theories of more general not necessarily renormalizable
high energy theories.
The condition of having low energy renormalizable field theories can be so stringent
that the corresponding Lagrangian may turn out to obey extra accidental symmetries, 
that were not symmetries of the higher energy theories~\cite{Weinberg:1995mt}.

This observation becomes important on the lattice.
If we discretize the continuum QCD action with Wilson fermions~\cite{Wilson:1974sk}, 
the Wilson term explicitly breaks chiral symmetry and the lattice action is not invariant anymore under
the field rotations mentioned before.
But we still have the freedom to choose the Wilson term and the mass term
to point in different relative ``directions'' in the Dirac and flavour space~\cite{Frezzotti:2000nk}. 
This freedom is the key to constrain the form of the cutoff effects induced by the Wilson term.

The observation that physical observables computed with the Wilson lattice action 
are automatically O($a$) improved in the ``infrared safe'' 
(i.e. with no spontaneous symmetry breaking)
chiral limit is relevant also for the renormalization properties of local operators that depend
on the breaking of chiral symmetry induced by the Wilson term.

To summarize: Wilson twisted mass QCD is a lattice regularization that allows automatic O($a$)
improvement only tuning one parameter. The bare untwisted quark mass $m_0$ has to be tuned
to the so called critical mass in order
to maximally disalign the Wilson term and the mass term.
In this approach the renormalization of local operators relevant for phenomenological
applications is significantly simplified with respect to the standard Wilson regularization.
The price to pay is the existence of O($a^2$) cutoff effects that break parity and flavour symmetry.
All these statements will be demonstrated explicitly in this report.

This is not the only review on twisted mass QCD. A set of lectures has been presented by S.Sint 
at the School "Perspectives in Lattice Gauge Theories"~\cite{Sint:2007ug}.
In these lectures a nice introduction on the basic setup, exceptional configurations, 
automatic O($a$) improvement together with few applications of Wtm is given.
Particular emphasis is also put in finite volume renormalization schemes with 
chirally twisted bounday conditions.
In our report we enlarge the topics covered by Sint, and we elaborate on the ones already there.
On the other side we touch only marginally finite volume renormalization schemes with 
chirally twisted bounday conditions. For this reason we believe that the present review is 
in many respects complementary to the one of Sint, and together they can be used as a complete
introduction to all the topics connected with twisted mass QCD.

The paper is organized as follows.
In section 2 I use classical considerations in continuum QCD to show the equivalence 
of twisted mass (tm) QCD and QCD. 
I also describe the rigorous theoretical properties of Wilson twisted mass QCD (Wtm QCD),
for degenerate and non-degenerate quarks.
In section 3 I discuss the O($a$) discretization errors of Wilson-like lattice actions.
In particular I show several proof of automatic O($a$) improvement, with a 
particular emphasis on the choice of the critical mass. Numerical results confirming this 
property will conclude the section.
In section 4 I derive again some of the results obtained in the previous sections using a different
fermion basis. Hopefully this could help the reader in a better understanding of the subject of this review.
In section 5 I analyse O($a^2$) parity and isospin violating discretization errors and O($a^2$) cutoff
effects responsible for the non trivial chiral phase structure of the lattice theory.
In section 6 I discuss selected numerical results obtained with Wtm QCD and 
show different methods to ease the renormalization of local operators.
In section 7 I make a short digression on algorithms to simulate light 
Wilson-like quarks. This is an important prerequisite for part of the numerical results
presented in this review.
Conventions, notations and more technical discussions are deferred to the Appendix.

\newpage
\section{Basic properties}
\label{sec:basic}

In this section we introduce twisted mass QCD in the continuum using classical 
arguments for a doublet of degenerate quarks.
This academic exercise allows the reader to get acquainted with twisted mass QCD and 
to learn how to relate correlation functions from QCD to twisted mass QCD.
To extend this concepts at a quantum level we discretize the theory on a
lattice using Wilson fermions. The resulting theory Wilson twisted mass (Wtm) QCD
is ultra-local, unitary, reflection positive and renormalizable to all orders
in perturbation theory.

\subsection{Twisted mass QCD in the continuum}
\label{ssec:tmQCD}

We first consider the continuum limit 
of the twisted mass (tm) QCD  action for $N_{\rm f} = 2$ degenerate quarks. 
We will always work in euclidean space and 
all the definitions and conventions are collected in app.~\ref{app:A}.
We call the set of fermion fields $\{\chi,\chibar\}$ the {\it twisted} basis.
In this basis the tm QCD action reads
\be
  S_{\rm F}[\chi,\chibar,G] =\int {\rm d}^4x\,\chibar\left(\gamma_\mu D_\mu
                +m_{\rm q}+i\mu_{\rm q}\gamma_5\tau^3\right)\chi, 
\label{eq:tmQCDcont}
\ee
where $D_\mu=\partial_\mu+G_\mu$ 
denotes the covariant derivative in a given gauge field $G_\mu$,
and $\tau^3$ is the third Pauli matrix acting in flavour space.
To better understand the structure of the action we write explicitly all the
indices. The fermion fields are 
\be
\chi_{A,\alpha,i}(x)
\ee
where $A$ is the colour index, $\alpha$ the Dirac index and $i$ the flavour
index. The action then reads
\bea
  S_{\rm F}[\chi,\chibar,G] =\int {\rm d}^4x\,
\chibar_{A,\alpha,i}(x)&& \left(  (\gamma_\mu)_{\alpha \beta}(D_\mu)_{AB}\delta_{ij} 
                +(m_{\rm q}) \delta_{AB}\delta_{\alpha \beta} \delta_{i j} \right.\nonumber \\
                 && + \left. (i\mu_{\rm q})(\gamma_5)_{\alpha \beta}(\tau^3)_{i j}
                \delta_{AB}
                \right)\chi_{B,\beta,j}(x). 
\label{eq:tmQCDcont2}
\eea
In the following we will use the compact notation in order to keep the reading
not too heavy.
The mass term of the tm action can be written as
\be
m_{\rm q}+i\mu_{\rm q}\gamma_5\tau^3 = M {\rm e}^{i\alpha \gamma_5 \tau^3}
\label{eq:massterm}
\ee
where
\be
M = \sqrt{m_{\rm q}^2+\mu_{\rm q}^2}
\label{eq:polarmass}
\ee
is the so called polar mass.
At the moment the tmQCD action is just a rewriting of standard QCD in a different basis.
In fact performing the following axial transformation 
\be
 \psi    =\exp(i \omega\gamma_5\tau^3/2)\chi,\qquad
 \psibar =\chibar\exp(i \omega\gamma_5\tau^3/2),
 \label{eq:axial}
\ee
the form of the action is left invariant, but 
the mass term transforms into
\be 
M {\rm e}^{i(\alpha - \omega)\gamma_5 \tau^3}.
\label{eq:transmass}
\ee
In particular, the standard QCD action for $N_{\rm f}=2$ degenerate quarks 
\be
  S_{\rm F}[\psi,\psibar,G] =\int {\rm d}^4x\,\psibar\left(\gamma_\mu D_\mu
                +M\right)\psi, 
\label{eq:QCDcont}
\ee
is obtained if $\omega = \alpha$ , i.e. if the {\it twist angle} $\omega$ satisfies the relation
\be
 \tan{\omega}=\mu_{\rm q}/m_{\rm q}.
 \label{eq:angle}
\ee
We call the {\it physical basis} $\{\psi,\psibar\}$ the basis where the continuum QCD action
takes the standard form. The two basis are related by the rotation~(\ref{eq:axial}) with $\omega$ 
satisfying eq.~(\ref{eq:angle}).

In the following we will mainly use the {\it twisted} basis since this is the basis used in the 
numerical simulations. Although the physical interpretation of the fermionic correlation functions
is most transparent in the {\it physical} basis the renormalization of gauge--invariant 
correlation functions, including those with insertions of local operators, 
looks simpler in the {\it twisted} basis.
This will become clear later when we will discretize the action~(\ref{eq:tmQCDcont}) 
using Wilson fermions~\cite{Wilson:1974sk,Wilson:1975id}.
In sect.~\ref{sec:phys} I derive again some results in the physical basis.

Twisted mass QCD~(\ref{eq:tmQCDcont}) and standard QCD~(\ref{eq:QCDcont})
actions are exactly related by the transformation~(\ref{eq:axial})
and therefore share all the symmetries. The symmetry transformations in the 
{\it twisted} basis are simply the transcription of the standard
symmetry transformations using eq.~\eqref{eq:axial}. 
These symmetry transformations will be from now on called {\it twisted} symmetries
and they are collected in app.~\ref{app:B}.
The twisted vector symmetry $SU_{\rm V}(2)_\omega$ defined in eq.~\eqref{eq:tv} 
is a symmetry of the tmQCD action ~\eqref{eq:tmQCDcont}, 
while the mass term $M$ breaks the twisted axial symmetry $SU_{\rm A}(2)_\omega$ 
defined in eq.~\eqref{eq:ta}. This is equivalent to the transformation properties
of the QCD action~\eqref{eq:QCDcont} under the standard $SU_{\rm V}(2)$ and
$SU_{\rm A}(2)$ symmetries transformations.

In order to give the correct physical interpretation of the interpolating
fields in the twisted basis it is important to have available the relations
between the currents in the two basis.
The axial and vector currents are
\be
  A_\mu^a = \chibar\gamma_\mu\gamma_5{{\tau^a}\over{2}}\chi,\qquad
  V_\mu^a = \chibar\gamma_\mu{{\tau^a}\over{2}}\chi,
  \label{eq:currents_local}
\ee
while the pseudoscalar and scalar densities are defined by
\be
  P^a  = \chibar\gamma_5{{\tau^a}\over{2}}\chi,\qquad
  S^0  = \chibar\chi.
\label{eq:densities}
\ee
The axial transformation~(\ref{eq:axial}) of the quark and anti-quark
fields induces a transformation of the composite fields. We indicate with a
calligraphic symbol the corresponding currents in the physical basis.
For example, the rotated axial and vector currents read,
\begin{alignat}{2}
  {\mcA}_\mu^a &\equiv \psibar\gamma_\mu\gamma_5{{\tau^a}\over{2}}\psi
  &=\;
  \begin{cases}
    \cos(\omega)A_\mu^a + \varepsilon^{3ab}\sin(\omega) V_\mu^b
     & \text{$(a=1,2)$},\\
    A_\mu^3 & \text{$(a=3)$},
  \end{cases} \label{eq:axial_current_rot}\\
  {\mcV}_\mu^a  &\equiv \psibar\gamma_\mu{{\tau^a}\over{2}}\psi &=\;
  \begin{cases}
    \cos(\omega)V_\mu^a + \varepsilon^{3ab}\sin(\omega) A_\mu^b
     & \text{$(a=1,2)$},\\
    V_\mu^3 & \text{$(a=3)$},
  \end{cases} \label{eq:vector_current_rot}   
\end{alignat}
and similarly, the rotated pseudo-scalar and scalar densities are
given by
\begin{alignat}{2}
  {\mcP}^a &\equiv \psibar\gamma_5{{\tau^a}\over{2}}\psi &=\;
  \begin{cases}
     P^a
     & \text{$(a=1,2)$},\label{eq:axial_density_rot}\\
    \cos(\omega) P^3+i\sin(\omega)\frac12 S^0 & \text{$(a=3)$},
  \end{cases}
\end{alignat}
\be
   {\mcS}^0 \equiv \psibar \psi = \cos(\omega) S^0 +2i\sin(\omega)P^3.
   \label{eq:scalar_density_rot}
\ee

The same procedure to change basis can be used for any local operator.
In app.~\ref{app:F} we give the example for a proton interpolating field.

The form of the Ward identities in the {\it twisted basis} is slightly different from
the standard form.
The local $SU_{\rm V}(2) \times SU_{\rm A}(2)$ chiral transformations of the fermionic
fields are defined as follows 
\be
\delta \chi(x) = i \big[ \alpha_V^a(x)\frac{\tau^a}{2} +
\alpha_A^a(x)\frac{\tau^a}{2} \gamma_5 \big]\chi(x)
\label{eq:deltachi}
\ee
\be
\delta \chibar(x) = i \chibar(x) \big[ - \alpha_V^a(x)\frac{\tau^a}{2} +
\alpha_A^a(x)\frac{\tau^a}{2} \gamma_5 \big] ,
\label{eq:deltachibar}
\ee
and the symmetry at the classical level ($\delta S =0$ where $\delta$ indicates
the variations on the fermion fields of eqs.~\ref{eq:deltachi}~-~\ref{eq:deltachibar}) 
gives the so-called partially conserved axial current (PCAC) 
and partially conserved vector current (PCVC) relations 
\be
  \partial_\mu A_\mu^a = 2m_{\rm q} P^a+i\mu_{\rm q} \delta^{3a}S^0, 
\label{eq:PCAC}
\ee
\be
  \partial_\mu V_\mu^a = -2\mu_{\rm q}\,\varepsilon^{3ab} P^b.
\label{eq:PCVC}
\ee
It is easy to verify that the rotated currents and densities 
satisfy the Ward identities in their standard form,
\be
  \partial_\mu {\mcA}^{a}_\mu = 2M {\mcP}^{a},\qquad
  \partial_\mu {\mcV}^{a}_\mu = 0,
\label{eq:WI_phy}
\ee
if $\omega$ is related to the mass parameters as in eq.~(\ref{eq:angle}).
The PCAC and PCVC masses that appear in eqs.~(\ref{eq:PCAC},\ref{eq:PCVC})
are the 2 components of the physical mass which is given by the polar mass $M$
(see eqs.~\ref{eq:polarmass} and~\ref{eq:WI_phy}).
A particular case is when one of the 2 masses vanishes: then the physical
mass is given by the non-vanishing mass.
We will see in the following that a very interesting case is when we work at
{\it full twist}, also called {\it maximal twist}. 
Full twist corresponds at the classical level to $m_{\rm q} = 0$ or equivalently to $\omega=\pi/2$. 
In this case the role of the physical mass is fully played by the twisted mass $\mu_{\rm q}$.

\subsection{Beyond the classical theory}
\label{ssec:beyond}

These classical considerations, based on the
possibility of performing the axial transformation~(\ref{eq:axial}) on the
fermion fields, show that there is a one-to-one correspondence 
between tmQCD and standard QCD.

Denoting the sum of the gauge and fermion QCD actions 
by $S = S_G + S_F$, the physical content of the quantum theory can be extracted 
from its $n$-point correlation functions
\be
  \langle {\mathcal O}(x_1,\ldots,x_n)\rangle = {\mcZ}^{-1}
  \int D[\psi,\psibar] D[U]\,\e^{\displaystyle -S} 
   {\mathcal O}(x_1,\ldots,x_n),
   \label{eq:func_int}
\ee
with 
\be
  {\mcZ}=\int D[\chibar,\chi] D[U]\,\e^{\displaystyle -S},
\label{eq:partition}
\ee
where ${\mathcal O}(x_1,\ldots,x_n)$ is a product of 
local gauge invariant composite fields which are localised
at the space time points $x_1,\ldots,x_n$. 
For the discussion of this subsection we concentrate on the dependence 
of the correlation functions on the fermionic fields and on the mass.
Formally in the functional integral~\eqref{eq:func_int} 
we can make the axial change of variables~\eqref{eq:axial}.
This change of variables is non anomalous, i.e. it does not change the integration measure.
As we have already discussed in the previous subsection only the mass term of the action is affected.
Then the relation between correlation functions in the two basis reads
\be
\langle \mathcal{O}[\psi,\psibar] \rangle_{(M,0)} = \langle O[\chi,\chibar] \rangle_{(m_{\rm q},\mu_{\rm q})}
\label{eq:continuum_rot}
\ee
where the index of the correlation function in the l.h.s indicates that it has been computed in
standard QCD with quark mass $M$ 
and the index of the correlation function in the r.h.s indicates that it has been computed in
tmQCD with quark masses $m_{\rm q}$ and $\mu_{\rm q}$.
The relation between the arguments of the correlation function is given exactly
by eq.~\eqref{eq:axial}.
An explicit example of eq.~\eqref{eq:continuum_rot} reads
\be
\langle \mcA_\mu^1(x)\mcP^1(y)\rangle_{(M,0)} = \cos \omega \langle A_\mu^1(x)P^1(y)\rangle_{(m_{\rm q},\mu_{\rm q})} + 
\sin \omega \langle V_\mu^2(x)P^1(y)\rangle_{(m_{\rm q},\mu_{\rm q})}.
\ee
Correlation functions in QCD can be written as a linear combination of correlation
functions computed in tmQCD at a given twist $\omega$.

This equivalence remains valid at finite lattice spacing for Wilson fermions up to discretization errors,
if the theory is correctly renormalized in a mass independent scheme~\cite{Frezzotti:2000nk}.

To carry over these formal considerations to a rigorous level, 
we have to regularize the theory with a regulator which 
preserves the chiral symmetry of the massless theory, i.e.
which preserves the axial symmetry containing 
the transformation~(\ref{eq:axial}).
A chiral invariant regularization is provided by 
Ginsparg-Wilson (GW) quarks~\cite{Ginsparg:1981bj} on the lattice.

With GW fermions we can repeat the same steps performed 
formally in the continuum theory. 
In particular with GW fermions eq.~\eqref{eq:continuum_rot} 
is valid in the bare theory, and analogously
the bare Ward identity masses coincide with the bare mass parameters of the action. 
Using then Ginsparg-Wilson fermions as a theoretical tool it is possible to
prove~\cite{Frezzotti:2000nk} that tmQCD and QCD are equivalent, i.e. given a lattice
regularization that preserves chiral symmetry, the one-to-one correspondence between tmQCD 
and standard QCD is preserved at {\it finite lattice spacing}: consequently they have the
same continuum limit.
There is only one condition that has to be fulfilled for this statement to be true:
all multiplicative renormalization constants have to be independent of the angle $\omega$,
not only up to cutoff effects~\cite{Frezzotti:2000nk}.
A simple example of a suitable renormalization scheme of this kind is a mass-independent scheme
which is obtained by renormalizing the theory in the chiral limit. 
Under this condition then the relations
between renormalized correlation functions take the analogous form 
given by classical considerations.
In other words in the continuum or with GW fermions, there is no reason to introduce a 
twisted mass term: it can be rotated away by a change of variable in the functional integral, 
because in both cases the massless theories are chirally symmetric.

Based on universality one then expects that tmQCD in other,
not necessarily chirally invariant, regularizations
can again be renormalized such that equivalence between renormalized correlation functions
computed in tmQCD and standard QCD is satisfied
up to cutoff effects~\cite{Frezzotti:2000nk}. 
In particular if the regulated theory breaks chiral symmetry even in the massless limit
the twisted mass term cannot be rotated away and the twist angle $\omega$
parametrizes a family of regularizations, which differ at finite lattice spacing, 
but have the same continuum limit.
It may hence be advantageous to choose the twist angle in a suitable way, e.g. to reduce
discretization errors.

The relations between local fields in the classical continuum theory, e.g. 
eqs.~(\ref{eq:axial_current_rot},\ref{eq:vector_current_rot}), thanks to the 
equivalence between tmQCD and QCD at the quantum level,
can be extended to the renormalized theory. 
Thus the classical theory may be used as a guide to establish the relations
between renormalized correlation functions.
This is a very important result and we will come back on it in sect.~\ref{ssec:corr}, 
in order to better understand which are the relations between tmQCD and QCD 
correlation functions.

\subsection{Wilson twisted mass QCD}
\label{ssec:Wtmaction}

We are interested to exploit the freedom given to us by the 
choice of the twist angle in order to improve the properties of chirally 
non-invariant Wilson regularization of lattice QCD.

The setup is a hypercubic infinite lattice, with lattice spacing $a$. 
Standard reference books for lattice field theories 
are~\cite{Creutz:1984mg,Montvay:1994cy,Smit:2002ug,Rothe:2005nw,DeGrand:2006zz}.
The gauge group is $SU(N_{\rm c})$ and the gauge field on the lattice is an 
$SU(N_{\rm c})$ matrix $U(x;\mu)$ that depends on the lattice point and on the
four directions. Fermion fields reside on the lattice sites and as we have explained 
in the previous section, they carry colour, Dirac and flavour indices.
The lattice action for $N_{\rm f} = 2$ degenerate flavours is of the form
\be
S[\chi,\chibar,U] = S_G[U] + S_F[\chi,\chibar,U] ,
\ee
where $S_G$ denotes the Wilson plaquette action
\be
S_G[U] = \frac{\beta}{6}\sum_{x;\mu\neq\nu} \tr\big\{1 - P^{1\times1}(x;\mu,\nu)\big\} , 
\label{eq:Wilson_gauge}
\ee
with $\beta=6/g_0^2$, $g_0$ being the bare gauge coupling and $P^{1\times1}(x;\mu,\nu)$ the parallel transporter 
around a plaquette $P$
\be
P^{1\times1}(x;\mu,\nu) = U(x;\mu) U(x+a\hat\mu;\nu) U(x+a\hat\nu;\mu)^{-1}U(x;\nu)^{-1} .
\label{eq:plaquette}
\ee
The sum runs over all the oriented plaquettes $P$ on the lattice.
With Wilson quarks the tmQCD Dirac operator is given by
\be
  S_{\rm F}[\chi,\chibar,U] =a^4\sum_x\chibar(x)\Big[D_{\rm W} + m_0 + i\mu_{\rm q}\gamma_5\tau^3\Big]\chi(x), 
\label{eq:WtmQCD}
\ee
where 
\be
D_{\rm W} = \frac{1}{2}\{\gamma_\mu(\nabla_\mu + \nabla^*_\mu) -a r \nabla^*_\mu\nabla_\mu\},
\label{eq:Wilson}
\ee
$\nabla_\mu$, $\nabla^*_\mu$ are the standard gauge covariant forward and backward
derivatives defined in app.~\ref{app:A}, $m_0$ and $\mu_{\rm q}$ are respectively 
the bare untwisted and twisted quark masses.
The parameter $r$ is the so called Wilson parameter and it is always set to 1 unless specified.
We will call the action~(\ref{eq:WtmQCD}) the Wilson twisted mass (Wtm) QCD action.
Sometimes Wilson actions are also written as
\bea
  S_{\rm F}[\chi,\chibar,U] &=& \sum_x \Big\{ \chibar(x)\Big[ 1 +
  i2\kappa a\mu_{\rm q} \gamma_5 \tau^3 \Big]\chi(x)
\nonumber \\
&-& {} \kappa \sum_{\mu=0}^3
  \chibar(x)\Big[ U(x;\mu)(1-\gamma_\mu)\chi(x+\hat{\mu})  \nonumber \\
&+& U(x-\hat{\mu};\mu)^{-1}(1+\gamma_\mu)\chi(x-\hat{\mu})\Big] \Big\}, 
\label{eq:WtmQCDhopping}
\eea
where we define the rescaled dimensionless fermion field
\be
\chi \rightarrow \frac{\sqrt{2 \kappa}}{a^{3/2}} \chi, \qquad x_\mu \rightarrow a x_\mu
\ee
and the hopping parameter
\be
\kappa = \frac{1}{8 + 2 am_0}.
\ee
The hopping parameter is an alternative way to label the bare untwisted quark mass, as $\beta$ defined before
is an alternative way to label the bare gauge coupling.

The first term of eq.~(\ref{eq:Wilson}) is a standard symmetric discretization
of a lattice derivative and the second term 
\be
a \chibar \nabla^*_\mu\nabla_\mu \chi
\label{eq:Wilson_term}
\ee
is the so-called Wilson term. It can be easily checked that this term is not
invariant under all the axial transformations, but it is needed in order to remove
from the spectrum of the theory other spurious particles, called doublers,
introduced by the lattice discretization. In the case of naive fermions,
without the Wilson term, the existence of the doublers is related to the so
called spectrum doubling symmetry~\cite{Karsten:1980wd}, that can be seen as
an exchange symmetry among the corners of the Brillouin zone in the reciprocal
momentum space. It can be shown that the doubling phenomenon is a more general
feature of ultra-local\footnote{
With ultra-local actions we think of actions where the interaction range is
spread over a finite number of points of the lattice.}
actions that goes under the name of Nielsen-Ninomiya 
theorem~\cite{Nielsen:1980rz,Nielsen:1981xu}.
We will not go further into this topic, but since we are here interested in
an action with a next-neighbour interaction (i.e. ultra-local) we need to
insert a Wilson term.

The Wilson term~\eqref{eq:Wilson_term} breaks explicitly the axial symmetry~\eqref{eq:axial}.
This implies that the twisted mass term cannot be rotated away by a chiral transformation
and the exact equivalence between the Wilson action with vanishing and non-vanishing 
twisted mass is lost: Wilson and Wilson twisted mass are different lattice regularization.
As we have discussed in sect.~\ref{ssec:beyond} the exact equivalence is recovered
only in the continuum limit.

\subsection{Symmetries}
\label{ssec:symmetries}

The introduction of a twisted mass and Wilson terms requires an analysis of
symmetries.
In sect.~\ref{ssec:tmQCD} we have already seen that at the classical level the symmetries
of tmQCD are the transcriptions of the standard QCD symmetry transformations via the axial
transformation~\eqref{eq:axial}.
They are collected in app.~\ref{app:B}.
In the regulated theory, the action given in (\ref{eq:WtmQCD}) breaks some of the symmetries 
of the classical action~\eqref{eq:tmQCDcont}. The Wilson term~\eqref{eq:Wilson_term} breaks
twisted parity $\mathcal{P}_\omega$~\eqref{eq:pomega}, twisted time reversal 
$\mathcal{T}_\omega$~\eqref{eq:tomega}, and twisted vector symmetry $SU_{\rm V}(2)_\omega$~\eqref{eq:tv}.
Wtm shares with standard Wilson fermions the following symmetries:
gauge invariance, lattice rotations, translations and charge conjugation $\mathcal{C}$
(see app.~\ref{app:B} for the definition).
The ordinary parity symmetry transformation
\be
\mathcal{P} \colon
\begin{cases}
U(x_0,\bx;0) \rightarrow U(x_0,-\bx;0), \quad 
U(x_0,\bx;k) \rightarrow U^{-1}(x_0,-\bx - a \hat{k};k), 
\quad k = 1, \, 2, \, 3 \\ 
\chi(x_0,{\bf x}) \rightarrow \gamma_0  \chi(x_0,-{\bf x}) \\
\chibar(x_0,{\bf x}) \rightarrow  \chibar(x_0,-{\bf x}) \gamma_0  
\end{cases}
\label{eq:parity}
\ee
is only a symmetry if combined either with a discrete
flavour rotation
\be
\mathcal{P}^{1,2}_F \colon
\begin{cases}
U(x_0,\bx;0) \rightarrow U(x_0,-\bx;0), \quad 
U(x_0,\bx;k) \rightarrow U^{-1}(x_0,-\bx - a \hat{k};k), 
\quad k = 1, \, 2, \, 3 \\ 
\chi(x_0,{\bf x}) \rightarrow i \gamma_0 \tau_{1,2} \chi_l(x_0,-{\bf x}) \\
\chibar(x_0,{\bf x}) \rightarrow -i \chibar(x_0,-{\bf x}) \tau_{1,2} \gamma_0  
\end{cases}
\label{eq:PF12}
\ee
or with a sign change of the twisted mass term 
\be
\widetilde{\mathcal{P}} \equiv \mathcal{P} \times 
[\mu_{\rm q} \rightarrow - \mu_{\rm q}]\,.
\label{eq:Ptilde}
\ee
The same holds for ordinary time-reversal
\be
\mathcal{T} \colon
\begin{cases}
U(x_0,\bx;0) \rightarrow U^{-1}(-x_0-a,\bx;0), \quad 
U(x_0,\bx;k) \rightarrow U(-x_0,\bx;k), 
\quad k = 1, \, 2, \, 3 \\ 
\chi(x_0,{\bf x}) \rightarrow i\gamma_0 \gamma_5 \chi(-x_0,{\bf x}) \\
\chibar(x_0,{\bf x}) \rightarrow  -\chibar(-x_0,{\bf x}) i \gamma_5 \gamma_0  
\end{cases}
\label{eq:time}
\ee
which is only a symmetry if combined either with a discrete
flavour rotation
\be
\mathcal{T}^{1,2}_F \colon
\begin{cases}
U(x_0,\bx;0) \rightarrow U^{-1}(-x_0-a,\bx;0), \quad 
U(x_0,\bx;k) \rightarrow U(-x_0,\bx;k), 
\quad k = 1, \, 2, \, 3 \\ 
\chi(x_0,{\bf x}) \rightarrow i \gamma_0 \gamma_5 \tau_{1,2} \chi_l(-x_0,{\bf x}) \\
\chibar(x_0,{\bf x}) \rightarrow -i \chibar(-x_0,{\bf x}) \tau_{1,2} \gamma_5 \gamma_0  
\end{cases}
\label{eq:PF12}
\ee
or with a sign change of the twisted mass term 
\be
\widetilde{\mathcal{T}} \equiv \mathcal{T} \times 
[\mu_{\rm q} \rightarrow - \mu_{\rm q}]\,.
\label{eq:Ptilde}
\ee
Incidentally this implies that $\mathcal{C}\mathcal{P}\mathcal{T}$ is a
symmetry of the lattice action~\eqref{eq:WtmQCD}.
Concerning the continuous symmetries the ordinary isovector $SU_{\rm V}(2)$
symmetry is broken explicitly by the $\mu_{\rm q}$ term down to the $U_{\rm v}(1)_3$
subgroup with diagonal generator $\tau_3$
\be
U_{\rm V}(1)_3 \colon
\begin{cases}
   \chi(x)    \rightarrow \exp(i \frac{\alpha_V}{2} \tau^3)\chi(x),\\
   \chibar(x) \rightarrow \chibar(x)\exp(-i\frac{\alpha_V}{2}\tau^3).
\end{cases}
\label{eq:u3}
\ee
which is a symmetry of the lattice action~\eqref{eq:WtmQCD} together with the $U(1)$ 
transformations associated with fermion number conservation.

\subsubsection{Chiral symmetry at $\omega={\pi \over 2}$}
\label{sssec:chiral}

In this section we analyze the case of full twist, i.e. $\omega = \pi/2$,
because for axial and vector transformations this is a special case.
We will discuss in sect.~\ref{ssec:corr} and extensively in
sect.~\ref{ssec:crit_mass} how to tune, in a non-perturbative way, the twist
angle $\omega$, i.e. the bare untwisted quark mass $m_0$. For what follows it
is enough to assume that $m_0$, i.e. $\omega$, has been tuned appropriately.
 
The twisted ``charged'' axial transformations for $a=1,2$ and $\omega = \pi/2$
read
\begin{displaymath}
\left[ U_{\rm A}(1)_{\pi \over 2}\right]_{1,2} : \left\{\begin{array}{lll}
   \chi(x)    & \longrightarrow & 
               \frac{1}{\sqrt{2}}\left(1 - i \gamma_5 \tau^3\right) 
\exp\left(i\frac{\alpha_A^{1,2}}{2}\gamma_5\tau^{1,2}\right)
 \frac{1}{\sqrt{2}}\left(1 + i \gamma_5 \tau^3\right) \chi(x),\\
   \chibar(x) & \longrightarrow & 
               \chibar(x)\frac{1}{\sqrt{2}}\left(1 + i \gamma_5 \tau^3\right)
\exp(i\frac{\alpha_A^{1,2}}{2}\gamma_5\tau^{1,2})
\frac{1}{\sqrt{2}}\left(1 - i \gamma_5 \tau^3\right) ,
\end{array} \right.
\end{displaymath}
hence
\be
\left[ U_{\rm A}(1)_{\pi \over 2}\right]_{1,2} \colon
\begin{cases}
   \chi(x)    \longrightarrow  
               \exp\left(\pm i\frac{\alpha_A^{1,2}}{2}\tau^{2,1}\right)\chi(x),\\
   \chibar(x)  \longrightarrow 
               \chibar(x) \exp(\mp i\frac{\alpha_A^{1,2}}{2}\tau^{2,1}) .
\end{cases}
\label{eq:uA12}
\ee
The twisted ``charged'' vector transformations for $a=1,2$ read
\be
\left[ U_{\rm V}(1)_{\pi \over 2}\right]_{1,2} \colon
\begin{cases}
   \chi(x)    \longrightarrow  
               \exp\left(\pm i\frac{\alpha_V^{1,2}}{2}\gamma_5\tau^{2,1}\right)\chi(x),\\
   \chibar(x)  \longrightarrow 
               \chibar(x) \exp(\pm i\frac{\alpha_V^{1,2}}{2}\gamma_5\tau^{2,1}) .
\end{cases}
\label{eq:uV12}
\ee
For $\omega={\pi \over 2}$ and for $a=1,2$ the form of the vector and axial
transformations is reversed compared with the ordinary transformations.
In the continuum this is really just a different transcription of the chiral
transformations, coming from a different choice of the fermionic basis.
On the contrary in the regulated theory different terms of the action break different sectors
of the axial and vector transformations.
In particular we observe that the Wilson term~\eqref{eq:Wilson_term} breaks
the ``charged'' twisted vector symmetry~\eqref{eq:uV12} and it is invariant
under the ``charged'' twisted axial symmetry transformation~\eqref{eq:uA12}.
The twisted mass term has effectively an orthogonal behaviour because it is
invariant under the ``charged'' twisted vector symmetry
transformation~\eqref{eq:uV12} 
and it breaks the ``charged'' twisted axial symmetry~\eqref{eq:uA12}.

To be more specific if we set the twisted mass to zero the Wilson theory is 
invariant under the group
\be
\widetilde{SU}_{\rm A}(2) \equiv \left[ U_{\rm A}(1)_{\pi \over 2}\right]_{1} 
\otimes \left[ U_{\rm A}(1)_{\pi \over 2}\right]_{2} 
\otimes \left[ U_{\rm V}(1)_{\pi \over 2}\right]_{3} ,
\label{eq:obl_grA}
\ee 
while the twisted mass term is invariant under the group
\be
\widetilde{SU}_{\rm V}(2) \equiv \left[ U_{\rm V}(1)_{\pi \over 2}\right]_{1} 
\otimes \left[ U_{\rm V}(1)_{\pi \over 2}\right]_{2} 
\otimes \left[ U_{\rm V}(1)_{\pi \over 2}\right]_{3} .
\label{eq:obl_grV}
\ee 
This means that for the ``charged'' 
axial transformations the lattice 
action~(\ref{eq:WtmQCD})\footnote{I acknowledge a very useful discussion with
  G.C. Rossi on this point.} 
has a continuum-like behaviour:
the twisted ``charged'' axial symmetry $\left[ U_{\rm A}(1)_{\pi \over 2}\right]_{1,2}$
is only softly broken by the mass term.
This exact symmetry of the massless theory protects the charged pion
from chiral breaking cutoff effects (see eq.~\ref{eq:mpisqGSM}).
This result is obviously independent on the choice of the basis and
we will discuss it again in sect.~\ref{sec:phys}. 
A consequence of this consideration is also that the 
charged vector current in the twisted basis is protected from renormalization 
and it is at the same time the current that defines the pseudoscalar decay constant for the charged pion
(see eqs.~\ref{eq:rotation_AP0},\ref{eq:rotation_AP},\ref{eq:fPS}).
The neat result being that at full twist the pseudoscalar decay constant for the charged pion does not need
to be renormalized (see sect.~\ref{sec:ren}).

This analysis also shows in which sense the Wilson and the mass term 
at full twist are maximally disaligned: they are maximally disaligned concerning
chiral symmetry.
While the Wilson term breaks twisted ``charged'' flavour
symmetry~(\ref{eq:uV12})
the mass term breaks as in continuum QCD the full axial group~(\ref{eq:ta}),
or to phrase it differently the Wilson term 
preserves a subgroup of twisted axial symmetry~(\ref{eq:uA12}) while the mass term does not.
What is relevant is that the Wilson and the mass term are ``orthogonal'' 
concerning the ``charged'' subgroup of chiral symmetry and this is achieved 
at full twist $\omega=\pi/2$.

\subsection{Tree-level}

To get some more insights on Wtm and to understand better 
the importance of disaligning mass term and Wilson term
we compute here the tree-level Wtm
propagator, that can be written as an integral over the first Brillouin zone of
\be
\widetilde{G}(p) =  \frac{-i \gamma_\mu
  \ppall_\mu + \mathcal{M}(p) - i \mu_{\rm q} \gamma_5 \tau^3}{\ppall_\mu^2 +
  \mathcal{M}(p)^2 + \mu_{\rm q}^2}
\label{eq:Wtmtree}
\ee
where we have defined 
\be
\ppall_\mu = \frac{1}{a}\sin(ap_\mu), 
\qquad \mathcal{M}(p) = m_0 +\frac{r}{2}a\pcap_\mu^2, \qquad \pcap_\mu =
\frac{2}{a}\sin(\frac{ap_\mu}{2}).
\ee
The poles of the propagator give us the spectrum of the theory.
If we now make an expansion for small lattice spacing $a$ neglecting all the
terms of O($a^2$) we obtain
\be
p^2 +m_0^2+am_0rp^2 + \mu_{\rm q}^2, \qquad {\rm where} \quad p^2 = p_\mu p_\mu .
\label{eq:disp}
\ee
The leading O($a$) discretization effects of the dispersion relation are
given by the term $am_0rp^2$.
We can already make some interesting remarks.
This term vanishes if we set $m_0=0$. This means that 
the chiral limit of the plain Wilson theory ($\mu_{\rm q} = 0$) does not have O($a$)
effects.
Of course we are only considering the tree-level and if we would switch on the
interaction the dynamics of the theory could change this result. We will see
that if we consider the theory in a finite volume with suitable boundary
conditions, where the mass dependence of the theory is smooth and there are no
phase transitions, this result is still true. 

From eq.~(\ref{eq:disp}) we also see that even if we set $m_0=0$ we can add a
mass to the theory without introducing O($a$) effects. To
understand how this can happen it is good to understand the origin of the
O($a$) term $am_0rp^2$. It comes from the cross term between the Wilson term and
the mass term in $\mathcal{M}(p)$. This cross term is absent with a twisted
mass because twisted mass and Wilson term point in different ``directions'' in
the chiral-flavour space.

Even if $m_0$ does not vanish but $m_0 = $ O($a$) this will not change our conclusion
since the term with $m_0$ will only change the O($a^2$) terms.

All these considerations are only at tree-level, but we will see in
sect.~\ref{sec:impro} that they remain true if we switch on the interaction
between quarks and gluons. In particular
setting the Wtm action at full twist allows to have physical observables that
are automatically O($a$) improved.

\subsection{Transfer matrix}
\label{ssec:transfer}

Necessary and sufficient conditions under which 
the physical content of the theory in Minkowski space can be reconstructed from 
Euclidean Green's functions are the so called Osterwalder-Schrader 
conditions~\cite{Osterwalder:1973dx,Osterwalder:1975tc}.
One of the required properties that does not obviously hold in a lattice theory
is physical positivity. This condition states that given a gauge invariant polynomial 
of positive time ($x_0 >0$) fundamental fields $O$ one should have
\be
\langle \Theta(O^\dagger)O\rangle \ge 0
\label{eq:OSpos}
\ee
where $\Theta$ denotes euclidean time reflection and $O^\dagger$ 
is the Hermitian conjugate of $O$.
An explicit expression for the transfer matrix that in particular
is strictly positive, i.e. all its eigenvalues are bigger than zero, 
was given for Wilson fermion and gauge actions in ref.~\cite{Luscher:1976ms}.
This allows to prove the positivity condition~(\ref{eq:OSpos}) for the Wilson 
action at finite lattice spacing.

In the app.~\ref{app:C} we briefly repeat the steps of the proof with the extension to Wtm,
because they become important for non-degenerate quarks.
Here we simply list the main result: adding a twisted mass term for 
degenerate quarks does not change~\cite{Frezzotti:2001ea} the constraint on $\kappa$ 
($|\kappa| < {1 \over 6}$) valid~\cite{Luscher:1976ms} for Wilson fermions.

\subsection{Renormalization}
\label{ssec:reno}

In perturbation theory, it has been  
shown that Wilson lattice QCD is 
renormalizable~\cite{Reisz:1987da,Reisz:1987pw,Reisz:1987px,Reisz:1987hx,Reisz:1988kk}, 
and we shall assume that this remains true beyond perturbation theory. 
Since the twisted mass term can be viewed 
as a super-renormalizable interaction term 
which does not modify the power counting, this result can be extended also to Wtm.

To understand the structure of the counterterms we use the symmetries of the Wtm 
lattice action~(\ref{eq:WtmQCD}), treating the masses as spurion fields which transform under these 
symmetries.
The counterterms to the action with dimension less or equal four are
\be
\tr\{F_{\mu \nu} F_{\mu \nu} \}, \quad \chibar \chi, \quad m_0 \chibar \chi, \quad
i \mu_{\rm q} \chibar \gamma_5 \tau^3 \chi , 
\label{eq:counter}
\ee
where $F_{\mu \nu}$ is the gluon field strength tensor.
The first counterterm gives a multiplicative renormalization of the bare gauge coupling.
The others enter in the renormalization of the bare quark masses.
The continuum renormalized quark action can then be written as
\be
S_0 =  S_G[A] + \int d^4 x \chibar(x) \Big[\gamma_\mu D_\mu + m_{\rm R} + i\mu_{\rm R} \gamma_5\tau^3\Big]\chi(x),
\label{eq:conttmQCD}
\ee
where $S_G$ now is the continuum gluon action. The renormalized parameters are given by
\be
  g_{\rm R}^2 =  g_0^2Z_{\rm g}(g_0^2,a\mu), 
\label{eq:gr}
\ee
\be
  m_{\rm R}   =  m_q Z_{\rm m}(g_0^2,a\mu), 
\label{eq:mr}
\ee
\be
  \mu_{\rm R}  =  \mu_{\rm q} Z_\mu(g_0^2,a\mu),
\label{eq:mur}
\ee
where $\mu$ denotes the renormalization scale dependence of the
renormalization constants $Z$, and 
\be
m_q = m_0 - m_{\rm cr}.
\ee
It is well known that due to the loss of chiral symmetry at finite lattice spacing, the bare untwisted
quark mass renormalizes also additively, with a linearly divergent (with the lattice spacing) counterterm
$m_{\rm cr}$. The critical line $m_{\rm cr}$ is the value of $m_0$ where the untwisted
quark mass vanishes. We will extensively discuss in sect.~\ref{ssec:crit_mass}
how to define non-perturbatively the critical mass.
For the moment we just assume that such a value exists.
Because of the $\widetilde{\mathcal{P}}$ symmetry defined in eq. (\ref{eq:Ptilde}) the flavour-parity
violating operator $\chibar\gamma_5\tau^3\chi$ comes with a coefficient odd in $\mu_{\rm q}$,
and opposite to what happens to the untwisted quark mass, the twisted mass is 
renormalized only multiplicatively. 
To state it differently, for zero twisted mass, parity is a symmetry of the bare action.
The residual $U_{\rm V}(1)_3$ symmetry~(\ref{eq:u3}) forbids 
bilinears containing flavour matrices $\tau^{1,2}$, and the parity 
flavour symmetry $\mathcal{P}_F^{1,2}$ requires that
parity and flavour are violated together, 
so it forbids flavour singlet parity violating terms 
$\chibar \gamma_5 \chi$ and $\epsilon_{\mu \nu \rho \sigma}F_{\mu\nu} F_{\rho\sigma}$, 
as well as the flavour violating, parity even, operator $\chibar \tau_3
\chi$. It is easy to check that all the dimension four operators which
violate parity or isospin or both are ruled out by $\mathcal{P}_F^{1,2}$.
The final continuum theory (\ref{eq:conttmQCD}) has now only an apparent
flavour and parity breaking.
As we have already explained in sect.~\ref{ssec:beyond}, provided that all the
renormalization constants are defined in a mass independent scheme, 
this theory is equivalent to standard $N_{\rm f} = 2$ degenerate
flavours QCD with a mass $M = \sqrt{m_{\rm R}^2 + \mu_{\rm R}^2}$, and no parity-flavour breaking.

\subsection{Correlation functions}
\label{ssec:corr}

In sects.~\ref{ssec:tmQCD} and \ref{ssec:beyond} we have seen  
that both in the classical theory, and in the quantum theory regularized in a chiral
invariant way, there is an exact equivalence between QCD and tmQCD. This
equivalence reflects itself in a correspondence between correlation functions
computed in the two theories. In particular eq.~\eqref{eq:continuum_rot} shows
the relation between correlators, valid in the bare lattice theory regularized with GW fermions.
Based on universality arguments we expect this equivalence to be true also
with Wtm fermions between renormalized correlation functions.

The equations which relate correlation functions in the two theories depend on
how the twist angle is defined.
The twist angle can be defined in the renormalized theory, analogously to the
continuum theory, by
\be
\tan \omega = \frac{\mu_{\rm R}}{m_{\rm R}} = \frac{Z_\mu}{Z_{\rm
      m}}\frac{\mu_{\rm q}}{m_0 - m_{\rm cr}} .
\ee
To tune the twist angle it is then necessary to compute the ratio of
renormalization constants $\frac{Z_\mu}{Z_{\rm m}}$ and the critical mass
$m_{\rm cr}$. We will discuss extensively in sect.~\ref{ssec:crit_mass} how
practically to determine the critical mass. One possible way to determine the
critical mass is using the PCAC relation
\be
m_{\rm R} = \frac{Z_A}{Z_P}m_{\rm PCAC} \quad {\rm with} \quad 
m_{\rm PCAC}  = \frac{\langle \partial_0 A_0^a(x) P^a(0)\rangle}
{2\langle  P^a(x) P^a(0)\rangle} \qquad a=1,2.
\label{eq:PCAC_mass}
\ee
Measuring directly the PCAC mass, the twist angle is obtained by
\be
\tan \omega = \frac{\mu_{\rm q}}{Z_A m_{\rm PCAC}} ,
\ee
where we have used the fact, implied by the PCVC~\eqref{eq:lat_PCVC}, that $Z_P = 1/Z_\mu$.

To tune the twist angle to $\omega=\pi/2$ it is not necessary to determine any
renormalization constant, but only the critical mass.

Now that we have renormalized the theory in a mass independent scheme and we
have determined the twist angle, using universality arguments, we can conclude that
the relation between correlation functions in QCD and tmQCD can be inferred
by the transformation of the integration variables~(\ref{eq:axial}), i.e.
\be
\langle \mathcal{O}[\psi,\psibar] \rangle_{(M_{\rm R},0)} = \langle O[\chi,\chibar]
\rangle_{(m_{\rm R},\mu_{\rm R})}
\label{eq:ren_rot}
\ee
and at finite lattice spacing is valid up to cutoff effects.
The index of the correlation function in the l.h.s indicates that it has been computed in
standard QCD with renormalized quark mass $M_{\rm R}$ and $\mu_{\rm R}=0$, 
and the index of the correlation function in the r.h.s indicates that it has been computed in
tmQCD with renormalized quark masses $m_{\rm R}$ and $\mu_{\rm R}$ satisfying 
\be
M_{\rm R}^2 = m_{\rm R}^2 +\mu_{\rm R}^2 .
\label{eq:masses}
\ee 
Hence a given standard correlation function in QCD can be written as a linear combination of correlation
functions computed in tmQCD at a given twist $\omega$.

To summarize the procedure:
\begin{itemize}
\item start with the QCD correlation function you are interested in;
\item perform the axial rotation that in the continuum brings the action from the physical basis 
to the twisted basis (see eq.~(\ref{eq:axial})) on the fields appearing in the correlation function;
\item compute the resulting correlation function with the Wtm 
lattice action in the twisted basis, with a choice of quark masses;
\item perform the continuum limit.
\end{itemize}
The final result will be exactly the desired QCD correlation function in the continuum 
with quark mass $M_{\rm R}$ given by eq.~
(\ref{eq:masses}).

We give here an explicit example, which is relevant for the extraction of the
pseudoscalar decay constants and will be analyzed again in
sect.~\ref{ssec:decay_constants}.

The pion decay constant $f_\pi$ can be determined in the standard Wilson case
from the correlation function
\be
\langle (\mathcal{A}_{\rm R})_0^1(x)
   P_{\rm R}^1(y)\rangle_{(M_{\rm R},0)} .
\label{eq:AP}
\ee
With Wilson fermions the axial current is not protected by chiral symmetry,
hence it needs to be renormalized by the scale independent renormalization
constant $Z_A$, which has to be determined to extract the decay constant
from~\eqref{eq:AP}.
If we want to compute the same correlation function in tmQCD we
perform first the axial rotation~\eqref{eq:axial} on the fermion fields
in~\eqref{eq:AP}, obtaining
\be
\langle (\mathcal{A}_{\rm R})_0^1(x)
   P_{\rm R}^1(y)\rangle_{(M_{\rm R},0)}
  =\cos(\omega)\langle (A_{\rm R})_0^1(x)P_{\rm R}^1(y)
                \rangle_{(m_{\rm R},\mu_{\rm R})}
+\sin(\omega)\langle (V_{\rm R})_0^2(x)P_{\rm R}^1(y)
                \rangle_{(m_{\rm R},\mu_{\rm R})}.
 \label{eq:rotation_AP0}
\ee
This relation is very useful because, as we will see in 
sect.~\ref{ssec:decay_constants}, there is a definition of
the vector current which is protected from any renormalization.
As a consequence at $\omega=\pi/2$, the decay constant can be computed with
Wtm without the computation of any renormalization constant.

\subsection{Exceptional configurations}
\label{ssec:except}

One of the historical reasons why a twisted mass term was introduced is the so called problem
of exceptional configurations that we are briefly going to explain.

We have seen that lattice QCD with Wilson quarks~\cite{Wilson:1974sk} 
breaks explicitly chiral symmetry.
To deal with this breaking we have to add suitable 
counterterms~\cite{Bochicchio:1985xa} in order to restore chiral symmetry in
the continuum limit. 
We have seen, e.g., in sect.~\ref{ssec:reno}, that the ordinary untwisted quark
mass is renormalized also additively.
As a consequence the value of $m_0$ which corresponds to physical 
light quark masses is typically negative.
This could have further practical consequences.
In fact it implies that the Wilson-Dirac operator 
is not protected against zero modes.
The Wilson operator $D_\mathrm{W}$~\eqref{eq:Wilson} fulfills 
the property $\gamma_5D_\mathrm{W}\gamma_5 = D_\mathrm{W}^\dagger$. 
We can then define the Hermitian Wilson operator
\be
  Q_W \equiv \gamma_5 (D_\mathrm{W} + m_0) \quad Q_W=Q_W^\dagger .
  \label{eq:Q}
\ee
$Q_W$ can have in general, for a given gauge configuration, 
a very small eigenvalue, even at values of $m_0$ which correspond to a not so
small quark mass. These modes are expected to disappear in the continuum
limit, but they can still be dangerous in numerical simulations

To understand this we write a pseudoscalar density propagator 
\be
C_P(x) = - \langle \psibar(x)\gamma_5\frac{\tau^1}{2}\psi(x)
\psibar(0)\gamma_5\frac{\tau^1}{2}\psi(0)\rangle
\ee
in terms of eigenfunctions and eigenvalues of $Q_W$.
Performing the functional integral (see eq.~\ref{eq:func_int})
over the fermion fields we obtain
\be
C_P(x) = \frac{1}{2}{\mcZ}^{-1}
  \int D[U]\, \det(Q_W^2)
  \tr\left[Q_W^{-1}(0,x)Q_W^{-1}(x,0)\right]\e^{\displaystyle -S_G} ,
   \label{eq:CP_exc}
\ee
This example shows the well-known fact that a functional integral
over Grassmann variables cannot diverge. 
In fact if we write the the r.h.s of eq.~\eqref{eq:CP_exc} in terms of
eigenfunctions $\phi_i(x)$ and eigenvalues $\lambda_i$ of $Q_W$ 
we obtain\footnote{Strictly speaking
  we are working in a finite volume where the spectrum of $Q_W$ is
  discrete. This is practically the case in all the numerical 
  simulations.}
\be
C_P(x) \propto \int D[U]\, \left[\prod_i \lambda_i^2\right]
\sum_{j,k}
\phi_j(0)\frac{1}{\lambda_j}\phi_j^*(x)\phi_k(x)\frac{1}{\lambda_k}\phi_k^*(0)
.
\ee
After integration 
over the quark fields, a small eigenvalue of the Wilson
operator appears both in the fermionic determinant and 
in the quark propagators entering the
correlation functions.  
If a small eigenvalue occurs in the course of the integration over the gauge
fields, the contributions from the denominators are always exactly compensated by the
same eigenvalues in the expression for the determinant.

A problem arises however in the so-called quenched 
model, which consists in neglecting the fermionic determinant. 
The contribution of a small eigenvalue to a fermionic 
correlator is then not balanced by the determinant, leading to 
large fluctuations in some of the observables which completely compromise
the ensemble average. 
Gauge field configurations where this happens are called
``exceptional''. 
The approach to the chiral limit in the quenched model
with ordinary Wilson quarks is then limited by exceptional configurations.

In refs.~\cite{Bardeen:1998dd,Schierholz:1998bq}, to solve the problem of exceptional configurations, 
was suggested to perform a chiral rotation of the mass term. But in ref.~\cite{Bardeen:1998dd}
was suggested to send the twisted mass to zero at the end of the computation, i.e. treating the twisted mass
as an external source not as the real mass term, and in both references
the axial rotation was anomalous, i.e. flavour singlet. Hence it would have not been possible
to use it as an alternativ discretization of lattice QCD.

Adding the twisted mass term~(\ref{eq:WtmQCD}) to the Wilson action solves
in a straightforward way the problem~\cite{Frezzotti:2000nk}.
For the 2 flavours Wtm operator $D = D_W + m_0 + i\mu_{\rm q} \gamma_5 \tau^3$ 
we have 
\be
Q = \gamma_5 D = Q_W + i\mu_{\rm q}\tau^3 \Rightarrow 
Q^\dagger Q = Q_W^\dagger Q_W + \mu_{\rm q}^2 = Q_W^2 + \mu_{\rm q}^2 \ .
\label{eq:QWtm}
\ee
This means that the Wtm operator does not have exceptionally small eigenvalues
for arbitrary gauge fields: they can only appear in the massless ($\mu_{\rm q}=0$) theory.

These considerations are true if we assume that the distribution of the
eigenvalues of $Q_W$ are either mass independent (like in the quenched model)
or the mass dependence is analytic near the chiral point. 
We will see in sect.~\ref{sec:asq} that in infinite volume
this is not true because of the non trivial chiral phase structure for Wilson fermions.
On the contrary, approaching the chiral limit at fixed finite volume could in principle 
be advantageous using Wtm.

\newpage
\section{Non-degenerate quarks}
\label{sec:nondeg}

In this section we show how twisted mass QCD can be generalized 
to a doublet of non-degenerate quarks.
We extend the topics covered in the previous section to the non-degenerate
case, emphasizing the main differences with the degenerate case.

\subsection{Continuum actions}
\label{ssec:ndegcont}

The continuum action we have discussed in sect.~\ref{ssec:tmQCD} describes two degenerate light quarks.
To add a further doublet of non-degenerate quarks in order to describe the heavier ($c$,$s$) quarks,
two proposals have been made~\cite{Frezzotti:2003xj,Pena:2004gb}. 
Both proposals can of course be used to describe a possible non-degeneracy also 
in the light sector. The first proposal~\cite{Frezzotti:2003xj} is based on a flavour off-diagonal splitting
\be
  S_{\rm F}[\chi,\chibar,G] =\int {\rm d}^4x\,\chibar\left(\gamma_\mu D_\mu
                +m_{\rm q}+i\mu_{\rm q}\gamma_5\tau^3 + \epsilon_{\rm q}\tau^1\right)\chi, 
\label{eq:ndtmQCDcont}
\ee
where we take $\mu_{\rm q} >0$ and $\epsilon_{\rm q} > 0$.
The off-diagonal splitting is particularly interesting because, as we will see, it retains all
the nice properties of tmQCD at full twist and it keeps the quark determinant real and positive
if $\sqrt{m_{\rm q}^2 + \mu_{\rm q}^2} > \epsilon_{\rm q}$ (see below and sect.~\ref{ssec:ndeglatt}).

In order to change from the {\it twisted basis} to the {\it physical basis} the following field 
transformations are needed.
First we need an isovector rotation of $\omega_2=\pi/2$
\be
 \chi'    =\exp(i \omega_2\tau^2/2)\chi |_{\omega_2=\pi/2} = \frac{1}{\sqrt{2}}(1+i\tau^2)\chi,
\label{eq:chiprime}
\ee
\be
 \chibar' =\chibar\exp(-i \omega_2\tau^2/2) |_{\omega_2=\pi/2} =
 \chibar\frac{1}{\sqrt{2}}(1-i\tau^2).
\label{eq:chibarprime}
\ee
This vector transformation leaves invariant the kinetic term and transforms the mass terms as
\be
 m_q + i \mu_q \gamma_5 \tau^3 + \epsilon_q \tau^1 \rightarrow m_q - i \mu_q \gamma_5 \tau^1 + \epsilon_q \tau^3.
 \label{eq:masses1}
\ee
Now we perform an axial rotation as before in the direction of the twisted mass term
\be
 \psi    =\exp(-i \omega_1\gamma_5\tau^1/2)\chi',\qquad
 \psibar =\chibar'\exp(-i \omega_1\gamma_5\tau^1/2).
 \label{eq:axial1}
\end{equation}
This transformation leaves the form of the action invariant, but 
transforms only the mass term with $\mu_{\rm q}$, and, if we want to have the standard action,
the rotation angle $\omega_1$ has to satisfy eq.~(\ref{eq:angle}).
The action now looks like 
\be
  S_{\rm F}[\psi,\psibar] =\int {\rm d}^4x\,\psibar\left(\gamma_\mu D_\mu
                +M + \epsilon_{\rm q}\tau^3\right)\psi, 
\label{eq:ndtmQCDcont_phys}
\ee
where $M = \sqrt{m_q^2 + \mu_q^2}$ is again the polar mass.
We remark here that these are the transformations needed given the action in eq.~(\ref{eq:ndtmQCDcont}).
Choosing a different basis for the action one is interested in 
(typically depending on the details of the lattice simulations),
will induce different field transformations in order to connect the initial basis with the physical one.

Analogously to the degenerate case we can derive the partial conservation laws
\be
  \partial_\mu A_\mu^a = 2m_{\rm q} P^a+i\mu_{\rm q} \delta^{3a}S^0 + \epsilon_{\rm q} \delta^{a1} P^0, 
\label{eq:PCACnd}
\ee
\be
  \partial_\mu V_\mu^a = -2\mu_{\rm q}\,\varepsilon^{3ab} P^b + 
i\epsilon_{\rm q}\varepsilon^{1ab} S^b ,
\label{eq:PCVCnd}
\ee
where 
\be
P^0 = \chibar \gamma_5 \chi , \qquad S^a = \chibar \frac{\tau^a}{2} \chi . 
\ee
If we have in mind to describe with this action the heavy doublet 
($c$,$s$) we will naturally associate the quark mass in the following way:
\be
m_c = M + \epsilon_q \qquad
m_s = M - \epsilon_q.
\ee
The fermion determinant will then be positive if $M > \epsilon_{\rm q}$.
This constraint will be reconsidered when introducing the Wilson term in the
lattice actions, and taking into account how the bare masses are renormalized.

Another way to extend the tmQCD action to four flavours has been proposed 
in ref.~\cite{Pena:2004gb}. In this proposal the continuum action reads
\be
  S_{\rm F}[\chi,\chibar,G] =\int {\rm d}^4x\,\chibar\left(\gamma_\mu D_\mu
                +\mathbf{m}+i\boldsymbol{\mu}\gamma_5 \right)\chi, 
\label{eq:ndSint}
\ee
where now $\chi$ collects four quark fields $\chi^T = (u,d,s,c)$ and the mass matrices have the form
\be
\small{
\mathbf{m} = 
\left( \begin{array}{ccccccc}
m_u &  \phantom{0} &0 &  \phantom{0} & 0   &\phantom{0} & 0 \\
0 &    \phantom{0} &m_d &\phantom{0} & 0   &\phantom{0} & 0 \\
0 &    \phantom{0} &0 &  \phantom{0} & m_s &\phantom{0} & 0 \\
0 &    \phantom{0} &0 &  \phantom{0} & 0   &\phantom{0} & m_c 
\end{array} \right) = 
 \left( \begin{array}{cccc}
M_u \cos \omega_l & 0 & 0 & 0 \\
0 & M_d \cos \omega_l & 0 & 0 \\
0 & 0 & M_s \cos \omega_h & 0 \\
0 & 0 & 0 & M_c \cos \omega_h
\end{array} \right) ,}
\label{eq:m_matrix}
\ee
\be
\small{
\boldsymbol{\mu} = 
\left( \begin{array}{ccccccc}
\mu_u & \phantom{0} &0     &\phantom{0} & 0     &\phantom{0} & 0 \\
0     & \phantom{0} &\mu_d &\phantom{0} & 0     &\phantom{0} & 0 \\
0     & \phantom{0} &0     &\phantom{0} & \mu_s &\phantom{0} & 0 \\
0     & \phantom{0} &0     &\phantom{0} & 0     &\phantom{0} & \mu_c 
\end{array} \right) = 
 \left( \begin{array}{cccc}
M_u \sin \omega_l & 0 & 0 & 0 \\
0 & -M_d \sin \omega_l & 0 & 0 \\
0 & 0 & M_s \sin \omega_h & 0 \\
0 & 0 & 0 & -M_c \sin \omega_h
\end{array} \right) ,}
\label{eq:mu_matrix}
\ee
Hence the theory has six independent parameters, 
namely the four polar quark masses $M_i$ ($i=u,d,s,c$),
with $M_i^2 = m_i^2 + \mu_i^2$, and the two twist angles 
$\omega_l,\omega_h$. In other words, the four standard mass parameters 
$m_i$ and the four twisted mass parameters $\mu_i$ are constrained by
\be
 \tan \omega_l = \frac{\mu_u}{m_u} = - \frac{\mu_d}{m_d},
 \qquad
 \tan \omega_h = \frac{\mu_s}{m_s} = -\frac{\mu_c}{m_c}.
 \label{eq:tan}
\ee
This framework extends the degenerate two-flavour tmQCD to non-degenerate quarks 
with the property that the quark mass terms remain flavour diagonal. 
At vanishing twist angles $\omega_l$ and $\omega_h$
one recovers the standard QCD action of four quark flavours,
while for $\omega_h=0$ and $M_u=M_d$ the two-flavour version 
of tmQCD in eq.~(\ref{eq:tmQCDcont}) is reproduced, 
with two additional untwisted quark flavours $s$ and $c$. 

Given the form of the mass matrices~(\ref{eq:m_matrix},\ref{eq:mu_matrix}) the rotation 
that has to be performed to go back to the {\it physical basis} is
\be
\chi    =\exp(-i \omega_l\gamma_5\tau_l^3/2 - i \omega_h\gamma_5\tau_h^3/2)\psi,\qquad
\chibar    =\psibar \exp(-i \omega_l\gamma_5\tau_l^3/2 - i \omega_h\gamma_5\tau_h^3/2),\qquad
\label{eq:axial_sint}
\ee
with
\begin{displaymath}
\tau^3_l = 
\left( \begin{array}{ccccccc}
1 & \phantom{0} & 0 & \phantom{0} & 0 & \phantom{0} & 0 \\
0 & \phantom{0} & -1 & \phantom{0} & 0 & \phantom{0} & 0 \\
0 & \phantom{0} & 0 & \phantom{0} & 0 & \phantom{0} & 0 \\
0 & \phantom{0} & 0 & \phantom{0} & 0 & \phantom{0} & 0 
\end{array} \right) \qquad 
\tau^3_h = \left( \begin{array}{ccccccc}
0 & \phantom{0} & 0 & \phantom{0} & 0 & \phantom{0} & 0 \\
0 & \phantom{0} & 0 & \phantom{0} & 0 & \phantom{0} & 0 \\
0 & \phantom{0} & 0 & \phantom{0} & 1 & \phantom{0} & 0 \\
0 & \phantom{0} & 0 & \phantom{0} & 0 & \phantom{0} & -1
\end{array} \right) ,
\label{eq:tau3}
\end{displaymath}
and $\omega_l$ and $\omega_h$ satisfying eqs.~(\ref{eq:tan}).
The four flavour QCD action now takes the standard form
\be
  S_{\rm F}[\psi,\psibar,G] =\sum_{i = u,d,s,c} \int {\rm d}^4x\,\psibar_i(x)\left(\gamma_\mu D_\mu
                + M_i \right)\psi_i(x). 
\label{eq:4fl_sint}
\ee
The rotation in eq.~(\ref{eq:axial_sint}) will also give the relations among local fields in the 2 basis.

For a generic four flavour QCD theory with non-degenerate quarks,
chiral and flavour symmetries are broken explicitly and this is expressed
by the PCAC and PCVC relations (for $i \ne j$)
\be
  \partial_\mu A_{\mu,ij} = (m_i+m_j)P_{ij} + i(\mu_i+\mu_j)S_{ij},
 \label{eq:PCAC_sint}
\ee
\be
  \partial_\mu V_{\mu,ij} = (m_i-m_j)S_{ij} + i(\mu_i-\mu_j)P_{ij}, 
  \label{eq:PCVC_sint}
\ee
where the bilinear fields are defined by
\be
\label{bilinears}
 S_{ij} =  \chibar_i \chi_j, \quad
 P_{ij} =  \chibar_i \gamma_5 \chi_j, \quad
 A_{\mu,ij} =  \chibar_i \gamma_\mu\gamma_5 \chi_j, \quad
 V_{\mu,ij} =  \chibar_i \gamma_\mu \chi_j.
\ee

\subsection{Lattice actions}
\label{ssec:ndeglatt}

We can now write also the fermionic action proposed in ref.~\cite{Frezzotti:2003xj} 
for two non-degenerate quarks in the {\it twisted basis}
\be
  S_{\rm F}[\chi,\chibar,U] =a^4\sum_x\chibar(x)\Big[D_{\rm W} + m_0 + i\mu_{\rm q}\gamma_5\tau^3 + 
\epsilon_{\rm q}\tau^1\Big]\chi(x), 
\label{eq:WtmQCDnondeg}
\ee
The introduction of the off-diagonal splitting leaves intact, for example, 
the symmetries $\widetilde{\mathcal{P}}$ and $\mathcal{P}^{1}_F$, while
$\mathcal{P}^{2}_F$ is a symmetry only if combined with a sign change of
$\epsilon_{\rm q}$
\be
\widetilde{\mathcal{P}}_F^2 \equiv \mathcal{P}_F^2 \times 
[\epsilon_{\rm q} \rightarrow - \epsilon_{\rm q}]\,.
\label{eq:PtildeF2}
\ee
In sect.~\ref{ssec:ndegcont} we have argued that the determinant is always real and positive provided 
the constraint $\mu_{\rm q} > \epsilon_{\rm q}$ is fulfilled\footnote{For simplicity we consider here
the full twist case.}. This is an important practical issue in order to perform dynamical simulations
with the currently available algorithms, since the fermionic determinant is usually included with the gauge action
to form an effective Boltzmann weight (see sect.~\ref{sec:algo} for a discussion on recent algorithmic developments).

To understand how the condition on the bare masses translates to the quantum theory,
we anticipate here that the renormalization factors of $\mu_{\rm q}$ and $\epsilon_{\rm q}$ are 
related to the renormalization factors of the pseudoscalar and scalar currents, 
i.e. 
\be
\mu_{\rm R} = Z_{\rm P}^{-1}\mu_{\rm q}, \qquad \epsilon_{\rm R} = Z_{\rm S}^{-1} \epsilon_{\rm q}.
\label{eq:muepsren}
\ee
It is then easy to show that the constraint $\mu_{\rm q} > \epsilon_{\rm q}$
augmented with the definitions~\eqref{eq:muepsren} gives 
\be
\frac{Z_{\rm P}}{Z_{\rm S}} > \frac{(\mu_{\rm c})_{\rm R} - (\mu_{\rm s})_{\rm R}}
{(\mu_{\rm c})_{\rm R} + (\mu_{\rm s})_{\rm R}},
\ee
where, having in mind phenomenological applications, we have defined
\be
(\mu_{\rm c})_{\rm R} = \mu_{\rm R} + \epsilon_{\rm R}, \qquad 
(\mu_{\rm s})_{\rm R} = \mu_{\rm R} - \epsilon_{\rm R}.
\ee
To give an example, fixing the values of the
renormalized {\it strange} and {\it charm} quark masses, gives the following constraints
\be
 (\mu_{\rm c})_{\rm R} \simeq 1.5 {\rm GeV} \qquad  (\mu_{\rm s})_{\rm R} \simeq 0.1 {\rm GeV}
  \Rightarrow {Z_{\rm P} \over Z_{\rm S}} \gtrsim 0.875 .
\ee

The fermionic action proposed in ref.~\cite{Pena:2004gb} for four
non-degenerate flavours reads
\be
  S_{\rm F}[\chi,\chibar,U] =a^4\sum_x\chibar(x)\Big[D_{\rm W} + \mathbf{m} + i\boldsymbol{\mu}\gamma_5 \Big]\chi(x), 
\label{eq:WtmQCDnondeg_sint}
\ee
with $\chi^T = (u,d,s,c)$, $\mathbf{m}$ and $\boldsymbol{\mu}$ defined 
in eqs.~(\ref{eq:m_matrix},\ref{eq:mu_matrix}).
To discuss the properties of the fermionic determinant with the action~(\ref{eq:WtmQCDnondeg_sint})
we assume that the light doublet is mass degenerate,
\be
    m_u=m_d=m_l,\qquad  \mu_u=-\mu_d=\mu_l.
\ee
Then the integration over the light fermion fields yields to
\be
 \det_{N_f=2}\left[\left(D_W + m_l\right){\mathbbm 1} + i\mu_l\gamma_5\tau^3\right]=
 \det_{N_f=1}\left[ \left(D_W + m_l \right)^\dagger \left(D_W + m_l \right) 
  + \mu_l^2 \right],
\ee
where the indices indicate in which flavour space the determinant is taken~\cite{Frezzotti:2000nk}.
Hence the determinant of the light doublet is positive at non-zero 
$\mu_l$, irrespective of the background gauge field. 
Integrating over strange and charm quarks one obtains
\be
 \det_{N_f=1}\Big[(D_W + m_s )^\dagger
 (D_W + m_c ) - \mu_s \mu_c + i \mu_c \gamma_5 (D_W + m_s)
 + i \mu_s \gamma_5 (D_W + m_c )  \Big],
\ee
which is real and positive only for degenerate strange and charm quarks.
The other possibility to ensure the reality of the determinant is to
employ untwisted strange and charm quarks, $\mu_s=\mu_c=0$,
as the fermion determinants for individual Wilson quark flavours are real.
Even if in this case the positivity of the determinant is not guaranteed, recent
numerical results~\cite{Kaneko:2003re}
indicate that this is indeed the case.

Practically, with the action~(\ref{eq:WtmQCDnondeg_sint}), there are two options: 
do not include the strange and the charm quarks in the determinant, or include them 
but without twist.
The first option would correspond to a partially quenched simulation with 
$N_f=2$ light dynamical quarks, i.e. the strange and the charm would remain quenched.
The second option would correspond to a $N_f=3,4$ dynamical simulation with two
light twisted quarks and two heavier non-degenerate untwisted quarks.

\subsection{Transfer matrix}
\label{ssec:transfer_ndeg}

In sect.~\ref{ssec:transfer} we have seen that provided $|\kappa| < {1 \over 6}$
Wtm fulfills physical positivity.
In app.~\ref{app:C} we briefly repeat the steps of the proof with the extension to Wtm,
because they are important for non-degenerate quarks.

In fact in app.~\ref{app:C} we show that if we add a non-degenerate doublet 
as in eq.~(\ref{eq:WtmQCDnondeg}) the constraint on $\kappa$ has to be changed into
\be
|\kappa| < \frac{1}{6+2a\epsilon_{\rm q}}, \qquad \epsilon_{\rm q} >0 .
\label{eq:knondeg}
\ee
This difference can be understood by the different Hermiticity property of the twisted mass term 
$i\mu_{\rm q} \gamma_5\tau^3$ and the splitting term $\epsilon_{\rm q}\tau^1$.

First simulations with non-degenerate twisted quarks~\cite{Chiarappa:2006ae}
indicate that suitable values of $\epsilon_{\rm q}$ are rather small and since,
in the continuum limit and for a value of the untwisted bare quark mass close to the critical mass,
$\kappa$ is near $1/8$, for practical purposes 
the constraint~(\ref{eq:knondeg}) should not cause any limitation.

To conclude Wtm, like the pure Wilson theory, is reflection positive 
for all the relevant values of the bare parameters.

\subsection{Renormalization}
\label{ssec:reno_ndeg}

Here we shortly discuss the structure of the counterterms for the non-degenerate action~(\ref{eq:WtmQCDnondeg})
and we refer to the original paper~\cite{Pena:2004gb} for the counterterm structure of the 
action~(\ref{eq:WtmQCDnondeg_sint}).
The only additional counterterm to those in eq.~(\ref{eq:counter}) 
allowed by the lattice symmetries is
\be
\epsilon_{\rm q} \chibar \tau^1 \chi.
\label{eq:counter2}
\ee
Because of the $\widetilde{\mathcal{P}}_F^2$ symmetry defined in eq. (\ref{eq:PtildeF2}), the flavour
violating operator $\chibar\tau^1\chi$ comes with a coefficient odd in $\epsilon_{\rm q}$,
and the twisted mass splitting is renormalized only multiplicatively. 
All the dimension four operators which violate parity or isospin or both are
absent because $\mathcal{P}_F^{1,2}$ are still symmetries when $\epsilon_{\rm q} = 0$.
The dimension three operators that are absent in
eqs.~(\ref{eq:counter},\ref{eq:counter2}) are all ruled out by $\mathcal{P}_F^1$
symmetry, with the exception of $\chibar \gamma_5 \tau^2 \chi$ that is ruled
out by charge conjugation $\mathcal{C}$ which is still a symmetry of the action~(\ref{eq:WtmQCDnondeg}).
To conclude, the continuum renormalized quark action for non degenerate quarks
is 
\be
S_0 =  S_G[A] + \int d^4 x \chibar(x) \Big[\gamma_\mu D_\mu + m_{\rm R} + 
i\mu_{\rm R} \gamma_5\tau^3 + \epsilon_{\rm R} \tau^1 \Big]\chi(x),
\label{eq:conttmQCDndeg}
\ee
where in addition to the degenerate case we just have to add
\be
  \epsilon_{\rm R}  =  \epsilon_{\rm q} Z_\epsilon(g_0^2,a\mu).
\label{eq:mur}
\ee

\newpage
\section{O($a$) improvement}
\label{sec:impro}

The continuum limit of lattice QCD is of fundamental importance
to relate numerical simulations with experimental results.
Practically to perform the continuum limit
it is crucial to simulate at several values of the lattice spacing
$a$, and it is also possible (and sometimes mandatory) 
to improve the rate of the discretization errors from
$a$ to $a^2$, using a suitable lattice QCD action.
A possibility is to apply Symanzik's improvement program
\cite{Symanzik:1981hc,Symanzik:1983dc,Symanzik:1983gh}, where the O($a$) cutoff effects in on-shell quantities are
cancelled by adding local O($a$) counterterms to the lattice action and to the
composite fields of interest \cite{Luscher:1984xn,Sheikholeslami:1985ij,Wohlert:1987rf,Luscher:1996sc,Heatlie:1990kg}. 
A technical difficulty is that the improvement coefficients multiplying these counterterms
are not known a priori. An alternative is to use Wtm at full twist. By this we mean Wtm with 
bare parameters tuned in order to have in the continuum limit a vanishing untwisted quark mass.
We will show that in this case physical observables will be automatically O($a$)
improved without the knowledge of any improvement coefficient. 
The only parameter tuning required is that of the 
bare untwisted quark mass to its critical value.

\subsection{Symanzik expansion}
\label{ssec:sym}

The form of the unimproved lattice action is 
\be
S[\chibar,\chi,U] = S_G[U] + S_F[\chibar,\chi,U] \ .
\label{eq:laction}
\ee
In this section we will analyse the O($a$) effects, so we leave unspecified the form 
of the gauge lattice action $S_G$, since it would only change the theory at O($a^2$). 
$S_F$ is the Wtm quark action defined in eq.~(\ref{eq:WtmQCD}), that we rewrite here for convenience
\be
  S_{\rm F}[\chi,\chibar,U] =a^4\sum_x\chibar(x)\Big[D_{\rm W} + m_0 + i\mu_{\rm q}\gamma_5\tau^3\Big]\chi(x) .
\label{eq:WtmQCD2}
\ee
Following the program of Symanzik, the long distance properties of Wtm close to the continuum limit
may be described in terms of a local effective theory with action
\be
S_{\rm eff} = S_0+aS_1+a^2S_2+\ldots
\label{eq:eff_action}
\ee
The key constraint is that each term of the Lagrangian 
has to be invariant under the symmetries of the regularized theory, i.e. the lattice theory.
The leading term, $S_0$, is the action of the target continuum theory
\be
S_0 =  S_G[A] + \int d^4 x \chibar(x) \Big[\gamma_\mu D_\mu + m_{\rm R} + i\mu_{\rm R} \gamma_5\tau^3\Big]\chi(x) .
\label{eq:conttmQCD2}
\ee
discussed already in sect.~\ref{ssec:reno}.
The remaining operators $S_k$ have to be interpreted as operator insertions in the continuum theory.
The continuum theory can be defined employing a lattice with spacing much smaller
than $a$, or using a chiral invariant regularization that fulfills the
Ginsparg-Wilson relation.
The terms in the effective action are of the form
\be
S_k=\int d^4y {\mathcal L}_k(y)
\label{eq:operators}
\ee
where the Lagrangians ${\mathcal L}_k(y)$ are linear combinations of local
composite fields of dimension $4+k$.

Cutoff effects come also from the local composite fields one is interested in.
A generic renormalized local gauge invariant field $\phi_R(x)$,  
constructed from quark and gluon fields on the lattice, 
is represented in the effective theory by an effective field of the form
\be
\phi_{\rm eff}(x) = \phi_0(x) + a\phi_1(x) + a^2\phi_2(x) + \ldots 
\ee
where the fields $\phi_k$ should have the appropriate dimension 
and should transform under symmetries as the lattice field.

All on-shell quantities in QCD can be extracted from correlation functions of local composite fields.
These correlation functions are needed at non-zero physical distance.
We take a generic connected lattice correlation function made of a multiplicatively renormalized multilocal field 
\be
G(x_1,\ldots,x_n) = \langle \phi_R(x_1) \cdots \phi_R(x_n) \rangle \equiv \langle \Phi \rangle
\label{eq:phi}
\ee
and we always consider $x_1 \neq x_2 \neq \cdots \neq x_n$.
In the effective theory up to order $a$ it will be described by
\be
\langle \Phi \rangle = \langle \Phi_0 \rangle_0 - a \int d^4y \langle \Phi_0
     {\mathcal L}_1(y) \rangle_0 + a \langle \Phi_1 \rangle_0 + {\rm O}(a^2)
\label{eq:sym_exp1}
\ee
where the expectation values on the right hand side are to be taken in the continuum theory
with action $S_0$ and 
\be
\langle \Phi_0 \rangle_0 \equiv \langle\phi_0(x_1) \cdots \cdots \phi_0(x_n) \rangle_0
\ee
\be
\langle \Phi_1 \rangle_0 \equiv \sum_{k=1}^{n}\langle\phi_0(x_1) \cdots \phi_1(x_k) \cdots \phi_0(x_n) \rangle_0
\ee
The second term in the r.h.s. of eq.~(\ref{eq:sym_exp1}) develops potentially divergent
contact terms whenever $y=x_k$. An important remark~\cite{Luscher:1996sc} is that these contact 
terms do not spoil the form of the expansion. Any contact term coming from $\phi_0(x){\mathcal L}_1(y)$ 
when $y \rightarrow x$ will be given by an operator that has the same dimension and symmetry properties
as $\phi_1(x)$. Since we leave the expression of $\phi_1$ unspecified, the way used to subtract 
the divergence from the contact term will not change the form of eq.~(\ref{eq:sym_exp1}).
The explicit $a$ dependence in eq.~(\ref{eq:sym_exp1}) is not the full $a$ 
dependence of the lattice correlation function: 
$\phi_1$ and ${\mathcal L}_1(y)$ are linear combinations of fields, the coefficients of which depend
on $a$ logarithmically, as shown in perturbation theory~\cite{Symanzik:1983gh}.
Additional O($a$) effects can arise if one integrates the lattice correlation functions over short distances,
and these cutoff effects will not be described by the effective theory.
We remind here that this is not a crucial restriction since hadron masses and matrix elements
are computed from correlation functions at non-zero distance.
This remark is also important because it allows further simplifications 
in determining the set of operators ${\mathcal O}_i$ contributing to ${\mathcal L}_1$.
In app.~\ref{app:D}, I briefly summarize how to construct ${\mathcal L}_1$
and I give as an example the operators contributing to $\phi_1$ for the currents
$A_\mu^a$, $V_\mu^a$ and $P^a$.

The result of this analysis gives as the leading term of the effective Lagrangian
\be
{\mathcal L}_1 = \sum_{i=1}^5 c_i {\mathcal O}_i
\label{eq:L1}
\ee
where
\be
{\mathcal O}_1 = i \chibar\sigma_{\mu\nu} F_{\mu\nu} \chi,
\label{eq:o1}
\ee
\be
{\mathcal O}_2 = m_{\rm q} \tr \{ F_{\mu\nu} F_{\mu\nu} \},
\label{eq:o2}
\ee
\be
{\mathcal O}_3 = m_{\rm q}^2\chibar \chi,
\label{eq:o3}
\ee
\be
{\mathcal O}_4=m_{\rm q}\mu_{\rm q} i \chibar\gamma_5\tau^3\chi,
\label{eq:o4}
\ee
\be
{\mathcal O}_5=\mu_{\rm q}^2\chibar\chi.
\label{eq:o5}
\ee

Now that we know the form of the leading corrections to the effective action
we can add to the Wtm Lagrangian the suitable counterterms in order to remove
the O($a$) terms from the lattice action. This will be already enough to improve all
the spectral quantities like the hadron masses. The counterterms to add to the Wtm action are
\be
a^5\sum_x \sum_{i=1}^5\hat{c}_i\hat{{\mathcal O}}_i \ ,
\ee
where the fields $\hat{{\mathcal O}}_i$ will
be some lattice representation of the continuum ${\mathcal O}_i$.
In general the form of the lattice fields $\hat{{\mathcal O}}_i$ is not fixed because 
this amounts to change the O($a^2$) terms of the theory.
These discretization ambiguities allow to
represent the gauge strength field and the local scalar density in the way
they already appear in the Wtm action. The O($a$) counterterms ${\mathcal O}_2$
to ${\mathcal O}_5$ amount then to a reparametrisation of the twisted and untwisted
quark masses together with the reparametrisation of the bare coupling $g_0$.
These reparametrisations are important if one chooses a mass independent
renormalization scheme. We have seen in sect.~\ref{sec:basic} that it is important
to make such a choice, in order to
have equivalence between twisted mass and standard QCD correlation functions. 
We assume now that a mass-independent renormalization scheme has been chosen.
The O($a$) improved action, also called clover action, is given by 
\be
S_{\rm impr}[\chibar,\chi,U] = S[\chibar,\chi,U] + \delta S [\chibar,\chi,U] \ ,
\label{eq:clovertm}
\ee
\be
\delta S [\chibar,\chi,U] = a^5\sum_x c_{\rm sw} \chibar(x)\frac{i}{4}\sigma_{\mu \nu}
\hat{F}_{\mu \nu}(x)\chi(x) \ ,
\label{eq:clover}
\ee
where $\hat{F}_{\mu \nu}$ is a lattice representation of the gluon field
strength tensor and $c_{\rm sw}$ is the so-called Sheikholeslami-Wohlert
parameter~\cite{Sheikholeslami:1985ij}. This parameter depends on the bare
gauge coupling and all the details of the lattice action (but not on the quark
masses) and has to be tuned in order to achieve on-shell O($a$) improvement.
To define, consistently with O($a$) improvement and a mass independent
renormalization scheme, the renormalized coupling and masses we have to define
the theory around the chiral point. We have seen in sect.~\ref{ssec:reno}, in the renormalization procedure,
that in the plane of the bare parameters the
massless point is given by $(m_0,\mu_{\rm q}) = (m_{\rm cr},0)$ where the critical line 
$m_{\rm  c}$ is the value of $m_0$ where the physical quark mass vanish.
Then is natural to introduce the subtracted bare quark mass 
\be
m_{\rm q} = m_0 - m_{\rm cr}
\ee
and to define the renormalized O($a$) improved masses and coupling
constant as
\be
  g_{\rm R}^2 =  \tilde{g}_0^2Z_{\rm g}(\tilde{g}_0^2,a\mu), 
\label{eq:gr}
\ee
\be
  m_{\rm R}   =  \tilde{m}_q Z_{\rm m}(\tilde{g}_0^2,a\mu), 
\label{eq:mr}
\ee
\be
  \mu_{\rm R}  =  \tilde{\mu}_{\rm q} Z_\mu(\tilde{g}_0^2,a\mu),
\label{eq:mur}
\ee
where $\mu$ denotes the renormalization scale dependence of the
renormalization constants $Z$, and the improved bare
parameters are given by
\be
  \tilde{g}_0^2       =  g_0^2(1 + b_{\rm g}am_{\rm q}), 
\label{eq:gtilde0}
\ee
\be
  \tilde{m}_q         =  m_q(1+b_{\rm m}am_{\rm q}) + \tilde{b}_{\rm m}a\mu_{\rm q}^2, 
\label{eq:mtildeq}
\ee
\be
  \tilde{\mu}_{\rm q}  =  \mu_{\rm q}(1 + b_{\rm \mu}a m_{\rm q}) \ .
\label{eq:mutildeq}
\ee
It easy to recognize the $4$ terms which correspond in the effective theory to
$\mathcal{O}_i$ with $i=2,3,4,5$.
To summarize, to achieve full O($a$) improvement it is necessary to tune
not only $c_{\rm sw}$, but also the $b$ and $\tilde{b}$ parameters defined in
eqs.~(\ref{eq:gtilde0}-\ref{eq:mutildeq}), and the improvement coefficients
related to the operators one is interested in.
To be specific the renormalized O($a$) improved axial current will look like
\be
(A_{\rm R})_\mu^a = Z_{\rm A}(1 + b_{\rm A} a m_{\rm q})\Big[A_\mu^a + a
c_{\rm A} \partial_{\mu} P^a + a \mu_q \tilde{b}_{\rm
  A}\epsilon^{3ab}V_\mu^b\Big] \ ,
\label{eq:Aimpr}
\ee
where we have eliminated the operator $(\mathcal{O}_8)^a_\mu$~(\ref{eq:O8app}) using 
the equations of motion.
In principle all these improvement coefficients would need to be computed
to have an O($a$) improved evaluation of an hadronic matrix element including an axial current.
At this point the introduction of a twisted mass has added the $b_\mu$ and
$\tilde{b}_m$ parameters together with some $\tilde{b}$ parameters for the improved
operators.
Still the number of improvement coefficients needed to be computed even
without twisted mass is not negligible,
especially if one considers non-degenerate quarks~\cite{Bhattacharya:2005rb}.
There are two technical remarks to make. The set of O($a$) counterterms that
have been introduced are slightly redundant~\cite{Frezzotti:2001ea}. 
This generic feature of tmQCD can
be traced back to the equivalence of correlation functions of tmQCD and
standard QCD in the continuum limit.
O($a$) improved Wtm is a one-parameter family of O($a$) improved theories: one
of the improvement coefficient can be chosen arbitrarily. The choice is
usually to set $\tilde{b}_{\rm m} = -1/2$ because with this choice almost all 
the other improvement coefficients vanish at tree-level of perturbation theory. 
The second remark is to note that the case $m_{\rm R} = 0$ is particularly
interesting.
In the most general case this corresponds to $m_{\rm q}$ being an O($a$)
(see eq. \ref{eq:mr}), and in the spirit of O($a$) improvement, where
O($a^2$) effects are neglected, from eqs.~(\ref{eq:gtilde0}-\ref{eq:mutildeq}) we infer that 
one remains with only one parameter $\tilde{b}_{\rm m}$ and moreover mass
dependent O($a$) effects in the bare coupling are absent. 
Given also the remark of the redundancy of the improvement coefficients Wtm at full twist 
is O($a$) improved just tuning the clover term and all the improvement
coefficients needed to improve the local operators.
The $\tilde{b}$ parameters are associated with
opposite parity operators and it is in principle 
possible to get rid of them in a quantum mechanical analysis (we will come
back to this topic in sect.~\ref{sec:asq}). 
So implementing the standard O($a$) Symanzik's improvement program with Wtm
at full twist ($m_{\rm R} = 0$) is cheaper in the number of improvement coefficients to compute, 
with respect to standard Wilson fermions.

We will see in sec.~\ref{ssec:auto} that the situation of full twist when $m_{\rm R} = 0$ is of
capital importance, and it has tremendous consequence in the cutoff effects of
correlation functions. It is already clear then that it becomes extremely important
to discuss the way of practically implementing on the lattice the condition $m_{\rm R} = 0$.

Before discussing the consequences of working at full twist, I briefly summarize some numerical results
obtained with Wtm adopting the Symanzik's improvement program just discussed.

\subsubsection{Numerical tests}

The Symanzik program requires the knowledge of a set of improvement coefficients, that we have
just discussed.
A first possibility is to compute them in perturbation theory. A one loop computation would leave physical
observables with O($ag_0^4$) discretization errors and if the gauge coupling (the lattice spacing)
is small enough it is reasonable to hope that the quantity of interest will numerically scale 
with O($a^2$) corrections. A further check could be to change by factors of O(1) the 
one loop improvement coefficients and check if the scaling violations change 
dramatically or not.
A better approach is to compute the improvement coefficients non-perturbatively.
Some of the improvement coefficients have been computed non-perturbatively, within the
ordinary Wilson framework, both in the quenched model~\cite{Luscher:1996ug,Bhattacharya:2000pn}, in the 
$N_f = 2$~\cite{Jansen:1998mx} and in the $N_f = 3$ theory~\cite{Aoki:2005et}. 

The improvement coefficients computed in a mass independent renormalization scheme 
do not depend on the form of the mass term and they can be used also with Wtm fermion. 
Only the additional improvement coefficients, specific to Wtm,
need to be determined in addition.
In \cite{Frezzotti:2001ea} it has been shown how to implement the standard Symanzik program for Wtm,
and many improvement coefficients have been computed at one loop in perturbation theory, especially
the improvement coefficients related to Wtm.

One is certainly interested in checking numerically if the O($a$) improvement has been successfully 
implemented and if the remaining O($a^2$) effects are small.
In particular quantities which have a finite continuum limit can be computed at 
several values of the lattice spacing to check the amount of {\it scaling violations}.
In particular these scaling studies have to be performed on a line of constant physics.
With this we mean that changing the lattice spacing the bare parameters have to be changed
keeping a number of (physical) quantities, corresponding to the number of bare 
parameters, fixed.

In the quenched model it is standard to tune the quark mass keeping fixed an hadronic mass and to tune
the gauge coupling keeping fixed the so called Sommer parameter $r_0$~\cite{Sommer:1993ce}. 
$r_0$ is an intermediate distance (usually fixed to be $0.5$ fm) where the force between two static 
quark is evaluated. While this quantity can be measured on the lattice very precisely, it has a rather 
uncertain phenomenological value. In the quenched model this is not a big problem, since the systematic 
error of neglecting the fermionic determinant is anyhow unknown, and the precise determination of the value
of the lattice spacing allows very careful analysis of scaling violations. In the results presented here
the values of $r_0/a$, where needed, are taken from~\cite{Guagnelli:1998ud}. 

A scaling test with O($a$) improved Wtm has been performed~\cite{DellaMorte:2001ys} in the quenched model
and in a finite volume
($L^3 \times T \simeq 0.75^3 \times 1.5$) ${\rm fm}^4$ with Schr\"odinger functional 
boundary conditions~\cite{Luscher:1992an,Sint:1993un,Sint:1995rb}.\footnote{These boundary conditions allow
to perform in intermediate volumes scaling studies of the lattice theory close to the continuum limit.
In particular this framework allows very precise determinations of several quantities 
which have a well defined continuum limit and can be used to study scaling violations.
Moreover these quantities become phenomenologically
relevant in the infinite volume limit.}
The renormalized quark masses were fixed in such a way that the ratio of untwisted to twisted mass was
\footnote{It is interesting to note that in this study, 
even if the untwisted quark mass is small, we are not at full twist.}
\be
\frac{m_{\rm R}}{\mu_{\rm R}} \simeq 0.131.
\ee
The outcome of the study is that even if some improvement coefficients are known only at
one loop in perturbation theory, for lattice spacings $a \ge 0.093$ fm 
the scaling behaviour of some renormalized and improved
quantities (which in large volume yield the mass and the
decay constant of pseudoscalar and vector mesons) is
consistent with O($a$) improvement, with residual
cutoff effects at $a =0.093$~fm ranging from $0.5\%$ to $9\%$.

The same setup and observables have then been employed
for a study in large volumes~\cite{DellaMorte:2001tu}: $L= 1.5$ to $2.2$~fm
and $T=(2-3) L$. This study was restricted to two lattice 
resolutions, $a=0.093$ and $0.068$~fm, and, for each of them,
three sets of quark mass parameters, which correspond to $|\omega| = \pi/2 + {\rm O}(a)$
and pseudoscalar meson masses $M_{\rm PS}$ in the range $1.85 \geq
(M_{\rm PS}/M_{\rm K^\pm})^2 \geq 0.85$. 

\begin{figure}[t]
\vspace{-10pt}
\begin{center}
\includegraphics[angle=0,width=15cm]{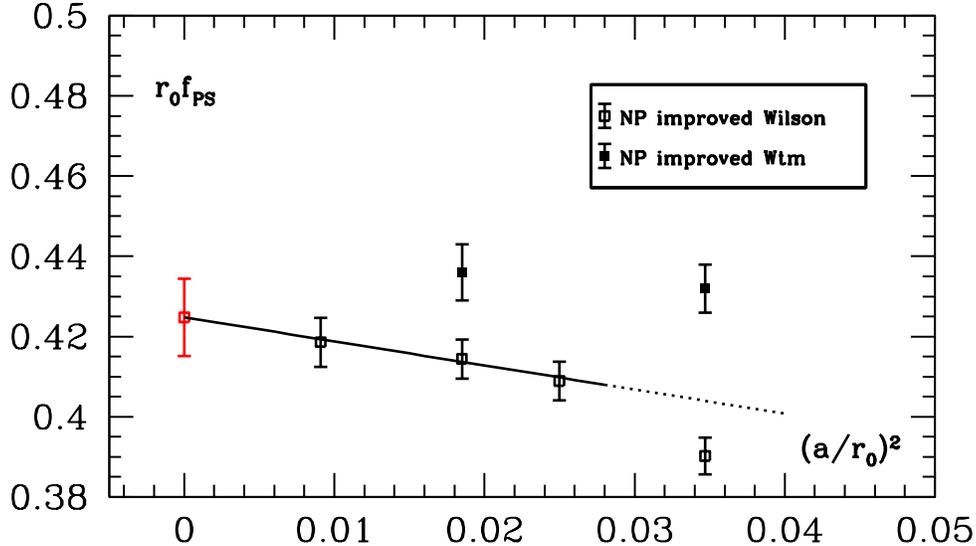}
\vspace{-7.0cm}
\end{center}
\caption{$f_{\rm PS}$ vs.\ $a^2$ at at a fixed value of the pseudoscalar mass 
$M_{\rm PS} \! \simeq \! 1.2 M_{\rm K^\pm}$ 
for non-perturbatively (NP) improved Wilson fermions ($\square$)~\cite{Garden:1999fg} and non-perturbatively (NP)
improved Wilson twisted mass ($\blacksquare$)~\cite{DellaMorte:2001tu}.
The continuum extrapolation by ref.~\cite{Garden:1999fg} is also shown.}
\label{fig:fPS_clover}
\end{figure}

The scaling behaviour of the pseudoscalar decay constant $f_{\rm PS}$ in large volume 
is presented in fig.~\ref{fig:fPS_clover} for the O($a$) improved Wilson
formulations with $|\omega| \simeq \pi/2$~\cite{DellaMorte:2001tu} ($\blacksquare$)
and $\omega =0$~\cite{Garden:1999fg} ($\square$). In the latter case four lattice
resolutions were considered to allow for continuum extrapolation.
The results for $f_{\rm PS}\,r_0$ 
that are obtained from the two lattice formulations
should agree in the continuum
limit: this seems to be the case within the statistical errors shown in the 
figures.
It is very interesting to note that Wtm at full twist, for this physical quantity, 
shows a much weaker $a^2$ dependence compared with O($a$) improved Wilson fermions.

\subsection{Automatic O($a$) improvement}
\label{ssec:auto}

We have shown that O($a$) improved Wtm requires the computation of less improvement coefficients
than Wilson fermions.
In particular we have seen that setting Wtm at full twist only $c_{\rm sw}$ has to be computed in order
to remove the O($a$) from the action and from the reparametrisation of the quark masses and gauge coupling.
While to improve operator matrix elements the standard improvement
coefficients are needed.
All the improvement coefficients depend on the details of the lattice action,
in particular they depend for example on the choice of the gauge action 
or on the way it is discretized the lattice derivative.

In a remarkable paper of Frezzotti and Rossi \cite{Frezzotti:2003ni} a step forward was made. 
It was proved that correlation functions made of parity even multiplicatively renormalizable
fields are free from O($a$) effects, and so no improvement coefficients are needed, 
if in the continuum limit $m_{\rm R} = 0$ (see eq.~\ref{eq:conttmQCD2} and eq.~\ref{eq:mr}),
i.e. if the physical quark mass is given solely by the twisted mass $\mu_{\rm R}$.

We will call this property {\it automatic O($a$) improvement}.
The first remark is that the theory itself is not improved, but the 
physical correlation functions are. To say it in another way, from 
all the possible sets of correlation functions that define the regularized theory, 
the O($a$) effects are all in those which vanish in the continuum limit. In particular the parity
odd correlators, which vanish in the continuum limit, will have a first contribution
of O($a$) while the parity even correlators will have as a first correction to the continuum value
an O($a^2$) error.

\subsubsection{Proof}

Automatic O($a$) improvement can be demonstrated in many ways, and many proofs 
appeared in the 
literature~\cite{Frezzotti:2003ni,Aoki:2004ta,Sharpe:2004ny,Frezzotti:2005gi,Shindler:2005vj,Sint:2005qz,Aoki:2006nv}.
The first proof was given in ref.~\cite{Frezzotti:2003ni}, which is 
sketched in app.~\ref{app:E}.
This proof is based on a set of spurionic symmetries of the lattice action.

Automatic O($a$) improvement can be proved in a different way just considering the 
symmetries of the continuum action. 
In the following I try to summarize the proof which emphasize the role of
the symmetries of the continuum action~\cite{Frezzotti:2005gi,Shindler:2005vj}, the automatic O($a$) 
improvement of the massless Wilson operator in a finite volume~\cite{Sint:2005qz},
and the usage of symmetries which are not spontaneously broken 
in infinite volume continuum QCD~\cite{Aoki:2006nv}.

Before going in details in the proof I would like to give an historical remark, 
in Minkowski space, that is not directly connected to twisted mass, but it helps
to clarify the nature of the cutoff effects in the Wilson theory.
The leading non-renormalizable corrections to QED
would be those interactions of dimension 5, which are suppressed by only one factor 
of $1/E$, where $E$ is some high energy scale.
According to Lorentz, gauge and $\mathcal{C}\mathcal{P}$ invariance there is only one term allowed: the Pauli term
proportional to $1/E \psibar \sigma_{\mu \nu} F^{\mu \nu} \psi$.
The contribution of this term to the magnetic moment of the electron or muon 
would give a constraint on the value of $E$.
It is well known, see for instance sec. 12.3 of~\cite{Weinberg:1995mt}, 
that this constraint can be strengthened using the following argument.
The QED Lagrangian is symmetric under the following chiral transformation
$\psi \rightarrow \gamma_5 \psi$ and $m \rightarrow -m$, were $m$ 
is the lepton mass. Then assuming that also the inclusion of a Pauli term should
respect this symmetry we see immediately that the Pauli term in the Lagrangian would have to
appear with an extra multiplicative factor $m/E$, i.e. $m/E^2 \psibar \sigma_{\mu \nu} F^{\mu \nu} \psi$. 
We will see now that this is essentially the mechanism responsible for automatic O($a$) improvement, 
because the Pauli term describes the leading discretization errors with an energy scale $E \sim 1/a$.

We first consider the massless Wtm action (eq.~\ref{eq:WtmQCD} with $m_0=m_{\rm cr}$, $\mu_{\rm q} = 0$) 
in a finite volume with suitable boundary conditions for all the fields. The choice to work in a finite volume 
is done to keep the mass dependence smooth and avoid any complications with 
possible phase transitions.

The Symanzik effective action reads
\be
S_{\rm eff} = S_0 + aS_1 + \ldots
\ee
and we are interested in a massless continuum target theory.
\be
S_0 = \int d^4x \chibar(x) \big [ \gamma_\mu D_\mu \big] \chi(x)
\label{eq:masslessQCD}
\ee
The correction terms in the effective action are given by
\be
S_1 = \int d^4y  {\mathcal L}_1(y) \qquad {\mathcal L}_1(y) = \sum_i c_i
  {\mathcal O}_i(y).
\ee
In the massless case the only operator contributing is
\be
{\mathcal O}_1 = 
  i \chibar\sigma_{\mu\nu}F_{\mu\nu}\chi,
\label{eq:sym_op}
\ee
which is the usual clover term. 
We consider now a general multiplicatively renormalizable multilocal field
that in the effective theory is represented by the effective field
\be
\Phi_{\rm eff} = \Phi_0 + a \Phi_1 + \ldots
\ee
A lattice correlation function of the field $\Phi$ to order $a$ is given by
\be
\langle \Phi \rangle = \langle \Phi_0 \rangle_0 - a \int d^4y \langle \Phi_0
     {\mathcal L}_1(y) \rangle_0 + a \langle \Phi_1 \rangle_0 + \ldots
\label{eq:sym_exp}
\ee
where the expectation values on the r.h.s are to be taken in the continuum
theory with action $S_0$.
The key point is that the continuum action (\ref{eq:masslessQCD}) is chirally symmetric, e.g.
the following discrete chiral symmetry
\be
\mathcal{R}^{1,2}_5 \colon
\begin{cases}
\chi(x_0,{\bf x}) \rightarrow i \gamma_5 \tau^{1,2} \chi(x_0,{\bf x}) \\
\chibar(x_0,{\bf x}) \rightarrow  \chibar(x_0,{\bf x}) i \gamma_5 \tau^{1,2}
\end{cases}
\label{eq:R512}
\ee
is a symmetry of the continuum action, while
all the operators in eq. (\ref{eq:sym_op}), of
the Symanzik expansion of the lattice action, are odd under the discrete chiral symmetry $\mathcal{R}^{1,2}_5$ 
of the continuum action.\footnote{Strictly speaking in the massless case the 
non trivial flavour structure of $\mathcal{R}^{1,2}_5$ it is not needed to prove automatic O($a$) 
improvement. It will become necessary in the massive case.}
If the operator $\Phi$ is a lattice representation of the continuum chirally even field $\Phi_0$,
then the second term in the r.h.s. of
eq. (\ref{eq:sym_exp}) vanishes. To show that also the $\Phi_1$ term vanishes we have to
show that an operator of one dimension higher than the original one but with the same lattice symmetries
has opposite chirality. To do this we introduce a symmetry that essentially counts the dimensions of the
operators~\cite{Frezzotti:2003ni}
\be
\mathcal{D} \colon
\begin{cases}
U(x;\mu) \rightarrow U^{\dagger}(-x-a\hat{\mu};\mu), \\ 
\chi(x) \rightarrow {\rm e}^{3 i \pi/2} \chi(-x) \\
\chibar(x) \rightarrow  \chibar(-x){\rm e}^{3 i \pi/2}.
\end{cases}
\ee
The gauge lattice action is invariant under $\mathcal{D}$ while in the fermion lattice action
the terms that break chiral symmetry are odd.
But in particular the lattice action is invariant under $\mathcal{R}^{1,2}_5 \times \mathcal{D}$.
So the operators in $\Phi_1$ will necessarily have opposite chirality to $\Phi_0$. Given the 
fact that the continuum action is chirally symmetric also $\Phi_1$ vanishes.
The conclusion is then: the chiral limit of the Wilson theory is automatically O($a$) improved, 
if we stay in a finite volume where no symmetry can be spontaneously broken~\cite{Sint:2005qz}.
We remark that the case of the Schr\"odinger functional is different 
since there the standard boundary conditions~\cite{Sint:1993un,Sint:1995rb}
break the chiral symmetry so automatic O($a$) improvement 
does not apply\footnote{Chirally twisted boundary conditions~\cite{Sint:2005qz} have been introduced in order 
to obtain a bulk automatic O($a$) improvement, with remaining O($a$) cutoff effects
stemming solely from the boundaries.}.

We add now a standard mass term $m_{\rm R}\chibar(x) \chi(x)$ to the action~(\ref{eq:masslessQCD}).
The dimension 5 operators contributing to ${\mathcal L}_1$
are now given by the operators in eqs.~(\ref{eq:o1}-\ref{eq:o3}). All these operators are odd under 
the spurionic symmetry~\footnote{Actually
also the operators in eqs.~(\ref{eq:o6app}) and~(\ref{eq:o7app}), 
which are eliminated using the equations of motion, are odd under the spurionic symmetry
$\widetilde{\mathcal{R}}^{1,2}_5$. This means that in principle the equation of motion
are not really needed to eliminate them in the context of automatic O($a$) improvement.}
\be
\widetilde{\mathcal{R}}^{1,2}_5 \equiv \mathcal{R}^{1,2}_5 \times (m_{\rm R} \rightarrow - m_{\rm R})
\ee
while the continuum action is even under $\widetilde{\mathcal{R}}^{1,2}_5$. 
This does not mean that the insertions of these operators in the Symanzik expansion
vanish, but since we are in finite volume
where the mass dependence is smooth this means 
that the cutoff effects are all of the kind O($a m_{\rm q}$).

It seems like a similar argument could be given in infinite volume. This indeed is not the case
because of the spontaneous breaking of chiral symmetry. Now the quark mass dependence around the chiral limit
does not need to be smooth and even if all the leading O($a$) effects are 
odd in the quark mass because of the spurionic symmetry $\widetilde{\mathcal{R}}^{1,2}_5$, a 
possible non-analyticity in the quark mass can generate pure O($a$) cutoff 
effects.~\footnote{A simple example is given by $a {\rm ~sign}(m_{\rm q})$ where
  the sign function is still odd under $m_{\rm q} \rightarrow -m_{\rm q}$ but
  it is non analytic in $m_{\rm q} = 0$.} 
To say it differently, the insertion of chirally odd operators does not vanish
in the chiral limit because of the spontaneous symmetry breaking of chiral symmetry. 

To obtain automatic O($a$) improvement in infinite volume 
including a mass term, we have to consider 
as a target continuum theory for the fermion fields 
\be
  \int d^4 x \chibar(x) \Big[\gamma_\mu D_\mu + i\mu_{\rm R} \gamma_5\tau^3\Big]\chi(x),
\label{eq:cWtm}
\ee
where the physical mass is given by the twisted mass term.
While in the massless case it is not possible to make a distinction between vector and axial
symmetries, in the massive case one usually associates the symmetry broken by the mass term with the
axial symmetry.
In the following we will refer generically to chiral symmetry, the form of which will depend on
the form of the mass term. We remind the reader that in the twisted basis the symmetry 
left unbroken by the mass term is the ``twisted'' vector symmetry~\eqref{eq:obl_grV}, 
and the symmetry broken by the mass term is the ``twisted'' axial symmetry~\eqref{eq:obl_grV}
(cf. sec.~\ref{sssec:chiral}).

We can repeat the same steps done in the first proof adding in $\mathcal{L}_1$ the term
\be
{\mathcal O}_5 =
  \mu_{\rm q}^2\chibar \chi.
\label{eq:mcO5}
\ee
The reason to have the physical mass term fully given by the twisted mass is that the 
continuum action~(\ref{eq:cWtm}) is still invariant under the discrete symmetry
$\mathcal{R}^{1,2}_5$, and if we now count the dimensions of the operators including the mass
\be
\widetilde{\mathcal{D}} =  \mathcal{D}\times [\mu_{\rm q} \rightarrow -\mu_{\rm q}]
\ee
then $\mathcal{R}^{1,2}_5 \times \widetilde{\mathcal{D}}$ is also still a
symmetry of the lattice action~\cite{Aoki:2006nv}.
We can then conclude that the second term in the r.h.s. of
eq. (\ref{eq:sym_exp}) vanishes, and $\Phi_1$, being of one dimension higher, 
is odd under a $\mathcal{R}^{1,2}_5$ transformation: for the same reason the third term in the r.h.s of
eq. (\ref{eq:sym_exp}) vanishes. Possible contact terms
coming from the second term amount to a redefinition of $\Phi_1$ as we have discussed 
in sec.~\ref{ssec:sym}, and so do not harm the proof.

The reason why now spontaneous breaking of chiral symmetry does not spoil 
the proof, is that the twisted mass term breaks chiral symmetry in an orthogonal direction compared to the
breaking of the Wilson term. Spontaneous chiral symmetry breaking is in the chiral flavour direction
of the mass term, while the Wilson term, and accordingly the relevant dimension five operators 
${\mathcal O}_1$ and ${\mathcal O}_5$ are ``orthogonal'', so are not affected by spontaneous
chiral symmetry breaking.

It is interesting to note that $\mathcal{R}^{1,2}_5$ corresponds to a
discrete twisted ``charged'' vector transformation defined in
eq.~(\ref{eq:uV12}). The actual symmetry used to prove automatic O($a$)
improvement is what would correspond in the physical basis to the standard
charged flavour symmetry which is well known not to be spontaneously broken in
continuum QCD~\cite{Vafa:1983tf}.

The same proof can be repeated identically using, instead of chiral symmetry, 
the twisted parity symmetry~\cite{Frezzotti:2005gi,Shindler:2005vj}
\be
\mathcal{P}_{\frac{\pi}{2}} \colon
\begin{cases}
U(x_0,\bx;0) \rightarrow U(x_0,-\bx;0), \quad 
U(x_0,\bx;k) \rightarrow U^{-1}(x_0,-\bx - a \hat{k};k), 
\quad k = 1, \, 2, \, 3 \\ 
\chi(x_0,{\bf x}) \rightarrow \gamma_0(i \gamma_5\tau^3)\chi(x_0,-{\bf x}) \\
\chibar(x_0,{\bf x}) \rightarrow  \chibar(x_0,-{\bf x})
(i\gamma_5\tau^3)\gamma_0 ,
\end{cases}
\label{eq:tmparity}
\ee
which is also not spontaneously broken in continuum QCD~\cite{Vafa:1984xg}.
The only difference would be that for matrix elements with non-vanishing momenta, the twisted parity
proof requires, an average of matrix elements computed with momentum $\mathbf{p}$ and $-\mathbf{p}$, while this
is not required using chiral symmetry.
It is then clear that in order to achieve automatic O($a$) improvement, the
continuum target theory must have a vanishing untwisted quark mass $m_{\rm
  R}$, otherwise the standard mass term $m_{\rm R} \chibar \chi$ 
will break the residual ``twisted'' vector symmetry $\mathcal{R}^{1,2}_5$ 
(or the twisted parity symmetry $\mathcal{P}_{\frac{\pi}{2}}$) of the continuum action. The
most natural way to achieve this on the lattice is
by setting the untwisted bare quark mass to
its critical value $m_0 = m_{\rm cr}$.
The proof also shows that a possible uncertainty of O($a$) in the critical
mass does not invalidate automatic O($a$) improvement since these uncertainties
are odd under  ``twisted'' vector symmetry (or twisted parity).

The careful reader may wonder what happens to automatic O($a$) improvement if
the critical mass is fixed such that the untwisted quark mass is of O($a$). 
We recall that the theory itself it is not improved, but only the physical correlators are.
In principle to obtain automatic O($a$) improvement we have seen it is necessary
to have $m_{\rm R}=0$, only in the continuum limit,
which means at finite lattice spacing at most $m_{\rm q} =$ O($a$)
(see eq.~(\ref{eq:mr},\ref{eq:mtildeq})).
This uncertainty can be described by a dimension 5 operator
\be
{\mathcal O}_0 = \Lambda^2\chibar \chi,
\label{eq:O_0}
\ee
where $\Lambda^2$ is some energy scale squared which depends on the way the critical mass is determined,
e.g. it could be of the order of the QCD scale $\Lambda_{\rm QCD}^2$, or it could 
be something proportional to $\Lambda_{\rm QCD}\mu_{\rm q}$.
In particular to have a massless continuum theory in general we have $m_{\rm q} = $ O($a$).
We can say that the operator ${\mathcal O}_0$ parameterizes O($a$) uncertainties in the critical mass.
This uncertainty does not harm automatic O($a$) improvement because is described by
an operator which is odd under the transformation
$\mathcal{R}^{1,2}_5$ which is symmetry of the continuum action~(\ref{eq:cWtm}).

We can conclude that correlators which are even under a twisted parity or twisted vector 
transformation, are automatically O($a$) improved without the knowledge of any improvement coefficient,
and by just tuning the critical mass such that the untwisted quark mass is at most $m_{\rm q} =$ O($a$).

From the proof apparently there are no constraints on the values of the quark
masses $\mu_{\rm q}$ where automatic O($a$) improvement is at work.
The presence of the Wilson term in the lattice action enforces us to perform a continuum limit
first at a fixed value of the renormalized quark mass, and then to study the
quark mass dependence, but it is important to understand how low we can go
with the quark mass at fixed lattice spacing.

To have a first guess we can take the polar mass
\be
M_{\rm R} = \sqrt{\mu_{\rm R}^2 + m_{\rm R}^2} = \sqrt{\mu_{\rm R}^2 + (\eta_1 a \Lambda^2)^2},
\label{eq:polar}
\ee
where the $\eta_1$ term parameterizes the mass independent O($a$) uncertainties in
the value of the untwisted quark mass $m_{\rm q}$ (see eqs. \ref{eq:mr} and \ref{eq:mtildeq}).
Expanding in powers of $a$ we have
\be
M_{\rm R} \simeq \mu_{\rm R}\Big[1+\frac{\eta_1 a^2\Lambda^4}{2\mu_{\rm R}^2} +
  O(a^4)\Big].
\label{eq:pole_exp}
\ee
We observe immediately that if numerically $\mu_{\rm R} < a\Lambda^2$, even if parametrically
O($a$) terms are absent in (\ref{eq:pole_exp}), there is a term of O($a^2$) with a
coefficient that tends to diverge as soon as $\mu_{\rm R}$ is made smaller and smaller.
From this example we can conclude that to have an effective automatic O($a$) 
improvement, without big O($a^2$) effects, with a generic choice of the critical mass, such that the
uncertainties in the untwisted quark mass are of order $a\Lambda^2$, we need to have the
constraint 
\be
\mu_{\rm R} > a\Lambda^2 .
\label{eq:constr1}
\ee
From a practical point a view this constraint can be very strong. If we take the
reasonable value $\Lambda = 300$ MeV and a lattice spacing $a=0.1$ fm then the
minimal quark mass that can be simulated without being affected by large
O($a^2$) effects is $\mu_{\rm R} = 45$ MeV corresponding, in the pseudoscalar sector,
roughly to the mass of a kaon made up by two degenerate quarks. 
It is then clear that in order to go closer
to the physical point corresponding to the {\it up} and {\it down} quark
masses the constraint has to be weakened.
From the example of the pole mass (\ref{eq:pole_exp}) we immediately
understand that the crucial issue is the determination of the critical line
and the understanding of its O($a$) uncertainties.

\subsection{The critical mass}
\label{ssec:crit_mass}

In this section, to let the interested reader have a general background, we
give the basic information on the two main theoretical frameworks used in the
discussion on the determination of the critical mass: Symanzik expansion and
Wilson chiral perturbation theory (W$\chi$PT). 
We then list the main results and discuss them, omitting some of the technical
details of their derivations.

The issue of the choice of the critical line was raised by the work of Aoki
and B\"ar~\cite{Aoki:2004ta} and by the numerical results obtained
in~\cite{Bietenholz:2004wv}. 
This problem has been further analyzed in several aspects in
\cite{Sharpe:2004ny,Frezzotti:2005gi,Sharpe:2005rq}.

We have seen that to obtain automatic O($a$) improvement the untwisted quark
mass has to be set to its critical value, i.e. to a value such that in the
continuum limit $m_{\rm R} = 0$. To understand how to impose this, it is
enough to understand which are the symmetries that are recovered in the
continuum if $m_{\rm R} = 0$ and to impose suitable identities on the lattice. 

The symmetries are the twisted parity defined in eq.~(\ref{eq:tmparity}) and
twisted vector symmetry in the isospin direction $1$ and
$2$~\eqref{eq:uV12}.
One way to impose the restoration of twisted vector symmetry is using the PCAC relation, 
i.e. determining the critical mass setting the PCAC quark mass to zero. 
To restore twisted parity it is enough to use a twisted 
parity violating matrix element like a correlator between charged axial and pseudoscalar currents,
and setting it to zero. 
We remark at this point that actually imposing the restoration of twisted parity 
automatically restores the twisted vector symmetry and vice versa.
This can be understood observing that the discrete version of twisted vector symmetry~\eqref{eq:R512}
times twisted parity~\eqref{eq:tmparity} is a symmetry of the lattice action, 
i.e. $\mathcal{P}^{1,2}_F$~\eqref{eq:PF12}.

All these options have been investigated both analytically and numerically.
There are two possible approaches to show that the restoration of twisted
parity or twisted vector symmetry is enough to ensure
automatic O($a$) improvement down to quark masses satisfying a weaker
constraint than eq.~(\ref{eq:constr1}). 
One is the analysis, using the Symanzik expansion, of the O($a$) effects in the correlators used to define
the critical mass. The second one is the use of a suitable modified chiral expansion in order to include
discretization errors in the effective Lagrangian describing pion interactions.

\subsubsection{Symanzik expansion}
\label{sssec:Sym_crit}

One possible way to restore twisted vector symmetry is to tune the bare untwisted quark mass
$m_0$ to a critical value $m_{\rm cr}$ such that the PCAC mass 
\be
m_{\rm PCAC} = \frac{\sum_{\bf x} \langle \partial_0 A_0^a(x) P^a(0)\rangle}
{2\sum_{\bf x} \langle  P^a(x) P^a(0)\rangle} \qquad a=1,2.
\label{eq:PCAC_lat}
\ee
vanishes for large euclidean times.

\begin{figure}[htb]
\epsfig{file=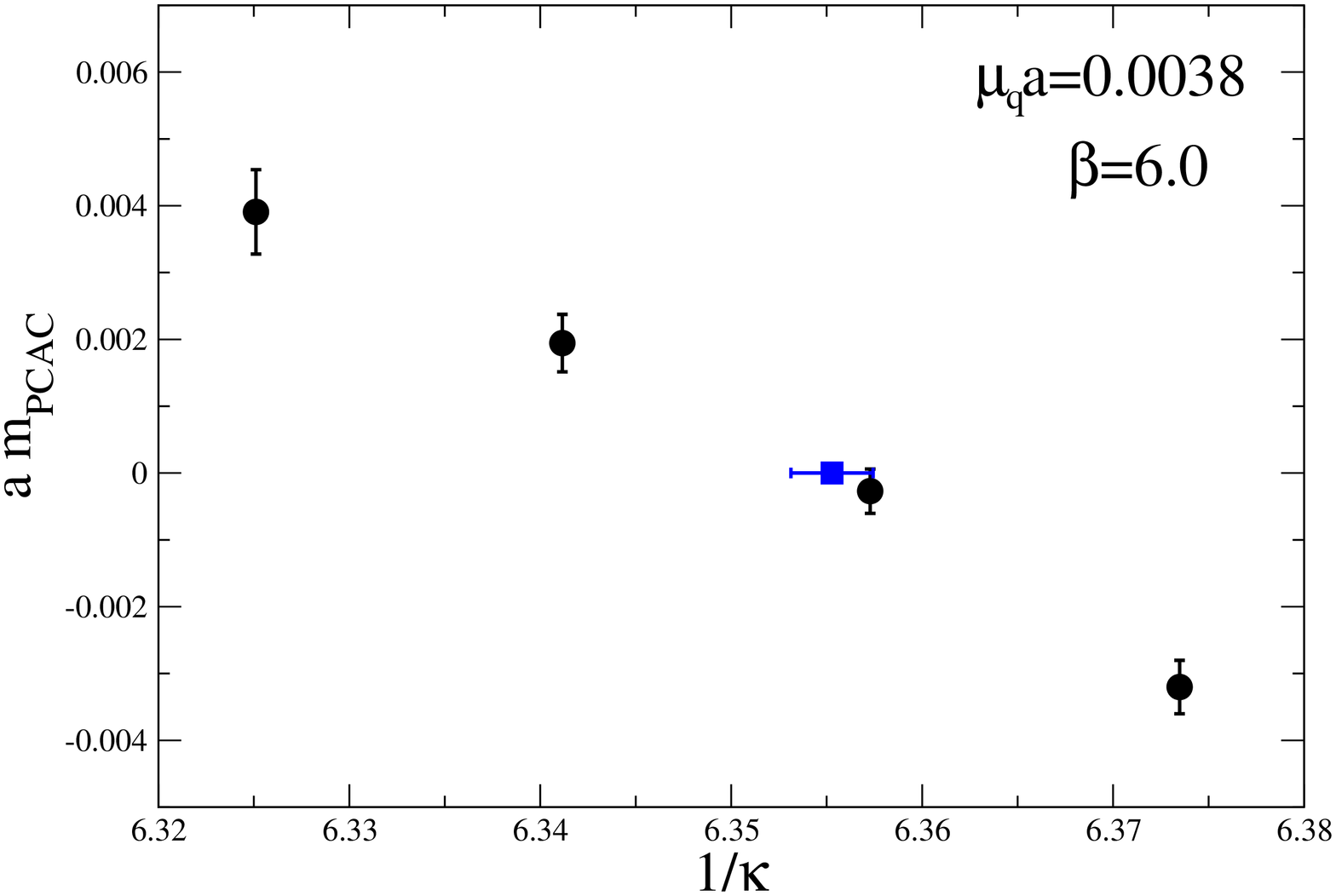,angle=270,width=0.52\linewidth}
\epsfig{file=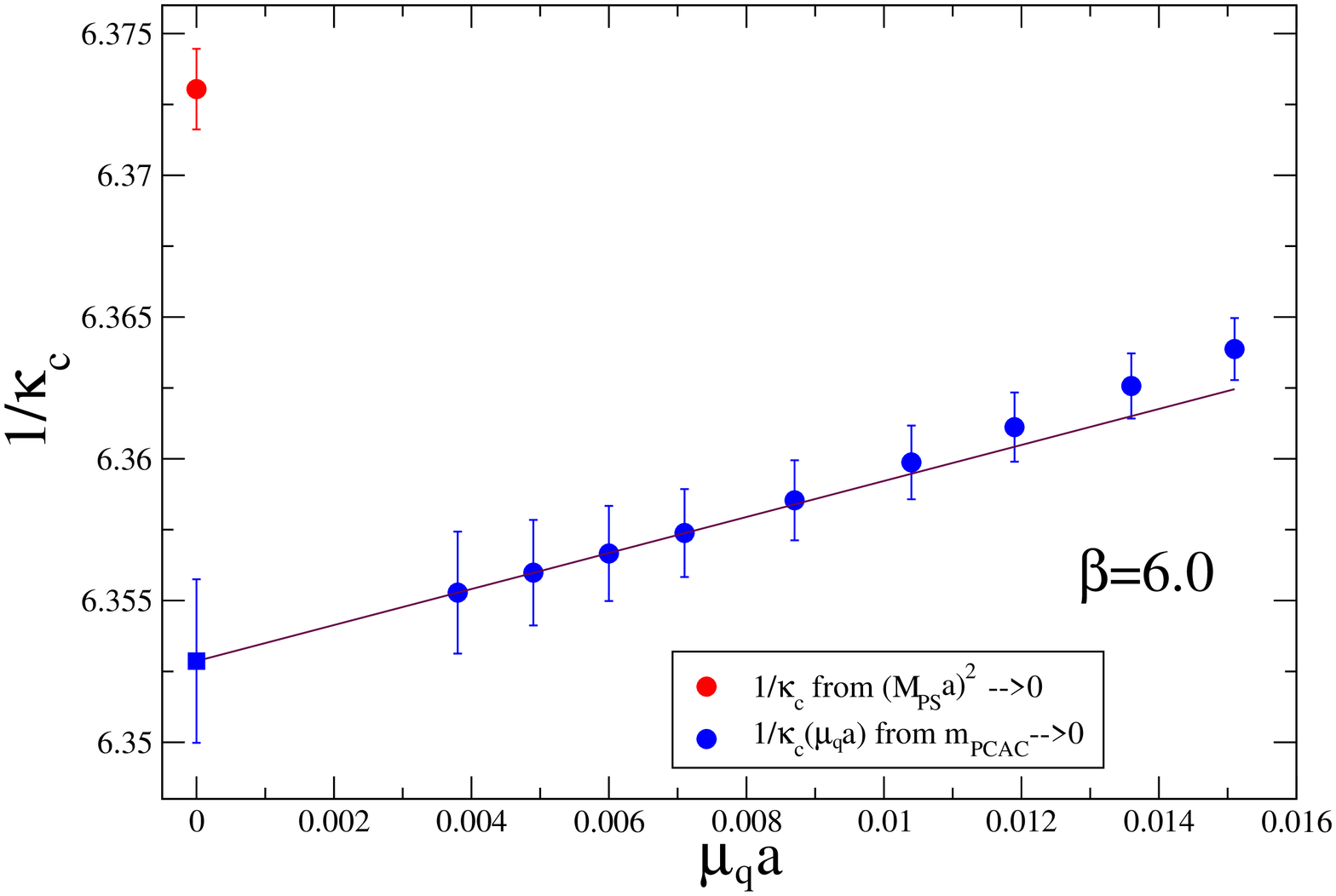,angle=270,width=0.52\linewidth}
\caption{Determination of the critical mass $m_{\rm c}$ ($\kappa^{-1} =
2am_0 +8$) for a given value of $\mu_{\rm q}$ at $a=0.093$ fm (left plot), 
and extrapolation to $\mu_{\rm q}=0$ (right plot). 
The red point in the right plot is the critical mass determined using the ``Wilson pion''
definition (see sect.~\ref{sssec:wchipt} and~\ref{ssec:mcsum} for details).}
\label{fig:kappa_c}
\end{figure}

It is possible to show that the cutoff effects of the PCAC mass 
in infinite volume can be schematically written as
\be
\eta_1 a \Lambda_{\rm QCD}^2 + \eta_2 a \mu_{\rm R}^2 + \eta_3 a \Lambda \mu_{\rm R} .
\label{eq:PCAC_cutoff}
\ee
We are here implicitly assuming that the only physical scales of the theory are
$\Lambda_{\rm QCD}$ and $\mu_{\rm R}$, i.e. we have already a rough estimate of the critical
mass such that $m_{\rm R}=$O($a\Lambda_{\rm QCD}$).
Practically at this point one has several possibilities.
For example~\cite{Jansen:2005gf,Jansen:2005kk} 
for each value of $\mu_{\rm q}$ it is possible to determine the critical mass $m_0=m_{\rm cr}(\mu_{\rm q})$ 
(see left panel of fig.~\ref{fig:kappa_c}) tuning the PCAC mass to zero, 
and then to extrapolate the obtained set of $m_{\rm cr}(\mu_{\rm q})$ to $\mu_{\rm q} = 0$ 
(see right panel of fig.~\ref{fig:kappa_c}).
This extrapolated value of $m_{\rm cr}$ (or equivalently $\kappa_{\rm c}$) can then be used to
perform simulations for all the values of $\mu_{\rm q}$. 
In fact from eq.~\eqref{eq:PCAC_cutoff} we observe that the critical mass has been tuned
such that for all the values of $\mu_{\rm q}$ the PCAC mass has at most O($a\mu_{\rm q}$).
The slope of the curve in the right plot of fig.~\ref{fig:kappa_c} is proportional, as it has been discussed in
\cite{Aoki:2005ii,Frezzotti:2005gi,Sharpe:2005rq}, to O($a$) cutoff effects related to the
discretization errors of the PCAC mass. In other words this slope is proportional to
the $\eta_3$ term in eq.~\eqref{eq:PCAC_cutoff}.
We remind that it is not surprising that the PCAC mass is not
automatically O($a$) improved since it is an odd quantity under the twisted parity
transformations~\eqref{eq:tmparity} and discrete symmetry~\eqref{eq:R512}.

Another possibility would be after the determination of the function $m_{\rm cr}(\mu_{\rm q})$
to use a different value of $m_{\rm cr}$ such that $m_{\rm PCAC}$ vanishes for each 
value of $\mu_{\rm q}$\footnote{In practice one could smoothly interpolate the curve 
$m_{\rm cr}(\mu_{\rm q})$ at the desired value of the twisted mass}.

This procedure has been used in~\cite{Abdel-Rehim:2005gz} but using a slightly different correlator.
One possible way to impose twisted parity restoration, is to tune $m_0$ to a critical value 
$m_{\rm cr}$ such that the twisted parity violating correlator
\be
a^3\sum_{\bf x} \langle A_0^1({\bf x},x_0) P^1(0) \rangle
\label{eq:crit_par}
\ee
vanishes for large euclidean times $x_0$.
The Symanzik expansion of the correlator~\eqref{eq:crit_par} is, in form, identical
to the one of the PCAC mass~\eqref{eq:PCAC_cutoff}. So the same considerations done for the PCAC
mass apply here. In particular if one performs an extrapolation of $m_{\rm cr}(\mu_{\rm q})$ to $\mu_{\rm q}=0$,
the cutoff effects of the critical mass would be at most of O($a\mu_{\rm q}$), while if 
$m_{\rm cr}$ is tuned to have twisted parity restoration for each value of $\mu_{\rm q}$ then the critical mass
is tuned such that $m_{\rm PCAC}=0$ for each lattice spacing and each twisted mass value.

Another possible way to fix the critical mass, especially practical for
expensive dynamical simulations, is to compute the critical mass, using the
PCAC relation at the smallest value of $\mu_{\rm q}=\mu_{\rm min}$, and then use this critical mass
for all the simulation points at heavier twisted masses.
This method has been used in a recent work~\cite{Boucaud:2007uk} where for the first time
large scale dynamical simulations have been performed with Wtm.
This is justified if at different lattice spacings the values of $\mu_{\rm R}$
where $m_{\rm cr}$ is computed are properly matched.
Using again eq.~\eqref{eq:PCAC_cutoff} one sees that for all the values of $\mu_{\rm q} > \mu_{\rm min}$
the PCAC mass has cutoff effects at most of O($a \mu_{\rm min}$).

We immediately observe that all these methods tune the critical mass such that 
the PCAC mass is identically zero or at most an O($a\mu_{\rm q}$).

To obtain the same result it is possible to use a clover term in the action
with a non-perturbatively tuned value for $c_{\rm sw}$. This will remove by definition
the O($a\Lambda$) in the PCAC mass, leaving again the PCAC mass with O($a\mu_{\rm q}$).
The critical mass can be determined also in the standard Wilson framework, with $\mu_{\rm q}=0$.
In this case a non-perturbatively tuned value for $c_{\rm sw}$, will again eliminate the 
O($a\Lambda$) errors, leaving only O($am_{\rm q}$) cutoff effects in the PCAC mass.
Performing then a chiral extrapolation at fixed lattice spacing would implicitly determine the critical
mass $m_{\rm cr}$ up to O($a^2$).
The only disadvantage of using this method with $\mu_{\rm q} = 0$ being that for quenched computations a 
long extrapolation to the chiral point is needed because of the occurrence of exceptional
configurations. 

The same considerations apply if the critical mass is determined in large volume
simu\-lations using Schr\"odinger Functional boundary conditions.
If the standard Schr\"o\-dinger Functional is used in a small volume ($L \lesssim 0.5$ fm),
the PCAC mass properly improved (computed with the proper values of $c_{\rm sw}$, $c_{\rm A}$)
can be used to determine the critical mass with residual discretization errors of 
O($a^2$)~\cite{Luscher:1996ug} without almost any extrapolation to the chiral limit. 
This is possible because the standard chirally breaking boundary conditions protect 
the spectrum of the Wilson operator from the appearence of very small eigenvalues, 
allowing simulations almost at the chiral point.

What is relevant is that the cancellation of the O($a\Lambda$) obtained with a 
properly tuned $c_{\rm sw}$ is mass independent.
This allows a tuning to full twist without a recomputation of the critical mass.
In fact in refs.~\cite{Dimopoulos:2006dm,Becirevic:2006ii,Dimopoulos:2007cn} old 
determinations\footnote{In~\cite{Dimopoulos:2006dm} $m_{\rm cr}$ has been recomputed only at one lattice spacing.}
\cite{Luscher:1996ug,Rolf:2002gu,Guagnelli:2004ga} of the critical mass with clover improved fermions
have been used.

To summarize, all the determinations of the critical mass based on correlators which violate
twisted parity and twisted vector symmetry, are such that the PCAC mass is affected
at most by O($a\mu_{\rm q}$) cutoff effects.

If we recall the example of the polar mass~\eqref{eq:pole_exp} 
we see that now we could relax the constraint~\eqref{eq:constr1}. 
It is possible to show~\cite{Frezzotti:2005gi} that this observation 
is actually true in general.

In~\cite{Frezzotti:2005gi}, it has been shown that the cutoff effects which
diverge at small quark masses previously discussed
(see eq.~\ref{eq:pole_exp}), 
so called infrared divergent (IR) cutoff effects, are a general property of
Wtm. 
In general a Wtm correlator will be automatically O($a$) improved at full
twist, but could suffer from numerically large O($a^2$) effects as soon as 
$\mu_{\rm R} \simeq a \Lambda^2$.
The result of ref.~\cite{Frezzotti:2005gi} can be summarized as follows: in
the Symanzik expansion of the lattice correlator $\langle \Phi \rangle$
defined in eq.~(\ref{eq:phi}) at order $a^{2k}$ ($k \ge 1$) there are terms of the
kind
\be
\left(\frac{a}{M_\pi^2}\right)^{2k}, \qquad k \ge 1
\label{eq:ampi2}
\ee
proportional to the matrix element
\be
|\langle \Omega | {\mcL}_1 | \pi^0 ({\bf 0}) \rangle_0|^{2k}.
\label{eq:xipi}
\ee
It is possible to recognize the $(a/\mu_{\rm R})^2$ of eq.~\eqref{eq:pole_exp}
as the $k=1$ case in eq.~\eqref{eq:ampi2} recalling that to a first approximation
$M_\pi^2 \propto \mu_{\rm R}$.
The terms in eq.~\eqref{eq:ampi2} are called leading infrared divergent cutoff effects, and they
come from continuum correlators where ${\mcL}_1$ is inserted $2 k$ times.
In the following, unless specified, we use the same notation for the dimension
5 fields and the corresponding operators.
A bit of notation is needed now: this is the continuum (see the index $0$)
matrix elements of the dimension $5$ Lagrangian that appears in the Symanzik
expansion of a lattice correlator defined by
\be
{\mcL}_1 = c_0 {\mcO}_0 + c_1 {\mcO}_1 + c_5 {\mcO}_5
\ee
where ${\mcO}_{0,1}$ are defined in eq.~(\ref{eq:sym_op}) and ${\mcO}_5$ in
eq.~(\ref{eq:mcO5}). Actually being proportional to $\mu_{\rm q}^2$ the
field ${\mcO}_5$ is not relevant for the following discussion. We also remind
that we work at full twist with an unspecified estimate of the critical mass
$m_{\rm cr}$.The dimension $5$ Lagrangian ${\mcL}_1$ has the quantum numbers of
a neutral pion field because we recall we are in the twisted basis at full twist. 
Then it has a non-zero matrix element between the vacuum $\langle \Omega|$ and
the neutral pion at rest $|\pi^0({\bf 0}) \rangle $ states.

To remove these dangerous cutoff effects, in ref.~\cite{Frezzotti:2005gi} 
it is proven that, setting the critical mass imposing the restoration of 
the twisted vector symmetry,
the leading infrared divergent cutoff effects
are removed. In particular in~\cite{Frezzotti:2005gi} it is
suggested to compute for each value of $\mu_{\rm q}$ the critical mass
imposing that the correlator in eq.~(\ref{eq:crit_par}) vanishes for large euclidean times,
and then to extrapolate to $\mu_{\rm q} = 0$. This leads to the result
\be
\lim_{\mu_{\rm q} \rightarrow 0} |\langle \Omega | {\mcL}_1 | \pi^0 ({\bf 0})
\rangle_0|^{2k} = 0.
\ee
Incidentally in the same paper it is also suggested 
that an alternative possibility would be to
use non-perturbatively improved clover fermions which would set 
the operator $\mathcal{O}_1$ to zero.
The conclusion of ref.~\cite{Frezzotti:2005gi} is that with an ``optimal'' choice
of the critical mass, obtained as just discussed, Wtm is automatically O($a$) improved for 
\be
\mu_{\rm  R} > a^2 \Lambda^3 .
\label{eq:constr2}
\ee

Before ending the section we want to briefly show with an example how even without an extrapolation 
to $\mu_{\rm q}=0$ the method of ref.~\cite{Boucaud:2007uk} does not imply the existence of infrared
divergent cutoff effects.

We set the PCAC mass to zero at a value $\mu_{\rm R} = \mu_1$ and as an example we
consider again the polar mass~(\ref{eq:polar}) up to O($a^2$) at a value
$\mu_{\rm R} = \mu_2 \neq \mu_1$ 
\bea
M_{\rm R} \simeq \mu_R \left[ 1 + \frac{1}{2} \eta_2^2 a^2 \mu_2^2 
\left(1- \frac{\mu_1^2}{\mu_2^2}\right)^2 \right. & + & \nonumber \\ 
+ \frac{1}{2} \eta_3^2 a^2 \Lambda^2 \left(1-
    \frac{\mu_1}{\mu_2}\right)^2 & + & \left.
\eta_2 \eta_3 a^2 \Lambda \left(\mu_2 + \mu_1\right) \left(1-
  \frac{\mu_1}{\mu_2}\right)^2
\right] .
\eea
It is then clear from this example that no dangerous infrared cutoff effects appear
if $\mu_2 > \mu_1$. Even if $\mu_2 \lesssim \mu_1$ no big
enhancement are visible and if $\mu_2 \gg \mu_1$ the cutoff effects take the
standard form as the PCAC mass would have been set to vanish at $\mu_{\rm q}
= 0$. 

All the methods we have discussed before show that all the determinations
of the critical mass imply at most O($a\mu_{rm q}$) cutoff effects in the PCAC mass.
We will see in the next section that actually from a purely theoretical point of view,
all the determinations of the critical mass will have this property.
While certainly the analysis of ref.~\cite{Frezzotti:2005gi} is correct, if we consider 
only O($a$) cutoff effects there is really no ``optimal'' choice of the critical mass, 
but they are all equivalent.
We will analyze further this point, in particular because from a practical point of view
this might not be the case.

\subsubsection{Wilson chiral perturbation theory}
\label{sssec:wchipt}

An alternative method to analyse cutoff effects in physical observables at low energies 
is to apply the methods of chiral perturbation theory to the Wtm action, i.e.
at non-zero lattice spacing.
The discretization errors can be included systematically in a combined expansion
in the lattice spacing and in the quark mass.

We recall first the basic principles of chiral perturbation theory ($\chi$PT) in the continuum
considering $N_f=2$ flavour QCD.
If the quark masses are set to zero the QCD action~(\ref{eq:tmQCDcont}) is symmetric under the chiral
group $SU_{\rm L}(2) \times SU_{\rm R}(2)$.
One then {\it assumes} that the theory spontaneously breaks this symmetry (see ref.~\cite{Gasser:1982ap} and
refs. therein) down to $SU_{\rm V}(2)$\footnote{Strictly speaking for $N_f=2$, 
chiral symmetry is spontaneously broken when $(m_u + m_d) \rightarrow 0$. In the following we will concentrate 
on degenerate light quarks and neglect physical isospin breaking effects.}.
The symmetry manifests itself in the occurrence of $2^2-1=3$ pseudoscalar
Goldstone bosons. In reality the QCD action contains a mass term which breaks the symmetry and this appears
in the non-conservation of the N\"other currents~(\ref{eq:PCAC},\ref{eq:PCVC}).
Since the masses of the $u$ and $d$ quarks are small, the low energy properties associated with chiral
symmetry should show a small deviation due to the quark masses.
A rough estimate of the size of these deviations should be given by the ratio between the quark masses
and the intrinsic QCD scale, giving a violation of a few percent.

The low energy structure of the correlation functions in QCD depends on the size of the quark masses.
Heavy quarks play a minor role because their degrees of freedom are frozen at low energies.
Here we consider only two flavours $u$ and $d$. This is suitable for our purposes where 
we consider a lattice action for two degenerate flavours. 
In this restricted framework we are able to discuss only the dependence of the correlation
functions on the $u$ and $d$ quark masses, and the remaining quark masses are fixed.
The method we are going to briefly review~\cite{Gasser:1983yg,Gasser:1985gg} is the extension
of the analysis carried by Weinberg~\cite{Weinberg:1978kz} for the S-matrix elements, to an expansion
of correlation functions in powers of the momenta and the quark masses.

The method consists in adding space dependent external fields to the QCD Lagrangian 
which transform accordingly in order to keep the Lagrangian
invariant under a local chiral transformation. Then using the assumption of spontaneous symmetry breaking,
it is possible to write a general low energy effective chiral Lagrangian where the pion fields, i.e. the fields
corresponding to the Goldstone bosons, are collected in a unitary matrix that automatically fulfills the 
requirements of chiral symmetry. The source terms of the QCD Lagrangian are collected in the effective
Lagrangian according to their symmetry transformations under chiral symmetry.
The behaviour at small momenta and quark masses of the QCD correlation functions can be recovered
{\it matching} them with the expansion of the chiral effective Lagrangian in powers of the
derivative of the external fields and the fields themselves.
We remark that this low-energy expansion is not a Taylor series: the pions generate poles
at small momenta. The correlation functions admit a Taylor expansion only if the momenta are
much smaller then the pion mass. So the power counting will be identified by the pion momentum
$p^2$ and the pion mass $M_\pi^2$ (or the quark mass). In particular they will be treated to be both small
but with the value of the ratio $p^2/M_\pi^2$ fixed and unconstrained.

The QCD Lagrangian with external fields reads
\be
\mathcal{L} =  \mathcal{L}_0 + \chibar(x)\Big[s(x) + i\gamma_5 p(x)\Big]\chi(x).
\label{eq:QCDwso}
\ee
For simplicity, since not needed in the following analysis, we neglect vector and axial external fields
and $\theta$-terms induced by the anomaly. $\mathcal{L}_0$ is the massless QCD Lagrangian 
that includes the gauge part,\footnote{
It could in principle also include the heavier quarks.} and the external fields
$s(x)$ and $p(x)$ are $2 \times 2$ Hermitean matrices in flavour space (we assume that $p(x)$ is traceless)
\be
s(x) = s^0(x) \mathbb{I} + s^a(x) \tau^a, \qquad p(x) = p^a(x) \tau^a.
\ee
The connected correlation functions are obtained performing functional
derivatives with respect to the sources $s(x)$ and $p(x)$ on the generating functional defined as 
\be
{\mcW}[s,p] =  \log {\mcZ}[s,p], \qquad 
  {\mcZ}[s,p] = \int D[\chibar,\chi] D[U]\,\e^{\displaystyle -S[s,p]}, \qquad S[s,p] = \int d^4 x \mathcal{L},
\ee
and then fixing them at their physical value:
in the twisted basis the correlation functions for massive quarks at full twist are obtained
expanding around $s=0$ and $p = \mu_{\rm q} \tau^3$.
For example 
\be
\langle P^a(x) P^b(y) \rangle  = \Big(\frac{-i}{2} \frac{\delta}{\delta p^a(x)}\Big)
\Big(\frac{-i}{2} \frac{\delta}{\delta p^b(y)}\Big) {\mcW}[s,p]{\Big{|}}_{s=0,p= \mu_{\rm q} \tau^3}.
\ee
The local $SU_{\rm L}(2) \times SU_{\rm R}(2)$ transformation of the fields is 
\be
\chi(x) \rightarrow V_R(x)\frac{1}{2}(1+\gamma_5)\chi(x) + V_L(x)\frac{1}{2}(1-\gamma_5)\chi(x)
\ee
\be
s(x) + i p(x) \rightarrow V_R(s(x) + i p(x))V_L^\dagger.
\ee

On the other side, the pion fields are collected in a unitary matrix $\Sigma$ which transforms according
to the linear representation
\be
\Sigma(x) \rightarrow V_R \Sigma(x) V_L^\dagger.
\ee
The singlet field is eliminated imposing $\det \Sigma = 1$, where $\det$ is applied in flavour space.

The effective chiral Lagrangian will be a function of the pion fields and their derivatives 
together with the external fields
\be
\mathcal{L}_{\chi} = \mathcal{L}_{\chi}(\Sigma, \partial_\mu \Sigma,s,p,...).
\ee
The order of the arguments reflects their low-energy dimensions: $\Sigma$ counts as 
a field of order $1$, $\partial_\mu \Sigma$ as order $p$ and $s(x)$,$p(x)$ as order $p^2$.
The general effective Lagrangian of order 1 is only a function of $\Sigma$ and since it has to be chiral
invariant it can only depend on $\det \Sigma$ or 
$\Tr\left[\left(\Sigma \Sigma^\dagger\right)^n\right]$ where $\Tr$ is applied in flavour space, 
i.e. an irrelevant constant. We conclude that chiral symmetry implies a leading derivative coupling.
The matrix $\Sigma$ that collects the pion fields can be written as
\be
\Sigma(x) = \Sigma_0 {\rm exp}\Big(i\frac{\pi^a(x) \tau^a}{f^2}\Big)
\ee
where $\pi^a(x)$ are the pion fields, the dimensionfull constant $f$ 
is the decay constant in the chiral limit normalized to $f_\pi = 93$ MeV, and $\Sigma_0$ is the 
vacuum expectation value of $\Sigma$, that breaks the chiral symmetry down to
$SU_{\rm V}(2)$.
The pion fields parametrize the fluctuations around $\Sigma_0$.

The most general form for the effective Lagrangian consistent with chiral symmetry
is given by
\be
\mathcal{L}_{\chi}^{(2)} = \frac{f^2}{4} \Big[
\langle \partial_\mu \Sigma(x)^\dagger \partial_\mu \Sigma(x) \rangle +
\langle \sigma(x)\Sigma(x)^\dagger + \sigma(x)^\dagger \Sigma(x) \rangle\Big],
\label{eq:Lchi2}
\ee
where the brackets $\langle \cdot \rangle$ here indicate the trace in flavour space.
The field $\sigma(x)$ collects the external field dependence of the effective
Lagrangian according to chiral symmetry
\be
\sigma(x) = 2B_0 \big[s(x) + ip(x)\big].
\ee
Analogously to what we have done for the quark Lagrangian~(\ref{eq:QCDwso}) we can do here for 
the effective chiral Lagrangian, defining the generating functional for connected correlation functions.
The strategy is then to equate the correlation functions obtained in the two theories.
This will express the QCD correlators as a function of the quark masses and the 
low energy constants (LEC) $f$ and $B_0$. Since we are expanding around massless QCD 
the only scale that can appear is $\Lambda_{\rm QCD}$, so one expects $f \sim B_0 \sim \Lambda_{\rm QCD}$.

To include the discretization errors in effective chiral theory one proceeds 
in two steps~\cite{Sharpe:1998xm}. 
First, one determines the continuum Symanzik action describing the interactions of quarks and gluons
with momenta much smaller than $\pi/a$. Discretization errors enter with explicit factors of $a$,
and are controlled by the symmetries (or lack thereof) of the underlying
lattice theory. Second, one uses standard techniques to develop a generalized 
chiral expansion for the Symanzik effective theory. We will call generically this expansion 
Wilson chiral perturbation theory (W$\chi$PT) having in mind a general form for the mass term
that includes also the twisted mass case. 

The form of the Symanzik effective Lagrangian including the external sources is given by
\be
{\mcL}_{\rm eff} = {\mcL} + a{\mcL}_1 + \ldots
\label{eq:cel}
\ee
with ${\mcL}$ given by eq.~(\ref{eq:QCDwso}) and ${\mcL}_1$ by eq.~(\ref{eq:L1}).
In order to ease the construction of the chiral Lagrangian in the following we keep all the terms in 
${\mcL}_1$ without using the freedom to eliminate them through a redefinition at O($a$) 
of the bare parameters.

At this point it is useful to anticipate the power counting scheme including the lattice spacing $a$.
This is 
\be
1 \gg m_{\rm R},\mu_{\rm R},p^2,a \gg 
m_{\rm R}^2,\mu_{\rm R}^2,p^4,a^2,m_{\rm R}\mu_{\rm R}, m_{\rm R}p^2, a m_{\rm R},
\mu_{\rm R}p^2,a \mu_{\rm R},a p^2 \gg \ldots
\label{eq:GSM}
\ee
The factors of $\Lambda$ (with $\Lambda$ a scale of order $\Lambda_{\rm QCD}$) 
necessary to make all these quantities dimensionless are implicit 
from now on unless specified.

We recall that this approach to the description of the lattice data does not
require a continuum extrapolation, hence the power counting scheme does not
imply that $\mu_{\rm R}$ goes to zero in the continuum limit but represents
only an order of magnitude equality. In other words the following formul\ae~
are useful in describing at fixed lattice spacing the quark mass dependence of
a given lattice correlator only in a region of quark masses appropriate given
the power counting scheme and the value of the lattice spacing.

The next step is to {\it match} the continuum effective Lagrangian~(\ref{eq:cel}) into a generalized
chiral Lagrangian. At LO the sole term that survives in ${\mcL}_1$ is the clover term 
(${\mcO}_1$ in eq.~\ref{eq:sym_op}), all the other terms being at least of NLO.
The key property that allows this matching was noticed by 
Sharpe and Singleton in ref.~\cite{Sharpe:1998xm} where they realized that the 
{\it clover term transforms under chiral symmetry exactly as the mass term}.
It is the possible to add to the chiral Lagrangian~(\ref{eq:Lchi2}) a simple source term in order to include
the leading discretization effects.
Namely the modified LO continuum chiral Lagrangian reads
\be
\mathcal{L}_{W\chi}^{(2)} = \frac{f^2}{4} \Big[
\langle \partial_\mu \Sigma(x)^\dagger \partial_\mu \Sigma(x) \rangle +
\langle \sigma(x)\Sigma(x)^\dagger + \sigma(x)^\dagger \Sigma(x) \rangle + 
\langle A(x)\Sigma(x)^\dagger + A(x)^\dagger \Sigma(x) \rangle\Big].
\label{eq:Wchi2}
\ee
At the end of the analysis the sources are set to their physical value
\be
\sigma(x) \rightarrow 2B_0(m_{\rm R} + i\tau^3 \mu_{\rm R}), \qquad A(x) \rightarrow 2 W_0 a
\ee
where $W_0$ is an unknown dimensionful constant which parametrizes the leading cutoff effects.
We remark that if the $c_{\small{\rm SW}}$ coefficient would be set to its ``correct'' 
non-perturbative value we would have $W_0 = 0$. This does not mean that all the O($a$) terms will disappear, 
because there will be O($a$) terms at NLO, that would have to be cancelled by other 
improvement coefficients.

Because of the chiral transformation properties of the clover term, the LO Lagrangian~(\ref{eq:Wchi2}) 
is unchanged from its continuum form if one shifts the external sources
\be
\sigma' \equiv \sigma + A.
\ee
This at the quark level corresponds to a redefinition of the untwisted quark mass
\be
m_{\rm R} \rightarrow m_{\rm R} + a W_0/B_0 \equiv m'.
\label{eq:mprime}
\ee
Recalling the definition of $m_{\rm R}$ in eq.~(\ref{eq:mr}), this
is equivalent to a shift in the critical mass.
This observation is rather important because it means that at LO all the
O($a$) effects in spectral quantities can be reabsorbed in the definition of the
quark mass through the offset, or shift, $a W_0/B_0$.
This shift is not measurable, however, since
$m_{\rm cr}$ is not known {\em a priori}. It must be determined
non-perturbatively from the simulation itself.
The traditional definition is that $m_{\rm cr}$ is the bare mass
at which $M_\pi^2\to 0$ on the Wilson axis, i.e. $\mu_{\rm q} = 0$.
Since at LO the chiral Lagrangian~(\ref{eq:Wchi2}) predicts $M_\pi^2\propto m'$,
we discover that this ``Wilson pion'' definition of $m_{\rm cr}$
{\em automatically} includes the shift in critical mass,
and chooses the untwisted quark mass to be $m'$.
This remark is important because firstly it tells us that with 
the standard numerical definition of $m_{\rm cr}$,
the pion and vacuum sectors are automatically
$O(a)$ improved at LO in W$\chi$PT {\it for any twist angle}, and secondly 
it indicates that from a theoretical point of view, at least at LO, there is 
no difference in the critical line computed in this way or by using other methods 
which set $m'=0$. We will come back later to this point when analysing numerical results.

Analogously to what we have done in the continuum, we define a polar mass and a twist angle
that includes this O($a$) shift
\be
M'{\rm e}^{i\omega'\tau^3} \equiv (m' + i \mu_{\rm R} \tau^3).
\label{eq:polar_LO}
\ee

In the continuum the PCAC quark mass~(\ref{eq:PCAC})
is the untwisted quark mass which appears in the Lagrangian.
If we consider now the lattice PCAC quark mass defined in (\ref{eq:PCAC_lat}),
evaluating the correlators at large distances in order to
let the single pion state dominate one finds
\be
m_{\rm PCAC} = m'.
\ee
This shows that this quantity automatically includes the O($a$) offset in the untwisted mass, exactly as 
the pion definition does.
We remark that at this order it does not matter if the critical line is computed on the Wilson axis
or with $\mu_{\rm q} \neq 0$.

At NLO we should add to the Symanzik expansion the O($a^2$) terms (see eq.~\ref{eq:GSM}). 
These can be collected in three categories.
First, those that are invariant under Euclidean and chiral symmetries, and they
simply modify, in the chiral Lagrangian, the leading order continuum results
by $a^2$, i.e. they lead to an $O(a^2)$ correction to the LEC $f$~\cite{Lee:1999zx}.
Second, there are
four-fermion operators which violate chiral symmetry.
In the chiral Lagrangian the corresponding operator is already present, having
been produced by two insertions of the clover term~\cite{Bar:2003mh}. 
This shows that what is relevant for matching are
the symmetries broken by the operators (here, chiral symmetry), 
and not their detailed form.
The four-fermion operators simply change the unknown low energy constant corresponding to a double insertion
of the clover term. This also means that 
using a non-perturbatively $O(a)$ improved quark action,
this low energy constant does not vanish.
Finally, there are the terms violating Euclidean symmetry.
These can be decomposed into Euclidean singlet and non-singlet
parts, and can be shown to be of higher order~\cite{Lee:1999zx}.

The detailed form of the Wilson chiral Lagrangian at NLO can be found in many 
papers and reviews (see for example~\cite{Sharpe:2006pu} and refs. therein).
Here we want to list the main 
results~\cite{Munster:2003ba,Scorzato:2004da,Aoki:2004ta,Sharpe:2004ny} 
and add some considerations.

The charged pion mass squared at NLO reads
\be
M_{\pi^{\pm}}^2 = 2B_0M'\left[1 + \frac{2B_0M'}{32 \pi^2 f^2}\log(2B_0M'/\Lambda_\pi^2)\right]
+ 2aB_0M' \cos\omega' { (2 \delta_W-\delta_{\widetilde W})}
+ 2 a^2 w' \cos^2\omega' .
\label{eq:mpisqGSM}
\ee
The first term in the r.h.s. is the continuum NLO result~\cite{Gasser:1983yg}
while remaining terms show the impact of a finite lattice spacing.
In particular $\delta_W$ and $\delta_{\widetilde W}$ parametrize NLO O($a$) 
effects of order $\Lambda_{\rm QCD}$. The $w'$ term parametrizes NLO O($a^2$) effects of order 
$\Lambda_{\rm QCD}^2$.
We have reabsorbed the scale dependence of the chiral logs in the LEC $\Lambda_\pi$,
and $2\delta_W-\delta_{\widetilde W}$ is scale invariant. 
Chiral logs do not contain discretization corrections at this order 
because the LO discretization errors can be absorbed into $\sigma'$.

The result (\ref{eq:mpisqGSM}) shows the different possibilities 
for removing $O(a)$ errors.
\begin{itemize}
\item
Non-perturbatively $O(a)$ improve the quark action,
in which case $2\delta_W-\delta_{\widetilde W}=0$ and the $O(a)$ term
vanishes.
\item
Use ``mass averaging''\cite{Frezzotti:2003ni}
in which one averages over $\omega'$ and $\omega'+\pi$
at fixed $M'$. This flips the sign of both
$m'$ and $\mu_{\rm R}$, and thus of $\cos\omega'$,
and cancels the $O(a)$ term. 
\item
Work at full twist, $\omega'=\pm\pi/2$.
This removes the $O(a)$ term
and, in this case though not in general, also the $O(a^2)$ term.
This can be traced back to the exact ``charged'' 
twisted chiral symmetry~(\ref{eq:uA12}) of the massless Wtm lattice action~(\ref{eq:WtmQCD}).
\end{itemize}

\begin{figure}[!ht]
\begin{center}
\includegraphics[angle=0,width=14cm]{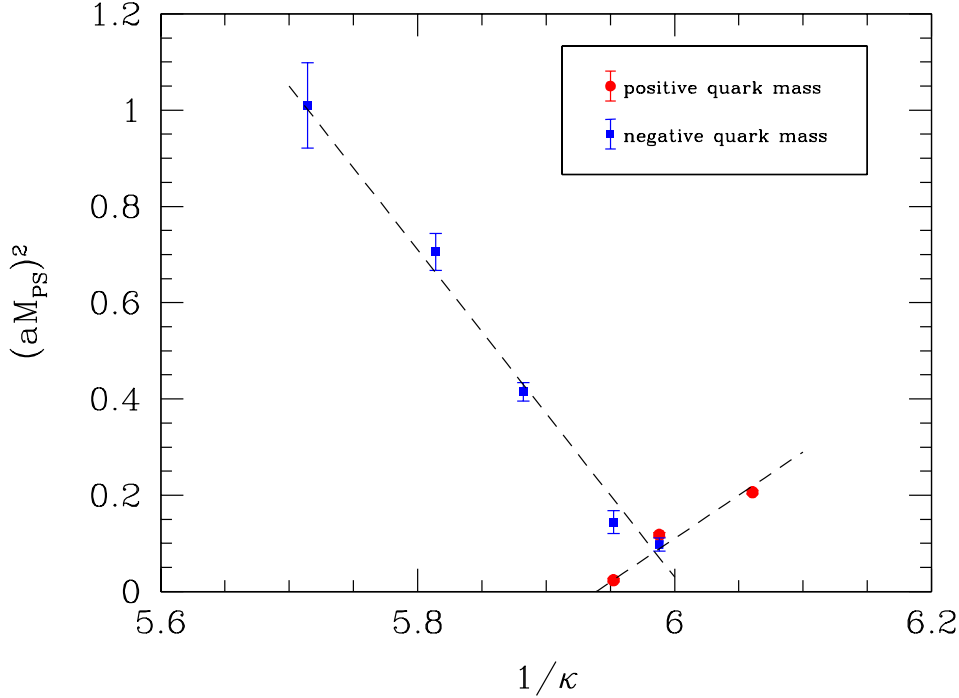}
\end{center}
\vspace{-3.0cm}
\caption{Unquenched results for $(aM_{\rm PS})^2$ as a 
function of $\kappa^{-1}= 2m_0+8$ 
for $\mu_{\rm q}=0$ and with $a \approx 0.2\;$fm\protect\cite{Farchioni:2004fs}.
Straight lines are to guide the eye.}
\label{fig:mpi_mu}
\end{figure}

There are two further features of the result (\ref{eq:mpisqGSM}). 
The $O(a)$ errors
are determined for every twist angle by the combination
$2\delta_W-\delta_{\widetilde W}$. In particular, on the Wilson axis, this term
predicts an asymmetry in the slopes on the two sides of $m_c$.
The asymmetry is defined by~\cite{Sharpe:2005rq} 
\be
\frac{M_\pi^2(m') - M_\pi^2(-m')}{M_\pi^2(m') + M_\pi^2(-m')}, \qquad {\rm with}~\mu_{\rm R}~{\rm fixed.}
\ee
Neglecting chiral logs and O($a^2$) terms and setting $\mu_{\rm R} = 0$ in
eq.~(\ref{eq:mpisqGSM}) 
the asymmetry is given by 
\be
a \frac{m'}{|m'|} \left( 2\delta_W-\delta_{\widetilde W} \right).
\ee
This asymmetry has been previously observed numerically~\cite{Farchioni:2004fs}, in the course 
of the initial studies of the properties of Wtm. I show an example of the
results in Fig.~\ref{fig:mpi_mu}.
The observed asymmetry in fig.~\ref{fig:mpi_mu} can be viewed by the
difference between the absolute values of the slopes of the two straight lines
that leads to a value for the asymmetry $\sim 0.3$~\cite{Sharpe:2005rq}.
This estimate is consistent with the expected size of
$(2\delta_W-\delta_{\widetilde W}) \sim \Lambda_{\rm QCD} \simeq 0.3$ GeV 
considering that $a^{-1}\approx 1$ GeV.

\subsubsection{Full twist}

The form of the $O(a)$ correction in the result for
$M_\pi^2$ in eq.~(\ref{eq:mpisqGSM}), is the generic structure of the O($a$)
terms in all the physical quantities, i.e. all the O($a$) cutoff effects
are proportional to $a \cos\omega'$.
This is the way to observe automatic O($a$) improvement in the framework of
W$\chi$PT.
To obtain automatic $O(a)$ improvement one 
needs $\cos\omega'=O(a)$ and thus $\omega'=\pi/2 + O(a)$.
In other words, full twist means at most ``up to
$O(a)$''. 
I want to briefly explain here in which sense the analysis carried out with
the Symanzik expansion and W$\chi$PT gives a consistent picture for the determination 
of the full twist setup.

In sect.~\ref{ssec:auto} and~\ref{ssec:crit_mass} we have already discussed in the framework of
the Symanzik expansion that to achieve automatic O($a$) improvement down to
masses obeying the constraint~(\ref{eq:constr2}) the untwisted quark mass has
to be tuned to be zero up to errors at most of O($a\mu_{\rm q}$).
In the power counting scheme~(\ref{eq:GSM}) we assume for the W$\chi$PT analysis, with $\mu_{\rm R} \sim a $,
an $O(a)$ accuracy in $\omega'$ requires $m'= O(a^2)$.
This is easily obtained considering that 
\be
\cos \omega' = \frac{m'}{M'} .
\ee
If we again recall our working power counting it is easy to see that $m'=
O(a^2)$  corresponds exactly to the requirement of having a vanishing PCAC quark mass up to O($a\mu_{\rm q}$) or
O($a^2$).
So the two analyses are consistent and the result can be summarized as
follows:
to achieve automatic O($a$) improvement down to a twisted mass which obeys the
constraint~(\ref{eq:constr2}) we have to tune the theory to full
twist, meaning that the critical mass $m_{\rm cr}$ has to be
tuned such that the PCAC quark mass vanishes up to O($a\mu_{\rm q}$) or O($a^2$).

The traditional definition of $m_{\rm cr}$
used with Wilson fermions is to extrapolate $m_0$ to the point
where $M_\pi^2=0$. We now consider the NLO expression for the squared charged
pion mass~(\ref{eq:mpisqGSM}) and set $\mu_{\rm q}=0$.
Eq.~(\ref{eq:mpisqGSM}) tells us that if one is able to perform a perfect chiral
extrapolation including the chiral log,
this method is in principle adequate to tune $m_{\rm cr}$ such that $m'= $
O($a^2$) in the regime $\mu_{\rm R} \sim a$.

Alternatively to obtain $m_{\rm R} = 0$ in the continuum limit,
it is enough to understand which are the symmetries that are recovered in the
continuum if $m_{\rm R} = 0$ and to impose suitable identities on the lattice.
For $m_{\rm R} \neq 0$ twisted parity~(\ref{eq:tmparity}) and twisted flavour~(\ref{eq:R512})
are broken symmetries.
Enforcing this restoration in particular correlators
for $a\ne 0$ gives a non-perturbative determination of the twist angle 
(or equivalently of $m_{\rm cr}$) to full twist.

A way to understand how to impose the restoration of twisted parity symmetry
in the continuum limit is to write a correlator which breaks parity in the physical basis. 
We have seen in sec.~\ref{sec:basic} which is
the relation between fields in the twisted and physical basis.
The idea is now to take either (\ref{eq:axial_current_rot},\ref{eq:vector_current_rot})
or (\ref{eq:axial_density_rot},\ref{eq:scalar_density_rot}) 
as a {\em definition} of $\omega$,
and enforce parity restoration in a particular correlator.
Two examples are the $\omega_A$ method~\cite{Farchioni:2004fs}
\be
\langle{\mcV}^2_\mu(x) {\mcP}^1(y) \rangle
\propto \langle 0| {\mcV}^2_\mu | \pi^1\rangle = 0,
\ee
and the $\omega_P$ method~\cite{Sharpe:2004ny}
\be
\langle {\mcS}^0(x) {\mcA}^3_\mu(y) \rangle
\propto \langle 0| {\mcS}^0 | \pi^3\rangle = 0 .
\ee
The correlators are to be evaluated for $x\ne y$ and large euclidean times
in order to isolate the single pion contribution.
Using (\ref{eq:axial_current_rot}-\ref{eq:scalar_density_rot}) one can
manipulate these criteria into results for the twist angle
in terms of correlators in the twisted basis:
\be
\tan\omega_A \equiv \frac{\langle V_\mu^2(x) P^1(y) \rangle}
                       {\langle A_\mu^1(x) P^1(y) \rangle} 
\,,\quad
\tan\omega_P \equiv \frac{i \langle S^0(x) A_\mu^3(y)\rangle}
                         {2 \langle P^3(x) A_\mu^3(y)\rangle}\,.
\label{eq:omegaAP}
\ee
Full twist occurs when the denominators vanish, i.e
\be
\omega_A=\pi/2 \Rightarrow \langle A_\mu^1(x) P^1(y) \rangle = 0\,,
\quad
\omega_P=\pi/2 \Rightarrow \langle P^3(x) A_\mu^3(y) \rangle = 0\,.
\label{eq:maxtwistAP}
\ee
The correlator in the ``$\omega_P$ method''
includes quark-disconnected contractions and is much more difficult to
calculate in practice. 
The ``$\omega_A$ method'' is used in practice.
One fixes $\mu_{\rm q}$
and varies $m_0$ until $\omega_A=\pi/2$.
The resulting $m_{\rm cr}(\mu_{\rm q})$ depends on the choice of
discretization of the axial current (e.g. $O(a)$ improved or not),
and, in general, upon the separation $x-y$.
In fig.~\ref{fig:kappa_c} a typical $\mu_{\rm q}$ dependence of the critical line
is shown.
At large distances, which are used in practice,
the pion contribution dominates and the resulting $m_{\rm cr}$ becomes
independent of separation.
At such distances the ``$\omega_A$ method'' is
equivalent to the vanishing of the PCAC mass~(\ref{eq:PCAC_lat}).
We remind that the determination of $\omega$ at full twist
does not require any computation of $Z$-factors~\cite{Frezzotti:2000nk}.

Both methods can be studied
in W$\chi$PT and at full twist we have~\cite{Sharpe:2004ny}:
\be
 \omega_A=\pi/2 \  \Rightarrow\ 
\omega' = \pi/2 + a{\delta_W}\,,
\quad
\omega_P=\pi/2 \ \Rightarrow\ 
\omega' =  \pi/2\,,
\ee
To summarize: all the known methods to compute the critical line
are theoretically equivalent in order to achieve automatic O($a$) improvement
provided $\mu_{\rm R} > a^2 \Lambda_{\rm QCD}^3$.
Practically the situation can be very different 
as we are going to discuss in the next section.

\subsubsection{The bending phenomenon: a closed chapter}

The discussion up to this point makes clear the importance of accurate
tuning to full twist. Initial studies of Wtm in the quenched model 
observed a phenomenon called ``bending''.
Although this is largely a closed chapter in the history of
Wtm, it is worth learning the appropriate lessons.
Here we assume that we work in a region of quark masses such that $\mu_{\rm R} \gg a^2$.
We will enter into the region where $\mu_{\rm R} \sim a^2$ in sect.~\ref{sec:asq}.
\begin{figure}
\begin{center}
\includegraphics[width=5cm]{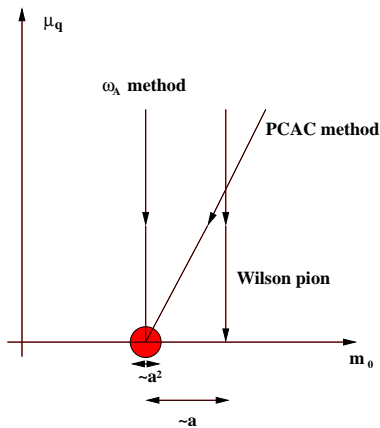}
\caption{Sketch of the different methods for working at full twist.
The arrows represent the direction one moves
to approach the chiral limit (which occurs at $\mu_{\rm q}=0$). The plot 
represents the regime in which $M_{\rm R}\sim a$. The regime
in which $M_{\rm R}\sim a^2$ is represented by the shaded region.
It is discussed in sec.~\protect\ref{sec:asq}. 
For the ``Wilson pion'' method the supposed behaviour given the
quenched numerical data is plotted. See text for discussion.}
\label{fig:crit_all}
\end{center}
\end{figure}

\begin{figure}[htb]
\begin{center}
\includegraphics[angle=0,width=14cm]{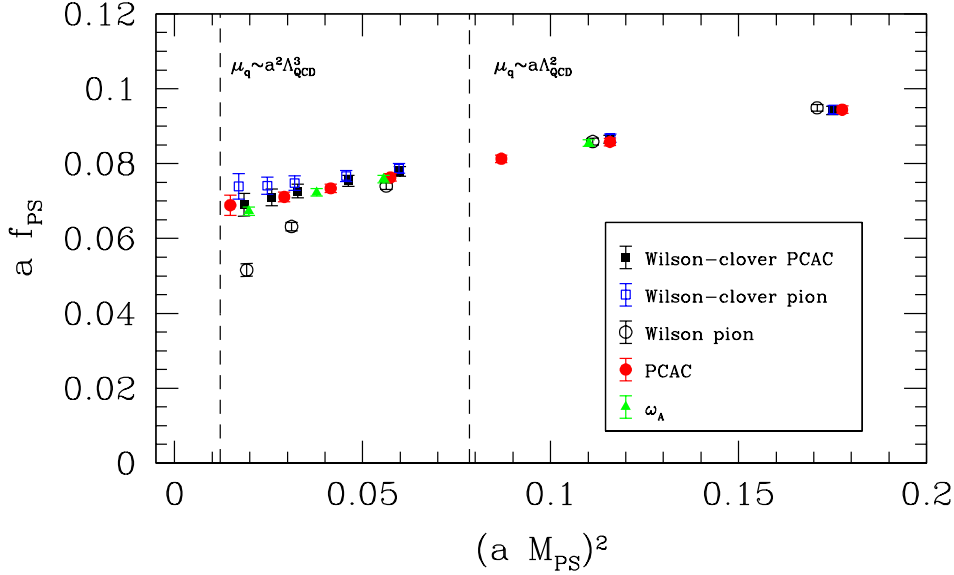}
\end{center}
\vspace{-5.0cm}
\caption{Chiral behaviour of the pseudoscalar decay constant at fixed lattice spacing $a = 0.093$ fm for several 
determinations of the critical mass $m_{\rm cr}$: ``Wilson-clover PCAC'' ({\tiny{$\blacksquare$}}) 
 and ``Wilson-clover pion'' ({\color{blue}{\tiny{$\square$}}}) \protect\cite{Becirevic:2006ii}, 
``Wilson pion'' ({\large{$\circ$}}) and PCAC ({\color{red}{\large{$\bullet$}}}) 
\protect\cite{Jansen:2005gf,Jansen:2005kk}, $\omega_A$ ({\color{green}{\small{$\blacktriangle$}}}) 
\protect\cite{Abdel-Rehim:2005gz}.}
\label{fig:bend}
\end{figure}

I show in fig.~\ref{fig:bend}, the numerical results for $af_{\rm PS}$ at a fixed value
of the lattice spacing ($a=0.093$ GeV) using several definitions 
for the critical mass (see the caption for more details).
A clear ``bending'' is observed at smaller quark masses with the ``Wilson pion'' definition.
This bending can be explained taking the LO expression in W$\chi$PT for the pseudoscalar 
decay constant
\be
f_{\rm PS} = f ~ \frac{\mu_{\rm R}}{\sqrt{\mu_{\rm R}^2 + m_{\rm PCAC}^2}} \simeq f ~ \left[ 1 -
\frac{1}{2} \left(\frac{m_{\rm PCAC}}{\mu_{\rm R}}\right)^2 + ~ ... \right]~,
\label{eq:fpi_exp}
\ee
where $f$ is the pion decay constant in the chiral limit and $\mu_{\rm R}$ and
$m_{\rm PCAC}$ represent the renormalized twisted and PCAC quark masses
respectively. 
We have already seen that from a theoretical point of view, independently
on the definition used for tuning $m_{\rm cr}$, $m_{\rm PCAC}$ vanishes or is
a quantity of O($a\mu_{\rm q}$) or O($a^2$).
The term $\left(\frac{m_{\rm PCAC}}{\mu_{\rm R}}\right)^2$ causes a deviation
of the expected chiral behaviour as soon as $\mu_{\rm R} \simeq m_{\rm PCAC}$.
The fact that for the ``Wilson pion'' definition a ``bending'' appears
for $\mu_{\rm q} \simeq a \Lambda_{\rm QCD}^2$, gives an indirect evidence that
$m_{\rm PCAC} = $ O($a$). 
In practice in a quenched simulation, a long extrapolation along the Wilson axis is 
needed  because of the occurrence of exceptional configurations(see sec.~\ref{ssec:except}).
It is conceivable that performing this long extrapolation without including the
chiral logs and in a region of quark masses where the
applicability of NLO W$\chi$PT is debatable, the critical mass could be 
determined in such a way that $m_{\rm PCAC}$ is of order $a$. 
If this happens, this could generate, as we have seen, large O($a^2$) corrections.
A possible way to resolve the issue would be to perform chiral fits using NLO W$\chi$PT,
including $m_{\rm PCAC}$ as a fit parameter for several lattice spacings and
then studying the lattice spacing dependence of $m_{\rm PCAC}$.

An alternative approach is to use the clover improved Wtm
action~(\ref{eq:clovertm}) which removes from the theory all the cutoff
effects of the kind $a\Lambda_{\rm QCD}^2$~\cite{Frezzotti:2005gi}, and to use
the equivalent pion and PCAC determination of $m_{\rm cr}$. This is
nicely shown in fig.~\ref{fig:bend} where both the ``Wilson-clover pion''
definition and the ``Wilson-clover PCAC'' definition give no evidence of
bending down to masses $\mu_{\rm q} \simeq a^2\Lambda_{\rm QCD}^3$~\cite{Becirevic:2006ii}.
 
The final answer is given by a study of the lattice spacing dependence of
several physical quantities, with several definitions of $m_{\rm cr}$.

The ``Wilson pion'' method was used 
in the first scaling study of Wtm 
in the quenched model~\cite{Jansen:2003ir}. This study was performed in a region
of quark masses where most probably $\mu_{\rm R} \gg a$. Therefore the ``Wilson pion'' 
definition was good enough to obtain a clear evidence of automatic O($a$)
improvement, and no enhanced O($a^2$) effects.

\begin{figure}[htb]
\vspace{-0.0cm}
\begin{center}
\epsfig{file=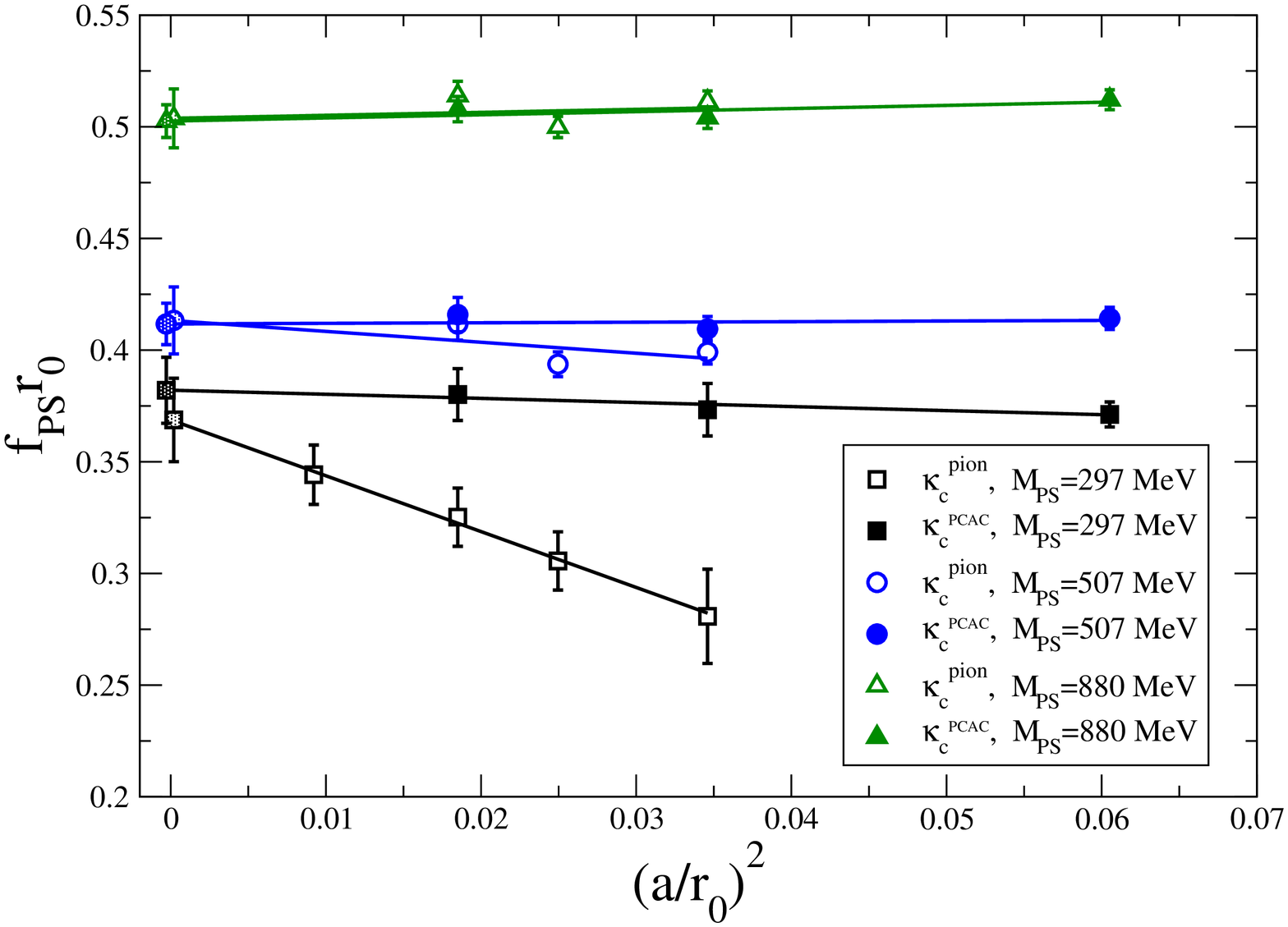,angle=270,width=0.8\linewidth}
\epsfig{file=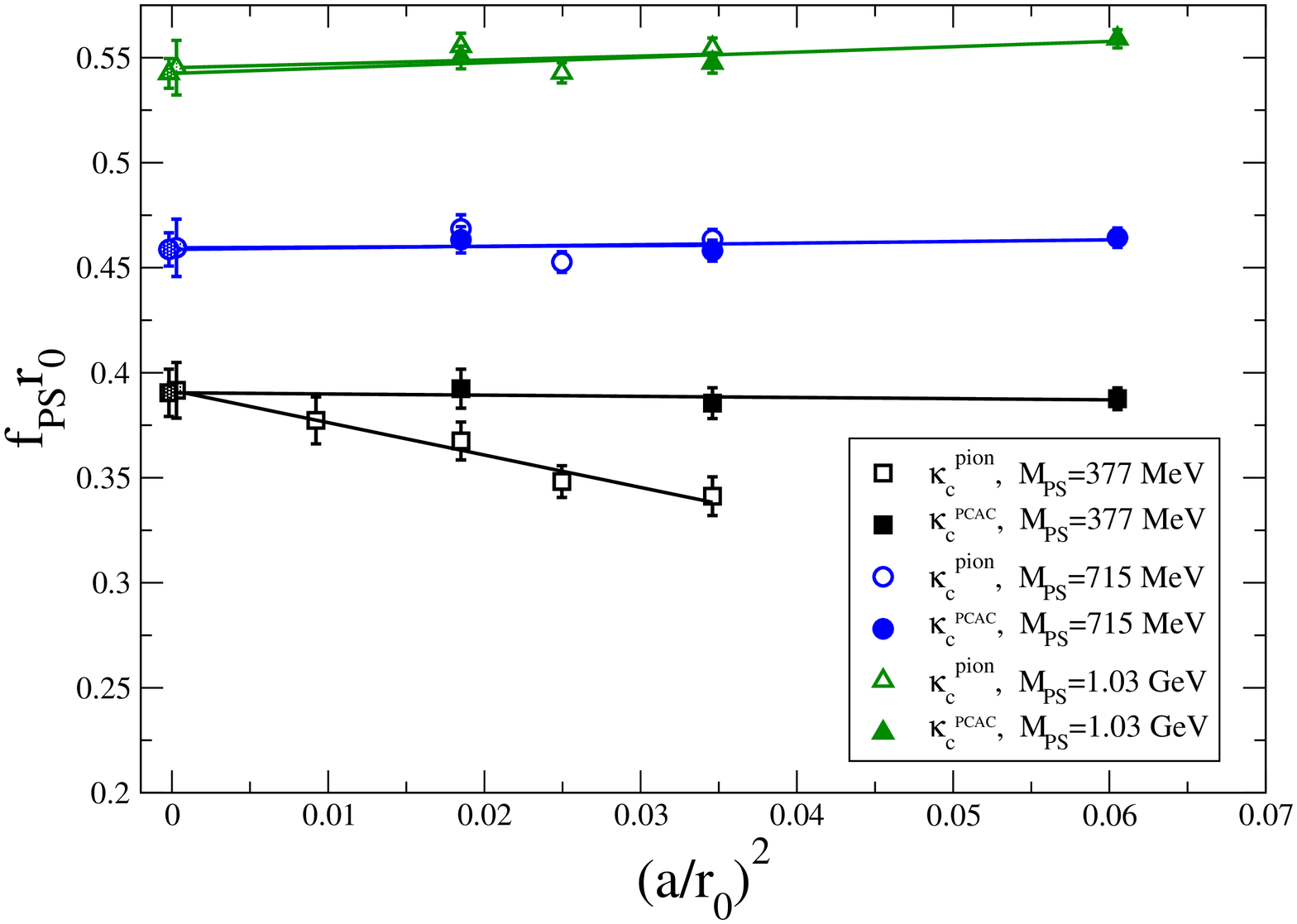,angle=270,width=0.8\linewidth}
\end{center}
\vspace{-0.0cm}
\caption{$r_0f_{\rm PS}$ as a function of $(a/r_0)^2$ using the 
``Wilson pion'' definition (open symbols) and the PCAC definition (filled symbols)
of the critical mass; fits are performed with a linear function in
$(a/r_0)^2$ separately for each set.
\label{fig:fpi}}
\end{figure}

In a further publication~\cite{Jansen:2005kk} the PCAC method and the ``Wilson pion'' method
have been compared, obtaining consistent results in the continuum limit but rather different O($a^2$) effects.
This is clearly seen in the fig.~\ref{fig:fpi}  
where $f_{\rm PS} r_0$ is plotted as a function of $(a/r_0)^2$ for pseudoscalar masses
in the range 297-1032 MeV, obtained with both definitions of $m_{\rm_c}$.
For small enough values of the lattice spacing, the values of $f_{\rm PS} r_0$ show, with both 
definitions of $m_{\rm cr}$, a linear behaviour in $(a/r_0)^2$. This nicely
demonstrates the $O(a)$ improvement for both definitions of the critical mass. 
However, for the pion definition we notice that the effects of
$O(a^2)$ are rather large, in particular at small pseudoscalar meson masses of 
$297$ MeV and $377$ MeV.
In contrast, the PCAC definition reveals an almost flat behaviour as a function
of $(a/r_0)^2$  even at these small pseudoscalar meson  masses.  
In order to take the continuum limit, one should identify the scaling
region where the data are well described by corrections linear in
$(a/r_0)^2$.
This turns out to start at $a=0.093$ fm with the pion definition of $m_{\rm cr}$ 
and at $a=0.123$ fm with the PCAC definition. The values in the
continuum limit are obtained separately by performing linear fits to the data
in these two regions. The results of these fits are shown in fig.~\ref{fig:fpi}
together with the simulation data. It is very reassuring that these independent linear
fits lead to completely consistent continuum values.

\subsection{The critical mass: summary}
\label{ssec:mcsum}

All this discussion about the critical mass deserves a summary.
In fig.~\ref{fig:crit_all} is sketched the approach to the chiral limit
according to the different definitions used for the critical mass $m_{\rm cr}$.
For the ``Wilson pion'' definition the supposed behaviour is shown given by the
quenched numerical data.
While they are all nominally at full twist: they could show different O($a^2$) scaling
behaviour depending on the value of the quark mass $\mu_{\rm R}$.
In table~(\ref{tab:mc_summary}) I collect all the definitions discussed 
\begin{table}[!th]
\begin{center}
\begin{tabular}{cc}
Method & Definition \\
\hline
Wilson pion & 
$\lim_{m_0 \rightarrow m_{\rm cr}} M_{\pi}^2 = 0$ \\
Wilson-clover pion & $\lim_{m_0 \rightarrow m_{\rm cr}} M_{\pi}^2 = 0$ \\
Wilson-clover PCAC & $\lim_{m_0 \rightarrow m_{\rm cr}} m_{\rm PCAC} = 0$ \\
$\omega_A$ & $\langle A_\mu^a(x) P^a(y) \rangle = 0 \quad a=1,2$ \\
$\omega_P$ & $\langle A_\mu^3(x) P^3(y) \rangle = 0 $ \\
PCAC & $m_{\rm cr} = \lim_{\mu_{\rm q} \rightarrow 0} m_{\rm cr}(\mu_{\rm q})$ \\
\hline
\end{tabular}
\label{tab:mc_summary}
\end{center}
\end{table}
and here are some considerations:
\begin{itemize}
\item ``Wilson pion'' : this is a valid definition to guarantee automatic O($a$) 
improvement down to $\mu_{\rm R} \sim a$. If this method is used in the quenched case
a long chiral extrapolation is needed because of the occurrence of exceptional
configurations.
The correct extrapolation to the chiral limit, including eventual
chiral logs, is crucial in order to have a definition of the critical mass that absorbes 
the O($a$) offset. If the extrapolation is not done correctly this definition can miss the critical line
in such a way that $m' = {\rm O(}a$). This is most probably the reason of the ``bending phenomenon''.
\item ``Wilson-clover pion'' : this definition has the advantage with 
respect to the previous one that, if one removes non-perturbatively the O($a$) effects in the 
lattice action, even if the extrapolation is needed it can miss the critical line only by O($a^2$).
The nice results of ref.~\cite{Becirevic:2006ii} indeed seem to indicate that this definition is good
enough to avoid the ``bending phenomenon''. This somehow confirms that the origin of the 
``bending phenomenon'' is the O($a$) effects present in the untwisted quark mass.
\item ``Wilson-clover PCAC'' : to this definition the same considerations 
apply as for the ``Wilson-clover pion'' method. 
\item $\omega_A$ : this definition (and the equivalent using the PCAC mass),
allows a clean determination of the critical line 
$m_{\rm cr}(\mu_{\rm q})$ for each value of $\mu_{\rm q}$. It can be safely used 
in the quenched case, because the twisted mass gives a sharp infrared cutoff 
to the lattice theory. It has the drawback that the unquenched case requires a tuning
for each value of $\mu_{\rm q}$ and this could be rather expensive from the computational side.
\item $\omega_P$ : this definition, theoretically very attractive, has the
drawback that it requires the computation of quark disconnected diagrams. 
These diagrams are usually more difficult to compute
so it has mainly an academic interest.
\item PCAC : this definition takes the one obtained with the $\omega_A$ method 
(or the equivalent with the PCAC mass) and then extrapolates $m_{\rm cr}(\mu_{\rm q})$ 
to $\mu_{\rm q} = 0$. The advantage of this definition is that one could perform the extrapolation
using a limited range of values of $\mu_{\rm q}$, and then the critical mass mass could be used for all
the values of $\mu_{\rm q}$ one is interested in. A possible cheaper, from a computational point of view,
alternative has been recently proposed
~\cite{Boucaud:2007uk} that uses the critical mass determined at the lowest value of $\mu_{\rm q}$ used in the
simulations. 
\end{itemize}

We can conclude this lengthy and technical discussion on the critical mass with the following statement:
provided we are away from the region $\mu_{\rm R} \sim a^2\Lambda_{\rm QCD}^3$ all the definitions given above 
are theoretically equivalent in order to achieve automatic O($a$) improvement. 
The ``Wilson pion'' definition has to be avoided in the quenched case because 
the long extrapolation to the chiral limit
can induce effective O($a$) cutoff effects in the untwisted quark mass.

\subsection{Numerical results}
\label{ssec:num_res}

The first numerical evidence of automatic O($a$) improvement was given in ref.~\cite{Jansen:2003ir},
where a first scaling test was performed at a fixed value of the pseudoscalar meson mass around 
$M_{\rm PS} \simeq 700$ MeV in the quenched model. 
In this paper, scaling violations of the pseudoscalar decay
constant and the vector meson mass have been studied, 
confirming that, without any improvement coefficient, Wtm at full twist
is consistent with automatic O($a$) improvement, with first indications that the remaining 
O($a^2$) effects should be small.
This very interesting result has triggered a set of quenched studies
\cite{Bietenholz:2004wv,Abdel-Rehim:2004gx,Abdel-Rehim:2005gz,Jansen:2005gf,Jansen:2005kk,Abdel-Rehim:2005qv} 
to further check the property of automatic O($a$) improvement~\cite{Frezzotti:2003ni},
and to gain experience with this formulation of lattice QCD.
\begin{figure}[htb]
\epsfig{file=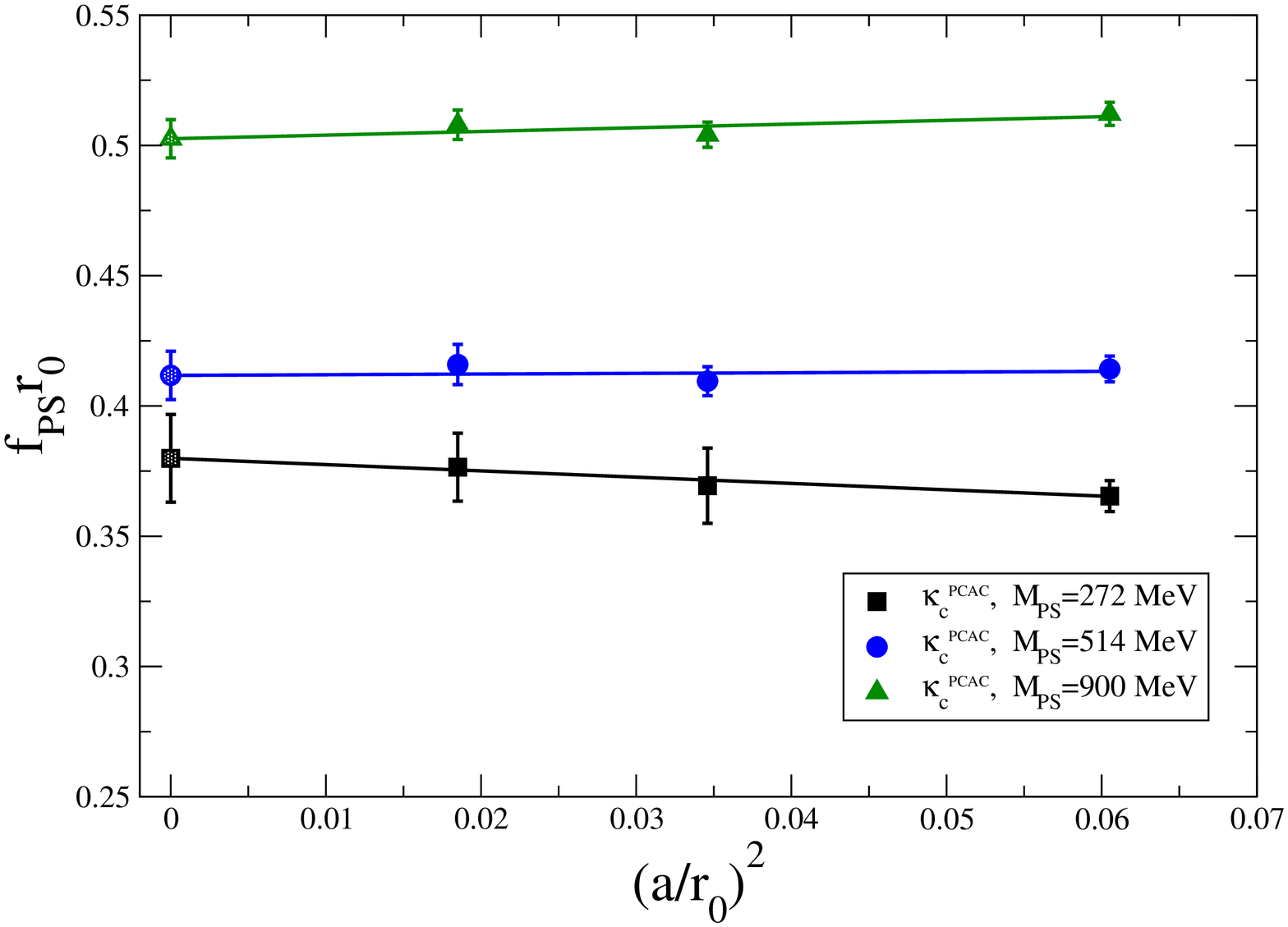,angle=270,width=0.5\linewidth}
\epsfig{file=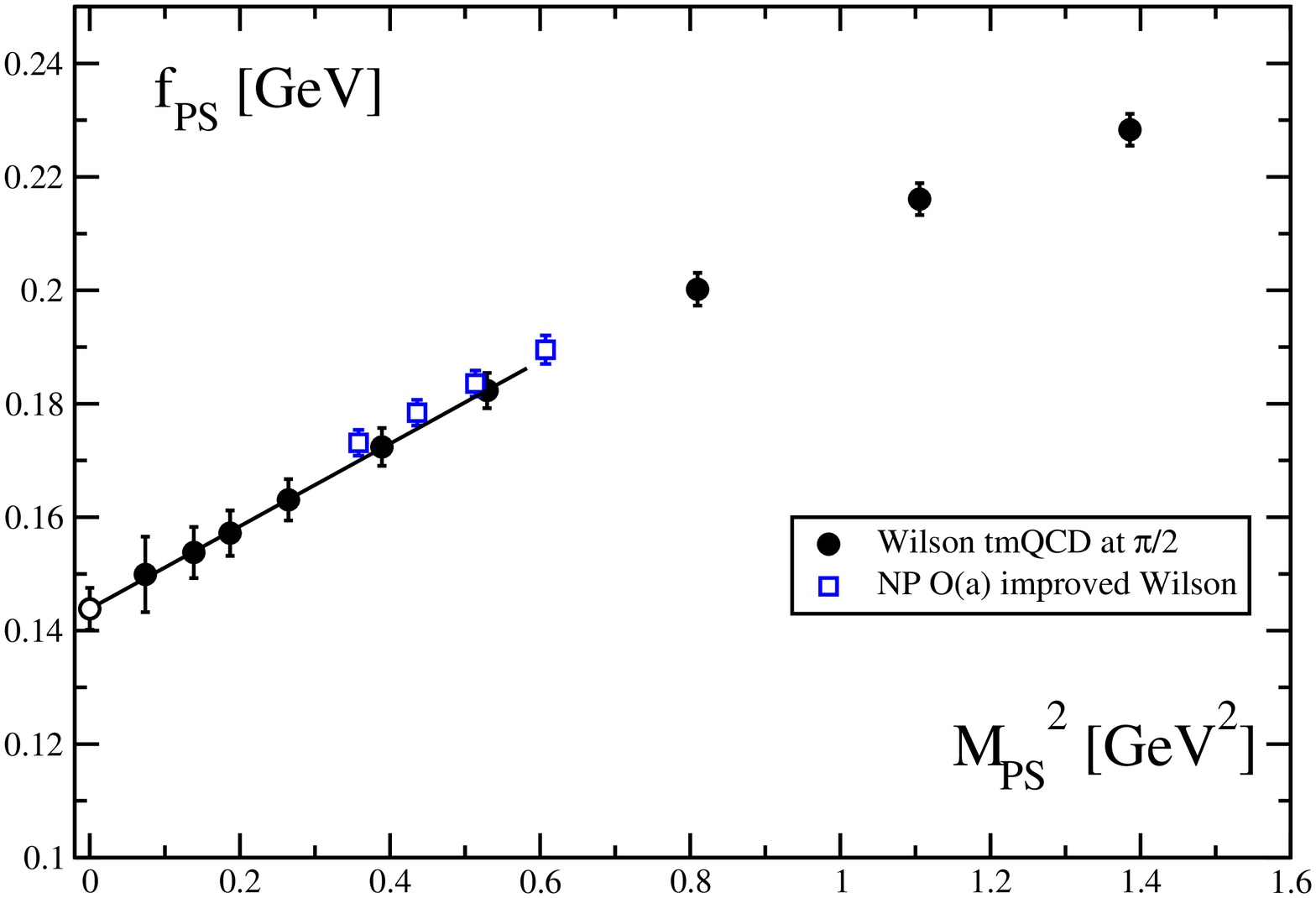,angle=270,width=0.5\linewidth}
\caption{Left plot: continuum extrapolation of $f_{\rm PS}$ as a function of $a^2$.
Right plot: continuum limit values for $f_{\rm PS}$
  as a function of $M_{\rm PS}^2$ in physical units. 
The empty squares are taken from \cite{Garden:1999fg}.}
\label{fig:fps}
\end{figure}
An interesting quantity to compute with Wtm is the pseudoscalar decay
constant $f_{\rm PS}$. As it was noted in
\cite{DellaMorte:2001tu,Frezzotti:2001du,Jansen:2003ir}, 
the computation of $f_{\rm PS}$ does not require any renormalization constant,
in contrast of ordinary Wilson fermions, and moreover, given automatic O($a$)
improvement, does not need the computation of any improvement coefficient. 
Thus the situation for this quantity is like with Ginsparg-Wilson fermions. 
The reason for this nice property is that the continuum Ward identity~(\ref{eq:PCVC}),
remains exact on the lattice 
\be
\langle\partial^*_\mu \widetilde{V}^a_\mu(x) O(0)\rangle= -2 \mu_{\rm q} \epsilon^{3ab}
\langle P^b(x) O(0)\rangle \quad a=1,2, 
\label{eq:PCVC_lat}
\ee
(where $\partial^*_\mu$ is the lattice backward derivative,
and $O$ is a local lattice operator)
if one uses a slightly modified point-split vector current 
\begin{eqnarray}
   \widetilde{V}^a_\mu(x) &=&  
 \frac12 \Bigl\{\chibar(x)(\gamma_\mu-1)\frac{\tau^a}{2}
 U(x,\mu)\chi(x+a\hat\mu) \nonumber\\
 && \hphantom{\frac12 } +\chibar(x+a\hat\mu)(\gamma_\mu+1)\frac{\tau^a}{2} 
 U(x,\mu)^{-1}\chi(x)\Bigr\}.
\end{eqnarray}
Given the fact that at full twist we have 
\be
\langle (\mathcal{A}_{\rm R})_0^1(x) P^1(0) \rangle_{(M_{\rm R},0)} = 
\langle (V_{\rm R})_0^2(x) P^1(0) \rangle_{(m_{\rm R},\mu_{\rm R})} ,
\label{eq:AP_VP}
\ee
inserting a complete set of states in a standard fashion on the r.h.s. of 
eq.~(\ref{eq:AP_VP}) and using the PCVC relation~(\ref{eq:PCVC_lat}) 
we obtain
\be
f_{\rm PS}=\frac{2\mu_{\rm q}}{M_{\rm PS}^2} | \langle
0 | \hat{P}^a | {PS}\rangle | \qquad a=1,2 
\label{indirect}
\ee
where $M_{\rm PS}$ is the charged pseudoscalar mass and $| {PS}\rangle$ denotes
the corresponding pseudoscalar state. 

In fig. \ref{fig:fps} (left plot) the
continuum limit of $r_0 f_{\rm PS}$, the critical mass being computed with the PCAC method, 
is shown, from ref.~\cite{Jansen:2005kk} as a function of $(a/r_0)^2$. The scaling is consistent with being of
O($a^2$), and moreover it is reassuring that the O($a^2$) effects are rather small for all the pseudoscalar
masses investigated down to $M_{\rm PS} = 272$ MeV. The right panel of
fig. \ref{fig:fps} shows the chiral behaviour of the continuum pseudoscalar
decay constant~\cite{Jansen:2005kk}, compared with the non-perturbatively O($a$) improved
data of~\cite{Garden:1999fg}. We remark that this comparison is purely
illustrative since it is in the quenched model, and the simulations
with clover fermions had to stop around $M_{\rm PS} \simeq 500$ MeV due to the
appearance of exceptional configuration.
Using a linear extrapolation of these data to the chiral limit 
(performed on the six
smallest masses) gives the values of the pion and
kaon decay constant $f_\pi$ and $f_K$ (the latter in the $SU(3)$
symmetric limit). The ratio of the two gives $f_K/f_\pi=1.11(4)$, which
is $10\%$ smaller than that obtained experimentally. This is however
consistent with what was observed in previous quenched 
calculations~\cite{Heitger:2000ay}.    

Another interesting quantity that has been computed is the vector meson mass.
Applying the standard axial rotation~(\ref{eq:axial}) we obtain for $\omega = \frac{\pi}{2}$
\be
\langle (\mathcal{V}_{\rm R})^{1,2}_\mu(x) (\mathcal{V}_{\rm R})^{1,2}_\mu(0) \rangle_{(M_{\rm R},0)} = 
\langle (A_{\rm R})^{2,1}_\mu(x) (A_{\rm R})^{2,1}_\mu(0) \rangle_{(m_{\rm R},\mu_{\rm R})} .
\ee
Another possible interpolating field for the vector channel is given by the following
component of the tensor current
\be
\mathcal{T}_k^a=\bar{\psi}\sigma_{0 k}\frac{\tau^a}{2}\psi .
\ee
This current is invariant under the rotation~(\ref{eq:axial}), so 
the vector meson mass can be extracted from the correlators
\be
C_A^a(x_0) = \frac{a^3}{3}\sum_{k=1}^3\sum_{\mathbf x} \langle A_k^a(x)A_k^a(0)\rangle 
\quad a=1,2 
\ee
\be
C_T^a(x_0) = \frac{a^3}{3}\sum_{k=1}^3\sum_{\mathbf x} \langle T_k^a(x)T_k^a(0)\rangle
\quad a=1,2 
\ee
One observation~\cite{Jansen:2005kk} is that the tensor correlator
systematically shows smaller statistical fluctuations.
 
\begin{figure}[h]
\epsfig{file=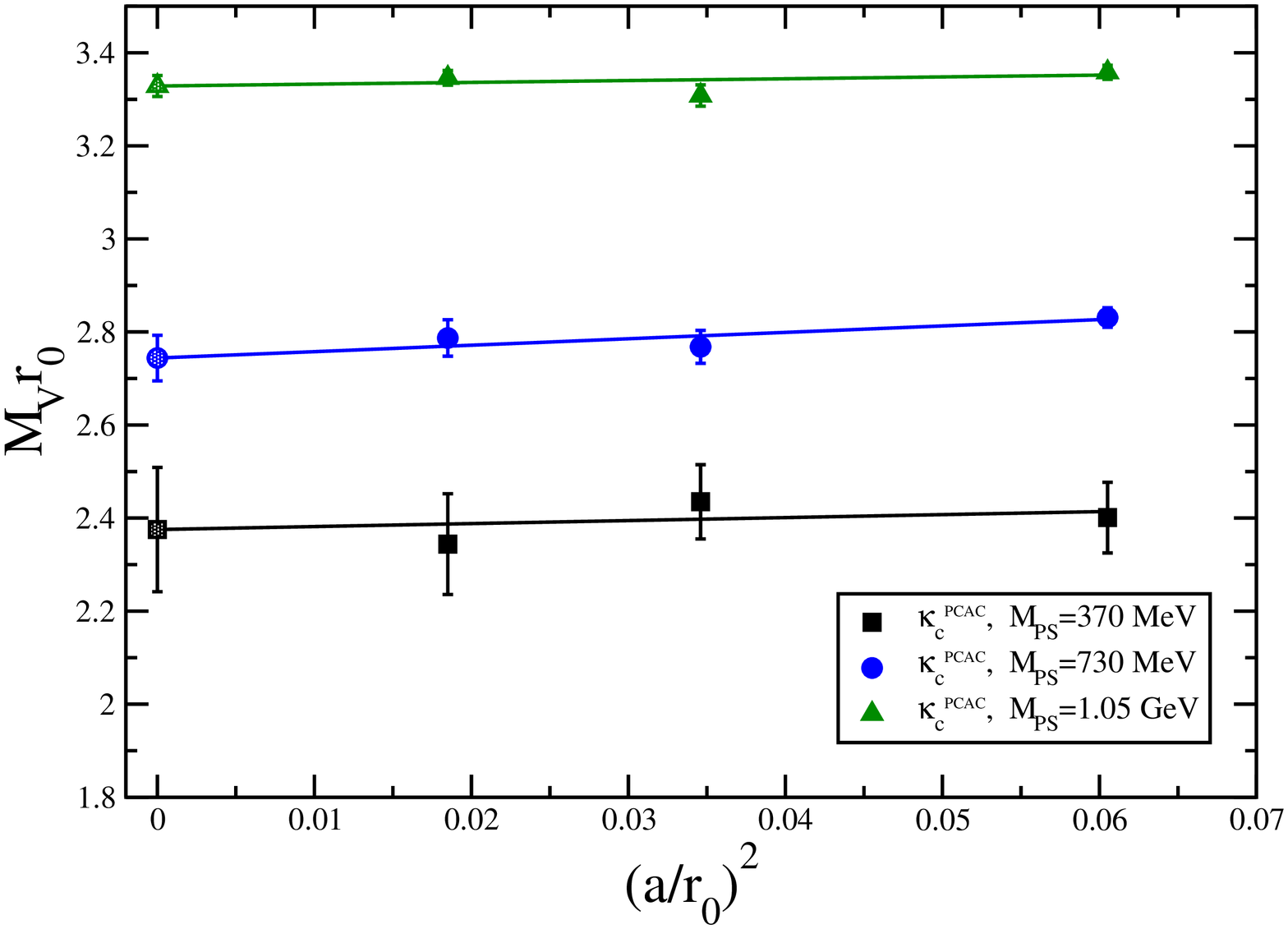,angle=270,width=0.5\linewidth}
\epsfig{file=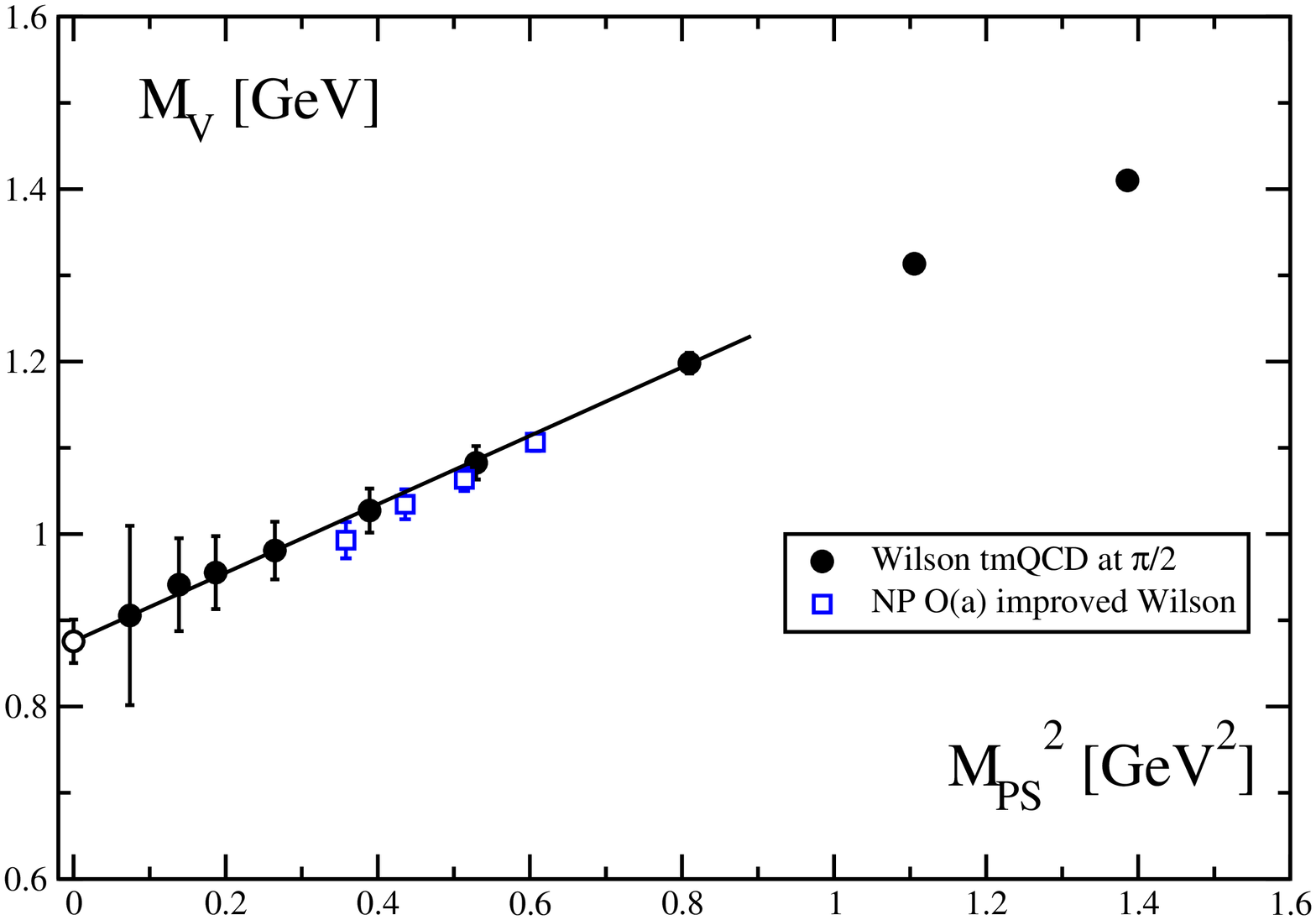,angle=270,width=0.5\linewidth}
\caption{Left plot: continuum extrapolation of $M_{\rm V} r_0$ as a function of
$(a/r_0)^2$. Right plot: continuum limit values of the vector meson mass (filled
  circles) as a function of the pseudoscalar meson mass squared. 
  The continuum limit vector masses obtained with non-perturbatively improved
  Wilson fermions (open squares)~\cite{Garden:1999fg} are also plotted.  
\label{fig:mvcont}}
\end{figure}

In the left plot of fig.~\ref{fig:mvcont} we show the results for the vector meson mass as
a function of $(a/r_0)^2$.
Again, even for small pseudoscalar meson masses, the behaviour of the 
vector meson mass is almost flat in $(a/r_0)^2$, indicating that $O(a^2)$
lattice artefacts  
are also small for this quantity. The lines in 
fig.~\ref{fig:mvcont} (left panel) represent linear fits of these data 
as a function of $(a/r_0)^2$. The continuum extrapolated values for the vector meson
mass are presented in the right plot of fig.~\ref{fig:mvcont}. As a function of the 
pseudoscalar meson mass squared, they show a linear behaviour.
In fig.~\ref{fig:mvcont} we also plot the
continuum values obtained with non-perturbatively improved Wilson 
fermions~\cite{Garden:1999fg}. 
The results of a linear extrapolation (performed with
the seven smallest masses) are used to compute the values of $M_\rho$ and of $M_{K^*}$
(the latter in the $SU(3)$ symmetric limit). As already observed in quenched
calculations where the scale is determined through $r_0$, these values turn out
to be $10-15\%$ larger than the experimental values.

\subsection{Conclusions}
\label{ssec:conclusions}

Despite the fact that Wtm is a relatively new lattice action, extended
scaling tests with mesonic quantities in large volumes
have shown beyond any reasonable doubt that automatic O($a$) improvement
is at work, i.e. at the price of tuning one parameter, namely the critical mass $m_{\rm cr}$,
the physical correlation functions exhibits only O($a^2$) scaling violations.
The issue of the choice of the critical mass can be summarized as follows: provided
the definition of the critical mass is theoretically and practically valid each definition leads
to automatic O($a$) improvement. This statement obviously does not include
regions of the bare parameters where phase transitions can occur. In these regions
the validity of the Symanzik expansion itself can be questioned.
The importance of using both the Symanzik expansion and its application with $\chi$PT
theory has been shown in all its power.
Obviously the remaining O($a^2$) contributions of the theory are not predictable, and extensive
non-perturbative simulations have to be performed in order to understand the amount
of the remaining scaling violations.
In the quenched model the remaining O($a^2$) scaling violations are very small
for a wide range of pseudoscalar masses ($300 {\rm MeV} \lesssim M_{\rm PS}
\lesssim 1 {\rm GeV}$). Preliminary
results~\cite{Jansen:2006rf,Shindler:2006tm} indicate that this is true also
for $N_f = 2$ dynamical simulations.

\newpage
\section{The physical basis}
\label{sec:phys}

In the previous two sections we have seen how renormalizability and O($a$) cutoff effects 
of Wtm can be analyzed using the {\it twisted basis}. 
In the twisted basis the Wilson term takes the standard form while the mass term takes a 
``twisted'' form.
In this section we want to rederive some of the results already presented using the so called
{\it physical basis}~\cite{Frezzotti:2003ni}.
The physical basis is obtained from the twisted basis
performing an axial rotation in the $\tau^3$ isospin directionin in the bare lattice theory.
If one rotates the bare fields in the action and in the correlation functions,
the physics is completely unchanged and correlation functions calculated before and after the rotation
are the same on every single gauge configuration.
This happens because what matters are the
relative angles between the flavour-Dirac directions of the mass term and the Wilson term.
These angles are not changed by a bare field rotation.

In the physical basis the Wilson term is chirally rotated while the mass term
takes the standard form.
The advantage of this basis compared to the twisted one is 
that the dictionary of the correlation functions
is unchanged since after performing the continuum limit, the continuum QCD action takes 
the standard form. On the other side the dictionary has to be changed from the standard one, 
in the renormalization process, because the Wilson term is now chirally rotated.

\subsection{Chirally rotated Wilson term}

The setup is identical to the one described in sect.~\ref{ssec:Wtmaction}.
The only difference is the basis in which the fermion action is written, namely in this section the 
{\it physical basis} given by the set of fermion fields $\{\psi,\psibar\}$.
For a $SU(2)$ flavour doublet of mass degenerate quarks in the physical basis the action
has the form
\be
  S_{\rm F}[\psi,\psibar,U] =a^4\sum_x\psibar(x)\Big[D_{\rm tW}(\omega) + M \Big]\psi(x), 
\label{eq:tWQCD}
\ee
where 
\be
D_{\rm tW}(\omega) = \frac{1}{2}\{\gamma_\mu(\nabla_\mu + \nabla^*_\mu) - 
a r {\rm e}^{-i\omega \gamma_5 \tau^3}\nabla^*_\mu\nabla_\mu\},
\label{eq:tWilson}
\ee
is the twisted Wilson (tW) operator, 
$\nabla_\mu$, $\nabla^*_\mu$ are the standard gauge covariant forward and backward
covariant derivatives defined in app.~\ref{app:A} and $M$ is the standard mass term.

To ease the discussion in this section we will concentrate on the full twist case 
$\omega=\pi/2$ and we set $r=1$.Then eq.~(\ref{eq:tWQCD}) becomes 
\be
  S_{\rm F}[\psi,\psibar,U] =a^4\sum_x\psibar(x)\Big[D_{\rm tW} + M\Big]\psi(x), 
\label{eq:fulltWQCD}
\ee
with 
\be
D_{\rm tW} \equiv D_{\rm tW}(\omega=\pi/2) =\frac{1}{2}\big[\gamma_\mu(\nabla_\mu + \nabla^*_\mu) +
a i \gamma_5 \tau^3\nabla^*_\mu\nabla_\mu\big], 
\label{eq:tWilsonpi2}
\ee

The choices $\omega=\pi/2$ (and equivalently $\omega=-\pi/2$) is particularly useful because 
as we have seen in sect.~\ref{sec:impro} it can be
shown~\cite{Frezzotti:2003ni} that, despite the fact that the theory is not fully 
O($a$) improved, cancellation of O($a$) effects in quantities of physical 
interest (like energies and operator matrix elements) is automatic. 
We will prove again this property in this section using the physical basis.

We shortly recall now the symmetries of the lattice action given in eq.~(\ref{eq:fulltWQCD}):
gauge invariance, lattice rotations and translations, charge conjugation $\mathcal{C}$
(see app.~\ref{app:B} for the definition), and the $U(1)$
transformations associated with fermion number. 
It differs from standard Wilson fermions in two important ways.  First, the
flavour $SU_{\rm V}(2)$
symmetry is broken explicitly by the Wilson term down to the $U_{\rm V}(1)_3$
subgroup with diagonal generator $\tau_3$
\be
U_{\rm V}(1)_3 \colon
\begin{cases}
   \psi(x)    \rightarrow \exp(i\frac{\alpha_V}{2}\tau^3)\psi(x),\\
   \psibar(x) \rightarrow \psibar(x)\exp(-i\frac{\alpha_V}{2}\tau^3).
\end{cases}
\label{eq:u3psi}
\ee
In sect.~\ref{sssec:chiral} we have seen that Wtm preserves a subgroup of chiral symmetry.
In the physical basis this property is even more transparent since the lattice 
action~(\ref{eq:fulltWQCD}) with zero mass ($M=0$) is invariant under the standard axial transformations
\be
U_{\rm A}(1)_{1,2} \colon
\begin{cases}
   \psi(x)    \rightarrow \exp(i\frac{\alpha_A}{2}\gamma_5\tau^{1,2})\psi(x),\\
   \psibar(x) \rightarrow \psibar(x)\exp(i\frac{\alpha_A}{2}\gamma_5\tau^{1,2}).
\end{cases}
\label{eq:u_12}
\ee
The massless theory (eq.~\ref{eq:fulltWQCD} with $M=0$) is invariant under the
group transformations
\be
U_{\rm A}(1)_1 \otimes U_{\rm A}(1)_2 \otimes U_{\rm V}(1)_3 .
\label{eq:exact}
\ee
The ``charged'' sector has a continuum like behaviour and the exact
symmetry~(\ref{eq:exact}) protects the charged pion from chirally breaking
cutoff effects.

The discrete version of this symmetry
\be
\mathcal{R}_5^{1,2} \colon
\begin{cases}
\psi(x) \rightarrow i \gamma_5 \tau^{1,2}  \psi(x) \\
\psibar(x) \rightarrow  i \psibar(x) \gamma_5 \tau^{1,2}
\end{cases}
\ee
is still a symmetry of the massive lattice action~(\ref{eq:tWilsonpi2}) 
if combined with a sign change of the mass
\be
\widetilde{\mathcal{R}}_5^{1,2} \equiv \mathcal{R}_5^{1,2} \times [M \rightarrow - M],
\label{eq:Rtilde}
\ee
analogously to what happens in continuum QCD.

To understand the structure of the counterterms, we can use the symmetries of the  
lattice action~(\ref{eq:fulltWQCD}) as we have done in the twisted basis.
The counterterms to the action with dimension less or equal four are
\be
\tr\{F_{\mu \nu} F_{\mu \nu} \}, \quad i\psibar \gamma_5 \tau^3 \psi, \quad M \psibar \psi.
\label{eq:counterphys}
\ee
We notice immediately that quantum corrections imply the necessity to add to the lattice action 
a counterterm $i\psibar \gamma_5 \tau^3 \psi$ with a linearly divergent coefficient.
This is not surprising since this is the term that corresponds to the critical mass 
introduced in sect.~\ref{sec:basic}, which is the linear divergence induced by the presence
of the Wilson term.

On the contrary because of the $\widetilde{\mathcal{R}}_5^{1,2}$ symmetry~(\ref{eq:Rtilde}),
the operator $\psibar \psi$ comes with a coefficient odd in $M$,
and the mass is renormalized only multiplicatively. 
To state it differently, for zero quark mass the theory still preserves a remnant of chiral symmetry.
The residual $U_{\rm V}(1)_3$ symmetry (\ref{eq:u3psi}) forbids 
bilinears containing flavour matrices $\tau^{1,2}$.
The parity flavour symmetry $\mathcal{P}_F^{1,2}$ defined in
eq.~(\ref{eq:PF12}) 
takes the same form in the two bases
and thus is still a symmetry of the lattice action~(\ref{eq:fulltWQCD}). 
This symmetry requires that
parity and flavour are violated together so it forbids flavour singlet parity violating terms 
$\psibar \gamma_5 \psi$ and $\epsilon_{\mu \nu \rho \sigma}F_{\mu\nu} F_{\rho\sigma}$, 
as well as the flavour violating, parity even, operator $\psibar \tau_3 \psi$. 
It is easy to check that all the dimension four operators which
violate parity or isospin or both are ruled out by $\mathcal{P}_F^{1,2}$.
The lattice theory has to be modified in order to include the power divergent subtraction, and to keep 
consistency between the basis we write the lattice action
\be
  S_{\rm F}[\psi,\psibar,U] =a^4\sum_x\psibar(x)\left\{\frac{1}{2}\left[\gamma_\mu(\nabla_\mu + \nabla^*_\mu) +
i \gamma_5 \tau^3\big(a \nabla^*_\mu\nabla_\mu - m_{\rm cr}\big)\right] + M\right\}\psi(x) .
\label{eq:fulltWQCD2}
\ee
One can easily check now that this action is completely equivalent 
from the action in the twisted basis~\eqref{eq:WtmQCD},
just performing the rotations~\eqref{eq:axial} with $\omega=\pi/2$ and identifying $m_0 = m_{\rm cr}$
and $M=\mu_{\rm q}$.
The final continuum theory has now the standard QCD form
\be
S_0 = \int d^4x \psibar(x) \Big[ \gamma_\mu D_\mu + M_{\rm R} \Big] \psi(x) ,
\qquad {\rm with} \qquad M_{\rm R} = Z_{\rm M}(g_0^2,a\mu) M
\label{eq:cQCD_phy}
\ee

To understand how the choice of a twisted Wilson term influences the 
Ward identities of the theory we rewrite some of them in the physical basis.
To deduce them it is enough to start from the Ward identities in the twisted basis, where
using a mass independent renormalization scheme, one can make a standard analysis 
to construct the lattice fields that are multiplicatively 
renormalizable and respect, up to cutoff effects, the chiral multiplet structure.
The final step is then to rotate all the 
quark fields back into the physical basis.

Renormalized vector and axial currents can be taken to be 
\be
(\mathcal{V}_{\rm R})_\mu^a=Z_A\,\mathcal{V}_\mu^a \qquad
(\mathcal{A}_{\rm R})_\mu^a=Z_V\,\mathcal{A}_\mu^a \qquad a=1,2 \, ,
\ee
\be
(\mathcal{V}_{\rm R})_\mu^3=Z_V\,\mathcal{V}_\mu^3 \qquad
(\mathcal{A}_{\rm R})_\mu^3=Z_A\,\mathcal{A}_\mu^3 ,
\ee
where the local currents in the physical basis are defined in 
eqs.~(\ref{eq:axial_current_rot},\ref{eq:vector_current_rot})
and the finite renormalization constants, $Z_V$ and $Z_A$, are those for the local 
vector and axial currents of standard Wilson fermions, respectively. Notice 
the switch between $Z_V$ and $Z_A$ for the currents with flavour $a=1,2$, 
due to the presence of the factor $\gamma_5\tau_3$ in front of the Wilson 
term in eq.~(\ref{eq:tWilsonpi2}).

The expressions for the renormalized densities are
\be
\mathcal{P}_{\rm R}^a=Z_P\,\mathcal{P}^a \qquad
\mathcal{P}_{\rm R}^3=Z_{S^0}\Big[\mathcal{P}^3 + \frac{i\rho_P(aM)}{a^3}\Big] ,
\ee
\be
\mathcal{S}_{\rm R}^0=Z_P\Big[\mathcal{S}^0 + \frac{M\rho_{S^0}(aM)}{a^2} \Big]
\label{eq:S0ren}
\ee
where $\rho_P(aM)$ is a polynomial in $aM$, while $\rho_{S^0}(aM)$ is a
polynomial with even powers of $aM$. The reason for this is because 
$\widetilde{\mathcal{R}}_5^{1,2}$ is a symmetry of the lattice action~(\ref{eq:fulltWQCD2}).
Formula~(\ref{eq:S0ren}) is rather interesting because it shows that the chiral 
order parameter is only affected by an $M/a^2$ power divergence, 
analogously to what happens with Ginsparg-Wilson fermions.

The non-singlet Ward identities~(\ref{eq:WI_phy}), once the local operators
are properly renormalized, are valid at fixed lattice spacing up to
cutoff effects
\be
\partial_\mu^* \langle (\mathcal{V}_{\rm R})_\mu^a(x) \mathcal{O}_{\rm R}(y)
\rangle =0 
\ee
\be 
\partial_\mu^* \langle (\mathcal{A}_{\rm R})_\mu^a(x) \mathcal{O}_{\rm R}(y)
\rangle = 2 M_{\rm R} \langle \mathcal{P}_{\rm R}^a(x) \mathcal{O}_{\rm R}(y) \rangle
\ee
with $\mathcal{O}$ being a generic multilocal operator and $x \neq y$.

\subsection{Automatic O($a$) improvement}

In this section we are going to prove again automatic O($a$) improvement 
in the physical basis in infinite volume.

The target continuum theory for the fermion fields will be now
\be
  S_0 = \int d^4 x \psibar(x) \Big[\gamma_\mu D_\mu + M_{\rm R} \Big]\psi(x),
\label{eq:cQCD_phy}
\ee
The Symanzik effective action in eq.~(\ref{eq:eff_action}) now has correction terms given by
\be
S_1 = \int d^4y  {\mathcal L}_1(y) \qquad {\mathcal L}_1(y) = \sum_i c_i
  {\mathcal O}_i(y),
\ee
with operators of the form
\be
{\mathcal O}_0 = i \Lambda^2\psibar \gamma_5 \tau^3 \psi, \qquad {\mathcal O}_1 = 
  i \psibar\gamma_5 \tau^3 \sigma_{\mu\nu}F_{\mu\nu}\psi, \qquad {\mathcal O}_5 =
  i M^2\psibar \gamma_5 \tau^3 \psi.
\label{eq:sym_op_phy}
\ee
where $\Lambda$ is an energy scale of the order of the QCD scale $\Lambda_{\rm QCD}$.
The operator ${\mathcal O}_1$ is the usual clover term in the physical basis,
where the Wilson term is chirally rotated. The operator
${\mathcal O}_0$ parametrizes the unavoidable mass independent O($a$) uncertainties 
in the critical mass.

We can now repeat the same steps followed in sect.~\ref{sec:impro}.
We consider a general multiplicatively renormalizable multilocal field
that in the effective theory is represented by the effective field
\be
\Phi_{\rm eff} = \Phi_0 + a \Phi_1 + \ldots
\ee
A lattice correlation function of the field $\Phi$ to order $a$ is given by
\be
\langle \Phi \rangle = \langle \Phi_0 \rangle_0 - a \int d^4y \langle \Phi_0
     {\mathcal L}_1(y) \rangle_0 + a \langle \Phi_1 \rangle_0 + \ldots
\label{eq:sym_exp2}
\ee
where the expectation values on the r.h.s are to be taken in the continuum
theory with action $S_0$.
The key point is that the continuum action (\ref{eq:cQCD_phy}) is symmetric
under isovector flavour transformations.
In particular the discrete flavour rotations 
\be
\mathcal{F}^{1,2} \colon
\begin{cases}
\psi(x_0,{\bf x}) \rightarrow i \tau^{1,2}  \psi(x_0,{\bf x}) \\
\psibar(x_0,{\bf x}) \rightarrow  -i \psibar(x_0,{\bf x}) \tau^{1,2}
\end{cases}
\label{eq:F12}
\ee
are symmetries of the continuum action.
On the contrary all the operators in eq. (\ref{eq:sym_op_phy}), of
the Symanzik expansion of the lattice action, are odd under the the discrete flavour
symmetry~(\ref{eq:F12}).
If the field $\Phi$ is a lattice representation of the flavour even field $\Phi_0$ then 
the second term in the r.h.s. of
eq. (\ref{eq:sym_exp2}) vanishes. To show that also the $\Phi_1$ term vanishes we have to
show that an operator of one dimension higher than the original one but the same lattice symmetries
has opposite discrete flavour number. 
This goes exactly in the same way as in the twisted basis (see
sect.~\ref{sec:impro}).
The gauge action is invariant under the symmetry~\cite{Frezzotti:2003ni}
\be
\widetilde{\mathcal{D}} =  \mathcal{D}\times [M \rightarrow -M]
\ee
where
\be
\mathcal{D} \colon
\begin{cases}
U(x;\mu) \rightarrow U^{\dagger}(-x-a\hat{\mu};\mu), \\ 
\chi(x) \rightarrow {\rm e}^{3 i \pi/2} \chi(-x) \\
\chibar(x) \rightarrow  \chibar(-x){\rm e}^{3 i \pi/2},
\end{cases}
\ee
while in the fermion action the terms that break flavour symmetry $\mathcal{F}^{1,2}$ symmetry are odd.
But in particular the lattice action is invariant under $\mathcal{F}^{1,2} \times \widetilde{\mathcal{D}}$.
So the operators in $\Phi_1$ will necessarily have opposite flavour number to $\Phi_0$. Given the 
fact that the continuum action is flavour symmetric also $\Phi_1$ vanishes.
Possible contact terms coming from the second term amount to a redefinition of
$\Phi_1$ as we have discussed 
in sec.~\ref{sec:impro} and so do not harm the proof.

The same proof can be repeated using the parity symmetry
$\mathcal{P}$ defined in eq.~(\ref{eq:parity}) instead of using the discrete flavour symmetry
$\mathcal{F}^{1,2}$.

\newpage
\section{O($a^2$) cutoff effects}
\label{sec:asq}

In this section I will be mainly concerned with the O($a^2$) cutoff effects of the
Wtm formulation and 
in particular on the cutoff effects induced by the breaking of flavour and parity
symmetry. Some of these require a formal description of the
quantum mechanical representation for a lattice correlator computed with the Wtm action.
The first part of this section will be devoted to a brief introduction of the basic
Hamiltonian formalism, while the second part will be concentrated on the typical
consequences of these breaking effects on physical observables. A third part will deal
with a possible strategy in order to attenuate these O($a^2$) effects.

\subsection{Hamiltonian formalism}

We assume we are in a finite volume $L$, which is large enough to consider
hadronic states as point-like. This is likely to be the situation for the actual
lattice simulations with $2 {\rm fm} \lesssim L \lesssim 4 {\rm fm}$.
The relation among euclidean correlation function,
hadron masses and matrix elements is given in the Hamiltonian formalism by
the transfer matrix of the lattice theory.
We have shown in sect.~\ref{sec:basic} that for Wtm the transfer matrix
exists and it is positive definite.
As a consequence of physical positivity a lattice correlation function admits a quantum mechanical
representation in term of states belonging to the Hilbert space of the theory.
This complete set of states can be chosen to be the eigenstates $|n,q \rangle $ of 
the Hamiltonian $\mathbb{H}$
\be
\mathbb{H} | n,q \rangle = E_n^{(q)} | n,q \rangle \ , \qquad \mathbbm{1} = \sum_{n,q} | n,q \rangle \langle n,q |
\ee
where $q$ indicates a set of the quantum numbers corresponding to the symmetries of the lattice
action and $n$ the energy level given a certain quantum number.
We choose here the following normalization
\be
\langle n',q' | n,q \rangle = 2E_n^{(q)}\delta_{n',n}\delta_{q',q}
\ee
The set of quantum numbers that it is possible to use to classify the states for Wtm are 
$q \equiv \{j,P_F^1,C,Q_3,{\bf p}\}$ corresponding to the symmetries of Wtm:
charge conjugation ${\mathcal C}$, 
parity combined with flavour exchange ${\mathcal P}_F^1$, charge associated to
the residual $U_{\rm V}(1)_3$ 
flavour symmetry $Q_3$, the representation of the $H(3)$ group of spatial discrete rotations $j$ 
and the spatial momentum ${\bf p}$\footnote{${\mathcal P}_F^2$ is not used
  because it is not independent 
from ${\mathcal P}_F^1$ and $Q_3$.}.

In the case of plain Wilson fermions (eq.~\ref{eq:WtmQCD} with $\mu_{\rm q} = 0$) we could replace
${\mathcal P}_F^1$ and $Q_3$ 
with standard parity and isospin symmetry ${\mathcal P}$, $I$ and $I_3$.
For Wtm the set of quantum numbers $q$ is smaller than in the continuum. This means that certain
states that in the continuum have different quantum numbers, cannot be disentangled anymore with Wtm.
A typical example is the neutral pion where we have $q_{\pi^0} = \{j,+,+,0,{\bf 0}\}$ 
that is certainly not distinguishable from the vacuum.
More generally in the quantum mechanical representation of the correlation
functions computed with Wtm, 
there will be parity violating matrix elements which vanish only in the
continuum limit. 
Even if technically this is slightly
more complicated that analyzing correlators in the plain Wilson case, the state of art for 
the analysis of correlation functions requires a multi-state fit, which is what we need to extract
the masses and matrix elements we are interested in.
Moreover given the fact that these matrix elements vanish in the continuum
limit, it could be that numerically these contributions are very small at
sufficiently fine lattices.
To be specific I will concentrate here on 2 examples in order to show how to proceed.

\subsection{Pion channel correlator}

We take here a finite 4--d lattice with spacing $a$.
We want to write the quantum mechanical decomposition of the following correlator
\be
C_{\rm PP}^{11}(x_0)  = - a^3\sum_{\bf x} \langle P^1(x_0,{\bf x}) P^1(0) \rangle \ .
\ee
Since we sum over ${\bf x}$, only states with vanishing spatial momentum contribute.
Moreover we concentrate on the trivial representation of H(3) and since we neglect electromagnetic interactions
it is more convenient to study eigenstates of charge conjugation $\mathcal{C}$.
The relevant quantum numbers are the quantum numbers of the charged and
neutral pion
\be
q_{\pi^1}=\{0,-,+,1,{\bf 0}\} \ , \qquad q_{\pi^0}=\{0,+,+,0,{\bf 0}\} \ .
\ee
The eigenstates of charge conjugation are
\be
|0,\pi^1({\bf p}) \rangle = \frac{1}{\sqrt{2}}(|0,\pi^+ ({\bf p}) \rangle +
|0,\pi^- ({\bf p}) \rangle) \, \qquad 
|0,\pi^2({\bf p}) \rangle = \frac{-i}{\sqrt{2}}(|0,\pi^+({\bf p}) \rangle -
|0,\pi^-({\bf p}) \rangle) \ .
\ee
These two states have the same mass because of the residual $U_{\rm V}(1)_3$ isospin symmetry.
Using the standard representation of a correlation function in terms of the local operators
$\hat{P}^1$, and inserting a complete set of states, we obtain (for simplicity
we consider the lattice time extent $T \rightarrow \infty$)
\be
C_{\rm PP}^{11}(x_0) = \frac{1}{\mathcal{Z}} 
\sum_{n,q} \langle 0,\Omega|\hat{P}^1 |n,q \rangle \frac{e^{-E_n^{(q)} x_0}}{2 E_n^{(q)}}
\langle n,q|\hat{P}^1|0,\Omega \rangle \ .
\label{eq:PP_dec}
\ee
A detailed analysis of the excited states can become rather involved. Here we
concentrate on the fundamental and the first excited state which is a specific contribution of
Wtm
\be
n = 0 \Rightarrow |0,\pi^1 ({\bf 0})\rangle
\ee
\be
n = 1 \Rightarrow |1,\pi^1({\bf 0}) \pi^0 ({\bf 0})\rangle .
\ee
We are here implicitly assuming that in the chosen volume $L$ the energy level of the 
two pion state with zero relative momentum is well separated from the energy levels of the
excited two pion state. From fig.~1 of ref.~\cite{Lellouch:2000pv} this seems to be certainly the case
for the volumes which are currently simulated.

Contrary to what happens with standard Wilson fermions, the first excited state
is given by a two pion contribution. The parity symmetry of the the continuum
QCD action implies that the matrix element $|\langle
0,\Omega|\hat{P}^1|1,\pi^1 \pi^0 \rangle|$ must vanish in the continuum limit.
This means it is at most an O($a$) term.
This implies that the amplitude $|\langle 0,\Omega|\hat{P}^1|1,\pi^1 \pi^0
\rangle|^2$ is an O($a^2$) effect independently of whether we set the untwisted
quark mass to zero or not
\footnote{I acknowledge very useful discussions with G. Herdoiza and R. Frezzotti on this point.}.
The first two terms of eq.~(\ref{eq:PP_dec}) read
\bea
&&|\langle 0,\Omega|\hat{P}^1|1,\pi^1({\bf 0}) \rangle|^2 
\frac{e^{-M_{\pi^1} x_0}}{2 M_{\pi^1}} \times \nonumber \\
&\times& \left\{1+
\frac{|\langle 0,\Omega|\hat{P}^1|1,\pi^1({\bf 0}) \pi^0({\bf 0}) \rangle|^2}
{\langle 0,\Omega|\hat{P}^1|1,\pi^1({\bf 0}) \rangle|^2}
e^{-\left(E_{2\pi}(L) - M_{\pi^1}\right) x_0}\frac{M_{\pi^1}}{E_{2\pi}(L)} + \ldots \right\},
\eea
where $E_{2\pi}(L)$ represents the lowest energy level for the two pion state.
At finite lattice spacing $a$ the excited state correction increases
when the quark mass goes to zero. In fact, up to finite size corrections, the gap is
proportional to the neutral pion mass which, as we will see in sects.~\ref{sssec:pion_splitting}
and~\ref{sssec:WCPT_phase}, vanishes at very small values of the twisted mass.
This could be in principle a problem for a
reliable extraction of the charged pion mass.
Practically the problem is absent because $|\langle 0,\Omega|\hat{P}^1|1,\pi^1 \pi^0 \rangle|^2$
is of O($a^2$) and, like all the other parity violating contributions 
to the correlator, it is highly suppressed at sufficiently small lattice spacing.
An additional suppression factor is given by $\frac{M_{\pi^1}}{E_{2\pi}(L)}$.
We remark that, independently on the lattice QCD formulation used, the gap between
the fundamental and the first excited state in the pion channel
will always go to zero in large volumes, because the first excited state will be at 
most a 3 pions state with vanishing momentum.

Since Wtm does not preserve parity at finite lattice spacing, it would be possible in principle to have as
excited state a charged scalar. But the scalar that can be created must have opposite isospin component 
because parity ``times'' isospin in direction $1$ and $2$ 
($\mathcal{P}^{1,2}_F$, defined in eq.~\ref{eq:PF12}) is a symmetry of the lattice action.
This isospin component of the scalar current has opposite charge conjugation.
The residual $U_{\rm V}(1)_3$ symmetry~(\ref{eq:u3}) excludes mixing with the remaining
isospin components of the scalar particle.

\begin{figure}[htb]
\begin{center}
\includegraphics[angle=0,width=14cm]{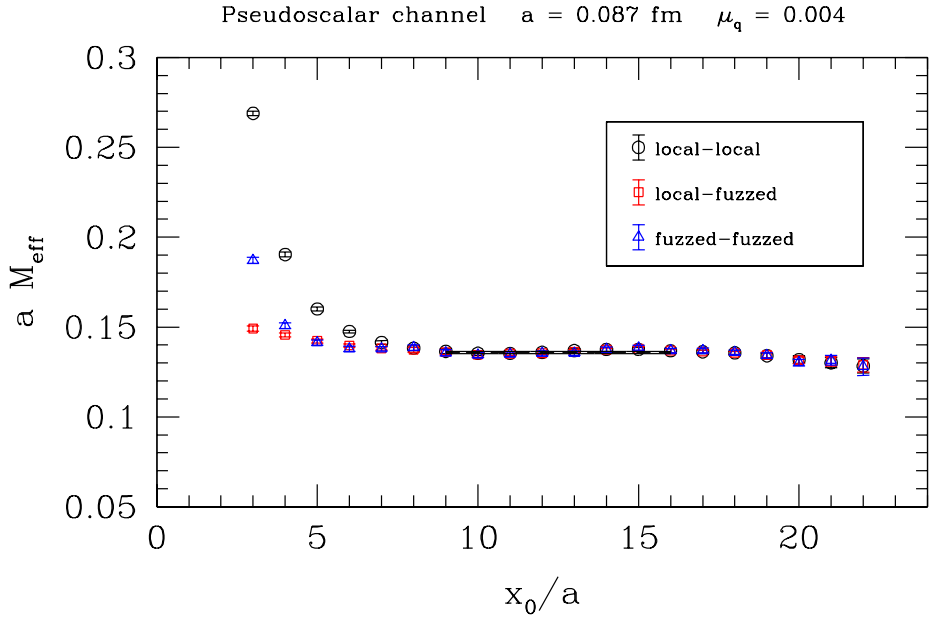}
\end{center}
\vspace{-5.0cm}
\caption{Effective mass for the pseudoscalar channel for a lattice spacing of $a=0.087$ fm and a corresponding
mass of $292$ MeV. The masses using 3 different interpolating operators are shown.}
\label{fig:eff_mass_PS}
\end{figure}

The conclusion of this discussion is that the first excited state
contributions typical of Wtm are either absent, like the charged scalar, or
highly suppressed like the two pion state.
This is nicely confirmed by the recent unquenched simulations performed by the
ETM Collaboration~\cite{ETMC2}.
In fig.~\ref{fig:eff_mass_PS}, I show a typical effective mass for $N_f=2$ dynamical quarks.
A long plateau in euclidean time is observed that allows a very precise determination of the pseudoscalar mass.
We conclude that Wtm does not seem to make more difficult extracting the pseudoscalar mass, 
compared with other lattice actions which preserve parity symmetry.
 
\subsection{Charged scalar channel correlator}

The same analysis can be repeated for the charged scalar correlator.
In particular we want to study the correlator for the charged $a_0$ scalar meson.
The relevant quantum numbers are
\be
q_{a_0^1}=\{0,+,+,1,{\bf 0}\} \ , \qquad q_{\eta}=\{0,-,+,0,{\bf 0}\}\ .
\ee
The $\eta$ we consider here is the pseudoscalar isosinglet in the $SU(2)$
framework, sometimes labelled as $\eta_2$.
The leading contributions of the charged scalar correlator
\be
C_{\rm SS}^{11}(x_0) = \frac{1}{\mathcal{Z}} 
\sum_{n,q} \langle 0,\Omega|\hat{S}^1 |n,q \rangle \frac{e^{-E_n^{(q)} x_0}}{2 E_n^{(q)}}
\langle n,q|\hat{S}^1|0,\Omega \rangle \ ,
\ee
are
\be
n = 0 \Rightarrow |0,a_0^1({\bf 0}) \rangle
\ee
\be
n = 1 \Rightarrow |1,a_0^1 ({\bf 0})\pi^0({\bf 0}) \rangle, \qquad  |1,\eta ({\bf 0}) \pi^1 ({\bf 0}) \rangle.
\ee

The $|1,\eta \pi^1  \rangle $ state is the
natural decay channel for the charged scalar particle. If the values of the
quark masses in the simulations are such that the decay threshold is not open,
this term contributes as an excited state with a gap from the fundamental
state which could be very small.

The parity violating term in the spectral decomposition is, analogously to the
pseudoscalar channel, proportional to 
\be
\frac{|\langle 0,\Omega|\hat{S}^1|1,a_0^1 ({\bf 0}) \pi^0 ({\bf 0}) \rangle|^2}{|\langle
  0,\Omega|\hat{S}^1|0,a_0^1 ({\bf 0}) \rangle|^2} \e^{-\left(E_{a_0\pi}(L) - M_{a_0}\right) x_0} ,
\label{eq:exc2}
\ee
with an amplitude squared of O($a^2$).

\begin{figure}[htb]
\begin{center}
\includegraphics[angle=0,width=14cm]{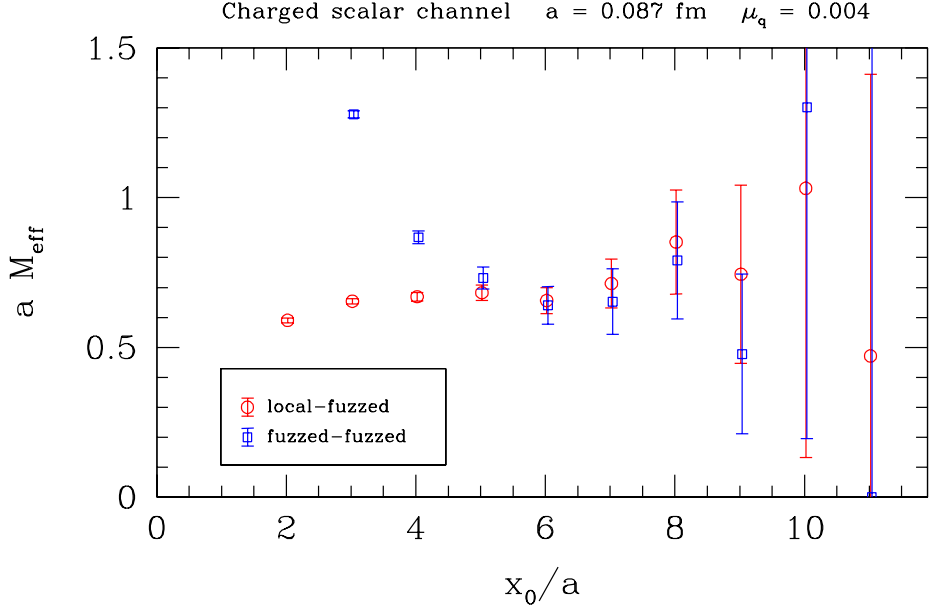}
\end{center}
\vspace{-5.0cm}
\caption{The same as fig.~\ref{fig:eff_mass_PS} but for the charged scalar channel and with the 
two best correlators.}
\label{fig:a0a0}
\end{figure}

In fig.~\ref{fig:a0a0} we see an example of a charged scalar correlator, from
a recent analysis of $N_f=2$ dynamical configurations~\cite{ETMC2}.

Using the same argument we have use in the previous section to exclude a charged scalar as
a possible excited state of the charged pseudoscalar, we can exclude the presence of
a single pion intermediate state.
As a confirmation of this fact there is no sign in fig.~\ref{fig:a0a0} 
that the correlator dips down to the value of the pseudoscalar mass. 

\subsection{Isospin violation}
\label{ssec:isoviol}

The fact that Wtm at finite lattice spacing breaks isospin symmetry, while retaining
the isovector subgroup $U_{\rm V}(1)_3$, 
allows a splitting between neutral and charged pions.
This is not a physical splitting and in the continuum limit of a 
$N_{\rm f}=2$ theory with degenerate quarks we will obtain again a degenerate triplet of pions
(if we neglect, as we do here, the electromagnetic interactions).
But at finite lattice spacing the masses of the charged and neutral pion are not the same.
This is a very important issue to understand and rather surprisingly only recently numerical simulations
started to investigate this issue~\cite{Jansen:2005cg,Becirevic:2006ii,Abdel-Rehim:2006ve}. 
Other examples of isospin violating cutoff effects are the splitting in the
kaon sector or in the $\Delta$-baryon multiplet investigated numerically
in~\cite{Abdel-Rehim:2005gz}. 
In general the isospin splitting is an O($a^2$) effect independently of the
value of the twist angle. 
This can be shown using the symmetry $\mathcal{P}_F^{1,2}$ (\ref{eq:PF12}) which is
the product of parity and flavour exchange. This is a symmetry of 
Wtm~\eqref{eq:WtmQCD} independently on the value of the twist angle.
This symmetry of the lattice action implies that all the dimension 5 counterterms
of the Symanzik effective action will have to be even under $\mathcal{P}_F^{1,2}$.
For parity conserving correlation functions this automatically implies that 
their O($a$) insertion will contribute in the same way to the correlation functions
independently on their flavour structure. 
Thus the isospin splitting has to be at least of O($a^2$) independently on the
value of the twist angle. In other words the O($a$) affecting the neutral
and charged pion correlation functions out of full twist are the same. 
This is confirmed by the results obtained in W$\chi$PT~\cite{Scorzato:2004da,Sharpe:2004ps}
and discussed in the following section.

\subsubsection{Pion splitting}
\label{sssec:pion_splitting}

To fix the notation we recall some basic definitions. The charged pseudoscalar
densities are given by 
\be
P^\pm(x) =\chibar(x)\gamma_5\tau^\pm\chi(x) \qquad  \tau^\pm = {\tau^1
  \pm {\rm i} \tau^2 \over 2}
\ee
and a possible interpolating field for the neutral pion, with $\omega = \pi/2$, is the scalar current
\be
\frac{1}{\sqrt{2}}S^0(x) = \frac{1}{\sqrt{2}}\chibar(x)\chi(x) .
\ee
The charged and neutral pseudoscalar masses can be extracted by the following correlators
\be
C_{\pi^{+}} (x_0) = -a^3\sum_{\mathbf x} \langle P^+(x) P^-(0) \rangle \qquad
C_{\pi^{0}} (x_0) = \frac{a^3}{2}\sum_{\mathbf x} \langle S^0(x)S^0(0) \rangle  .
\label{eq:pion_corrs}
\ee
If we separate the flavour components and we perform the fermionic
contractions we obtain
\bea
C_{\pi^{0}} (x_0) &=& \frac{a^3}{2}\sum_{\mathbf x} \big\{ \left\langle - \tr 
\big[ G_u(0,x) G_u(x,0) \big] - \tr \big[ G_d(0,x) G_d(x,0) \big] \right. \nonumber \\
&& + \left. \tr \big[G_u(x,x)\big] \tr \big[G_u(0,0)\big]  + \tr \big[G_u(x,x)\big] \tr
\big[G_d(0,0)\big] \right. \nonumber \\
&& + \left. \tr \big[G_d(x,x)\big] \tr \big[G_u(0,0)\big]  + \tr \big[G_d(x,x)\big] \tr
\big[G_d(0,0)\big]  
\right\rangle\big\}
\label{eq:corrpi0}
\eea
where $G_{u,d}(x,y)$ are the fermionic propagators for the the {\it up} and
{\it down} quarks.

The neutral pion correlator~(\ref{eq:corrpi0}) contains fermionic disconnected
diagrams (the last four terms in eq.~\ref{eq:corrpi0}), which are notoriously more difficult to compute on the lattice.
One way to circumvent this problem is to consider only the fermionic 
connected part (the first two terms of eq.~\ref{eq:corrpi0}). 
The question would be now if it is still
possible to interpret the fermionic connected piece as coming from a
correlation function between local operators which represent a pion field.
The answer is positive if we consider the valence quarks of the theory
discretized with the so called Osterwalder-Seiler (OS) action~\cite{Osterwalder:1977pc,Frezzotti:2004wz}.
To be specific we consider now the following model: the sea quarks are
described by the standard $N_f=2$ degenerate Wtm action~(\ref{eq:WtmQCD})
while the valence quarks, that we call $u$ and $d'$ have the following actions
\be
S_{\rm OS}^{(u)} = a^4\sum_x \big\{\bar{u}(x)\left[D_{\rm W} + m_{\rm cr}
+i\mu_u^v\gamma_5\right] u(x)\big\} .
\label{eq:OSu}
\ee
\be
S_{\rm OS}^{(d')} = a^4\sum_x \big\{\bar{d'}(x)\left[D_{\rm W} + m_{\rm cr}
+i\mu_{d'}^v\gamma_5\right] d'(x)\big\} .
\label{eq:OSdprime}
\ee
These two quarks have the same mass $\mu_u^v = \mu_{d'}^v$, thus they have identical actions,
differing only because they have different flavours.
We can then define as interpolating field for the neutral pion
\be
(\pi^0)'(x) = \frac{1}{\sqrt{2}}\left[ \bar{u}(x) u(x) - \bar{d'}(x) d'(x)
\right] .
\label{eq:pi0prime}
\ee
It is easy to see that, performing the axial rotations~(\ref{eq:axial})
corresponding to the actions~(\ref{eq:OSu},\ref{eq:OSdprime})
\be
\begin{cases}
   u(x)     \longrightarrow 
               \exp(-i\frac{\omega}{2}\gamma_5) u(x) \\
   \bar{u}(x)  \longrightarrow 
               \bar{u}(x)\exp(-i\frac{\omega}{2}\gamma_5) ,
\end{cases} \qquad 
\begin{cases}
   d'(x)     \longrightarrow 
               \exp(-i\frac{\omega}{2}\gamma_5) d'(x) \\
   \bar{d'}(x)  \longrightarrow 
               \bar{d'}(x)\exp(-i\frac{\omega}{2}\gamma_5) ,
\end{cases}
\label{eq:udprimerot}
\ee
with $\omega=\pi/2$, this is a valid interpolating field for the neutral pion.
Performing standard Wick contractions we obtain
\be
C_{(\pi^{0})'} (x_0) = a^3\sum_{\mathbf x} \big\{ \left\langle - \tr 
\big[ G_u(0,x) G_u(x,0) \big] - \tr \big[ G_{d'}(0,x) G_{d'}(x,0) \big] 
\right\rangle\big\} ,
\label{eq:corrpi0prime}
\ee
where the disconnected terms cancel because the $u$ and $d'$ quarks have identical propagators.
Since parity ``times'' flavour exchange is a symmetry of the lattice action we
have
\be
\gamma_0 G_d(0,x_{\rm P}) \gamma_0 = G_u(0,x) = G_{d'}(0,x), \qquad {\rm with} \quad x_{\rm
  P} = (x_0,-{\bf x}) .
\ee
The connected part of the neutral pion correlator
with Wtm~(\ref{eq:corrpi0}) is equal to the correlation
function~(\ref{eq:corrpi0prime}) for a model with OS valence quarks.
This argument tells us that the connected part of eq.~(\ref{eq:corrpi0}) is
not the $\pi^0$ propagator with Wtm, but it is still a $\pi^0$ propagator with
valence quarks having the same OS actions~(\ref{eq:OSu},\ref{eq:pi0prime}). 
In the continuum limit the two neutral pion correlators will coincide.

The OS action can be used only as a discretization for valence quarks
because a parity violating and isospin singlet counterterm like $F \widetilde{F}$ could be
otherwise generated in the process of renormalizing the theory
(see sect.~\ref{sec:basic}).
This freedom of choosing a different lattice action between valence and sea
quarks will be further used to ease the renormalization of local operator (see
sect.~\ref{sec:ren}).
Here we just use this freedom as a tool to interpret the fermionic connected
part as a theoretically valid expression of a correlation function between
local operators.
We remark that the fermionic connected part is {\it not} the neutral pseudoscalar meson of Wtm,
but it is an interesting quantity to study with precise data on its own, in view
of a possible use of mixed actions (the OS action for the valence quarks and
Wtm for the sea quarks).
Moreover in the quenched model the fermionic determinant is neglected anyhow,
and this puts the two models (quenched OS and quenched Wtm) on the same footing.

We also note that in the same model we could define a charged pion made up of
a $u$ and $d'$ quark. In this case the correct interpolating field would be
$\bar{u} d'$ that has identical correlation function with the neutral pion
field $(\pi^0)'$. This leads to a model with NO isospin splitting among the pions.
The drawback of this choice would be, apart from O($a^2$) unitarity
violations, that this charged pion
field is not the field that is protected by the exact lattice Ward
identity~(\ref{eq:lat_PCVC}).

The choice of valence actions different from the sea action allows freedom to solve
some problems which appear in the unitary formulation. We will see in sect.~\ref{sec:ren} how
this freedom can simplify the renormalization of local operators.
It remains to be seen how the O($a^2$) unitarity violations affect the theory 
and the approach to the continuum limit.
This is an open question which could be answered only performing detailed scaling tests.

In~\cite{Jansen:2005cg}, to which I refer for all the technical details,
a pilot quenched study has been performed to study flavour breaking 
effects with Wtm.
In this paper the scaling behaviour is studied in the quenched model of the
pion splitting using either only the connected correlator
($(M_\mathrm{PS}^0)_{\rm OS}^2-(M_\mathrm{PS}^+)^2$), or including also the
disconnected terms 
($(M_\mathrm{PS}^0)^2-(M_\mathrm{PS}^+)^2$)\footnote{In both computations the ``PCAC method'' is used for
the determination of the critical mass.}.
The results show O($a^2$) scaling
violations for both the pion splittings with indications that 
the neutral pseudoscalar meson for Wtm (with the 
inclusion of the disconnected correlator) has 
reduced splitting with the charged meson, within the rather large statistical errors. 
It is possible to give a very rough estimate of the pion splitting
$r_0^2((M_\mathrm{PS}^0)^2-(M_\mathrm{PS}^+)^2)\simeq c(a/r_0)^2$ with
$c \simeq 10$ (with large errors).
Comparing to a quenched simulation for na\"ive staggered fermions with Wilson
gauge action \cite{Ishizuka:1993mt}, one finds a similar size of the flavour
splitting encountered for the pion mass at a similar lattice spacing with a
value $c \simeq 40$. For dynamical improved staggered fermions
a value of $c \simeq 10$ has been found \cite{Aubin:2004wf}.

The pion splitting has been investigated in the quenched model in~\cite{Becirevic:2006ii}, where a
comparison between the impact on the methods used to determine the critical mass has been
made. The outcome of this study is that if one considers only the
fermionic connected diagrams (OS neutral pion) the inclusion of the clover
term into the Wtm lattice action~(\ref{eq:clovertm})
reduces the pion splitting in comparison with the Wtm action without clover
term. 
This result is rather interesting but not conclusive, in fact
the addition of the disconnected terms can have a tremendous impact as we have
just seen in the quenched model. Moreover recent dynamical
simulations~\cite{Boucaud:2007uk} indicate that the inclusion of the disconnected terms
changes the sign of the pion-splitting. To be precise the outcome is that
a first analysis at a value the pseudoscalar mass $M_{\rm PS} \simeq 300$ MeV,
taking the disconnected contribution in the neutral
channel fully into account, shows that the neutral pseudoscalar
meson is about $20\%$ lighter than the charged one.
Expressed differently one obtains
$r_0^2((M_\mathrm{PS}^0)^2-(M_\mathrm{PS}^+)^2)=c(a/r_0)^2$ with $c=-4.5(1.8)$.
This coefficient is, in absolute value, a factor of 2 smaller than the value found in
quenched investigations~\cite{Farchioni:2005hf}. If one considers only
the connected term, i.e. considering OS valence quarks, the charged pion is
lighter or in other words the sign of $c$ is positive.

From the theoretical side a tool
to address and understand this problem is again W$\chi$PT.
Adopting the same power counting scheme used in sect.~\ref{sec:impro}
given in eq.~(\ref{eq:GSM}) the result~\cite{Scorzato:2004da,Sharpe:2004ps} at NLO
order is 
\be
M_{\pi^0}^2 -  M_{\pi^{\pm}}^2 = -2a^2 w'\frac{\mu_{\rm R}^2}{m'^2 + \mu_{\rm R}^2} \,.
\label{eq:pionsplit}
\ee
The splitting is as expected an O($a^2$) effect and vanishes 
on the Wilson axis ($\mu_{\rm R} = 0$) as it should.
It is moreover necessarily even in $\mu_{\rm q}$, from the combined violation of parity
and isospin argument given in sect.~\ref{ssec:isoviol}.
The splitting is maximized when $m'=0$ as it should be since this is the
condition that minimizes the parity violation.
The sign of the LEC $w'$ tells us which pion is heavier. We will see in
sect.~\ref{ssec:phase} that $w'$ also determines the nature of the vacuum structure of
the theory around the chiral point.

Given the non-degeneracy among the pion multiplet, several issues would have
to be investigated in the future. 
The usage of a mixed action approach with OS valence quarks (or another
isospin conserving action like the overlap discretization) eliminates the
problem of isospin splitting at the valence level.
Of course the isospin breaking cutoff effects of the sea Wtm action
will still generate a splitting, but only among the virtual pions. One
possible consequence of this phenomenon is the change of the finite size
effects in a physical quantity induced by the pion cloud: the dominant
contribution comes from the lightest pion of the theory at finite lattice
spacing, that, as we have seen, is not necessarily the charged one  
\footnote{I thank C. Michael for bringing also my attention to this issue.}.

\subsubsection{Kaon splitting}

With an interest in the phenomenology of hadrons built of $u$, $d$ and $s$
quarks, it is useful to explore Wtm including heavier quarks.
We have already seen in sect.~\ref{sec:basic} that there is no unique way to
introduce the $s$ quark into the calculation. The method investigated in
ref.~\cite{Abdel-Rehim:2006ve} considers a pair of quark doublets $(u,d)$ and
$(``c",s)$, following the proposal of ref.~\cite{Pena:2004gb} discussed in sect.~\ref{ssec:ndeglatt}. 
In particular the partner of the $s$ quark does not play an active role and should not be
thought of as the physical charm quark; moreover with this method no mass splitting is 
introduced within either doublet. 

To be specific one introduces
Wtm with two degenerate quark doublets,
\begin{equation}\label{defdoublets}
\chi_l=\begin{pmatrix} u \\ d \end{pmatrix} \,, \qquad
\chi_h=\begin{pmatrix} c \\ s \end{pmatrix} \,,
\end{equation}
referred to as the light and heavy doublets respectively. 

As we have seen in sect.~\ref{sec:basic} this construction can be extended to
include non-degenerate quarks in a single doublet~\cite{Pena:2004gb}.
Even if with this approach the fermion determinant 
is not real, and so it would not be practical to perform dynamical simulations,
this action could still be useful in the spirit of using different
lattice actions for the valence and the sea quarks. 

The two-doublet lattice 
action is simply a block-diagonal version of two copies of the one-doublet 
theory~(\ref{eq:WtmQCD}).
\be
S^L_F = a^4\sum_{x} \chibar(x)
\Big[\frac{1}{2} \sum_{\mu} \gamma_\mu (\nabla^\star_\mu + \nabla_\mu)
- \frac{a}{2} \sum_{\mu} \nabla^\star_\mu \nabla_\mu 
+ \boldsymbol{m} + i \gamma_5 \boldsymbol{\mu} \Big] \chi(x) \,,
\label{eq:can_action}
\ee
where $\nabla_\mu$ and $\nabla^\star_\mu$ are the usual covariant forward and
backward lattice derivatives respectively, and
\begin{equation}
\chi =
\begin{pmatrix}
\chi_l \\
\chi_h
\end{pmatrix} \,, \qquad
\boldsymbol{m} = 
\begin{pmatrix}
m_{l,\,0}\mathbbm{1}_2 & 0 \\
0 & m_{h,\,0}\mathbbm{1}_2
\end{pmatrix} \,, \qquad
\boldsymbol{\mu} = 
\begin{pmatrix}
\mu_{l,\,{\rm q}}\tau_3 & 0 \\
0 & \mu_{h,\,{\rm q}}\tau_3
\end{pmatrix} \,.
\end{equation}

The method used in~\cite{Abdel-Rehim:2006ve} to fix the critical mass is the
``$\omega_A$ method'', with the untwisted bare quark mass tuned so as
to achieve full twist at each
particular twisted mass value.  

\begin{figure}[!htb]
\begin{center}
\includegraphics[width=14cm]{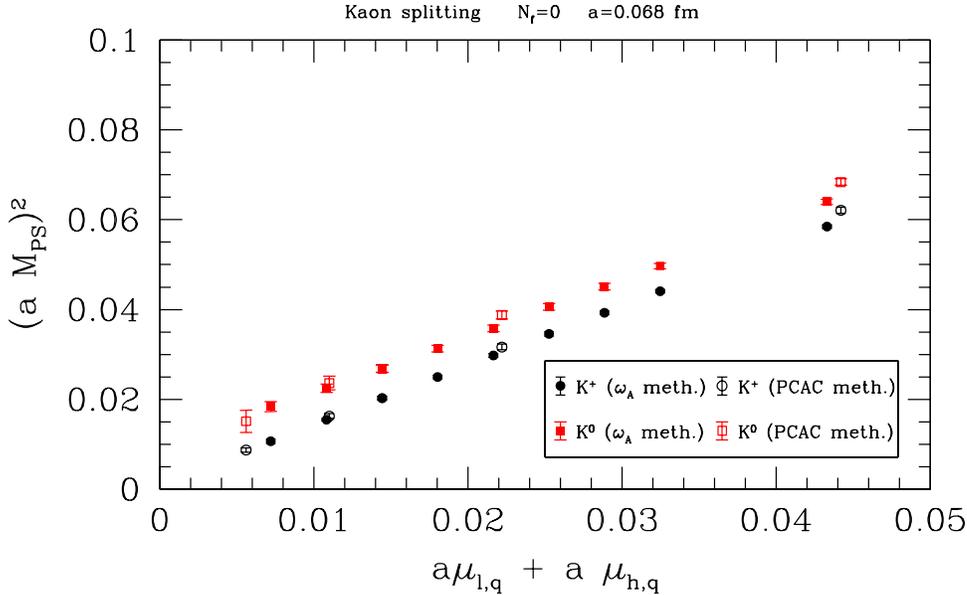}
\vspace{-5.0cm}
\caption{Pseudoscalar meson mass squared as a function of the sum of quark and
  antiquark twisted mass parameters.  Subscripts $l$ and $h$ indicate
  the light and heavy doublets. Results are taken from ref.~\protect\cite{Abdel-Rehim:2006ve}
  which uses the $\omega_A$ method and and from ref.~\protect\cite{Jansen:2005cg}
  which uses the PCAC method. The results from ~\protect\cite{Jansen:2005cg} 
  have equal masses for the quark and anti-quark.}
\label{fig:mPSvsmumu}
\end{center}
\end{figure}

\begin{figure}[!htb]
\begin{center}
\scalebox{0.45}{\includegraphics*[0cm,11mm][255mm,19cm]{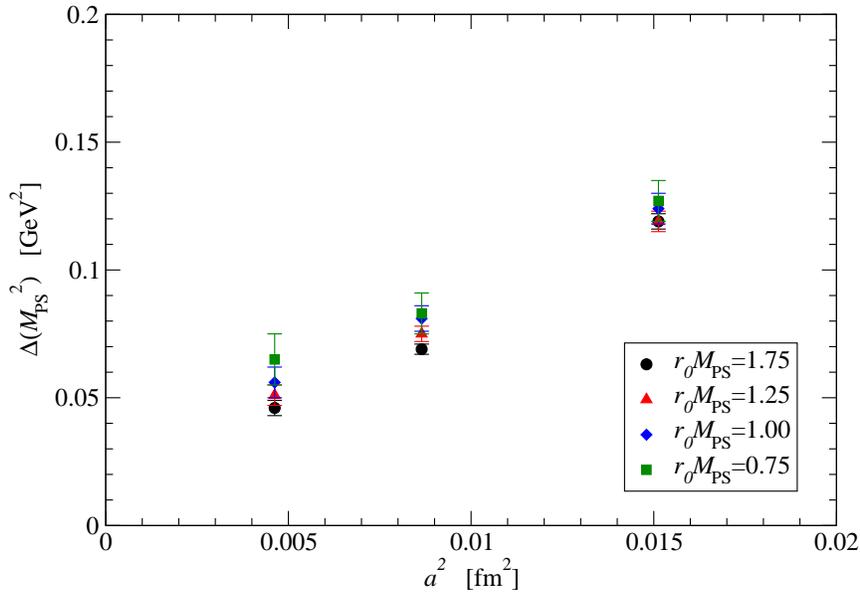}}
\caption{The difference between charged and neutral squared pseudoscalar meson
         masses as a function of squared lattice spacing, for selected
         values of the charged meson mass. Plot taken from ref.~\protect\cite{Abdel-Rehim:2006ve}}.
\label{fig:mKsqvsasq}
\end{center}
\end{figure}

NLO W$\chi$PT~\cite{Abdel-Rehim:2006ve} predicts a formula for $M_{{\rm
    K}^0}^2 - M_{{\rm K}^\pm}^2$ very similar to the formula for the pion
splitting~(\ref{eq:pionsplit}). In particular this formula, analogously to the pion
case, predicts for the kaon splitting at NLO no dependence on the twisted
quark mass if the twist angles for the light and the heavy sector are $\omega =
\pi/2$ for all the values of the twisted mass (``$\omega_{\rm A}$ method'').
The same formula, again analogously to the pion case, predicts that the kaon
splitting should depend linearly in $a^2$ and it should vanish
in the continuum limit.  Modulo the unknown higher order effects,
figs.~\ref{fig:mPSvsmumu} and~\ref{fig:mKsqvsasq} are in reasonable agreement with these expectations.
In fig.~\ref{fig:mPSvsmumu} are plotted the masses squared of the charged and neutral kaons,
{\it i.e.} the ground state pseudoscalar mesons containing one $s$ quark from
the heavy doublet and one $u$ or $d$ quark from the light doublet, as a
function of the sum of the valence twisted quark masses.
Fig.~\ref{fig:mKsqvsasq} shows the lattice spacing dependence of the squared
mass differences for four mass values. 
To summarize: the approximate mass independence of the kaon splitting is evident from
fig.~\ref{fig:mPSvsmumu}, and the
dependence on $a^2$ in fig.~\ref{fig:mKsqvsasq} is approximately linear, though a linear fit misses the
massless prediction at $a=0$ by a few (statistical) standard deviations.

It should be noted from fig.~\ref{fig:mKsqvsasq} that even at the smallest
lattice spacing, the mass splitting of $m_{K^0}-m_{K^\pm}\sim50$ MeV is
significant relative to the kaon mass itself. However, in
terms of the difference of mass squared, these results are consistent 
with the pseudoscalar meson mass splittings discussed in ref.~\cite{Jansen:2005cg}.

The splitting in the Kaon system that we have just discussed involves only twisted quarks.
Different O($a^2$) cutoff effects appear if one studies the splitting between 
pseudoscalar correlators computed with twisted and untwisted quarks.
One can consider a twisted $(u,d)$ doublet, and flavour-singlet
Wilson strange (and charm) quarks. This approach could be a viable one for doing
full dynamical simulations. 
On the other side automatic O($a$) improvement would be lost and there would be the need
to introduce suitable improvement coefficients.
In ref.~\cite{Dimopoulos:2007cn} a precise investigation has been performed, in the quenched model,
of this particular splitting between pseudoscalar correlators induced 
by the fact that twisted and untwisted actions have been used for the quarks.
The twisted and untwisted actions of~\cite{Dimopoulos:2007cn} are both clover improved, and 
all the relevant improvement coefficients have been used to improve the correlators.
The twisted and untwisted quark masses are always matched in order to have the same
continuum QCD theory with meson masses around
the Kaon mass. 
The critical mass $m_{\rm cr}$ used in~\cite{Dimopoulos:2007cn} has been taken from 
refs.~\cite{Luscher:1996ug,Rolf:2002gu,Guagnelli:2004ga} and 
it has been recomputed at one lattice spacing. 
The quantities studied in this paper are the pseudoscalar masses $M_{\rm tt}$, 
$M_{\rm tW}$, $M_{\rm WW}$ and decay constants $F_{\rm tt}$, 
$F_{\rm tW}$, $F_{\rm WW}$ for mesons which are respectively made of a doublet of twisted quarks $tt$,
a twisted and untwisted quark $tW$ and two untwisted quarks $WW$. 
The meson made of twisted quarks is the charged pseudoscalar correlator of 
eq.~\eqref{eq:pion_corrs}.
The outcome of this precise quenched study is that the splitting between
pseudoscalar meson masses and decay constants ranges, at the coarsest lattice spacing
$a=0.093$ fm, between 5-13\% and it goes
to zero in the continuum limit. The splitting seems to depend also on the value of the
improvement coefficient $c_{\rm A}$ which is needed to improve the axial current
with Wilson fermions.
In this paper the issue of the pseudoscalar splitting induced by the twisted mass
term is not addressed.

An exploratory dynamical study with two light degenerate quark and two heavier
non-degenerate quarks ($N_f=2+1+1$) has been carried out in
ref.~\cite{Chiarappa:2006ae}, following the approach explained in
sect.~\ref{sec:basic} with off-diagonal flavour
splitting~\cite{Frezzotti:2003xj}.
As we have already explained this approach has the big advantage of being well
suited for dynamical simulation. Moreover it has been noted
in~\cite{Chiarappa:2006ae} that with this formulation
the masses in the kaon doublet (and D-meson doublet)
are exactly degenerate.
This follows from an exact symmetry of the lattice action defined in
eq.~(\ref{eq:WtmQCDnondeg}).
To understand how this works, we first write two interpolating fields for the
kaon doublet in the physical basis (see sect.~\ref{sec:phys}):
\be
K^+(x) = \bar{u}(x) \gamma_5 s(x) , \qquad K^0(x) = \bar{d}(x) \gamma_5 s(x) .
\ee
Performing the rotations~(\ref{eq:chiprime},\ref{eq:axial1}) we obtain in the
twisted basis
\be
K^+(x) = -\frac{i}{\sqrt{2}} \bar{u}(x) c(x) + \frac{1}{\sqrt{2}}\bar{u}(x) \gamma_5 s(x) , \qquad K^0(x) =
-\frac{1}{\sqrt{2}} \bar{d}(x) \gamma_5 c(x)  - \frac{i}{\sqrt{2}} \bar{d}(x) s(x) .
\ee
Both the action for degenerate $u$ and $d$ quarks in eq.~(\ref{eq:WtmQCD}) and
for non degenerate $s$ and $c$ quarks in eq.~(\ref{eq:WtmQCDnondeg}) are
symmetric under parity ``times'' flavour exchange $\mathcal{P}_F^1$ defined in
eq.~(\ref{eq:PF12}).
Under this symmetry transformation the single flavoured quarks transform as
\be
\mathcal{P}^1_F \colon
\begin{cases}
u(x_0,{\bf x}) \rightarrow i \gamma_0 d(x_0,-{\bf x}) \\
d(x_0,{\bf x}) \rightarrow i \gamma_0 u(x_0,-{\bf x}) \\
c(x_0,{\bf x}) \rightarrow i \gamma_0 s(x_0,-{\bf x}) \\
s(x_0,{\bf x}) \rightarrow i \gamma_0 c(x_0,-{\bf x}) 
\end{cases}, \qquad 
\begin{cases}
\bar{u}(x_0,{\bf x}) \rightarrow -i \bar{d}(x_0,-{\bf x}) \gamma_0 \\
\bar{d}(x_0,{\bf x}) \rightarrow -i \bar{u}(x_0,-{\bf x}) \gamma_0 \\
\bar{c}(x_0,{\bf x}) \rightarrow -i \bar{s}(x_0,-{\bf x}) \gamma_0 \\
\bar{s}(x_0,{\bf x}) \rightarrow -i \bar{c}(x_0,-{\bf x}) \gamma_0 
\end{cases}.
\label{eq:PF1_single}
\ee
The effect of this {\em exact} symmetry transformation is 
\be
K^+(x_0,{\bf x}) \rightarrow K^0(x_0,-{\bf x}) , \qquad K^0(x_0,{\bf x})
\rightarrow K^+(x_0,-{\bf x}) .
\ee
Hence the equality of the masses within kaon doublets
follows. It is straightforward to repeat the same argument to show that also
the D-doublet is mass degenerate.

\subsubsection{Baryon splitting}

In principle O($a^2$) isospin breaking cutoff effects can appear also in
baryon correlators.
Here we concentrate on the spin-parity $(1/2)^{\pm}$ and $(3/2)^{\pm}$ baryons 
and we always consider Wtm with two degenerate quarks as specified by the action in eq.~(\ref{eq:WtmQCD}).

\begin{figure}[!htb]
\begin{center}
\includegraphics[width=15cm]{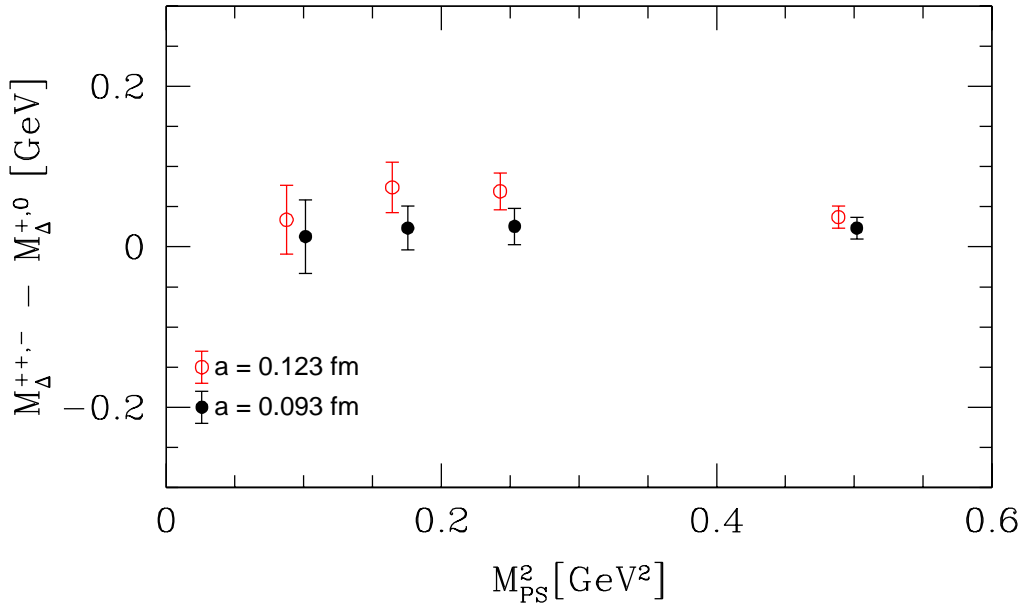}
\vspace{-6.5cm}
\caption{Flavour splitting within the $\Delta(1232)$ multiplet as a function
         of the pseudoscalar meson mass squared, for two lattice spacings:
         $a=0.123$ fm ({\color{red}\large{$\circ$}}) and $a=0.093$ fm ({\large{$\bullet$}}).
         The critical mass was fixed using the ``Wilson pion'' method.}
\label{fig:flavourwilson}
\end{center}
\end{figure}

\begin{figure}[!htb]
\begin{center}
\includegraphics[width=15cm]{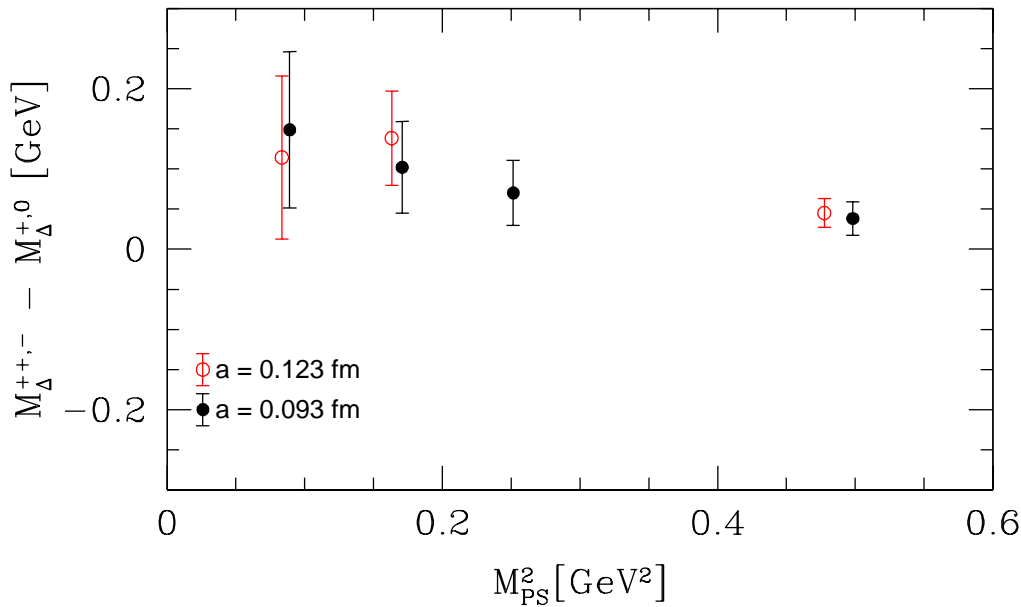}
\vspace{-6.5cm}
\caption{Same as fig.~(\protect\ref{fig:flavourwilson}) but with the critical
  mass fixed using the ``$\omega_A$ method''.}
\label{fig:flavourparity}
\end{center}
\end{figure}

We write now the interpolating fields for the baryons in the {\it physical basis}
\bea
\mathscr{P} &=& \epsilon_{ABC}[u_A^TC^{-1}\gamma_5d_B]u_C, \label{eq:proton} \\
\mathscr{N} &=& \epsilon_{ABC}[d_A^TC^{-1}\gamma_5u_B]d_C, \\
\Delta_k^{++} &=& \epsilon_{ABC}[u_A^TC^{-1}\gamma_ku_B]u_C, \\
\Delta_k^+ &=& \frac{2}{\sqrt{3}}\epsilon_{ABC}[u_A^TC^{-1}\gamma_kd_B]u_C
             + \frac{1}{\sqrt{3}}\epsilon_{ABC}[u_A^TC^{-1}\gamma_ku_B]d_C, \\
\Delta_k^0 &=& \frac{2}{\sqrt{3}}\epsilon_{ABC}[d_A^TC^{-1}\gamma_ku_B]d_C
             + \frac{1}{\sqrt{3}}\epsilon_{ABC}[d_A^TC^{-1}\gamma_kd_B]u_C, \\
\Delta_k^- &=& \epsilon_{ABC}[d_A^TC^{-1}\gamma_kd_B]d_C, \label{eq:Delta}
\eea
where $C$ is the charge conjugation matrix defined in app.~\ref{app:B}.
To get the corresponding interpolating fields in the {\it twisted basis} it is enough to apply the rotation 
in eq.~(\ref{eq:axial}) to the quark fields. In app.~\ref{app:F}, I show how this works
with the proton field.
Here I just want to prove that the proton and the neutron with Wtm are degenerate even at finite 
lattice spacing. This was shown in ref.~\cite{Abdel-Rehim:2005gz} and here we give a slightly different
argument based on the symmetries of the lattice action.
In particular we know that parity ``times'' flavour exchange $\mathcal{P}^1_F$, 
defined in eq.~(\ref{eq:PF12}) is a symmetry of Wtm. This symmetry is
independent on the basis of choice, in fact also the action in the physical
basis~(\ref{eq:fulltWQCD}) is invariant under the $\mathcal{P}^1_F$ symmetry transformation.
Applying this symmetry to the baryon fields we obtain
\be
\mathscr{P}(x_0,{\bf x}) \rightarrow -i\gamma_0 \mathscr{N}(x_0,{\bf -x})
\ee
\vspace{-0.8cm}
\be
\Delta_k^{++}(x_0,{\bf x}) \rightarrow -i\gamma_0 \Delta_k^-(x_0,{\bf -x})
\ee
\be
\Delta_k^+(x_0,{\bf x}) \rightarrow -i\gamma_0 \Delta_k^0(x_0,{\bf -x}).
\ee
\vspace{0.3cm}
If we consider a baryon field $\mathscr{B}$ and the corresponding correlation function
\be
a^3\sum_{\bf x} \langle \mathscr{B}(x_0,{\bf x}) \bar{\mathscr{B}}(0,{\bf 0}) \rangle , 
\ee
we conclude that proton and neutron are degenerate as are $\Delta_k^{++}$ with $\Delta_k^-$ and 
$\Delta_k^+$ with $\Delta_k^0$. There is on the other hand an O($a^2$) 
splitting between $\Delta_k^{++}$ and $\Delta_k^+$.
This splitting it is also numerically simpler to study than the pion splitting since it does not
require computation of disconnected diagrams. 
This was in fact noticed and studied in ref.~\cite{Abdel-Rehim:2005gz} in the quenched model.
Figs.~\ref{fig:flavourwilson} and \ref{fig:flavourparity} show the results for
the splitting in physical units in the $\Delta$ channel, as obtained respectively from
the ``Wilson pion'' definition of the critical mass and the ``$\omega_A$ method''.
No significant difference
is found between the two definitions of full twist for this observable. Based on
the larger statistics of fig.~\ref{fig:flavourwilson} (1000 configurations
rather than only 300), there is some evidence that
$m_{\Delta^{++,-}}-m_{\Delta^{+,0}}$ decreases as $a \rightarrow 0$, as expected.

The extension of W$\chi$PT, with a twisted mass term, to include the baryons 
has been worked out in ref.~\cite{WalkerLoud:2005bt}.
In this work, beside showing how to include the baryons in the effective theory, 
several interesting results have been obtained.
It has been confirmed automatic O($a$) improvement at full twist for baryon masses,
it has been discussed the splittings among the delta-resonances and a mass splitting formula has been given
for a genric twist angle $\omega$.

\subsubsection{Conclusions}

Isospin breaking cutoff effects are most probably the most delicate issue concerning Wtm.
In the pseudoscalar sector they are similar in magnitude with the staggered taste
violating cutoff effects, and they are even smaller if we include 
the disconnected diagrams in the pion sector.
The possibility to use the OS action for the valence quarks partially
mitigates this problem. 
The mixed action approach, while removing completely the isospin breaking at
the valence level, cannot cure completely the problem.
For physical quantities which involve 
pion scattering, isospin breaking cutoff effects are an issue which has to be
further investigated.
In the baryon sector isospin breaking cutoff effects seem to be under control.
Moreover choosing as a lattice action for the heavy non-degenerate quarks the action
proposed in~\cite{Frezzotti:2003xj} the splitting in the K and D sector is absent.

\subsection{Phase structure of Wilson fermions}
\label{ssec:phase}

In the first part of this sect.~\ref{sec:asq}, we have discussed the O($a^2$) 
isospin breaking cutoff effects. Now we want to analyze the impact of 
the O($a^2$) discretization errors the shape of the chiral phase diagram.
The O($a^2$) chirally breaking cutoff effects change the form of the chiral 
phase transition. Thus the knowledge of the shape of the chiral phase diagram
is an important prerequisite to perform large scale simulations with a given
lattice action.
In this section I show that both
numerically and analytically there is strong evidence that Wilson type
fermions show a peculiar phase diagram around the chiral point.
The structure of such a phase diagram is due to lattice artifacts.
We will see in sect.~\ref{sec:algo} that the approach to the chiral and continuum
limit in numerical simulations shows a substantial
slowing down in algorithmic performance, 
so in general a careful continuum limit and
chiral extrapolation is required in order to compare lattice results with
experimental measurements. The knowledge of the position of the phase
transition point as a function of the quark mass and the lattice spacing, becomes then
very important, given also the fact that it is not clear whether HMC-like
algorithms can correctly sample the configuration space in such extreme
situations.
In particular the shape of the phase structure is such that there is a minimal pion
mass that can be simulated at fixed lattice spacing. This bound depends on the details of
the Wilson-like action used in the simulations (e.g. on the gauge action).
Lattice simulations are performed in a finite volume and strictly speaking
there are no phase transitions in a finite box, but if the lattice size is
large enough, the shadow of the would-be phase transition 
could still be visible in the numerical simulations.

\subsubsection{W$\chi$PT analysis}
\label{sssec:WCPT_phase}

\begin{figure}[htb]
\centering
\psfrag{m}{$\mathfrak{m}$}
\psfrag{n}{$\mathfrak{n}$}
\includegraphics[width=2.5in]{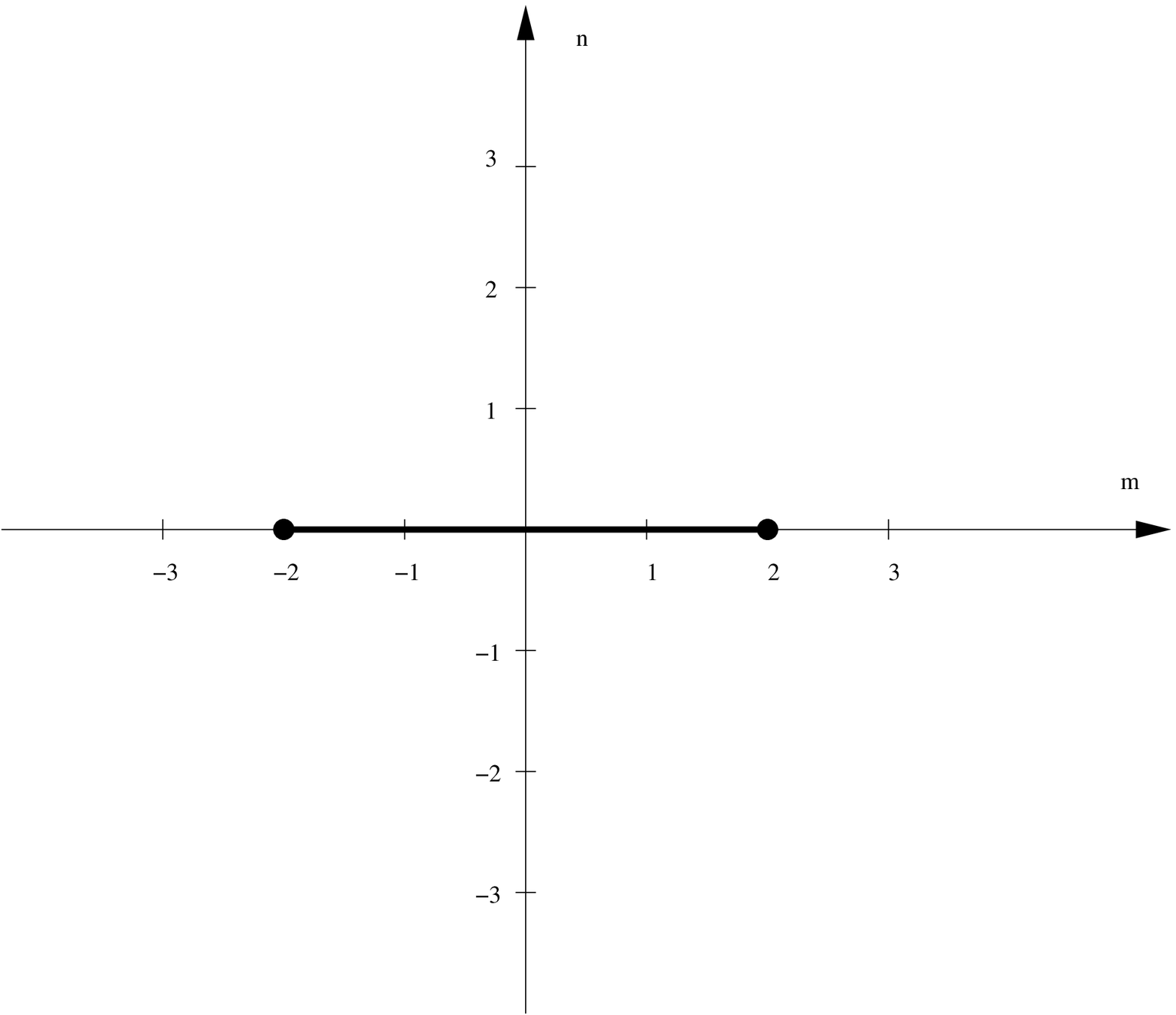}
\includegraphics[width=2.5in]{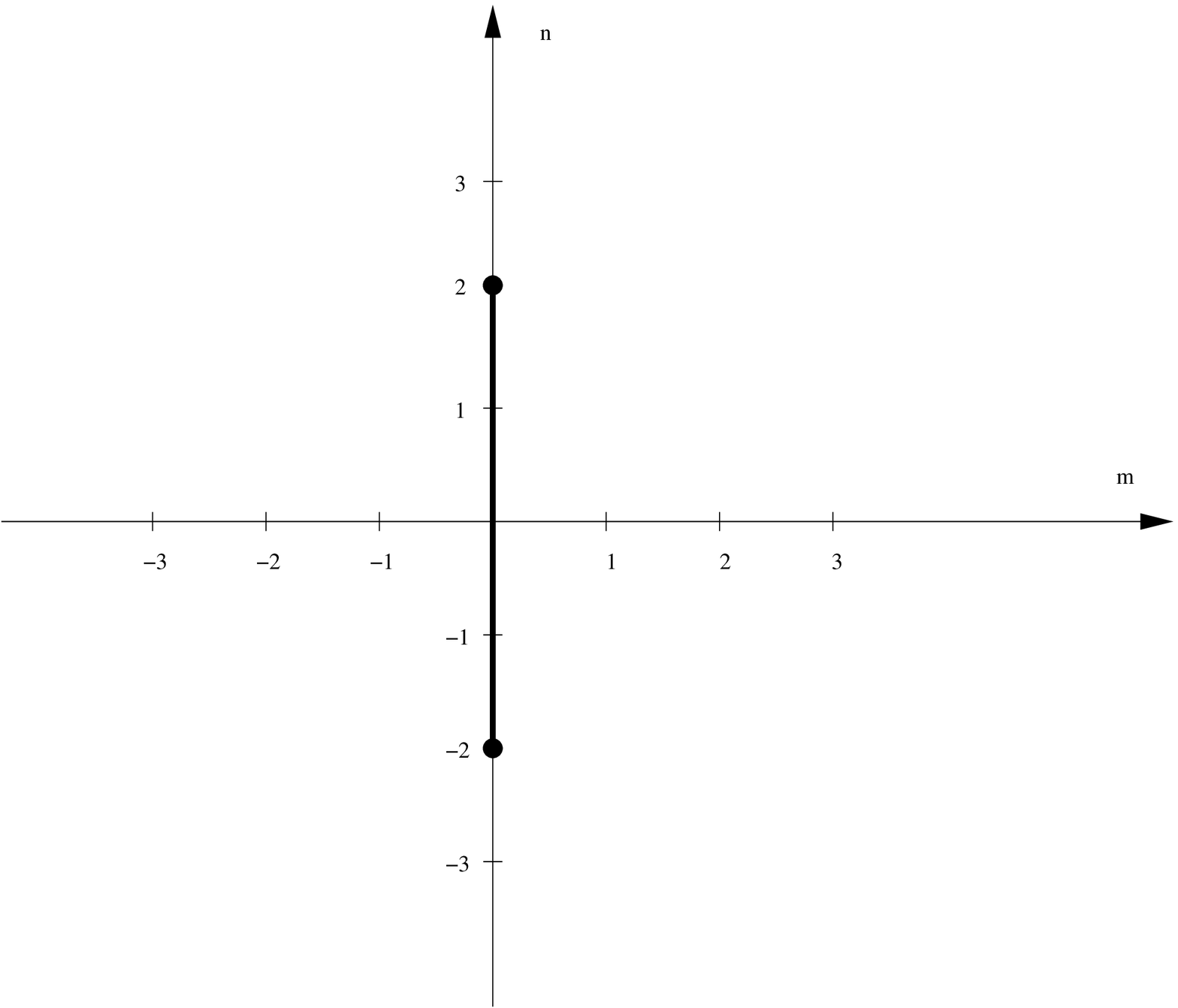}
\caption{Left panel: the phase diagram of Wilson diagram according to W$\chi$
  PT for $c_2>0$. Right panel: as in the left panel but for $c_2<0$. The x-axis
  is proportional to $m'/a^2$ and the y-axis to $\mu_{\rm R}/a^2$~(\ref{eq:mresc},\ref{eq:muresc}).}
\label{fig:tmphase}
\end{figure}

Before going into the study of the phase diagram of Wilson-like fermions
let us recall which is the situation in continuum QCD. In the continuum chiral
symmetry is spontaneously broken in the massless limit, and the order
parameter of such a transition is the chiral condensate. 
The structure of the phase diagram is thus simply: a singular point at the origin 
of the ``mass plane'' where we plot the untwisted quark mass along the horizontal axis 
and the twisted mass along the vertical axis (see fig.~\ref{fig:tmphase}).
The aim of this section is to understand how this picture is modified by the
inclusion of lattice artifacts in the chiral Lagrangian.
We have already discussed in sect.~\ref{sssec:wchipt} the construction of the chiral
Lagrangian including lattice spacing effects. In particular our starting
point was the power counting scheme specified in eq.~(\ref{eq:GSM}).
We have seen in sect.~\ref{sssec:wchipt} that at LO the inclusion of O($a$) lattice
artifacts can be fully reabsorbed in a shifted quark mass $m'$, defined in
eq.~(\ref{eq:mprime}). We can then immediately conclude that if we are in the
power counting scheme given by eq.~(\ref{eq:GSM}) the phase structure is
continuum like. To start to be sensitive to the modifications of the phase structure in
presence of lattice artifacts we have to further lower the quark masses at
fixed lattice spacing. In particular the appropriate power counting is given
\be
1 \gg m',\mu_{\rm R},p^2,a^2 \gg \ldots
\label{eq:aoki}
\ee
This region of quark masses is usually called the Aoki region because it is
the region where a possible Aoki phase appears~\cite{Aoki:1984qi}.
The Aoki phase, as we will see below, is characterized by a region in the bare parameters
space where parity and flavour are spontaneously broken.

Since we are changing the power counting scheme, the careful reader could
wonder if this does not imply a breakdown of the expansion.
The reason why this is not the case is that there is a rearrangement
in the ordering of the coefficients of the chiral Lagrangian.
It is possible to show~\cite{Sharpe:1998xm} that the neglected
terms of the expansion are at most of order $\sim a^3$,
and thus suppressed by one power of $a$.
Loop corrections are also suppressed, since they are
quadratic in $m'$ and $\mu_{\rm R}$ (up to logarithms) and thus $\sim a^4$.
The reordering of the expansion is possible only because
the leading order discretization error has exactly the
form of a mass term and so can be completely absorbed
into $m'$, to all orders in the chiral expansion~\cite{Sharpe:1998xm}.

Upon neglecting the derivative interaction terms, as we are interested in the vacuum state,
the potential of the effective chiral Lagrangian at
LO~\cite{Munster:2004am,Scorzato:2004da,Sharpe:2004ps,Aoki:2004ta} 
is given by
\be
V_{\chi} = 
-\frac{c_1}{4} \langle \Sigma + \Sigma^{\dagger} \rangle 
+ \frac{c_2}{16} \langle \Sigma + \Sigma^{\dagger} \rangle^2
+ \frac{c_3}{4} \langle i(\Sigma - \Sigma^{\dagger}) \tau_3
\rangle
\label{eq:chipot}
\ee
where the explicit forms of the coefficients,
and their sizes in our power-counting scheme, are
\begin{align} 
c_1 &= 2B_0f^2m' \sim m', \notag \\ 
c_2 &= - f^2 w' a^2 \sim a^2 , \notag \\
c_3 &= 2B_0f^2 \mu_{\rm R} \sim \mu_{\rm R}, \,.
\label{eq:ChLcoefs}
\end{align}
All the $c_i$ coefficients are independent at non-zero lattice spacing and
the only extra term introduced by the twisted mass is the one proportional to $c_3$.

In the Aoki region the coefficient $c_2$ is of
the same size as the coefficients $c_{1,3}$; all are of $O(a^2)$.
This implies a competition of the mass terms and the
O($a^2$) term in the shape of the potential that causes a non-trivial vacuum
structure.
Minimizing the potential~(\ref{eq:chipot}) gives rise to two possible 
scenarios~\cite{Sharpe:1998xm,Munster:2004am,Scorzato:2004da,Sharpe:2004ps} 
for the phase diagram of Wilson-like fermions: \\
\vspace{-0.6cm}
\begin{itemize}
\item the Aoki scenario \cite{Aoki:1984qi}; \\
\item the Sharpe-Singleton scenario~\cite{Sharpe:1998xm}.
\end{itemize}
Note that {\em both} $m'$ and $\mu_{\rm R}$ must be of $O(a^2)$ in order 
for such competition to occur;
if either $m'$ or $\mu_{\rm R}$ is of $O(a)$ then one
is in the continuum-like region.
I would like to remark that the presence of two possible phase structures was
foreseen by Creutz~\cite{Creutz:1996bg}.

The order parameter of the continuum chiral phase
transition is the chiral condensate.
The condensate in W$\chi$PT is determined by the minima of the potential energy~(\ref{eq:chipot}). 
We parametrize the chiral field in the standard way:
 $\Sigma = \Sigma_0 + i\Sigma_a \cdot \tau^a$ with 
real $\Sigma_0$ and $\Sigma_a$ satisfying $\Sigma_0 \cdot \Sigma_0 + \Sigma_a \cdot \Sigma_a = 1$,
so that  $\Sigma_0, \, \Sigma_a \in [-1,1]$.
Similarly, the condensate (the value of $\Sigma$ at the minimum
of the potential) is written 
$\Sigma^{(m)}_0 = \Sigma_0^{(m)} + i\Sigma^{(m)}_a \cdot \tau^a$.
In the Aoki region the potential is then
\be
V_{\chi} = -c_1 \Sigma_0 - c_3 \Sigma_3 + c_2 \Sigma_0^2\,.
\label{eq:pot}
\ee
We have now to compute the values of $\Sigma_0^{(m)}$ and $\Sigma_3^{(m)}$.
We will not enter in the details of the computations that are nevertheless
rather simple.
One finds that there two possible solutions depending on the sign of
$c_2$ (or $w'$). Thus it is the sign of $c_2$ (or $w'$) that determines the
possible scenario for the chiral phase structure of Wilson-like fermions. 
We summarize here the results illustrating them with plots~\cite{Sharpe:2004ps}.
For convenience we define the following rescaled mass variables
\be
\mathfrak{m} = \frac{c_1}{|c_2|} = 
\frac{2B_0 m'}{a^2 |w'|}
\sim m'/a^2 
\label{eq:mresc}
\ee
\be
\mathfrak{n} = \frac{c_3}{|c_2|} = 
\frac{2 B_0 \mu_{\rm R}}{a^2 |w'|} \sim \mu_{\rm R}/a^2 \,.
\label{eq:muresc}
\ee
They are of O(1) in the region of interest.

In fig.~\ref{fig:tmphase} I plot the lines corresponding to a first order
phase transition which have second order phase transitions end points,
where the x-axis is $\mathfrak{m}$ and the
y-axis is $\mathfrak{n}$. 

I first discuss the structure of the phase diagram for $c_2$ positive or negative
and then I will describe the physics one finds in the two situations.

In the left panel of fig.~\ref{fig:tmphase} the Aoki scenario
that appears if $c_2 >0$ ($w'<0$) is depicted. In this scenario along the Wilson axis
$\mathfrak{m}$ there is a region (indicated by the thick line) where parity 
and flavour are spontaneously broken: this is the Aoki phase~\cite{Aoki:1984qi}. 
At the endpoints of this line the three pions are massless. 
Inside the Aoki phase the charged pions stay massless while 
the neutral one becomes massive.
A non-zero value of the twisted mass washes out the Aoki
phase introducing an explicit breaking of flavour and parity symmetry. 

In the right panel of fig.~\ref{fig:tmphase} the
Sharpe-Singleton scenario that appears if $c_2 <0$ ($w'>0$) is depicted. 
In this scenario the first order phase transition line extends into
the twisted direction. The transition ends
with a second order phase transition point, where the neutral pion mass vanishes.

\bigskip
We go now to study the mass dependence of the condensate and
of the pion masses inside the Aoki region.
\begin{figure}
\centering
\psfrag{A0}[][]{\large $\Sigma^{(m)}_0 $}
\psfrag{a}{\large $\mathfrak{m}$}
\psfrag{b0}{\tiny $\mathfrak{n} = 0$}
\psfrag{b1}{\tiny $\mathfrak{n} = 1$}
\psfrag{b2}{\tiny $\mathfrak{n} = 2$}
\psfrag{b3}{\tiny $\mathfrak{n} = 3$}
\includegraphics[width=3in]{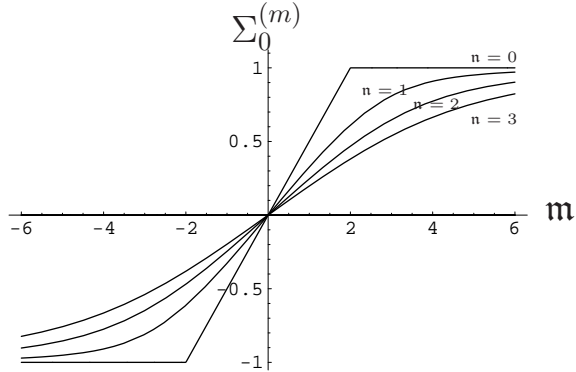}
\caption{\label{fig:c2gt0Am} The global minimum of the potential,
  $\Sigma^{(m)}_0 $, as a function of $\mathfrak{m}$, for $c_2 > 0$ and $\mathfrak{n} =
  0,\,1,\,2,\,3$.}
\end{figure}

\begin{figure}
\centering
\subfigure[Mass of $\pi_1$ and $\pi_2$]{
\psfrag{m2f2c2}[][]{\large $M^2_{\pi^{1,2}} / a^2 |w'|$}
\psfrag{a}{\large $\mathfrak{m}$}
\psfrag{b0}{\small $\mathfrak{n} = 0$}
\psfrag{b1}{\small $\mathfrak{n} = 1$}
\psfrag{b2}{\small $\mathfrak{n} = 2$}
\psfrag{b3}{\small $\mathfrak{n} = 3$}
\label{fig:c2gt0mpi:a}
\includegraphics[width=3in]{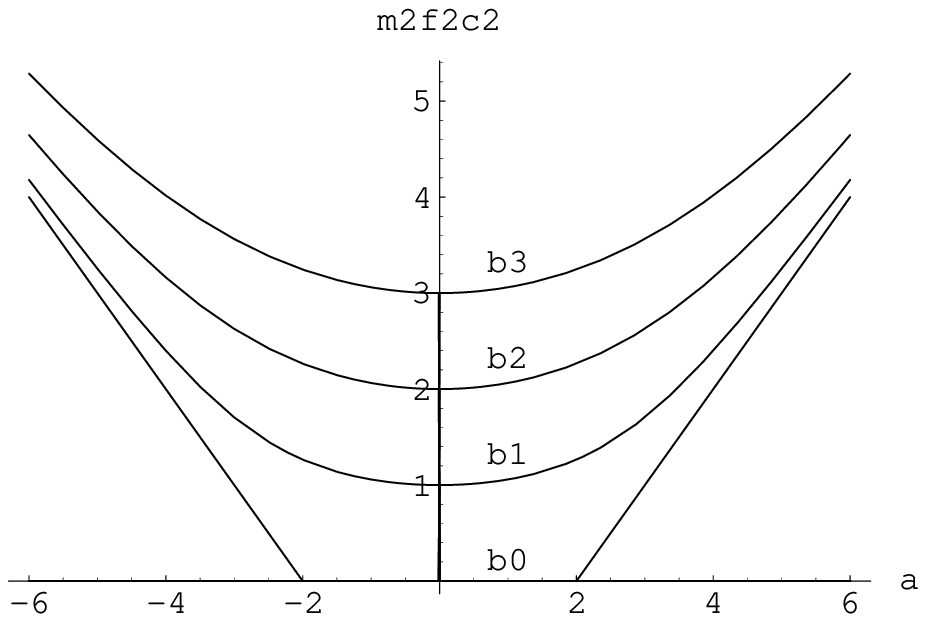}}
\hspace{0.1in}
\subfigure[Mass of $\pi_3$]{
\psfrag{m2f2c2}[][]{\large $M^2_{\pi^3} / a^2|w'|$}
\psfrag{a}{\large $\mathfrak{m}$}
\psfrag{b0}{\small $\mathfrak{n} = 0$}
\psfrag{b1}{\small $\mathfrak{n} = 1$}
\psfrag{b2}{\small $\mathfrak{n} = 2$}
\psfrag{b3}{\small $\mathfrak{n} = 3$}
\label{fig:c2gt0mpi:b}
\includegraphics[width=3in]{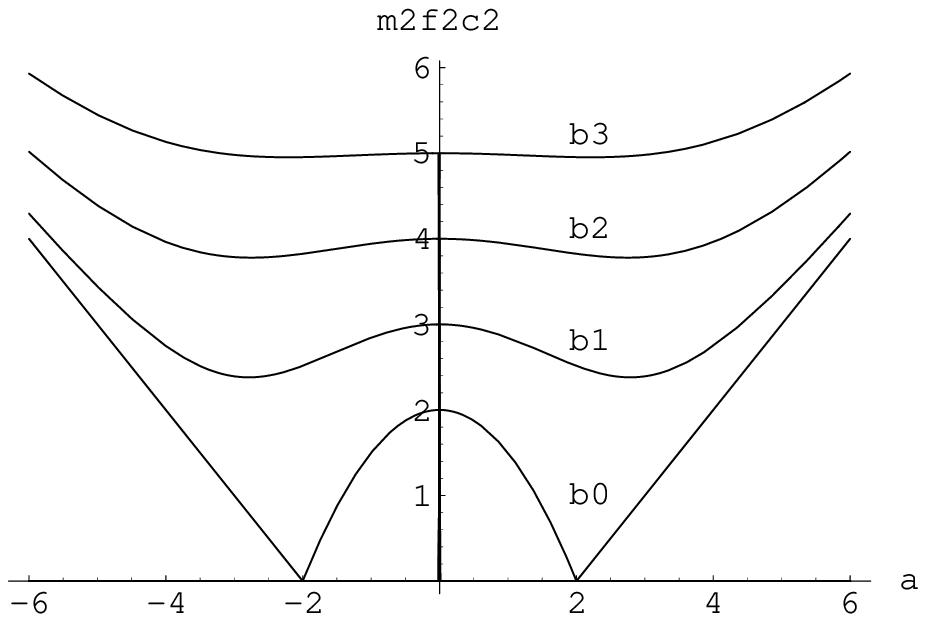}}
\caption{\label{fig:c2gt0mpi} Mass of the pions as a function of
  $\mathfrak{m}$, for $c_2 > 0$ and $\mathfrak{n} = 0,\,1,\,2,\,3$.} 
\end{figure}

We begin with results for $c_2 > 0$. 
Fig.~\ref{fig:c2gt0Am} shows the behaviour of the 
identity component of the condensate, $\Sigma^{(m)}_0 =\langle
\Sigma^{(m)}\rangle /2$,
for $\mathfrak{n}=0$, $1$, $2$ and $3$ as a function of $\mathfrak{m}$.
The corresponding pion masses are shown in fig.~\ref{fig:c2gt0mpi}.
In the untwisted theory ($\mathfrak{n}=0$)
there are second order transitions at $m' = \pm \frac{a^2 |w'|}{B_0}$,
as shown by the discontinuity in the derivatives respect to $\mathfrak{m}$ 
in $\Sigma^{(m)}_0$ and the vanishing of the
pion masses. In the Aoki phase, with $\Sigma^{(m)}_3 \ne 0$,
the $SU_{\rm V}(2)$ vector flavour symmetry is spontaneously broken into $U_{\rm V}(1)_3$,
and correspondingly there are two Goldstone bosons 
between these second-order points.
Once $\mathfrak{n}$ is non-vanishing, however, the transition is smoothed
out into a crossover, and the pion masses are always non-zero.
Flavour is broken for all $\mathfrak{m}$, with the charged pions lighter
than the neutral pion by $O(a^2)$, as given by eq.~(\ref{eq:pionsplit}).
For the special case of $\mathfrak{m}=0$ (full twist), $\Sigma^{(m)}_0$ vanishes
and $(\Sigma^{(m)}_3)^2=1$, so the mass-squared splitting is 
$2 a^2|w'|$ for all $\mathfrak{n}$ (which becomes a difference of
2 in the units in the plots).

If we keep $-2 \le \mathfrak{m} \le 2$ fixed, and we lower $\mathfrak{n}$,
it is possible to pass through the Aoki phase.
When we change the sign of twisted mass, i.e. passing through $\mathfrak{n} = 0$, 
there is a first-order phase transition, since $\Sigma^{(m)}_3$ 
jumps from $+\sqrt{1-(\Sigma^{(m)}_0)^2}=\sqrt{1-\mathfrak{m}^2/4}$ 
to $-\sqrt{1-(\Sigma^{(m)}_0)^2}=-\sqrt{1-\mathfrak{m}^2/4}$.

\bigskip

\begin{figure}
\centering
\subfigure[Global minimum, $\mathfrak{n} = 0$]{
        \psfrag{A0}[][]{\small $\Sigma^{(m)}_0$}
        \psfrag{a}{\small $\mathfrak{m}$}
        \label{fig:c2lt0:a}
        \includegraphics[width=2.2in]{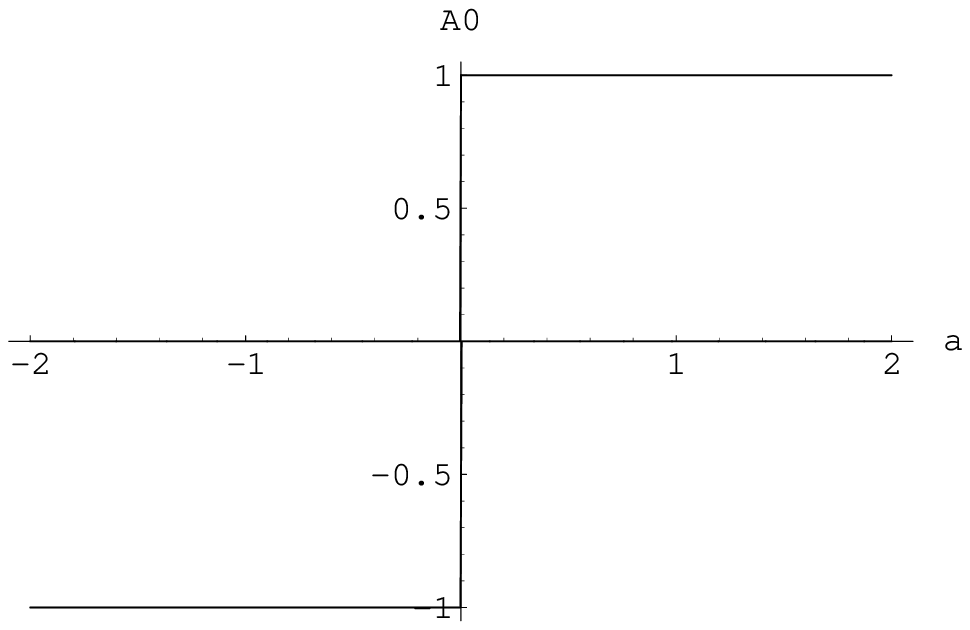}}
\hspace{0.1in}
\subfigure[Pion masses, $\mathfrak{n} = 0$]{
        \psfrag{m2f2c2}[][]{\small $M^2_\pi / a^2|w'|$}
        \psfrag{a}{\small $\mathfrak{m}$}
        \label{fig:c2lt0:b}
        \includegraphics[width=2.2in]{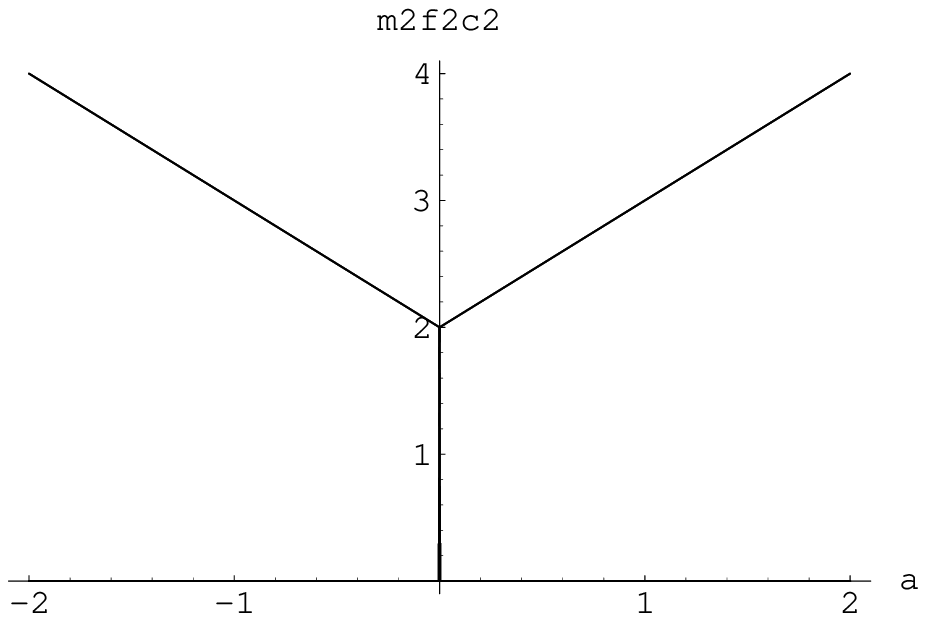}}

\subfigure[Global minimum, $\mathfrak{n} = 1$]{
        \psfrag{A0}[][]{\small $\Sigma^{(m)}_0$}
        \psfrag{a}{\small $\mathfrak{m}$}
        \label{fig:c2lt0:c}
        \includegraphics[width=2.2in]{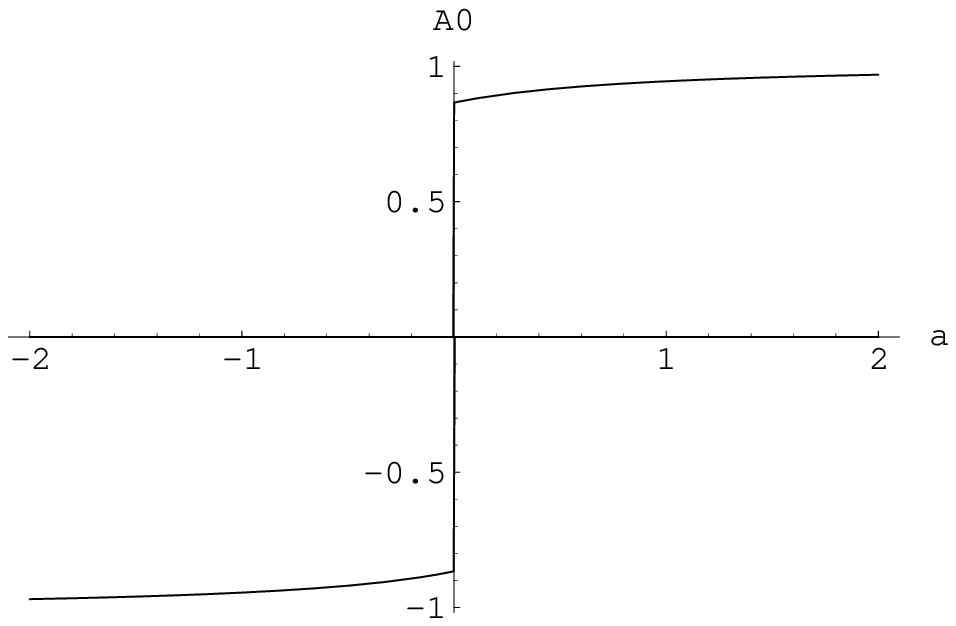}}
\hspace{0.1in}
\subfigure[Pion masses, $\mathfrak{n} = 1$]{
        \psfrag{m2f2c2}[][]{\small $M^2_\pi / a^2|w'|$}
        \psfrag{a}{\small $\mathfrak{m}$}
        \label{fig:c2lt0:d}
        \includegraphics[width=2.2in]{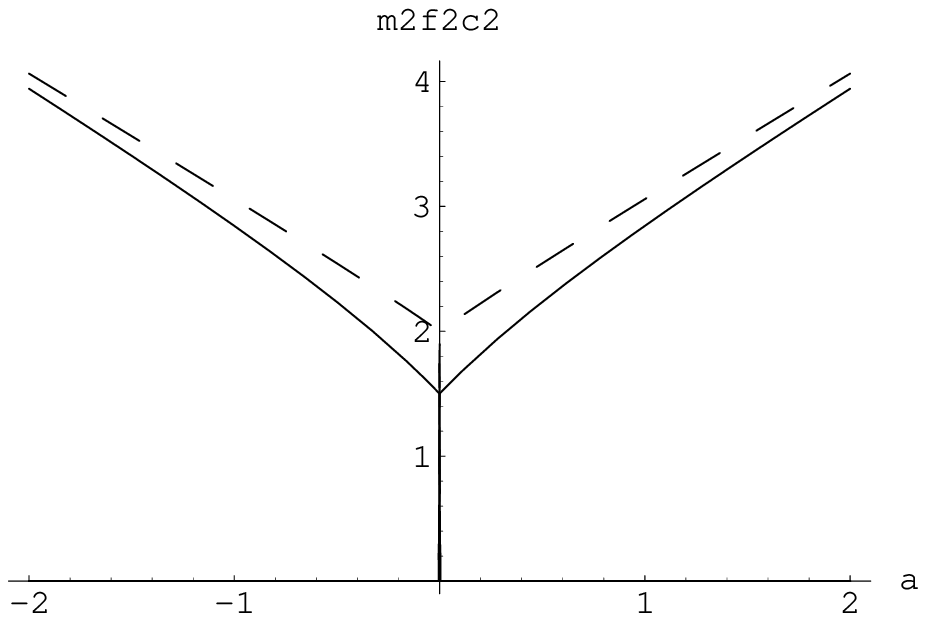}}

\subfigure[Global minimum, $\mathfrak{n} = 2$]{
        \psfrag{A0}[][]{\small $\Sigma^{(m)}_0$}
        \psfrag{a}{\small $\mathfrak{m}$}
        \label{fig:c2lt0:e}
        \includegraphics[width=2.2in]{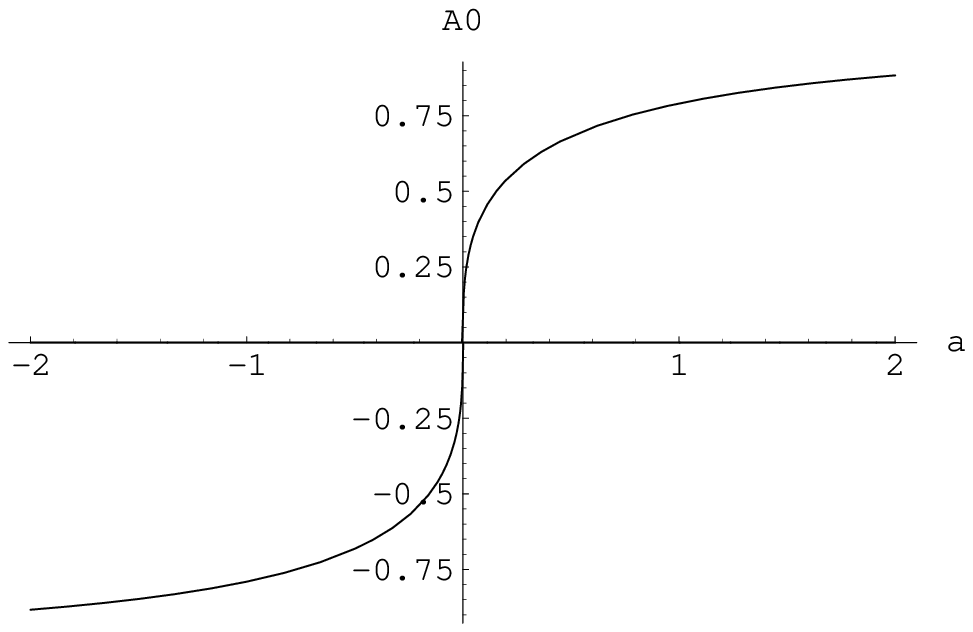}}
\hspace{0.1in}
\subfigure[Pion masses, $\mathfrak{n} = 2$]{
        \psfrag{m2f2c2}[][]{\small $M^2_\pi / a^2|w'|$}
        \psfrag{a}{\small $\mathfrak{m}$}
        \label{fig:c2lt0:f}
        \includegraphics[width=2.2in]{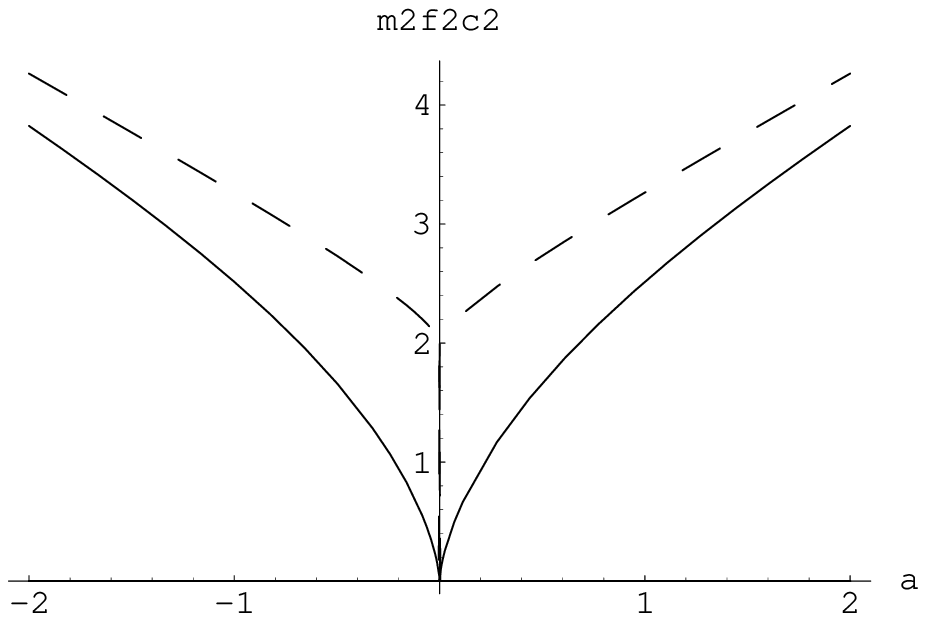}}

\subfigure[Global minimum, $\mathfrak{n} = 3$]{
        \psfrag{A0}[][]{\small $\Sigma^{(m)}_0$}
        \psfrag{a}{\small $\mathfrak{m}$}
        \label{fig:c2lt0:g}
        \includegraphics[width=2.2in]{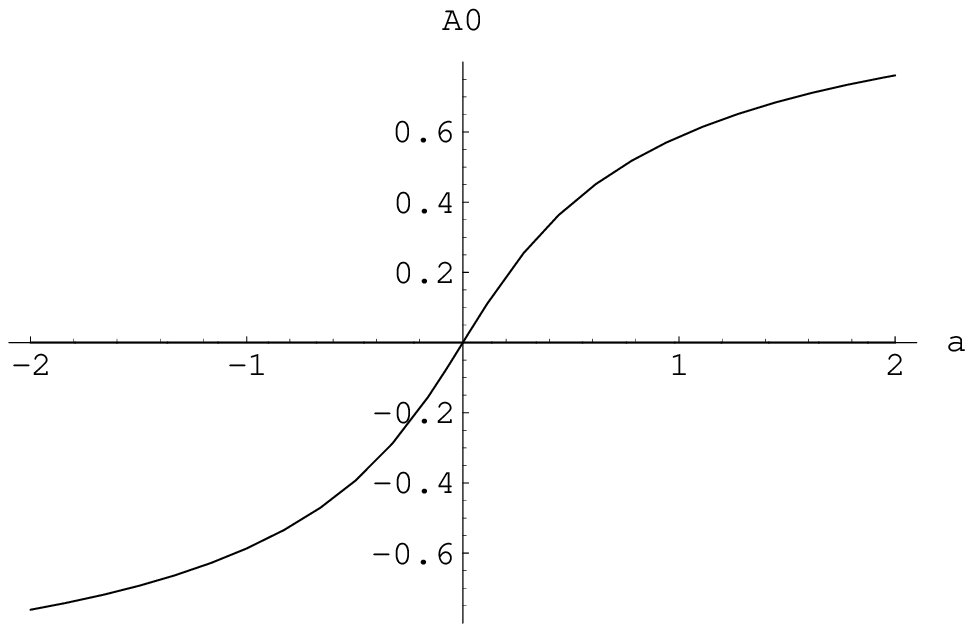}}
\hspace{0.1in}
\subfigure[Pion masses, $\mathfrak{n} = 3$]{
        \psfrag{m2f2c2}[][]{\small $M^2_\pi / a^2|w'|$}
        \psfrag{a}{\small $\mathfrak{m}$}
        \label{fig:c2lt0:h}
        \includegraphics[width=2.2in]{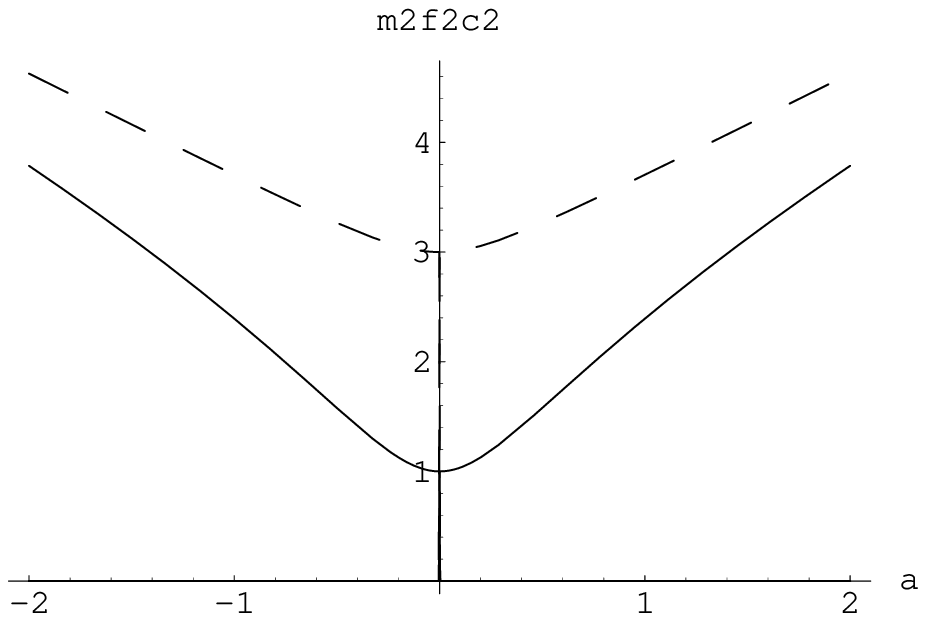}}
\caption{\label{fig:c2lt0} Global minimum $\Sigma^{(m)}_0$ and pion masses as a
  function of $\mathfrak{m}$, for $c_2 < 0$ and $\mathfrak{n} = 0,\,1,\,2,\,3$. The
  dashed lines are for $\pi_{1,2}$ and the solid lines are for $\pi_3$.}
\end{figure}

We now consider the case where $c_2$ is negative.
Fig.~\ref{fig:c2lt0} shows $\Sigma^{(m)}_0$ and the pion masses as
a function of $\mathfrak{m}$ for fixed values of $\mathfrak{n}$.
Fig.~\ref{fig:c2lt0:a} and fig.~\ref{fig:c2lt0:b} show the results 
for $\mu_{\rm R}=0$. The condensate
jumps from $\Sigma_0 = 1$ (and thus $\Sigma^{(m)}_0=1$, $\Sigma^{(m)}_a=0$) for $\mathfrak{m} > 0$ 
to $\Sigma_0=-1$ (and thus $\Sigma^{(m)}_0 = -1$, $\Sigma^{(m)}_a=0$) for $\mathfrak{m} < 0$. 
This is a first order transition without flavour breaking, so all pions remain massive and degenerate.

The rest of fig.~\ref{fig:c2lt0} shows what happens at non-zero
twisted mass. The effect of $\mu_{\rm R}$ is to twist the condensate, so that
there is a non-zero $\tau_3$ component $\Sigma^{(m)}_3$.
There is, however, still a first order transition at which $\Sigma^{(m)}_3$
flips sign between $\pm (1-\mathfrak{n}/2)$ (assuming $\mathfrak{n}>0$).
The neutral pion is now lighter than the charged pions due to
explicit flavour breaking. The neutral pion has a mass
$M_{\pi_3}^2= 2a^2 |w'|(1-\mathfrak{n}/2)^2$ at the transition, 
while, as noted above, the charged pions have
a $\mathfrak{n}$ independent mass given by $m_{\pi_{1,2}}^2=2 a^2|w'|$.
The transition weakens as $|\mathfrak{n}|$ increases, and ends with
a second order transition point at $\mathfrak{n}=\pm 2$,
at which the neutral pion is massless [see Fig.~\ref{fig:c2lt0:f})]. 
For larger $\mathfrak{n}$ the transition is smoothed out.
Note that, once away from the transition, for $\mathfrak{m}=0$ the
mass-squared splitting between charged and neutral pions is still 
$2 a^2 |w'|$ and independent from $\mathfrak{n}$.

\begin{figure}
\centering
\subfigure[Mass of $\pi_1$ and $\pi_2$]{
\psfrag{m2f2c2}[][]{\large $M^2_\pi / a^2|w'|$}
\psfrag{b}{\large $\mathfrak{n}$}
\psfrag{a0}{\small $\mathfrak{m} = 0$}
\psfrag{a1}{\small $\mathfrak{m} = 1$}
\psfrag{a2}{\small $\mathfrak{m} = 2$}
\psfrag{a3}{\small $\mathfrak{m} = 3$}
\label{fig:c2lt0mpi:a}
\includegraphics[width=3in]{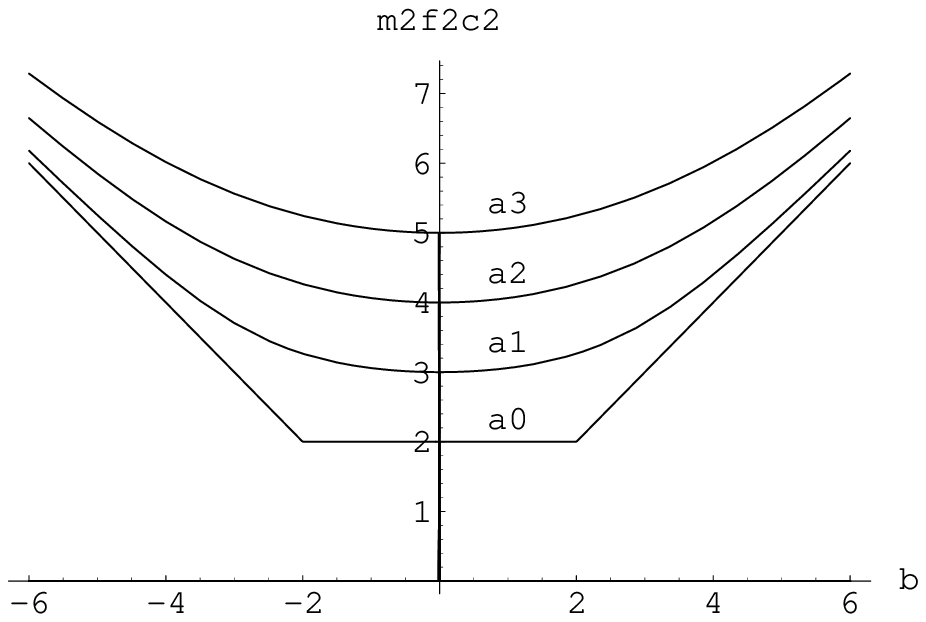}}
\hspace{0.1in}
\subfigure[Mass of $\pi_3$]{
\psfrag{m2f2c2}[][]{\large $M^2_\pi / a^2|w'|$}
\psfrag{b}{\large $\mathfrak{n}$}
\psfrag{a0}{\small $\mathfrak{m} = 0$}
\psfrag{a1}{\small $\mathfrak{m} = 1$}
\psfrag{a2}{\small $\mathfrak{m} = 2$}
\psfrag{a3}{\small $\mathfrak{m} = 3$}
\label{fig:c2lt0mpi:b}
\includegraphics[width=3in]{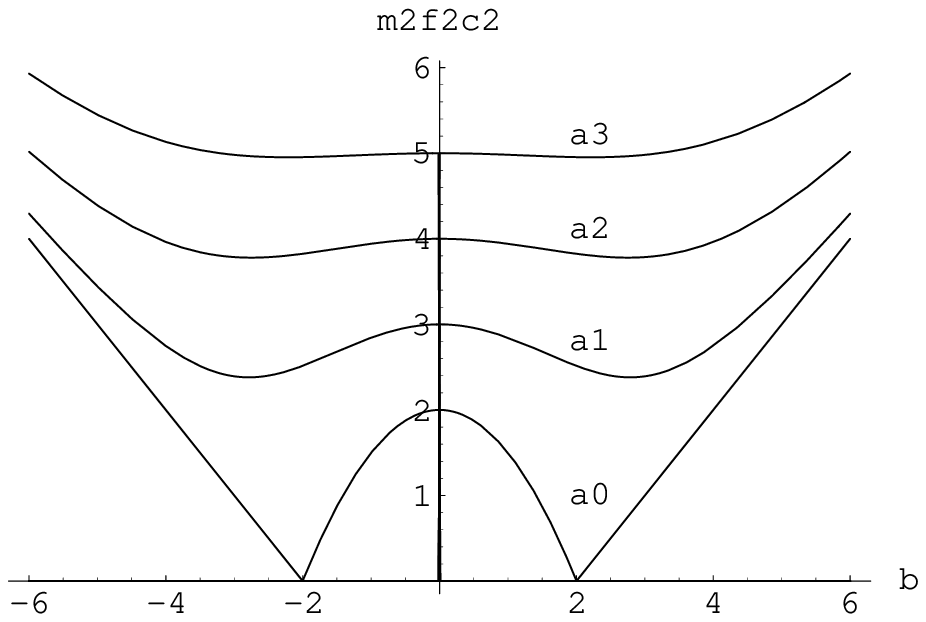}}
\caption{\label{fig:c2lt0mpi} Mass of the pions as a function of
  $\mathfrak{n}$, for $c_2 < 0$ and $\mathfrak{m} = 0,\,1,\,2,\,3$.} 
\end{figure}

These plots illustrate the general fact that
the $c_2<0$ case can be obtained from that with $c_2>0$ by a $\pi/2$
rotation and appropriate redefinitions. Indeed, the results for
$c_2>0$ can alternatively be viewed as the plots for $c_2<0$ at fixed
$\mathfrak{m}$ with $\mathfrak{n}$ varying, and {\em vice versa}, with the
exception of the charged pion masses, which differ by a constant
offset of $2 |c_2|/f^2$. To illustrate this latter point I
plot, in fig.~\ref{fig:c2lt0mpi}, the pion masses for $c_2<0$ as a
function of $\mathfrak{n}$ for fixed values of $\mathfrak{m}$.
Comparing to fig.~\ref{fig:c2gt0mpi}, we see the equality of
the neutral pion masses and the constant offset in the charged pion masses.

\subsubsection{Numerical results}

The analysis carried out with W$\chi$PT is very predictive but cannot choose
between the two possible shapes of the potential. 
The occurrence of one of the two scenarios depends on the sign of the coefficient
$c_2$ proportional to the O($a^2$) terms of the chiral Lagrangian, and 
this coefficient $c_2$, similarly to the other LEC, it is not determined by chiral symmetry, 
but it depends on the details of the lattice action one is using, like the gauge action, the
presence in the lattice action of the clover term and of course it depends on
the value of the bare gauge coupling, i.e. of the lattice spacing.

The lattice spacing dependence of the phase structure can be studied in detail
using lower dimensional models, allowing a better analytical control over the computations. 
An analysis with Wilson fermions of the two dimensional Gross-Neveau model
\cite{Izubuchi:1998hy} indicates that indeed both the scenarios describe
the phase structure of Wilson fermions depending on the value of the 
couplings of the model. The analysis shows that at strong coupling there is an Aoki phase
while at weak coupling the Sharpe-Singleton scenario sets in.
This analysis has been recently extended for the twisted mass case
\cite{Nagai:2005mi}, indicating even more complicated structures, like a coexistence
of the two scenarios for certain values of the bare couplings.

In QCD the situation is more involved.
The prediction of the existence of an Aoki-phase was made years before the
analysis of W$\chi$PT~\cite{Aoki:1984qi,Aoki:1985jj,Aoki:1986xr}.
The existence of a region of bare parameters where parity and flavour were
spontaneously broken was a possible mechanism to generate a massless pion at
finite lattice spacing.
Quenched studies~\cite{Aoki:1989rw,Aoki:1990ap,Aoki:1992nb} found evidence of such a phase structure.
Recently a quenched computation~\cite{Jansen:2005cg} of the charged and neutral 
pseudoscalar masses using Wtm has shown that the charged pion is lighter than the neutral, 
indicating that in the quenched case we are in an Aoki scenario.

On the other hand it is
interesting to revisit some old controversial results about the chiral phase
structure of Wilson fermions, 
using the standard Wilson gauge action~(\ref{eq:Wilson_gauge}).
In \cite{Aoki:1995yf,Aoki:1996pw} from a finite
temperature study there was an indication of difficulties in observing a phase
with spontaneous breaking of flavour and parity symmetry (Aoki phase) at
$\beta > 4.8$.
In \cite{Blum:1994eh} the MILC collaboration found a surprising bulk first
order phase transition at $\beta \simeq 4.8$.
In a recent investigation~\cite{Ilgenfritz:2003gw} of the Aoki phase 
for values of the coupling $\beta <5$ the authors confirm
evidence for an Aoki phase but only for values $\beta < 4.6$. For values of
$\beta > 5$ with Wilson plaquette gauge action a systematic investigation was
missing.

This gap has been filled in a set of publications that have thoroughly
investigated the chiral phase structure of Wilson-like fermions.
In \cite{Farchioni:2004us} the first study of Wtm with $N_{\rm f} = 2$
dynamical fermions was performed and rather surprising results were found:
the existence of the Sharpe-Singleton scenario.

The action used in this study is Wilson
gauge action combined with Wilson fermions with and without twisted mass.
In particular at a lattice spacing of $a\approx 0.16$ fm, strong evidence of a first
order phase transition was found for a rather large range of values of twisted
masses going from zero twisted mass to $\mu_{\rm q} \simeq 100$ MeV. This study
reveals also that the phase transition tends to disappear on increasing the value
of $\mu_{\rm q}$, it persists for $\mu_{\rm q}=0$ and it is volume independent.

\begin{figure}[hbt]
\begin{center}
\includegraphics[width=7.0cm]{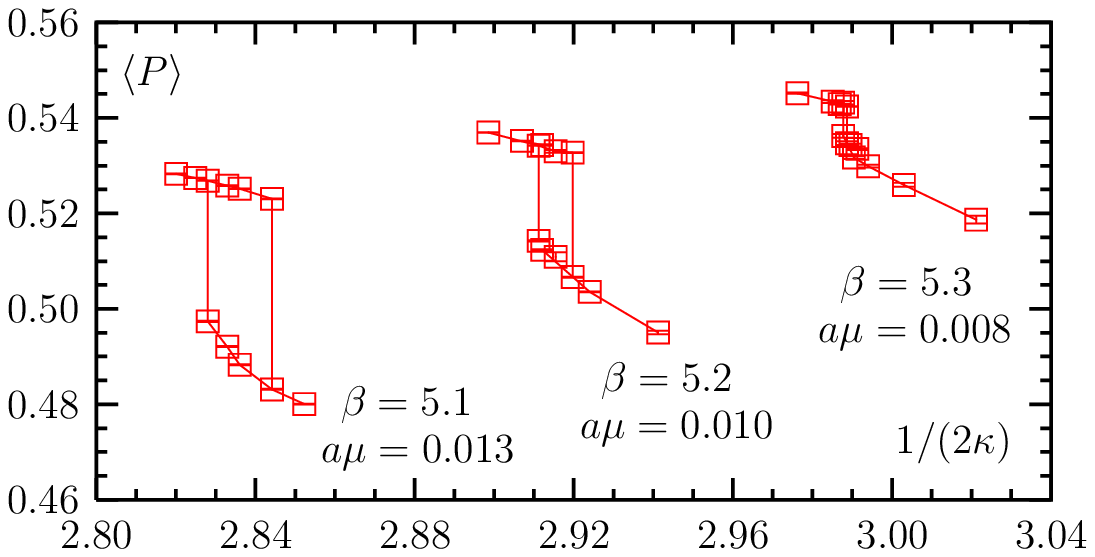}
\includegraphics[width=7.0cm]{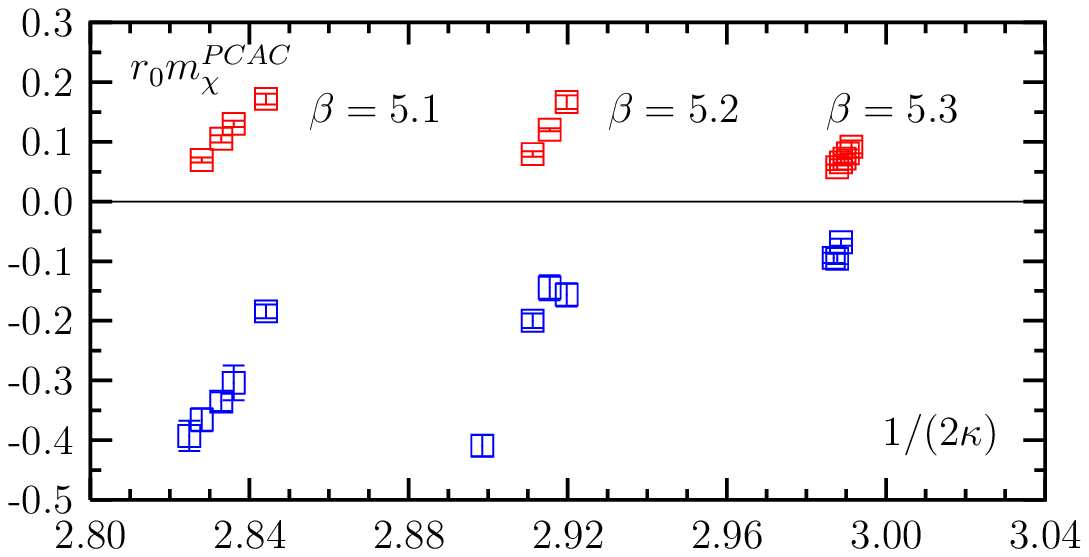}
\end{center}
\caption{Average plaquette and $m_{\rm PCAC}$ plotted vs.
$(2\kappa)^{-1}=m_0+4$ for fixed $\mu_{\rm q}$\protect\cite{Farchioni:2005tu}.}
\label{fig:firstorderresults}
\end{figure}

This study was extended in~\cite{Farchioni:2005tu}. 
In fig.~\ref{fig:firstorderresults} I show scans of the average plaquette and $m_{\rm PCAC}$ 
at fixed twisted mass (with $\mu_{\rm q}$ roughly fixed in physical units)
for three decreasing lattice spacings (at increasing $\beta$ values).
The plaquette and $m_{\rm PCAC}$ both have a discontinuity,
and show hysteresis. Both effects 
decrease as one approaches the continuum limit,
qualitatively consistent with expectations. 
The fact that $m_{\rm PCAC}$ has a minimum away from zero
is a manifestation of the non-zero minimum in the 
pion masses.

\begin{figure}[htb]
\vspace{-0.0cm}
\begin{center}
\epsfig{file=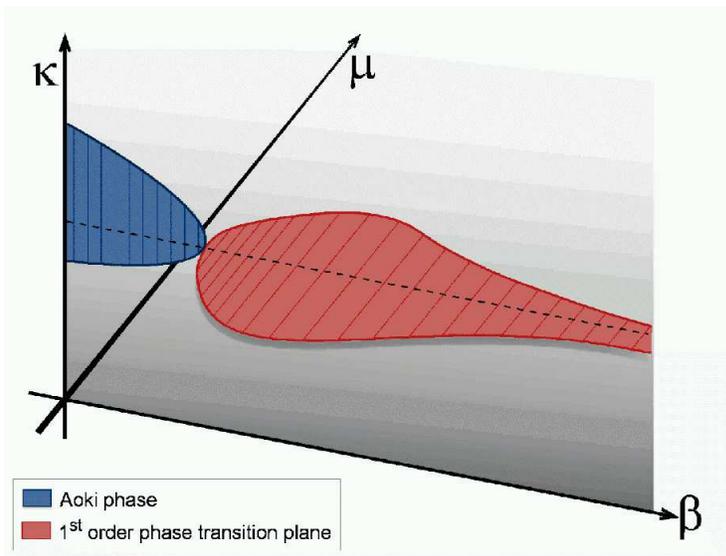,angle=0,width=0.65\linewidth}
\caption{Current knowledge of the Wilson lattice QCD phase diagram as function of the inverse
gauge coupling $\beta\propto 1/g^2$, the hopping parameter $\kappa = (8 + 2 a m_0)^{-1}$ and the
twisted mass parameter $\mu_{\rm q}$.}
\label{fig:phase}
\end{center}
\end{figure}

Our present understanding of the lattice QCD phase diagram with Wilson gauge action and Wilson-like fermions
can be summarized as follows.
For values of the lattice spacing much coarser than $a=0.15\
\mathrm{fm}$, there is an Aoki phase~\cite{Aoki:1984qi,Ilgenfritz:2003gw,Sternbeck:2003gy}. 
For smaller values of the lattice spacing, a first order phase transition
appears (Sharpe-Singleton scenario)~\cite{Farchioni:2004us,Farchioni:2004fs,Farchioni:2004ma,Farchioni:2005tu}
that separates the positive quark mass from the negative quark mass
phase, and extends into the twisted direction. 
This first order phase transition is reminiscent of the continuum 
phase transition when the quark mass is changed from positive to negative values
with the corresponding jump of the scalar condensate which is the order parameter of
spontaneous chiral symmetry breaking. 
The generic phase structure of lattice QCD 
is illustrated in fig.~\ref{fig:phase} and discussed in
refs.~\cite{Farchioni:2004us,Farchioni:2004fs,Farchioni:2004ma}.

\subsubsection{The gauge action}

In the Sharpe-Singleton scenario, the pseudoscalar mass
$m_{\rm PS}$ cannot be made arbitrarily small.
When lowering the quark mass from the twisted direction there is a minimal pion mass 
accessible with numerical simulations given directly by the
extension of the first order phase transition line, even if the twisted mass
provides a sharp infrared cutoff in the sampling performed by the 
simulation algorithm.

An important result of W$\chi$PT is that
both the size of the phase boundaries and the isospin splitting
for pions are determined by the same parameter, $w'$.
It thus makes sense
to try and tune the gauge and fermion actions to reduce $|w'|$.
Note that this tuning is not the same as a systematic improvement
program, but it is nevertheless very important.

In \cite{Farchioni:2005tu} the lattice spacing dependence of
the first order phase transition with Wilson gauge action has been studied, 
taking, as a measure of its strength, the
gap  between the two phases in the plaquette expectation value and in the PCAC
quark mass.
The qualitative estimate for the lattice spacing, where a minimal pion mass
$m_\pi \simeq 300 $ MeV could be reached without being affected by the first
order first transition is 0.07-0.1 fm.

It is suggestive to interpret, at the microscopic level, the occurrence of the
Sharpe-Singleton scenario with a massive rearrangement 
of the small eigenvalues of the Wilson-Dirac around the first order phase
transition line. This rearrangement could be
suppressed by the use of a renormalization group improved or O($a^2$) improved
gauge action, and indeed results from \cite{Aoki:2004iq} indicate that
metastabilities in the average plaquette observed for 
$N_{\rm f} = 3$ dynamical Wilson fermions with a clover term 
can be suppressed replacing the Wilson gauge action with the Iwasaki
action \cite{Iwasaki:1985we}. 

The dependence of the phase diagram on the gauge action used and on the
lattice spacing has been studied in a set of papers 
\cite{Farchioni:2004us,Farchioni:2004fs,Farchioni:2004ma,Farchioni:2005tu}
(see also \cite{Farchioni:2005ec} for a detailed summary of these results).
The gauge actions studied so far can be parametrized by
\be
    S_{\rm G} = \frac{\beta}{6} \big[b_0\sum_{x;\mu\neq\nu}\tr(1-P^{1 \times
    1}(x;\mu,\nu)) + b_1 \tr(1- P^{1 \times 2}(x;\mu,\nu)) \big]
\ee
with the normalization condition $b_0 = 1-8b_1$.
In particular three gauge actions have been investigated:
\begin{itemize}
\item Wilson action \cite{Wilson:1974sk} ~~$\Rightarrow b_1 = 0 $
\item tree-level Symanzik action \cite{Weisz:1982zw} ~~$\Rightarrow b_1 = - {1
    \over 12}$
\item DBW2 action \cite{Takaishi:1996xj} ~~$\Rightarrow b_1 = - 1.4088$
\end{itemize}

\begin{figure}[htb]
\begin{center}
\includegraphics[width=0.6\linewidth]{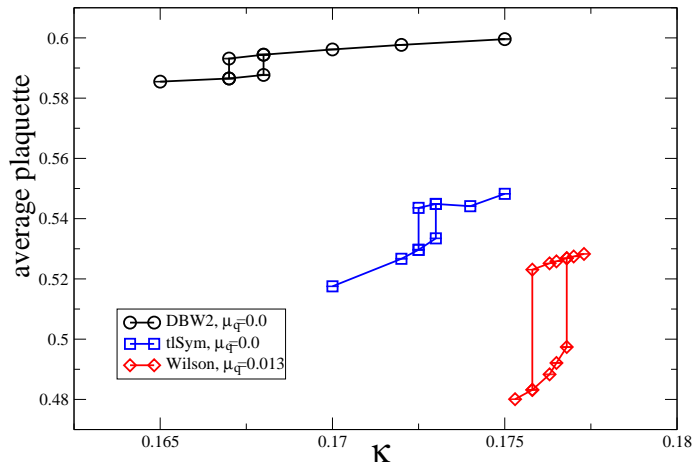}
\end{center}
\caption{Hysteresis of the average plaquette value as $\kappa$ is moved across
  the critical point, for Wilson, tlSym and DBW2 gauge action at $a\sim 0.17$~fm.
\label{fig:comp_plaq_wilson_dbw2_tlsym}}
\end{figure}

The result of these studies is summarized in fig.~\ref{fig:comp_plaq_wilson_dbw2_tlsym}
which shows that the discontinuities
in the plaquette are significantly reduced going from the Wilson to the DBW2 gauge action.

By varying the coupling $b_1$ which multiplies the 
rectangular plaquette term, one can
interpolate between various actions and this allows to understand in
more detail the properties of the phase structure, in particular how the
strength of the transition depends on the additional term and how
this influences the approach to the continuum limit.

That even a small value of $b_1$ can already
have a large impact on the phase structure is illustrated in 
fig.~\ref{fig:comp_plaq_wilson_dbw2_tlsym}
which shows the average plaquette value as a function of the hopping
parameter $\kappa$ for three different actions, i.e.~values of $b_1$, namely
$b_1=0$ (Wilson), $b_1=-1/12$ (tlSym) and $b_1=-1.4088$ (DBW2).
As one moves $\kappa$ from the negative or positive side across the critical
point, where the PCAC quark mass vanishes, a hysteresis in the average
plaquette value develops whose size and width are indicators of
the strength of the phase transition. Both the width and the
size of the gap in the plaquette value decreases considerably as we switch on
$b_1$ to $b_1=-1/12$ (tlSym action). Decreasing $b_1$ further down to
$b_1=-1.4088$ (DBW2 action) still seems to reduce the size of the gap, but the
effect is surprisingly small despite the large change in $b_1$.\footnote{
In fact, the reductions are greater than appears, as the
results with the Wilson action are for $\mu_{\rm q}\ne0$, 
where the transition is weaker than for $\mu_{\rm q}=0$,
which is the value used for the other actions.}
The results in fig.~\ref{fig:comp_plaq_wilson_dbw2_tlsym} are for a lattice
spacing $a\sim 0.17$~fm that is roughly the same for all three actions.
I remark also that the strength of the first order phase transition 
weakens rapidly when the lattice spacing is made finer.
This is illustrated in the left plot of fig.~\ref{fig:firstorderresults}
for the case of the Wilson plaquette gauge action. 

A satisfactory setup for dynamical simulations with, say, $N_f=2$ flavours of
quarks  would be to reach pion masses of about $250-300$~MeV and a box size
of $L>2$~fm. At the same time, one should stay at full twist to realize 
automatic O($a$) improvement.
From all the results presented here it is conceivable that for the tlSym action this can be
achieved with a reasonable computer time at $a \simeq 0.1$ fm on $L/a=24$ lattices. 
Although for the DBW2 action the situation might be somewhat better, the 
advantages of the tlSym action, such as good convergence of perturbation 
theory and not big scaling violations as found in~\cite{Necco:2003vh} for the DBW2 gauge action,
suggests the tlSym gauge action as the action of choice. 

Of course other options could be advocated like sticking to the Wilson gauge
action but simulating maybe at an even smaller lattice spacing, or
modifying some parts of the fermionic action like the form of the covariant
derivatives or by adding the clover term.
In~\cite{Becirevic:2006ii} it has been found that using the clover improved fermion
action reduces the pion isospin splitting and thus 
$w'$. This result cannot be considered however
conclusive for two reasons: the disconnected diagrams were neglected in
the computation of the neutral pion; it was a quenched study.
We have seen that these are critical issues because going from quenched to
unquenched simulations and/or including the disconnected diagrams can have
dramatic impacts in the phase diagram like even changing the sign of $w'$, i.e.
greatly affecting the neutral pion mass.

\newpage
\section{Renormalization and weak matrix elements}
\label{sec:ren}

In the previous sections we have extensively discussed the issue of the continuum limit
and we have analyzed cutoff effects of order $a$ and $a^2$ with Wtm.

Renormalization is necessary in order to perform the continuum
limit and correctly evaluate hadronic matrix elements.
Here we will not discuss the way how the renormalization is performed but only
the mixing patterns of relevant physical quantities according to the lattice
action used, i.e. according to the way the theory is regularized.
Throughout the section we will assume that a mass independent renormalization
scheme has been used, and that the scheme allows for a non-perturbative
renormalization.
Examples of such schemes are the so called RI-MOM
scheme~\cite{Martinelli:1994ty}, and the Schr\"odinger functional
(SF)~\cite{Luscher:1992an,Sint:1993un}. 

Whatever is the chosen strategy, the renormalization is a difficult problem,
that becomes even more difficult if one considers four-quark operators which
appear in the effective weak Hamiltonian.
The details of the renormalization patterns strongly depend on the symmetries 
preserved by the regularization adopted, i.e. on the lattice action.
Regularizations which preserve chiral symmetry are clearly 
advantageous~\cite{Capitani:2000da,Capitani:2000bm,Aoki:2000ee}, 
and are increasingly used despite 
their high computational costs~\cite{Bietenholz:2004wv}. 
If computationally cheaper quarks of the Wilson type are 
used instead, the operator renormalization 
is complicated, mainly due to the presence of 
power divergences \cite{Bochicchio:1985xa,Maiani:1986db} 
and/or mixings induced by the explicit chiral symmetry violation.
It is then more than welcome to make use of the possibility to ease the renormalization
pattern of relevant physical quantities using Wilson-like quarks.

In this section I will give two examples of how Wtm can be used to reach this
goal: the pseudoscalar decay constant and $B_K$.
This result can be obtained using the relative freedom of choosing 
the twist angle depending on the flavour of the quark field 
appearing in the four-fermion operators.
This freedom has been used in different ways in~\cite{Pena:2004gb,Frezzotti:2004wz}
leading to milder renormalization patterns for the four-fermion operators relevant for 
the $\Delta I = 1/2$ rule. The basic ideas on how to achieve these simplifications
can be understood in an easier way by analyzing in detail the renormalization of $B_K$.
This is presented in sect.~\ref{ssec:BK}.
A detailed treatment of the $\Delta I = 1/2$ rule goes beyond the scope of this report.
We thus refer to the original papers~\cite{Pena:2004gb,Frezzotti:2004wz} and 
references therein for a detailed analysis on this particular process.

\subsection{Decay constants}
\label{ssec:decay_constants}

The determination of the Cabibbo-Kobayashi-Maskawa (CKM) matrix elements 
is of phenomenological and theoretical interest 
and to test the unitarity of the CKM matrix, it is crucial to have a precise
determination of the matrix element $|V_{us}|$ that parametrizes the coupling
of the $u \rightarrow s$ transition in the Standard Model (SM).

Lattice QCD computations can address this issue determining in a very precise
way the decay constants of the pion and the kaon.

In fact the knowledge from first principles of $f_K/f_\pi$ allows for an 
accurate determination of $|V_{us}|/|V_{ud}|$ from the ratio of leptonic kaon
decay rates~\cite{Marciano:2004uf}:
\be
\frac{\Gamma(K\to\mu \bar\nu_\mu(\gamma))}{\Gamma(\pi\to\mu\bar\nu_\mu (\gamma))
} = K \frac{|V_{us}|^2 f^2_K m_K \left(1-\frac{m^2_\mu}{m^2_K}\right)^2} 
{|V_{ud}|^2 f^2_\pi m_\pi \left(1-\frac{m^2_\mu}{m^2_\pi}\right)^2} \,,
\label{eq:vusvud} 
\ee
where the prefactor $K$ on the r.h.s. takes into account radiative corrections. 
By combining the experimental result $\Gamma(K\to\mu \bar
\nu_\mu(\gamma))/\Gamma(\pi\to\mu\bar\nu_\mu (\gamma))$ with the 
Lattice QCD determination of $f_K/f_\pi$ and the experimental value  
of $|V_{ud}|$ one obtains $|V_{us}|$.

At the moment, the lattice errors dominate the uncertainty in the determination
of $|V_{us}|$, thus it is important to have available a lattice action that can
minimize possible sources of systematic uncertainty.

With Wtm it is possible to determine the pseudoscalar decay constant without
any computation of renormalization factors and without the knowledge of any
improvement coefficient, still with only O($a^2$) cutoff effects. 
It is only necessary to correctly tune the untwisted
bare quark mass to full twist as extensively detailed in sec.~\ref{sec:impro}.

Due to the exact flavour symmetry of massless Wilson quarks,
the situation with the vector
Ward identity is the same as with Ginsparg-Wilson
fermions, i.e. the classical PCVC relation~(\ref{eq:PCVC}) holds exactly on the lattice 
\be
\partial_\mu^* \langle \widetilde{V}^a_\mu(x) O(0) \rangle= -2 \mu_{\rm q} \epsilon^{3ab}
\langle P^b(x) O(0)\rangle 
\label{eq:lat_PCVC}
\ee
(where $\partial^*_\mu$ is the lattice backward derivative)
with the point-split vector current
\be
   \widetilde{V}^a_\mu(x) =  
 \frac12 \Big\{\chibar(x)(\gamma_\mu-1)\frac{\tau^a}{2}
 U(x,\mu)\chi(x+a\hat\mu)
 +\chibar(x+a\hat\mu)(\gamma_\mu+1)\frac{\tau^a}{2} 
 U(x,\mu)^{-1}\chi(x)\Big\},
\ee
and the local pseudoscalar density. This implies that 
the multiplicative renormalization constants of
composite fields which belong to the same isospin multiplet 
must be identical in order to respect the vector Ward identities. 
An example is the renormalized pseudoscalar
density which has the structure
\be
  P_{\rm R}^a =
     Z_{\rm P} (P^a+\delta^{a3}  \frac{c_{\rm P}}{a^3}).
\ee
The mixing with the identity operator appears because standard parity
is not a symmetry of the massive lattice theory, but while $P^{1,2}$
are still protected by the symmetries $\mathcal{P}_F^{1,2}$ defined
in eqs.~(\ref{eq:PF12}), $P^3$ is not. The vector Ward
identity~(\ref{eq:lat_PCVC}) here implies that $Z_{\rm P}$ is the same for all flavour
components, and
\begin{equation}
  Z_{\rm P}=\frac{1}{Z_\mu}.
  \label{eq:Zmu}
\end{equation}
Moreover since standard parity combined with a change of the sign of the twisted mass 
$\widetilde{\mathcal{P}}$ defined in eq.~(\ref{eq:Ptilde}) is a symmetry of the lattice action, the 
mixing pattern of $P^3$ is actually only with a quadratically divergent term {\large{${\mu_{\rm q} \over a^2}$}}
similarly to what happens with overlap fermions.

The existence of the point-split vector current 
can be used as a tool to determine the finite renormalization constant 
$Z_{\rm V}$ that normalizes the local vector current.
One possibility would be to determine $Z_{\rm V}$ through
\be
Z_{\rm V} = \lim_{\mu_{\rm q} \rightarrow 0} 
\frac{\langle \widetilde{V}^a_\mu (x) O(0) \rangle}{\langle V^a_\mu (x) O(0) \rangle}
\label{eq:ZV_ratio}
\ee
with the local vector current defined in eq.~(\ref{eq:currents_local}).
This method requires the computation of a correlation function 
involving the point-split current.
This can be bypassed substituting
$\widetilde{V}^a_\mu$ with $Z_V V^a_\mu$ in 
eq.~(\ref{eq:lat_PCVC}) and the relation
is now valid up to order $a^2$ lattice artifacts.
This relation allows for a determination of $Z_V$ through 
\be
Z_V=\lim_{\mu_{\rm q}\rightarrow 0} \frac{-2 \mu_{\rm q} \epsilon^{3ab} \sum_{\mathbf x}\langle P^b(x) 
P^b(0)\rangle} {\sum_{\mathbf x}\langle \tilde\partial_\mu V^a_\mu(x) P^b(0)\rangle} \qquad a=1,2 \,,
\label{eq:Z_V}
\ee
where $\tilde\partial_\mu$ is the symmetric lattice derivative.

\begin{figure}[!htb]
\begin{center}
\epsfig{file=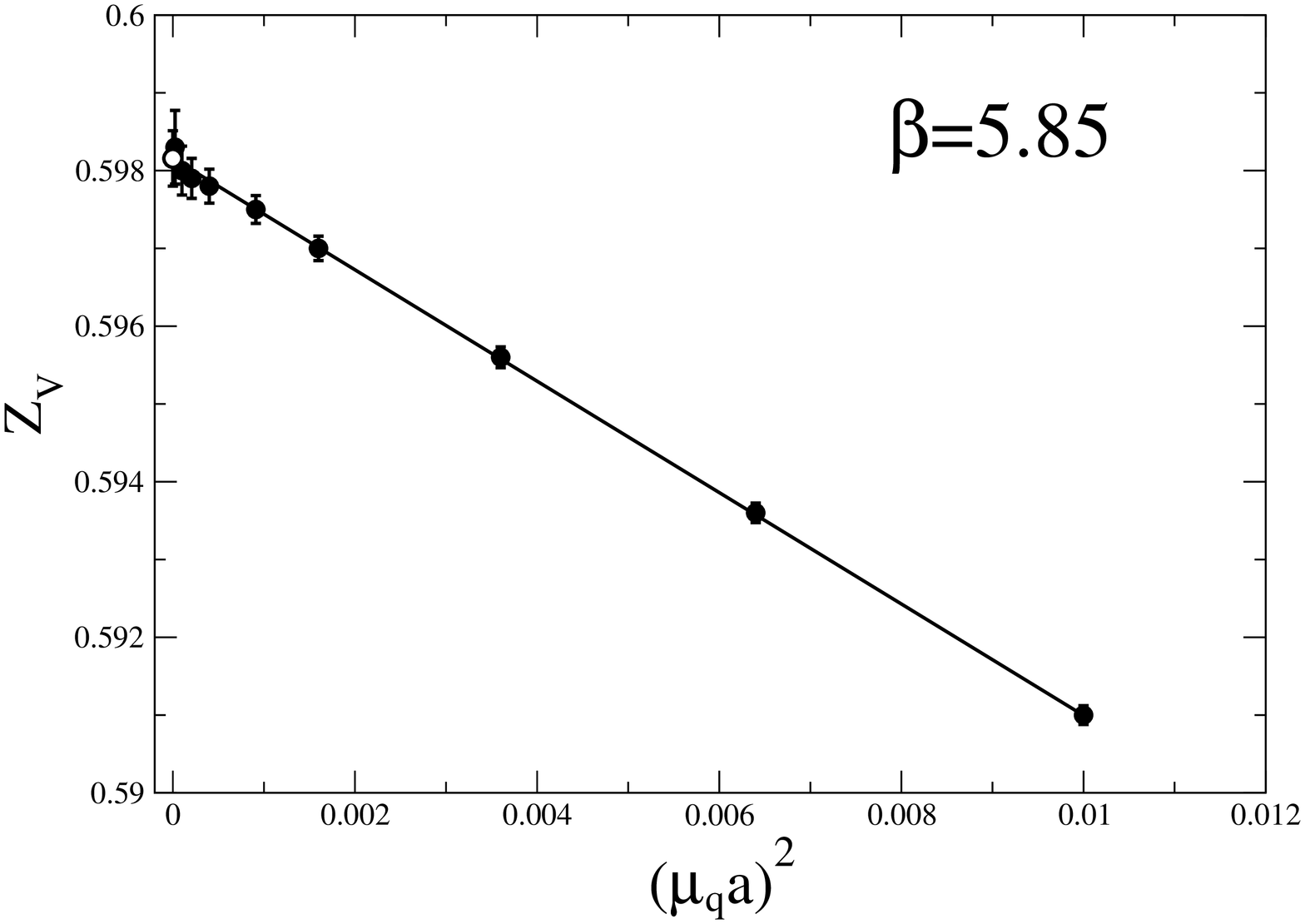,angle=270,width=0.8\linewidth}
~\\[-7.8cm]
\hspace*{4.1cm}
\epsfig{file=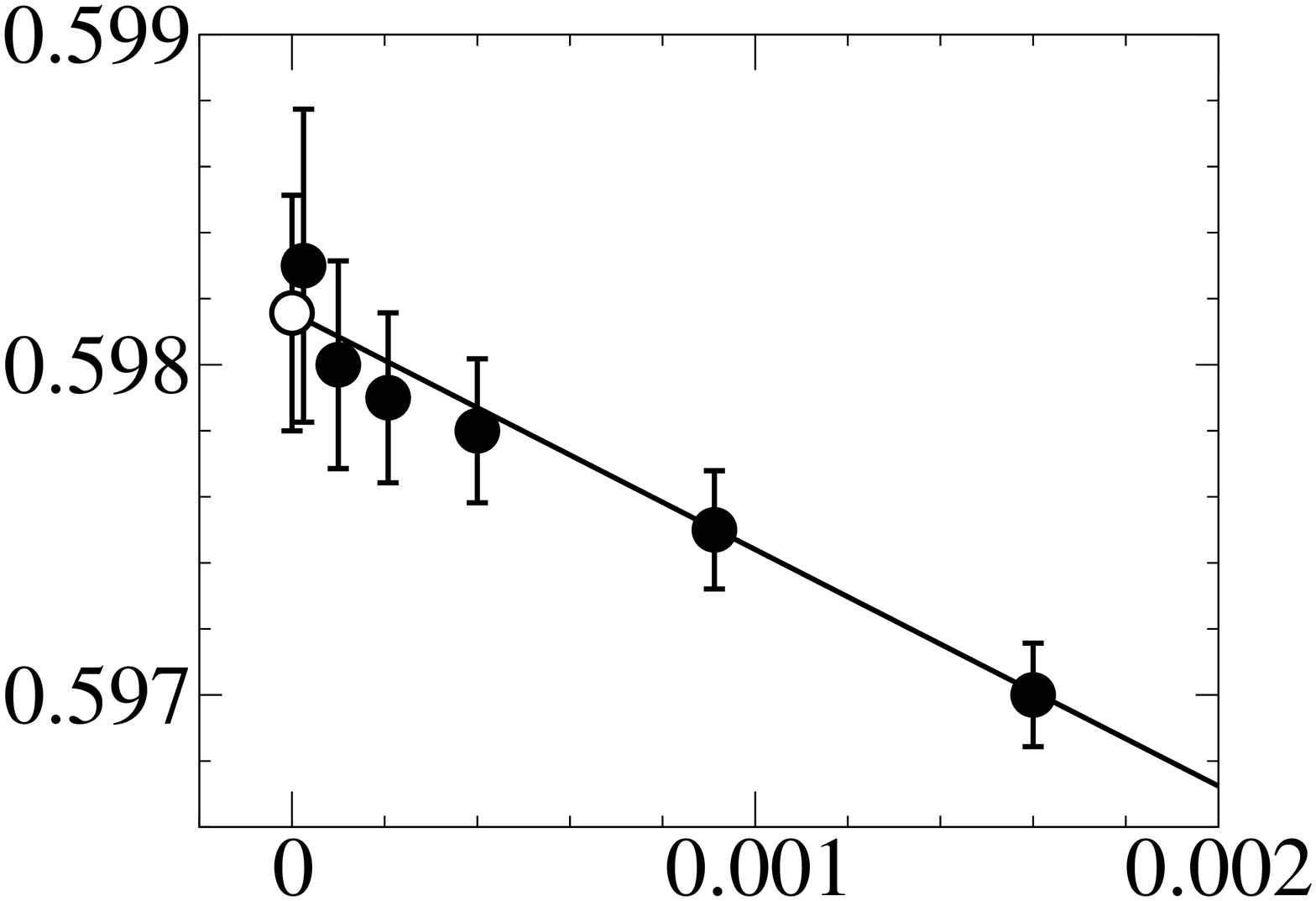,angle=270,width=0.31\linewidth}
\vspace*{4cm}
\end{center}
\caption{An example of the extrapolation of $Z_V$ to zero quark mass 
for a quenched simulation~\cite{Jansen:2005kk} at a lattice spacing of $a=0.123$ fm. 
The data show a linear behaviour in $(a\mu_{\rm q})^2$. 
\label{fig:zetav585}}
\end{figure}

In fig.~\ref{fig:zetav585} there is an 
example of $Z_V$ as a function of the quark mass for a quenched simulation~\cite{Jansen:2005kk}
determined using eq.~(\ref{eq:Z_V}).
It turns out that $Z_V$ is to a good approximation linear in $(a\mu_{\rm q})^2$. 

In contrast, the axial Ward identity does not hold in the
bare theory. Axial Ward identities therefore provide 
normalization conditions which determine finite renormalization
constants such as $Z_{\rm A}$, or finite 
ratios of scale dependent renormalization constants,
such as $Z_{\rm S}/Z_{\rm P}$~\cite{Bochicchio:1985xa}. \\

As Wtm and standard Wilson quarks are
not related by a lattice symmetry, the
counterterm structure for composite fields with the
same physical interpretation depends upon the twist angle $\omega$, and 
a particular choice of $\omega$ can lead to substantial simplifications.
A typical example is the computation of the pseudoscalar decay constant
from the axial coupling to the pion. We have argued in sec.~\ref{sec:basic} 
that for renormalized correlation functions, the relation between physical and twisted basis
can be inferred from the corresponding classical relations if the renormalization
scheme adopted is mass independent. With Wilson fermions usually the pseudoscalar
decay constant is extracted from the correlation function
\be
a^3 \sum_{\bf x} \langle (\mathcal{A}_{\rm R})_0^a(x) P_{\rm R}^a(0) \rangle .
\label{eq:WfPS}
\ee
This requires first the determination of the renormalized axial current and eventually all the
relevant improvement coefficients.
Performing the axial rotation~(\ref{eq:axial}) we obtain that in the continuum limit 
we should have (we set the isospin component $a=1$)
\be
\langle (\mathcal{A}_{\rm R})_0^1(x)
   P_{\rm R}^1(y)\rangle_{(M_{\rm R},0)}
  =\cos(\omega)\langle (A_{\rm R})_0^1(x)P_{\rm R}^1(y)
                \rangle_{(m_{\rm R},\mu_{\rm R})} 
+\sin(\omega)\langle V_0^2(x)P_{\rm R}^1(y)
                \rangle_{(m_{\rm R},\mu_{\rm R})}.
 \label{eq:rotation_AP}
\ee
If we set $\omega=\pi/2$ the r.h.s of eq.~(\ref{eq:rotation_AP}) only
contains the vector current which is protected against
renormalization.
Moreover using the exact PCVC relation on the lattice~(\ref{eq:lat_PCVC}) it is
easy to show~\cite{Frezzotti:2001du,DellaMorte:2001tu,Jansen:2003ir} that 
\be
f_{\rm PS} = \frac{2 \mu_{\rm q} | \langle \Omega | \hat{P}^a | {\rm PS} \rangle|}{M_{\rm PS}^2}
\qquad a=1,2 ,
\label{eq:fPS}
\ee
where $M_{\rm PS}$ is the charged pseudoscalar mass and $| {\rm PS}\rangle$ denotes
the corresponding pseudoscalar state. 
As we see, the neat result is that Wtm at full twist allows the determination
of an automatically O($a$) improved pseudoscalar decay constant without the knowledge 
of any renormalization constant and any improvement coefficient.

\begin{figure}[htb]
  \begin{center}
    \epsfig{file=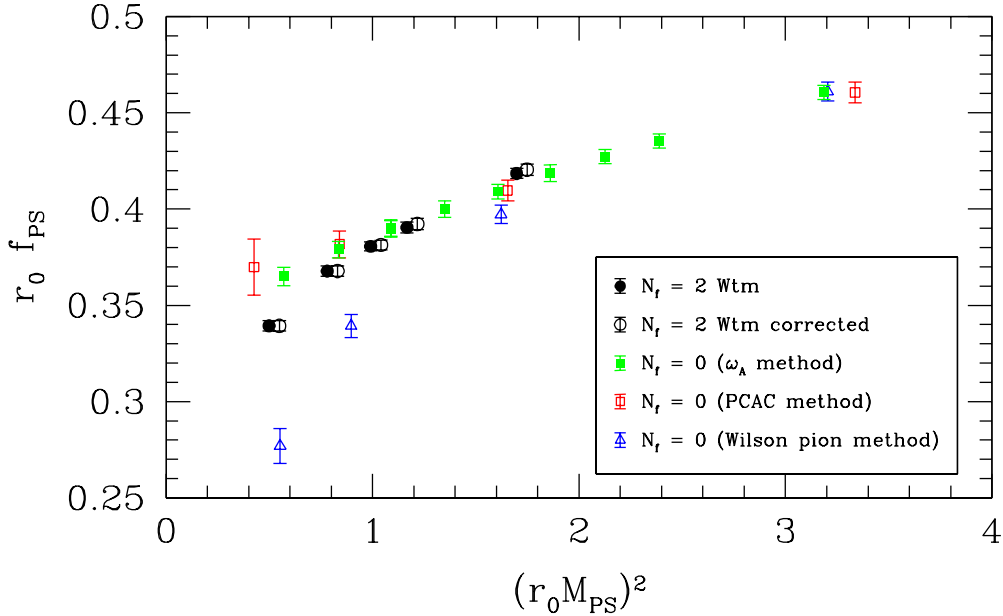,angle=0,width=1.0\linewidth}
  \end{center}
  \vspace{-6.0cm}
  \caption{Summary of the Wtm numerical results at $\omega = \pi/2$
    for the pseudoscalar decay constant 
$f_\mathrm{PS}$ as a function of $M_{\rm PS}^2$. The bending phenomenon is visible with the 
``Wilson pion'' definition of the critical mass for $N_f=0$.
 The corrected data for $N_f=2$ (slightly displayed for clarity) include the effect of the non vanishing
 $m_{\rm PCAC}$~(\ref{eq:fpi_corr}). 
 The perfect consistency of the data dispels all the doubts about the possible
 origins of the mass dependence for $f_{\rm PS}$.}
  \label{fig:fpi_all}
\end{figure}

In fig.~\ref{fig:fpi_all}, I summarize Wtm results coming from 
quenched~\cite{Jansen:2005kk,Abdel-Rehim:2006ve} ($N_f =0$) 
and unquenched~\cite{Boucaud:2007uk} ($N_f =2$) simulations, for the pseudoscalar 
decay constant over a range of pseudoscalar masses around $[300-700]$ MeV.
\begin{figure}[htb]
  \begin{center}
    \epsfig{file=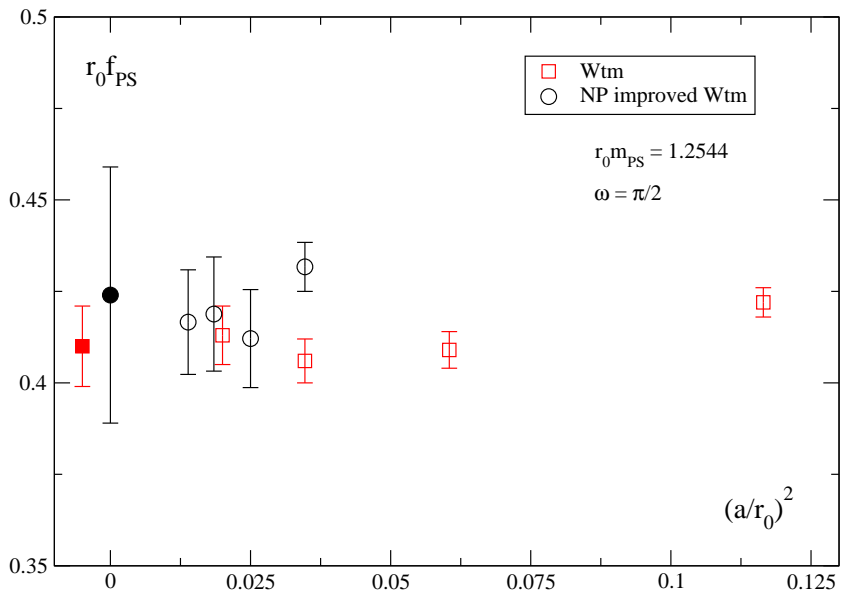,angle=270,width=0.9\linewidth}
  \end{center}
  \caption{Comparison of
 the continuum limit (filled symbols) at $r_0M_{\rm PS} =1.2544$ obtained in
 Wtm at full twist without~\cite{Jansen:2005kk} (red squares) and
 with~\cite{Dimopoulos:2007cn} (black circles) improvement coefficients.}
  \label{fig:q_comp}
\end{figure}
We recall that for Wtm, to produce such a plot, no renormalization constant are needed.
Even if the lattice spacing for the quenched ($a = 0.093$ fm) and the unquenched ($a = 0.087$ fm)
simulations are not exactly the same, it is interesting to notice the effect of the dynamical quarks in the
mass dependence of the pseudoscalar decay constant.
Moreover I also plot for the dynamical case the pseudoscalar decay constant corrected using 
eq.~(\ref{eq:fpi_exp}) and using the PCAC quark mass determined in the same set of simulations
\be
f_{\rm PS}^{\rm corr} = f_{\rm PS} ~ \frac{\sqrt{\mu_{\rm q}^2 + m_{\rm PCAC}^2}}{\mu_{\rm q}} .
\label{eq:fpi_corr}
\ee
In the statistical errors the two results are perfectly consistent showing that the mass dependence 
of $f_{\rm PS}$ for $N_f=2$ is a genuine $\mu_{\rm q}$ quark mass dependence and and the curvature
is not artificially induced by a 
non exactly zero PCAC mass. 
As a comparison I show also the $N_f=0$ results obtained with the 
``Wilson pion'' definition of the critical mass, 
where, on the contrary, the quark mass dependence is 
largely given by the cutoff effects, as discussed 
in sect.~\ref{sec:impro}.

In ref.\cite{Jansen:2005kk,Dimopoulos:2007cn} the continuum limit of the
quenched data is performed  at $\omega=\pi/2$ producing consistent results.
In fig.~\ref{fig:q_comp} we see an example of the two continuum
limits for a pseudoscalar mass fixed at the kaon mass ($r_0M_{\rm PS} =
1.2544$).
While in ref.~\cite{Jansen:2005kk} no improvement coefficients were used, in
ref.~\cite{Dimopoulos:2007cn} the whole set of improvement coefficients were used ($c_{\rm SW}$,
$c_{\rm A}$,...) in the computation of the decay constant from various lattice definitions.
Some of the improvement coefficients are known only at one loop, but
obviously at full twist they are actually irrelevant.
The data shown in fig.~\ref{fig:q_comp} corresponds to a definition involving
only $c_{\rm SW}$ and $\tilde{b}_{\rm V}$ (this is the coefficient multiplying the
O($a$) counterterm in eq.~\eqref{eq:O15}).
From the plot it is evident how automatic O($a$) improvement makes the usage
of the improvement coefficients not relevant for the cancellation of the
O($a$) discretization errors.
In ref.~\cite{Jansen:2005kk} the continuum limit is performed over a wide
range of quark masses as discussed in sect.~\ref{ssec:num_res} of this review
(e.g. see fig.~\ref{fig:fps}).
In ref.~\cite{Dimopoulos:2007cn} several definitions of the decay constant
allow to perform a constrained continuum limit with a final result  which has
a relative error smaller than $2\%$. 

The very precise unquenched data down to pseudoscalar masses around $300$ MeV~~\cite{Boucaud:2007uk}
allow a discussion of whether $\chi$PT formul{\ae} can reproduce the mass dependence 
for $am_\mathrm{PS}$ and $af_\mathrm{PS}$.
One possible source of systematic uncertainty is the finite size effects, and this can be
taken into account using $\chi$PT. In particular the lowest two quark masses
turn out to be significantly affected.
Preliminary results at a smaller lattice spacing~\cite{Jansen:2006rf,Shindler:2006tm}
show small discretization errors, indicating that discretization
errors are under control. Therefore one could try to 
use continuum $\chi$PT to describe consistently  the dependence of the data both on the 
finite spatial size ($L$) and on the quark mass $\mu_{\rm q}$.

The fit to the raw data for $M_\mathrm{PS}$ and $f_\mathrm{PS}$
has been performed simultaneously with the appropriate ($N_f=2$) $\chi$PT
formul{\ae}~\cite{Gasser:1986vb,Colangelo:2005gd}
\begin{equation}
  M_\mathrm{PS}^2(L) = 2B_0\mu_{\rm q} \, \left[
    1 + \frac{1}{2}\xi \tilde{g}_1( \lambda )    \right]^{2} \, \left[ 1 +
    \xi \log ( 2B_0\mu_{\rm q}/\Lambda_3^2 ) \right] \, ,
  \label{eq:chirfo1}
\end{equation}
\begin{equation}
  f_\mathrm{PS}(L) = F \, \left[
    1 - 2 \xi \tilde{g}_1( \lambda )    \right] \, \left[ 1 -
    2 \xi \log ( 2B_0\mu_{\rm q}/\Lambda_4^2 ) \right] \, ,
  \label{eq:chirfo2}
\end{equation}
where
\begin{equation}
  \xi = 2B_0\mu_{\rm q}/(4\pi F)^2 \, , \qquad
  \lambda = \sqrt{2B_0\mu_{\rm q} L^2}\ .  
\end{equation}
The finite size correction function $\tilde{g}_1(\lambda)$ was first
computed by Gasser and Leutwyler in ref.~\cite{Gasser:1986vb} and is
also discussed in ref.~\cite{Colangelo:2005gd}.
In eqs.~(\ref{eq:chirfo1}) and~(\ref{eq:chirfo2}) 
NNLO $\chi$PT corrections are assumed to be negligible.
The formul{\ae} above depend on four unknown parameters, 
$B_0$, $F$, $\Lambda_3$ and $\Lambda_4$, which will be determined by the fit. 
For details about the data analysis I refer to the original paper~\cite{Boucaud:2007uk}.

\begin{figure}[t]
  \hspace{-1.0cm}\subfigure[ \label{fig:f1}]
  {\includegraphics[width=0.44\linewidth,angle=270]{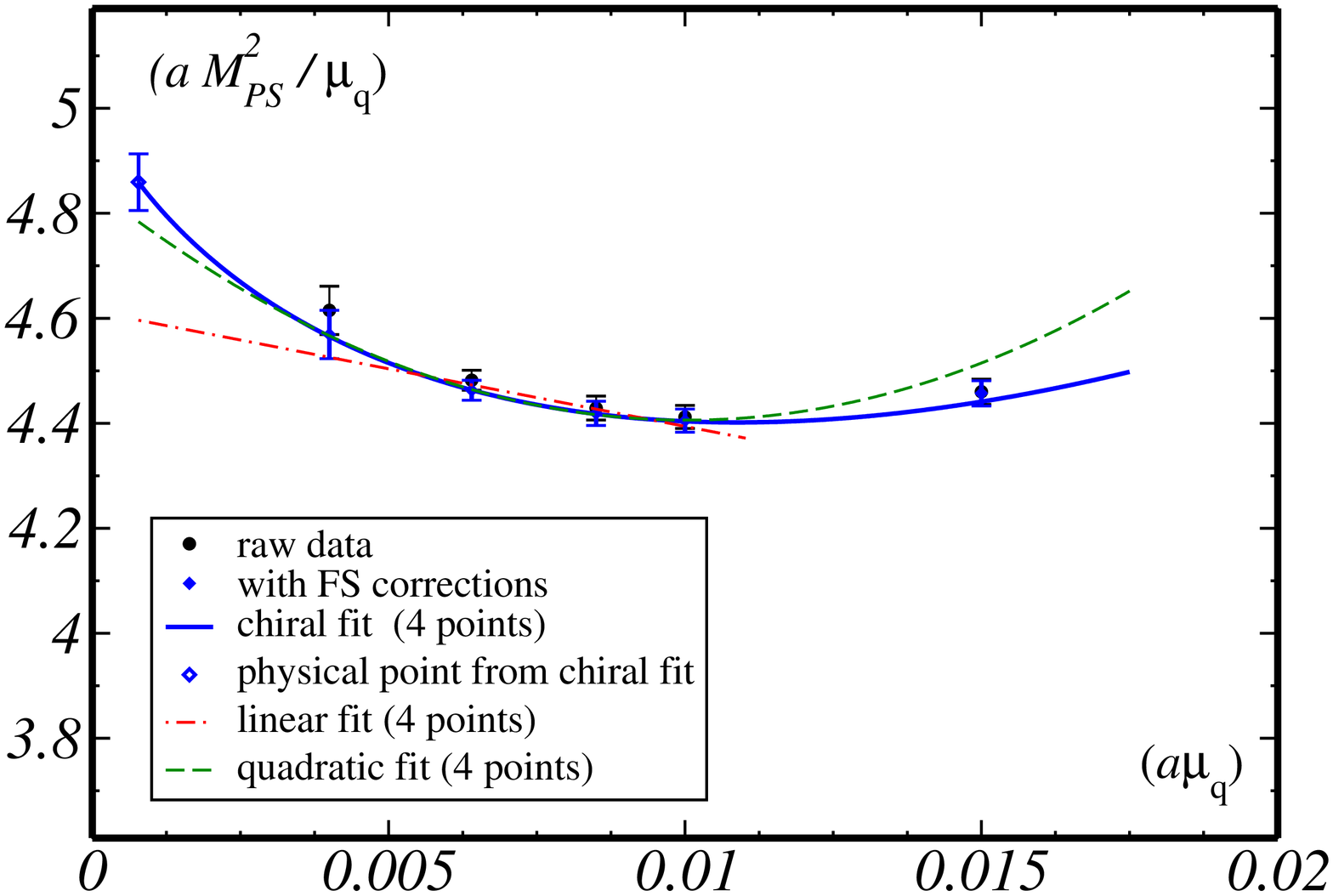}}
  \hspace{-1.0cm}\subfigure[ \label{fig:f2}]
  {\includegraphics[width=0.44\linewidth,angle=270]{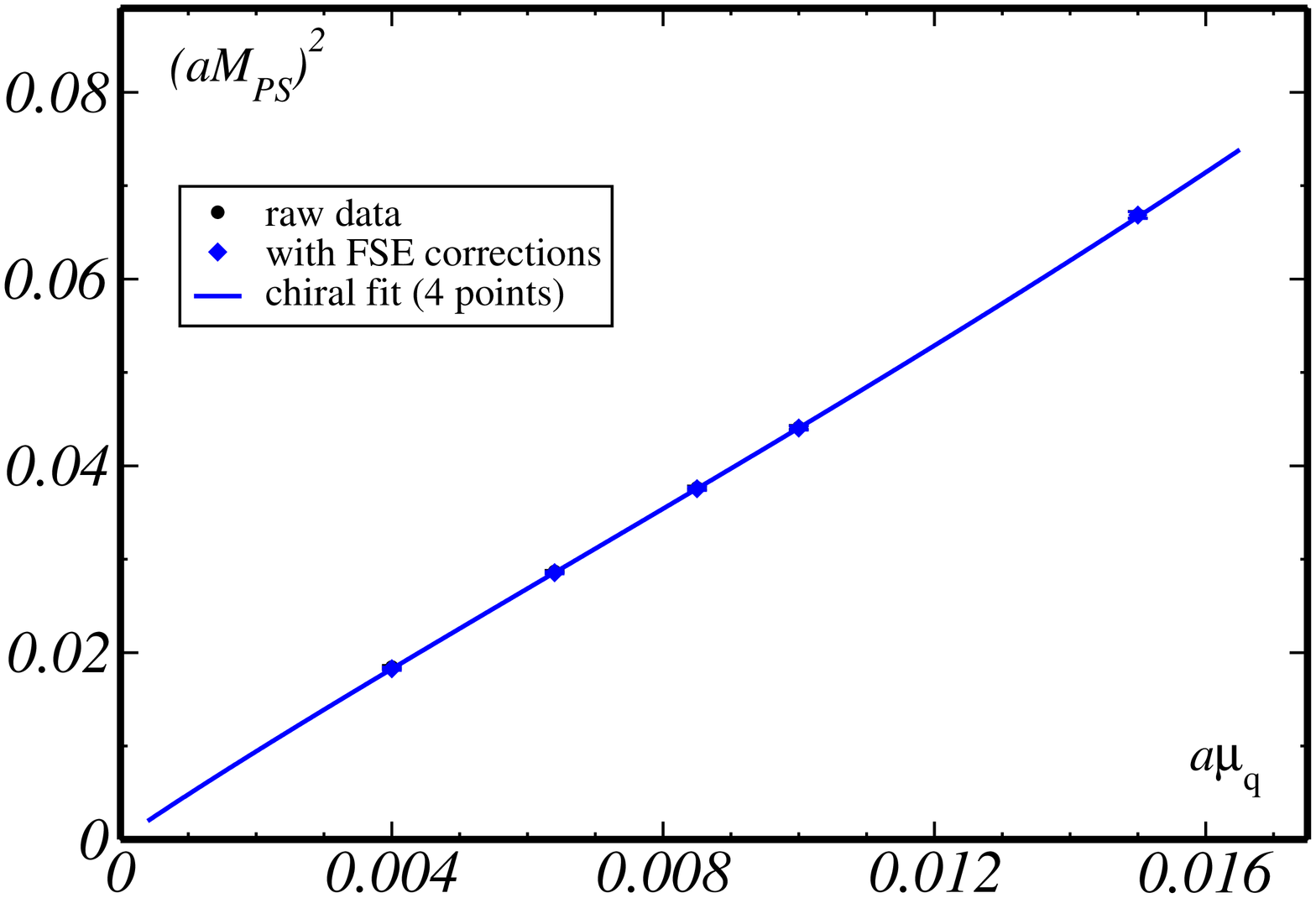}}
  \vspace{-0.5cm}\caption{In (a) I plot $(aM_\mathrm{PS})^2/(a\mu_{\rm q})$ as a function of
    $a\mu_{\rm q}$. In addition the $\chi$PT fit with
    eq.~(\ref{eq:chirfo1}) to the data from the 
    lowest four values of $\mu_{\rm q}$ is compared with linear and quadratic fits.
    In (b) I plot $(aM_\mathrm{PS})^2$ as a
    function of $a\mu_{\rm q}$ with the corresponding chiral fit.
    In both figures (a) and (b) the raw and the finite size
    corrected ($L\to\infty$) data are plotted.}\vspace{+0.5cm}
  \label{fig:mps}
\end{figure}

\begin{figure}
  \centering
  \includegraphics[width=.8\linewidth,angle=270]{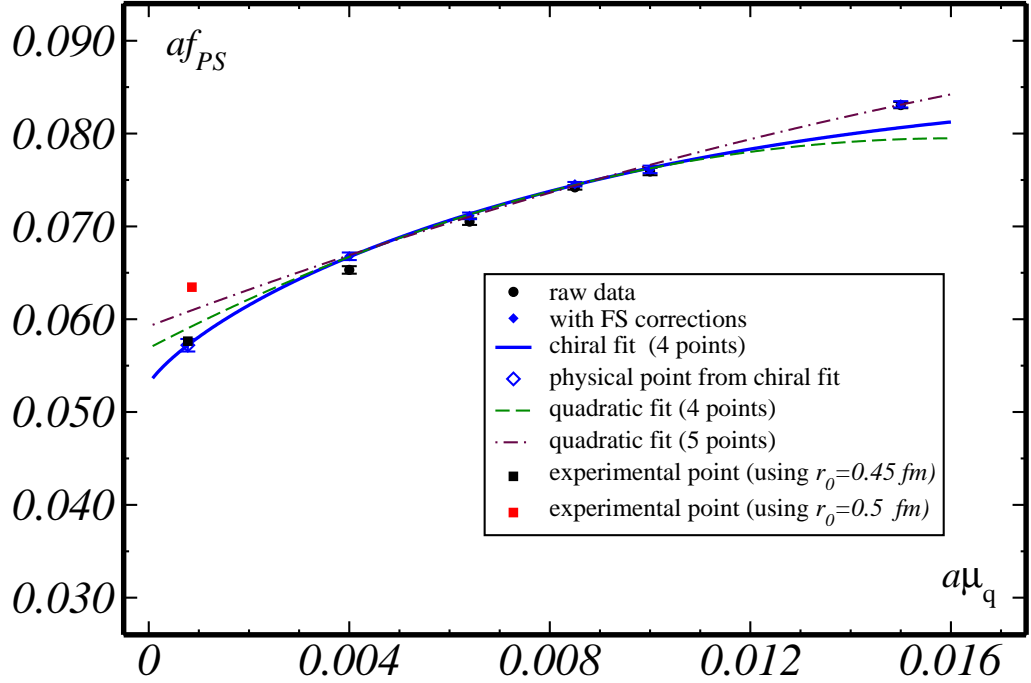}
   \vspace{-0.5cm}\caption{Plot of $af_\mathrm{PS}$ (before and after finite size corrections) 
    as a function of $a\mu_{\rm q}$ together
    with a fit to $\chi$PT formula~(\ref{eq:chirfo2}). This fit is compared with a quadratic fit
    including all the data points and excluding the heaviest mass.
    A comparison with the impact of different physical values for $r_0$ is also shown.}\vspace{+0.5cm}
  \label{fig:f3}
\end{figure}

For the lightest four values of $a\mu_{\rm q}$, an excellent fit to the
data on $f_\mathrm{PS}$ and $M_\mathrm{PS}$ is found (see figures \ref{fig:mps}
and \ref{fig:f3}). The fitted values of the four parameters are
\be
    2aB_0  =-4.99(6)\, , \nonumber 
\ee
\be
    aF =-0.0534(6)\, , \nonumber
\ee
\be
    \log(a^2\Lambda_3^2) = -1.93(10)\, , \nonumber \\
\ee
\be
    \log(a^2\Lambda_4^2) = -1.06(4)\ . 
    \label{eq:bestfit}
\ee
The data are clearly sensitive to $\Lambda_3$ as visualized in figure~\ref{fig:f1}. 

The value of $a\mu_{\rm q}$, $a\mu_\pi$, at which the pion 
assumes its physical mass, is determined~~\cite{Boucaud:2007uk} requiring that 
the ratio $[\sqrt{ [M_\mathrm{PS}^2(L=\infty)]
}/f_\mathrm{PS}(L=\infty)]$ takes the value $(139.6/130.7) = 1.068$.
From the knowledge of $a\mu_\pi$ one can evaluate \\ $\bar{l}_{3,4} \equiv
\log(\Lambda_{3,4}^2/M_{\rm PS}^2)|_{M_{\rm PS} = m_\pi}$ 
\be
  a\mu_\pi = 0.00078(2),\qquad\bar{l}_{3} = 3.65(12), \qquad
  \bar{l}_{4} = 4.52(06) .
  \label{eq:lec34}
\ee
These results compare nicely with other determinations (for a review see
ref.~\cite{Leutwyler:2006qq}). 
The inclusion of the results from $a\mu_{\rm q}=0.0150$ in the fit gives an 
acceptable description of $M_\mathrm{PS}^2$ but misses the data for 
$f_\mathrm{PS}$, as shown in figures~\ref{fig:f2} and~\ref{fig:f3}.
Note, however, that in Eqs.~(\ref{eq:chirfo1}, \ref{eq:chirfo2}), and
thus in the fit results~(\ref{eq:bestfit}, \ref{eq:lec34}), a number of
systematic errors, as discussed below, are not included.

The values presented here should hence be taken as a first estimate, 
the validity of which has to be checked in the future. 
Nevertheless, the statistical accuracy achieved implies
that there is a very good prospect of obtaining accurate and
reliable values for the low-energy constants from Wtm fermion simulations.

Based on the physical value of $f_{\pi}$, one gets 
\begin{equation}
  a = 0.087(1)\ \mathrm{fm}\, .
\end{equation}
Using the value of $r_0/a$  reported in ref.~\cite{Boucaud:2007uk},
this lattice calibration method yields  $r_0 = 0.454(7)\ \mathrm{fm}$.

We now discuss the possible sources of systematic error.
This analysis is based on lattice determinations of properties of 
pseudoscalar mesons with masses in the range 300 to 500 MeV 
on lattices with a spatial size slightly above 2~fm. 
Systematic errors can arise from several sources:

\noindent(i) Finite lattice spacing effects.
Preliminary results at a smaller value of the lattice spacing that were 
presented in refs.~\cite{Jansen:2006rf,Shindler:2006tm}  
suggest that $\mathcal{O}(a)$ improvement is nicely at work and that 
residual $\mathcal{O}(a^2)$ effects are small. 

\noindent (ii) Finite size effects (FSE). Even if the results presented here
are obtained from a box of physical size $L \gtrsim 2$ fm, 
to check that next to leading order (continuum) $\chi$PT adequately describes the FSE,
a run on a larger lattice is required.

\noindent (iii) Mass difference of charged and neutral pseudoscalar 
meson.  
In the appropriate W$\chi$PT  power counting for our values of the lattice spacing and quark masses,  
i.e.\ $\mu_{\rm R} \sim p^2 \sim a $~(\ref{eq:GSM}), the pion mass splitting is a NLO effect
$(M_{\pi^\pm})^2-(M_{\pi^0})^2 = 
\mathcal{O}(a^2\Lambda^4_{\mathrm{QCD}})$, from
which it follows that to the order we have been working the effects of
the pion mass splitting do not affect, in particular, 
the finite size correction factors
for $M_{\mathrm PS}$ and $f_{\mathrm PS}$. In spite of these formal remarks,
it is possible, however,  that the fact that the neutral pion is lighter
than the charged one (by about 20\% at $a\mu_{\rm q} = 0.0040$, see ref.~\cite{Boucaud:2007uk})
makes inadequate the continuum $\chi$PT description
of finite size effects adopted in the present analysis. This caveat 
represents a further
motivation for simulations on larger lattices, which will eventually resolve
the issue. 

\noindent (iv) Extrapolation to physical quark masses. We are assuming that 
$\chi$PT at next to leading order for the $N_f=2$ case is appropriate 
to describe the quark mass dependence of $M_{\mathrm PS}^2$ and $f_{\mathrm PS}$
up to $\sim 450$--$500$~MeV. The lattice data are consistent with this, 
but it would be useful to include higher order terms in the $\chi$PT fits 
as well as more values of $a\mu_{\rm q}$ to check this assumption. 
The effect of heavier quarks in the sea should also be explored, 
and preparatory studies with $N_f = 2+1+1$~\cite{Chiarappa:2006ae} dynamical quarks
show that the inclusion of heavier quarks in dynamical simulations is accessible using
current algorithms and machines.

\subsection{$B_K$}
\label{ssec:BK}

Indirect CP violation in $K\to\pi\pi$ decays is measured by the parameter
$\varepsilon_K$, defined in terms of kaon decay amplitudes as
\be
\varepsilon_K = \frac{T(K_L\to(\pi\pi)_{I=0})}{T(K_S\to(\pi\pi)_{I=0})} \, ,
\ee
where $I$ is the total isospin of the two-pion state. 
The long distance non-perturbative QCD contribution to $|\varepsilon_K|$ is provided by the following
matrix element
\be
B_K = \frac{\langle \bar K^0|O_{\rm R}^{\Delta S=2}| K^0 \rangle}{\frac{8}{3}f_K^2 M_K^2} \, ,
\ee
where $O_{\rm R}^{\Delta S=2}$ is the effective four-quark interaction renormalized operator, 
with bare operator
\be
O^{\Delta S=2} = (\bar s \gamma_\mu^{\rm L} d)(\bar s \gamma_\mu^{\rm L} d) \, ,
\ee
where $\gamma_\mu^{\rm L}=\gamma_\mu(\mathbbm{1}-\gamma_5)$.
$B_K$ largely dominates the uncertainty on the standard model (SM) value for 
$|\varepsilon_K|$ and improving the accuracy of $B_K$ is essential in order
to derive stringent bounds on the amount of non-SM CP violation in kaon decay.

One of the important sources of uncertainty in lattice QCD computations of $B_K$ with Wilson fermions 
arises from operator renormalization. The operator $O^{\Delta S=2}$ 
is usually split into parity-even and parity-odd parts as
\be
O^{\Delta S=2} = O_{\rm VV+AA} - O_{\rm VA+AV} ,
\ee
where with obvious notation we have
\be
O_{\rm VV+AA} = (\bar s \gamma_\mu d)(\bar s \gamma_\mu d) + 
(\bar s \gamma_\mu \gamma_5 d)(\bar s \gamma_\mu \gamma_5 d) \, ,
\label{eq:OVVpAA}
\ee
\be
O_{\rm VA+AV} = (\bar s \gamma_\mu d)(\bar s \gamma_\mu \gamma_5 d) + 
(\bar s \gamma_\mu \gamma_5 d)(\bar s \gamma_\mu d) \, .
\label{eq:OVApAV}
\ee
Since parity is a QCD symmetry, the only
contribution to the $K^0$--$\bar K^0$ matrix element comes from $O_{\rm VV+AA}$.
In regularizations which respect chiral symmetry,
the latter operator is multiplicatively
renormalizable. If chiral symmetry is not preserved,
$O_{\rm VV+AA}$ mixes with four other dimension-6
operators~\cite{Martinelli:1983ac,Bernard:1987rw,Donini:1999sf} with positive parity:
\begin{eqnarray}
(O_{\rm R})_{\rm VV+AA}(\mu) = Z_{\rm VV+AA}(g_0, a\mu) \Big [ O_{\rm VV+AA} (g_0)
+ \sum_{i = 1}^4 \Delta_i(g_0) O_i(g_0) \Big ] .
\end{eqnarray}
The operators $ O_i(g_0)$ belong to different chiral
representations than $O_{\rm VV+AA}$. The mixing coefficients
$\Delta_i(g_0)$ are finite functions of the bare coupling,
while the renormalization constant $Z_{\rm VV+AA}$
diverges logarithmically in $a\mu$, where $\mu$ is the renormalization scale.

There have been several proposals to eliminate this operator mixing with Wilson fermions,
all based on the observation~\cite{Bernard:1987pr,Donini:1999sf}
that, even in the absence of chiral symmetry, the
operator $O_{\rm VA+AV}$ is protected from finite operator mixing
by discrete symmetries, and thus it renormalizes multiplicatively
\be
(O_{\rm R})_{\rm VA+AV}(\mu) = Z_{\rm VA+AV}(g_0, a\mu) O_{\rm VA+AV} (g_0) \,.
\ee

In order to show this, it is convenient to work with massless Wilson fermions,
also having in mind a mass independent renormalization scheme.

First of all we note that $O^{\Delta S=2}$ cannot mix with operators of lower
dimensionality, because such operators do not
have the four-flavour content of the original one.
Thus $O^{\Delta S=2}$
can mix with any other dimension-six operator, provided it has
the same quantum numbers, i.e. with any operator which has the symmetries of 
$O^{\Delta S=2}$ and of the lattice action.
The generic QCD Wilson lattice action with 4 massless quarks
is symmetric under parity $\mathcal{P}$ defined in eq.~(\ref{eq:parity}), 
and charge conjugation $\mathcal{C}$ (see app~\ref{app:B} for the definition).
We define the generic four fermion operators
\be
O_{\Gamma^{(1)}\Gamma^{(2)}} =
(\chibar_1\Gamma^{(1)}\chi_2)(\chibar_3\Gamma^{(2)}\chi_4)
\label{eq:4fermiop}
\ee
for all $\Gamma^{(1)}$ and $\Gamma^{(2)}$
combinations of interest.
Moreover, there are other useful (flavour) symmetries of the action,
namely the switching symmetries $\mathcal{S}'$ and $\mathcal{S}''$ defined by~\cite{Bernard:1987pr}
\be
\mathcal{S}' \colon
\begin{cases}
\chi_1 \leftrightarrow \chi_2 , \\
\chi_3 \leftrightarrow \chi_4
\end{cases}
\label{eq:sprime}
\ee
\be
\mathcal{S}'' \colon
\begin{cases}
\chi_1 \leftrightarrow \chi_4 , \\
\chi_2 \leftrightarrow \chi_3 .
\end{cases}
\label{eq:ssecond}
\ee
In Table \ref{tab:g12} I classify the operators
$O_{\Gamma^{(1)} \Gamma^{(2)}}$, for all $\Gamma^{(1)}$ and $\Gamma^{(2)}$
with negative parity, according to the discrete symmetries
$\mathcal{C}$, $\mathcal{S}'$ and $\mathcal{S}''$, with notation
\be
O_{\Gamma^{(1)}\Gamma^{(2)} \pm \Gamma^{(2)}\Gamma^{(1)}} = 
O_{\Gamma^{(1)}\Gamma^{(2)}}\pm O_{\Gamma^{(2)}\Gamma^{(1)}} .
\ee
Parity violating operators, for which
$\mathcal{CS}''$ is not a symmetry, have been symmetrized or antisymmetrized in order to
obtain eigenstates of $\mathcal{CS}''$.
\begin{table}
\centering
\begin{tabular}{|r|r|r|r|r|r|}
\hline
$O_{\Gamma^{(1)} \Gamma^{(2)}}$ & $\mathcal{P}$ & $\mathcal{CS}'$ & $\mathcal{CS}''$ &
$\mathcal{CPS}'$ & $\mathcal{CPS}''$ \\ \hline \hline
$O_{VA+AV}$ & $-1$ & $-1$ & $-1$ & $+1$ & $+1$ \\
$O_{VA-AV}$ & $-1$ & $-1$ & $+1$ & $+1$ & $-1$ \\
$O_{SP-PS}$ & $-1$ & $+1$ & $-1$ & $-1$ & $+1$ \\
$O_{SP+PS}$ & $-1$ & $+1$ & $+1$ & $-1$ & $-1$ \\
$O_{T \tilde T}$ & $-1$ & $+1$ & $+1$  & $-1$ & $-1$  \\
\hline
\end{tabular}
\caption{Classification of parity violating four-fermion operators
$O_{\Gamma^{(1)} \Gamma^{(2)}}$ according to useful products of 
discrete symmetries $\mathcal{P}$, $\mathcal{C}$, $\mathcal{S'}$ and $\mathcal{S''}$. 
Note that $O_{\tilde T \tilde T} = O_{TT}$ and $O_{T \tilde T} = O_{\tilde T T}$.}
\label{tab:g12}
\end{table}

The parity violating four- fermion operators listed in
table~\ref{tab:g12} do not all have identical $\mathcal{CPS}'$ and $\mathcal{CPS}''$
values. It is straightforward to see that the operator $O_{VA + AV}$ cannot mix with the 
other parity violating operators\footnote{It could still mix with the Fierz transformed 
in Dirac space operator $O_{VA + AV}^F$. It turns out that $O_{VA + AV}^F = O_{VA + AV}$ 
(see ref.~\cite{Donini:1999sf} for details).}.

We are going now to discuss two possible approaches~\cite{Guagnelli:2002xz,Frezzotti:2004wz}
which use Wtm to extract the renormalized
$\langle \bar K^0 \vert O_{\rm VV+AA} \vert K^0 \rangle$ by relating it to the
matrix element of a parity violating operator.
As the Wtm action differs from the standard Wilson fermion action by the mass term, 
the renormalization properties of composite operators
in mass independent renormalization schemes are not modified.
In particular, $O_{\rm VA+AV}$ remains multiplicatively renormalizable,
with the same renormalization constant and running as with
Wilson fermions. Thus finite subtractions are avoided
in the Wtm determination of $B_K$.
In order to disentangle the two approaches we will call the proposal made
in~\cite{Guagnelli:2002xz,Dimopoulos:2006dm} the ``mixed twist'' method while
the proposal in~\cite{Frezzotti:2004wz} the ``mixed action'' method.
Hopefully the reasons for these names will become clear in the following.

\subsubsection{``Mixed twist'' method}

The first variant of the ``mixed twist'' method consists in choosing the following fermionic
action
\be
S_{\rm F}^{(\pi/2)} = a^4 \sum_x \,\, [ \chibar(x) (D_{\rm w,sw} 
+ m_l + i\gamma_5 \tau^3 \mu_l )\chi(x) \,+\,
\bar s(x) (D_{\rm w,sw} + m_s ) s(x) ] .
 \label{eq:tmQCD_action2}
\ee
where $D_{\rm w,sw}$ is the clover improved Wilson operator~(\ref{eq:clovertm},\ref{eq:clover})
and $\chi$ collects the light doublet
\begin{displaymath}
\chi =
\left( \begin{array}{c}
u \\
d
\end{array} \right) .
\end{displaymath}
The label $\pi/2$ refers to the choice of the twist angle that can be obtained
setting $m_{{\rm R},l} = 0$.
We see that while the twist angle of the first doublet is $\pi/2$, the twist
angle of the strange quark is zero, hence the name ``mixed twist''.

The second variant is based on the action
\be
 S_{\rm F}^{(\pi/4)} = a^4 \sum_x \,\, [ \bar u(x) (D_{\rm w,sw} + m_u ) u(x) \,+\,
 \chibar(x) (D_{\rm w,sw} 
+ m_l + i\gamma_5 \tau^3 \mu_l )\chi(x)
]\,.
 \label{eq:tmQCD_action4}
\ee
where now $\chi$ collects a doublet made of a $down$ and a $strange$ quark
\begin{displaymath}
\chi =
\left( \begin{array}{c}
s \\
d
\end{array} \right) .
\end{displaymath}
To set now the twist angle to $\pi/4$ requires $\mu_{{\rm R},l} = m_{{\rm
    R},l}$.
The variant with action~(\ref{eq:tmQCD_action4}) assumes {\em a priori} that the
$s$ and $d$ quarks have degenerate physical masses. This is necessary if one
would like to extend this variant to unquenched simulations: a non degenerate
diagonal twisted doublet would lead to a complex determinant~\cite{Pena:2004gb}.
As long as this action is used in quenched QCD this restriction is not needed
but all the computations carried out with it are performed with degenerate $s$
and $d$ quarks.
The action in eq.~(\ref{eq:tmQCD_action2}), on the other hand, is perfectly well suited for an
unquenched computation. 

It is important to stress that in both the variants not all the quark flavours
are fully twisted, i.e. automatic O($a$) improvement~\cite{Frezzotti:2003ni}
does not apply.
To have full O($a$) improvement of the matrix element it would be
necessary to subtract a number of dimension-seven counterterms from the
four-fermion operator. Such a procedure is highly impractical, and has
not been pursued. Hence leading cutoff effects in $B_K$ with the ``mixed
twist'' methods are expected to be linear in $a$.

We have already discussed in sect.~\ref{sec:basic}, that to relate
renormalized correlation functions computed with tmQCD to QCD, it is enough to
perform the needed change of variables (axial rotations).
This explains the choice of the twist angles for the two variants. In fact performing the
rotation in eq.~(\ref{eq:axial}) for the two actions, we obtain in both cases
\be
\langle K^0 \vert \,\, (O_R)_{\rm VV+AA} \,\, \vert \bar K^0 \rangle_{\rm QCD} =
- i \langle  K^0 \vert \,\, (O_R)_{\rm VA+AV} \,\, \vert \bar K^0 \rangle_{\rm tmQCD} \,,
\ee
which holds in the continuum limit for the two versions of Wtm under
consideration. From this identity, $B_K$ can be extracted from a $K^0$--$\bar K^0$
matrix element of the multiplicatively renormalizable operator $O_{\rm VA+AV}$.

Both these variants has been used to compute $B_K$ with quenched fermions.

In particular the non-perturbative renormalization has been performed in the 
SF scheme~\cite{Guagnelli:2005zc}, and the matrix element has been computed
using both the variants in quenched 
simulations~\cite{Dimopoulos:2006dm,Dimopoulos:2007cn}.
The twist angle has been tuned to $\omega=\pi/2$ setting $m_{{\rm R},l} = 0$
in the clover improved theory. This corresponds to the full twist tuning
discussed in sect.~\ref{sssec:Sym_crit}.
To tune the twist angle to $\pi/4$, the untwisted quark mass has been tuned
such that $\mu_{{\rm R},l} = m_{{\rm
    R},l}$, which via eqs.~(\ref{eq:mr},\ref{eq:mur}) 
and eqs.~(\ref{eq:mtildeq},\ref{eq:mutildeq}) translates into
\be
a m_{q,l} = \dfrac{Z_\mu}{Z_m} a \mu_l \big \{ 1 +
\big [ \dfrac{Z_\mu}{Z_m} (b_\mu - b_m ) - \frac{Z_m}{Z_\mu} \tilde b_m \big ]
a \mu_l \big \}
\ee
For a given choice of $a \mu_l$, $\kappa = (2am_l + 8)^{-1}$ is tuned so that 
$a m_{q,l}$ satisfies one of the two above relations. This requires the
knowledge of all the renormalization constants and improvement coefficients. 
The final result for the renormalization group invariant matrix element
$\hat{B}_K$ being
\be
\hat{B}_K = 0.735(71) .
\label{eq:BKquen}
\ee
The only systematic uncertainty that affects this computation is the usage of
quenched fermions. 

\begin{figure}
  \centering
  \includegraphics[width=.8\linewidth,angle=270]{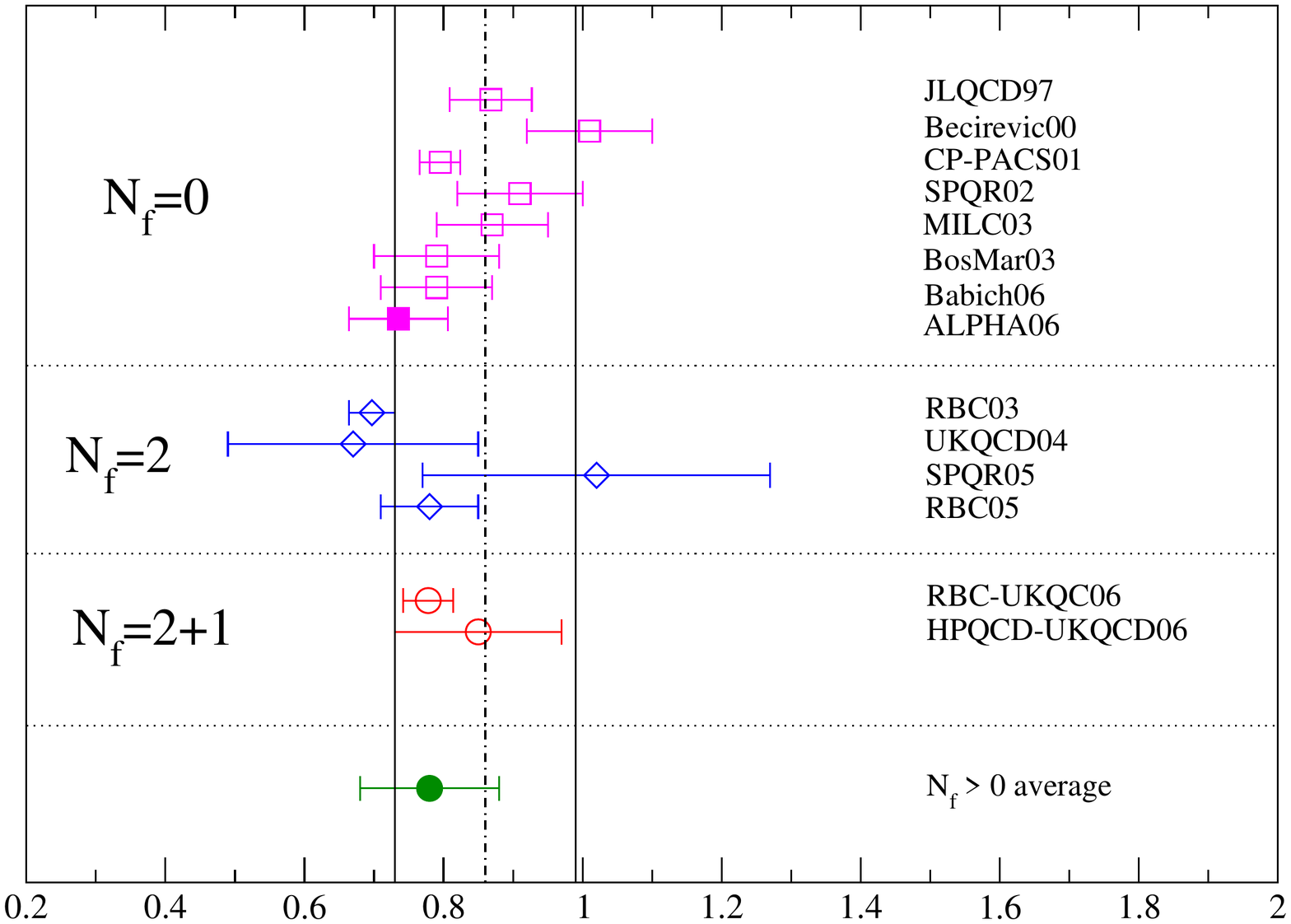}
   \vspace{-0.5cm}\caption{Summary plot of the results for $\hat{B}_K$ obtained by several collaborations
using different lattice actions. From the top to the bottom the refs. 
are~\cite{Aoki:1997nr,Becirevic:2000ki,AliKhan:2001wr,Becirevic:2004aj,DeGrand:2003in,Garron:2003cb,Babich:2006bh,Dimopoulos:2007cn,Aoki:2004ht,Flynn:2004xc,Mescia:2005ew,Dawson:2005za,Antonio:2007pb,Gamiz:2006sq}. The $N_f=0$ 
Wtm result~\cite{Dimopoulos:2007cn} is shown with a filled square.}\vspace{+0.5cm}
  \label{fig:BK_sum}
\end{figure}

In fig.~\ref{fig:BK_sum} we summarize results for $\hat{B}_K$ obtained by several collaborations
using different lattice actions.
In particular we observe that Wtm allows a very precise determinations of $\hat{B}_K$ 
which is competitive with the determination obtained with other
discretizations.
The main drawback of the ``mixed twist'' approach is the fact that the 
leading cutoff effects in $B_K$ are expected to be linear in $a$.

\subsubsection{``Mixed action'' method}

The second method we are going to discuss has all the quark flavours at full
twist, thus retaining automatic O($a$) improvement, but it makes use of
different lattice actions for valence and sea quarks. In particular the
valence actions are chosen in order to remove the unwanted mixings for $O_{\rm
  VA+AV}$.
To obtain this goal the number of valence quarks has to be extended, i.e. some
of the valence quarks involved in the correlation function under investigation
have to be doubled with a slightly different lattice action.
The resulting lattice theory is not unitary thus the approach 
to the continuum limit will be affected by O($a^2$) effects
coming from unitarity violations. With the ``mixed action''
a detailed understanding of these particular discretization 
errors have to be achieved in order to correctly perform the continuum limit.
This does not pose any problem of principle because once the renormalized
quark masses, or equivalently the corresponding pseudoscalar meson masses,
are matched in the continuum limit, one obtains the desired and correct
renormalized matrix element.\footnote{This statement could be put at a more formal level 
using a Ginsparg-Wilson regularization.}

For the light ($l$) and the heavy ($h$) sea doublet quarks the actions
are~\footnote{ In principle it is possible to choose also a non-degenerate quark action for
the light sector.}
\be
S_{\rm Wtm}^{(l)} = a^4 \sum_x \chibar_l(x) \Big[D_{\rm W} + m_0 + i\mu_l\gamma_5\tau^3\Big]\chi_l(x), 
\label{eq:WtmQCDl}
\ee
\be
S_{\rm Wtm}^{(h)} = a^4 \sum_x \chibar_h(x) \Big[D_{\rm W} +
m_0 + i\mu_h\gamma_5\tau^3 + \epsilon_h \tau^1\Big]\chi_h(x), 
\label{eq:WtmQCDh}
\ee
For the particular case of $B_K$ we introduce the following valence
quarks: $d$, $s$ and $s'$. The $d$ and the $s$ valence quarks have the Osterwalder-Seiler (OS) action
\be
S_{\rm OS}^{(f)} = a^4 \sum_x \chibar_f(x) \Big[D_{\rm W} + m_0 + i\mu^v_f\gamma_5\Big]\chi_f(x), 
\label{eq:OS1}
\ee
with $f=d,s$ while the $s'$ quark has the same OS action
\be
S_{\rm OS}^{(s')} = a^4 \sum_x \bar{s}'(x) \Big[D_{\rm W} + m_0 - i\mu^v_{s'}\gamma_5\Big]s'(x), 
\label{eq:OS2}
\ee
but with an opposite sign for the mass term. While this certainly does not
change the sign of the physical quark mass, it changes the leading discretization errors
of the actions, in such a way that to extract $B_K$ it is possible to use a
specific correlation function that renormalizes multiplicatively and it is
automatically O($a$) improved.

The OS action differs from Wtm by the fact that it does not violate
isospin. This action is not used for dynamical fermions since, as we have
discussed in sect.~\ref{sec:basic}, it could generate an $F\widetilde{F}$
in the renormalization process through vacuum polarization diagrams. Because 
of the peculiar flavour structure, this does not happen for Wtm, but anyhow the
OS action can be still used for valence quarks.
To correctly define the theory we would need to introduce for each valence
quark, the corresponding ghost action which cancels exactly the contribution
to the determinant~\cite{Sharpe:2001fh}.
We will assume that this it has been done. All the considerations which
follow will not involve correlation functions of ghost fields, thus the
standard Symanzik expansion will apply~\cite{Golterman:2005ie}, and all the
symmetries we are going to use will be naturally extended to the ghost fields.

We reiterate that the valence and sea quark masses have to be correctly
matched to obtain the correct continuum limit. 
We also recall that to obtain
automatic O($a$) improvement the untwisted quark mass has to be tuned to the
critical value, and so the physical quark mass is totally given by the twisted quark
mass.
This is obtained with the following matching prescription
\be
(\mu_d)_{\rm R} = (\mu_d^v)_{\rm R} ,  \nonumber
\label{eq:match1}
\ee
\be
(\mu_s)_{\rm R} = (\mu_s^v)_{\rm R} = (\mu_{s'}^v)_{\rm R} .
\label{eq:match2}
\ee

With the choice of the valence quark actions specified
in eqs.~(\ref{eq:OS1},\ref{eq:OS2}), it is easy to show, performing the usual
axial rotations, that the original matrix element
\be
\langle \overline{K}^0| O_{\rm R}^{\Delta S=2} |K^0 \rangle_{\rm QCD} =
\frac{8}{3}M_K^2 f_K^2 B_K
\label{eq:BKQCD}
\ee
is equivalent to the matrix element
\be
\langle \overline{K}'^0| 2 (\mathcal{O}_{\rm R})_{VA+AV} |K^0 \rangle =
\frac{8}{3}M_K M_{K'} f_K f_{K'} B_K
\label{eq:BKmodel}
\ee
where this second matrix element on the lattice is computed with the model
specified before.
In particular the bare operator reads
\be
\mathcal{O}_{VA+AV} = i \left[(\bar{s} \gamma_\mu d)(\bar{s}' \gamma_\mu \gamma_5 d) + 
(\bar{s} \gamma_\mu \gamma_5 d) (\bar{s}' \gamma_\mu d) \right]
\label{eq:OVAAVprime}
\ee
and the interpolating fields for the kaons are
\be
K^0(x) = i \bar{d}(x) s(x) , \qquad \overline{K}'^0(x) = \bar{s}'(x)\gamma_5 d(x).
\label{eq:kaons}
\ee
The matrix element~(\ref{eq:OVAAVprime}) can then be extracted by the
asymptotic behaviour of the correlation function
\be
C_{K'OK}(x_0,y_0) = 2 a^6 \sum_{{\bf x},{\bf y}} \langle (\bar{d} \gamma_5
s')(x) \mathcal{O}_{VA + AV} (0) (\bar{d}  s)(y) \rangle , 
\label{eq:KOK}
\ee
with $y_0 < 0$ and $x_0 > 0$. The values of $M_K$,$M_{K'}$,$f_K$ and $f_{K'}$, can be extracted in a
standard fashion from the 2-point correlation functions involving the
interpolating fields in~(\ref{eq:kaons}).
We repeat here briefly the reason why $\mathcal{O}_{VA + AV}$ is
multiplicatively renormalized.
Mixing with lower dimensional operators is excluded by flavour symmetry of
the massless lattice actions, in particular $\mathcal{O}_{VA + AV}$
can mix only with operators such that $\Delta s= \Delta s' = -\Delta d/2
= 1$. 
Parity symmetry of the action in the chiral limit forbids 
mixing with operators of the opposite parity.
Possible dimension 6 operators of negative parity that could mix with
$\mathcal{O}_{VA + AV}$ are
\be
O_{VA - AV}  = (\bar{s} \gamma_\mu d)(\bar{s}' \gamma_\mu \gamma_5 d) -
(\bar{s} \gamma_\mu \gamma_5 d) (\bar{s}' \gamma_\mu d)
\label{eq:OVAprime}
\ee
\be
O_{SP \pm PS}  = (\bar{s}  d)(\bar{s}' \gamma_5 d) \pm
(\bar{s} \gamma_5 d) (\bar{s}'  d)
\label{eq:OSPprime}
\ee
\be
O_{T \tilde{T}}  = \epsilon_{\mu \nu \lambda \rho}(\bar{s}
\sigma_{\mu \nu} d)(\bar{s}' \sigma_{\lambda \rho} d)
\label{eq:OTTprime}
\ee 
To rule out these operators is enough to use $\mathcal{CPS}'$ and $\mathcal{CPS}''$
symmetries defined in eqs.~(\ref{eq:sprime},\ref{eq:ssecond}), as we have done previously.

The main advantages of this approach is that all the quarks are at full twist
so to prove automatic O($a$) improvement it is enough to generalize the form of the 
{\it twisted parity} used in sect.~\ref{sec:impro} 
\be
\mathcal{P}_{\frac{\pi}{2}} \colon
\begin{cases}
U(x_0,\bx;0) \rightarrow U(x_0,-\bx;0), \quad 
U(x_0,\bx;k) \rightarrow U^{-1}(x_0,-\bx - a \hat{k};k), 
\quad k = 1, \, 2, \, 3 \\ 
\chi_{l,h}(x_0,{\bf x}) \rightarrow \gamma_0(i \gamma_5\tau^3)\chi_{l,h}(x_0,-{\bf x}) \\
\chibar_{l,h}(x_0,{\bf x}) \rightarrow  \chibar_{l,h}(x_0,-{\bf x})
(i\gamma_5\tau^3)\gamma_0 \\
q_f(x_0,{\bf x}) \rightarrow  \gamma_0(i \gamma_5)q_f(x_0,-{\bf x}) \\
\bar{q}_f(x_0,{\bf x}) \rightarrow  \bar{q}_f(x_0,-{\bf x}) (i \gamma_5)
\gamma_0 ,
\end{cases}
\label{eq:tmparity2}
\ee
where $q_f$ generically indicates all the valence fields.
Moreover non-degenerate quarks can be introduced retaining the nice property
of a real and positive definite determinant~\cite{Frezzotti:2003xj}.
This approach requires care in the matching procedure between valence and sea
quarks.

An obvious alternative to avoid renormalization problems consists
in using regularizations with exact chiral symmetry. However, the computational
costs involved make it difficult to perform continuum limit extrapolations and
study finite volume effects. A first attempt using Wtm sea and overlap valence
quarks has been presented at the last lattice conference~\cite{Bar:2006zj}.

We reiterate that one drawback of the mixed action approach 
is that the resulting theory is not unitary at non-zero lattice spacing. 
This introduces peculiar O($a^2$) cutoff effects, that nevertheless can be described 
by mixed action chiral perturbation theory~\cite{Bar:2002nr,Bar:2005tu}.
This cutoff effects have been investigated thoroughly in refs.~\cite{Chen:2006wf,Chen:2007ug},
and formul\ae, valid up to O($a^2$), have been given for correlators computed with valence
overlap fermions on a set of different sea quark actions.

The preliminary results of ref.~\cite{Bar:2006zj} indicate that the whole procedure 
is limited by the statistical accuracy achievable on a large volume with overlap fermions,
in order to correctly perform the matching procedure. A possible way to solve this problem is
to match the current quark masses in a finite volume.

\newpage
\section{Algorithms for dynamical fermions}
\label{sec:algo}

At present the only practical way to simulate numerically a 4--D
euclidean quantum field theory with fermions is to perform the Grassmann
integral on the fermion fields analytically and then to apply Monte Carlo methods in the resulting
effective bosonic theory.

After integrating out the fermion fields $\chi,\chibar$ the
partition function~(\ref{eq:partition}) of Wtm for $N_f = 2$ degenerate flavors is given by
\be
  \mathcal{Z} \propto \int\mathcal{D}U \det(Q^\dagger Q)
  e^{-S_\mathrm{G}[U]} = \int\mathcal{D}U e^{-S_\mathrm{eff}[U]} \ ,
\label{eq:ZQQ}
\ee
with
\be
  S_{\rm eff}[U] = S_{\rm G}[U] - \Tr \log Q^\dagger Q \ .
  \label{eq:eff_act}
\ee
The reality of the effective action is only guaranteed by the positivity of
the quark determinant. 

\subsection{Hybrid Monte Carlo}

The determinant $\det(Q^\dagger Q)$ 
can be expressed in terms of the so called pseudofermion complex
fields $\phi$ 
\be
  \det(Q^\dagger Q) \propto \int\mathcal{D}\phi^\dagger\mathcal{D}\phi
  \exp\Bigl(-S_{\mathrm{PF}}[U,\phi^\dagger,\phi]\Bigr)\, ,
  \label{eq:1}
\ee
where $S_{\mathrm{PF}}[U,\phi^\dagger,\phi] = (Q^{-1}\phi,Q^{-1}\phi)$ is the
pseudofermion action with standard scalar product $(u,v)$. The pseudofermion fields $\phi$ are formally
identical to the fermion fields $\chi$, but follow the statistics of
bosonic fields. 

A way to obtain a ``good'' global update with the non-local effective bosonic action 
$S_{\mathrm{PF}}$ is to simulate a {\it microcanonical ensemble} with a suitably
defined Hamiltonian. 
The Hamilton equations have to be integrated with a suitable integration
scheme, and it is then possible to correct for the the finite step errors with
a final stochastic accept-reject step. This is schematically what is called
the Hybrid Monte Carlo (HMC) algorithm~\cite{Duane:1987de}. To be more specific,
the $\phi$ version of the HMC algorithm is based on
the Hamiltonian
\begin{equation}
  \label{hmc:hamiltonian}
  H(\Pi, U, \phi, \phi^\dagger) = \frac{1}{2}\sum_{x,\mu}\Tr[\Pi(x,\mu)^2] + S_\mathrm{G}[U] +
  S_{\mathrm{PF}}[U,\phi,\phi^\dagger]\, ,
\end{equation}
where we introduced traceless hermitian momenta $\Pi(x,\mu)$.
The HMC algorithm is then composed by the following steps
\begin{itemize}
\item Global heat-bath for momenta and pseudofermion fields: \\
the initial momenta are randomly chosen according to a Gaussian distribution
$\exp(-\Pi^2/2)$, and the random fields $R$ are produced from a distribution
like $\exp(-R^\dagger R)$with the initial pseudofermions computed as $\phi =
Q \cdot R$.
\item Molecular dynamics evolution: \\
propose a gauge configuration $U'$ and a momentum $\Pi'$ integrating the
Hamilton equations of motion for the gauge field $U$ and the momentum $\Pi$ at
fixed pseudofermion field $\phi$.
\item Metropolis accept/reject step: \\
The proposals $U'$ and $\Pi'$ are accepted with probability ${\rm
  min}\{1,\exp(-\Delta H)\}$, where $\Delta H = H(\Pi',U',\phi,\phi^\dagger) -
H(\Pi,U,\phi,\phi^\dagger)$.
This step is needed because of the numerical inexact
integration of the equations of motion.
\end{itemize}
It possible to prove that the HMC algorithm satisfies the {\it detailed
  balance} condition~\cite{Duane:1987de} and hence the
configurations generated with this algorithm correctly represent the
intended ensemble. Since the Hamiltonian is conserved up to finite
step errors, the integration can be set up such that the gauge configurations
are globally updated keeping the acceptance rate high.

\subsection{Molecular dynamics evolution}

In the molecular dynamics part of the HMC algorithm the gauge fields
$U$ and the momenta $\Pi$ need to be evolved in a fictitious computer
time $\tau$. 
Differentiation on a compact Lie group is defined, given a generic function G, by
\be
G[U{\rm e}^{i\omega_aT^a}] = G[U] + \omega_a\delta_{\omega}^aG[U] + O(\omega^2) \ ,
\ee
where $U$ is group element, $\omega$ is an infinitesimal vector with dimension 
equal to that of the adjoint representation and $T^a$ are the group generators.
Since the variation $\delta_{\omega}^aG$ takes values in the Lie algebra it is 
natural to let the momenta $\Pi(x,\mu)$ be an element of the Lie algebra.
With respect to $\tau$, Hamilton's equations of motion read
\be
  \frac{d}{d\tau}U(x,\mu) = \Pi(x,\mu) U(x,\mu) \, , \qquad 
  \frac{d}{d\tau}\Pi(x,\mu) = -F(x,\mu) \ ,
  \label{eq:hmc1}
\ee
where the force $F(x,\mu)$ is obtained by differentiation with respect to the gauge links
\be
(\omega,F) = \delta_\omega S \ , \qquad \delta_\omega U(x,\mu) = \omega(x,\mu)U(x,\mu)
\label{eq:deltaS}
\ee
and $S = S_\mathrm{G} + S_{\mathrm{PF}}$. Since
analytical integration of the former equations of motion is normally
not possible, these equations must in general be integrated with a
discretized integration scheme that is area preserving and reversible.
The discrete update with integration
step size $\dtau$ of the gauge field and the momenta can be defined as 
\begin{equation}
  \begin{split}
    T_\mathrm{U}(\dtau)&:\quad U\quad\to\quad U' = E\left( i\dtau \Pi(x,\mu) \right)
      U\, ,\\
    T_\mathrm{S}(\dtau)&:\quad \Pi\quad\to\quad \Pi' = \Pi - i\dtau F\, ,\\
  \end{split}
  \label{eq:hmc2}
\end{equation}
where $E(\cdot)$ stands for the $SU(3)$ exponential function.
To integrate the equation of motion~(\ref{eq:hmc2}) it is possible to use many
different integrators. One of the simplest is the so called leap-frog integrator.
Given eq.~(\ref{eq:hmc2}) one basic time evolution step of the leap-frog reads 
\begin{equation}
  T = T_\mathrm{S}(\dtau/2)\ T_\mathrm{U}(\dtau)\ 
  T_\mathrm{S}(\dtau/2)\, ,
\label{eq:leap-frog}
\end{equation}
and a whole trajectory of length $\tau$ is achieved by a number of molecular dynamics 
$N_\mathrm{MD}=\tau/ \dtau$ successive applications of the transformation $T$.

\subsection{Preconditioning}

For each step along a trajectory of length $\dtau$ the force $F$ has to be computed. 
The computation of the force $F$ is the most expensive part in the HMC
algorithm since the inversion of the Wilson-Dirac operator is
needed.
In order to improve the integration (i.e. to increase the step size $\dtau$ or equivalently 
to reduce the number of steps $N_{\rm MD}$) it would be better to have small forces along the trajectory.
In general this is achieved by preconditioning the HMC algorithm.
A good strategy would then be to split the forces in such a way that those
which require more computer time to be computed 
are those with the smallest magnitude. 
When the chiral limit is approached,
the quark forces tend to increase, 
which requires the step sizes to be adjusted accordingly~\cite{Namekawa:2004bi},
and the choice of the preconditioner can have an
influence on this behaviour.
A parameter that cannot be predicted in a 
dynamical simulation is the autocorrelation
time of the quantity one is interested in.
Extended simulations are needed in order to understand if the autocorrelation
time is under control or not.

In general, preconditioning is always associated with
a factorization of the quark determinant 
into the determinants of certain different operators. 
The number and the type of operators depend on the chosen preconditioner.
A possible efficient preconditioning is obtained using a domain decomposition~\cite{Luscher:2005rx} (DD).
The preconditioner we are going to discuss is the so called 
Hasenbusch acceleration or mass preconditioning
\cite{Hasenbusch:2001ne}. From now on I will discuss only the case of the plain hermitian 
Wilson operator $Q_W$, because all the conclusions can be easily extended to the Wtm operator.
It was realized in ref.~\cite{Hasenbusch:2001ne} that using the identity 
\begin{equation}
  \label{eq:factorization}
  \det Q_W^2 = \det\left(Q_W^2+\rho^2\right)\ \det\left(\frac{Q_W^2}{Q_W^2 +\rho^2}\right) ,
\end{equation}
with an adjustable mass shift $\rho$, can speed up the HMC algorithm. Each of the
two determinants on the r.h.s. of eq.~(\ref{eq:factorization}) is
treated by a separate pseudofermion field $\phi_i$ and a
corresponding pseudofermion action $S_{\mathrm{PF}_i}$.
The Hamiltonian can be written as 
\be
  H = \frac{1}{2}\sum_{x,\mu}\Tr[\Pi(x,\mu)^2] + S_G + S_{\mathrm{PF}_1} + S_{\mathrm{PF}_2} \, .
  \label{eq:generalH}
\ee
where $S_{\mathrm{PF}_1}$ and $S_{\mathrm{PF}_2}$ are the pseudofermion actions respectively related to
$\det\left(Q_W^2+\rho^2\right)$ and $\det\left(\frac{Q_W^2}{Q_W^2 +\rho^2}\right)$.

In ref.~\cite{Hasenbusch:2002ai,DellaMorte:2003jj} it was argued that
the optimal choice for 
$\rho$ is given by $\rho^2 = \sqrt{\lambda_{\rm M}\lambda_{\rm m}}$. Here 
$\lambda_{\rm M}$ ($\lambda_{\rm m}$) is the maximal (minimal)
eigenvalue of $Q_W^2$. The reason for that choice is obtained from minimizing 
the sum of the condition numbers $K = K_1 + K_2$\footnote{
The condition number is the ratio of the maximal over the minimal eigenvalue of a given operator.}
for the operators appearing in $S_{\mathrm{PF}_1}$ and $S_{\mathrm{PF}_2}$.
With the optimal $\rho_{\rm opt}^2 =
\sqrt{\lambda_{\rm M}\lambda_{\rm m}}$ the two condition numbers $K_1$ and $K_2$ are
equal to $\sqrt{\lambda_{\rm M}/\lambda_{\rm m}}$, both of them being
smaller than the condition number of $Q_W^2$ which is
$\lambda_{\rm M}/\lambda_{\rm m}$. 

Since the force contribution in the molecular dynamics evolution is
supposed to be proportional to some power of the condition number, the
force contribution from the pseudofermion part in the action is
reduced and therefore the step size $\dtau$ can be increased, in
practice by about a factor of $2$ \cite{Hasenbusch:2001ne,Hasenbusch:2002ai}.

This preconditioning can be very easily adapted to the Wtm operator
observing that $Q_W^2 + \rho^2 = Q^\dagger Q$ with $\rho$ being the twisted mass
and $Q$ being $\gamma_5$ ``times'' the Wtm operators~(\ref{eq:QWtm}).

\subsection{HMC with multiple time scale integration and mass preconditioning}

In the paper by L\"uscher~\cite{Luscher:2005rx} impressive acceleration factors were 
obtained with a DD preconditioned HMC compared with a plain HMC.
One of the reasons for this impressive result is the observation that the
resulting forces after the preconditioning 
have a magnitude which decreases as the computer time needed to invert the corresponding operator
increases.
It is then beneficial to integrate each 
single force with different time steps.
This hierarchy of forces is obtained with a preconditioner that provides a strong 
infrared cutoff and separates low- and high-frequency modes of the system.
In~\cite{Urbach:2005ji} the idea was explored of using as an infrared cutoff 
the mass in the Hasenbusch acceleration method, which could also give a similar hierarchy of forces
that could then be combined with a suitable integrator.
Therefore it might be advantageous to change the point of view: instead
of tuning $\rho$ {\`a} la
refs.~\cite{Hasenbusch:2001ne,Hasenbusch:2002ai}, i.e. minimizing the condition number of the operators
appearing in the pseudofermion action, rather to exploit the
possibility of arranging the forces by the help of mass
preconditioning with the aim to reach a situation in which a
multiple time scale integration scheme is favorable, i.e. tuning $\rho$ to achieve
a hierarchy of forces as with the DD-HMC.

The idea to combine a multiple time step integrator with a separation of
infrared and ultraviolet modes was already proposed in
ref.~\cite{Peardon:2002wb}. This idea was applied to mass
preconditioning by using only two time scales in
refs.~\cite{AliKhan:2003mu,AliKhan:2003br} in the context of clover
improved Wilson fermions. However, a comparison of results
presented in the next section to the ones of
refs.~\cite{AliKhan:2003mu,AliKhan:2003br} is not possible, because
volume, lattice spacing and masses are different. The gain reported in~\cite{AliKhan:2003mu}
compared to the performance of the method described 
in~\cite{Hasenbusch:2001ne,Hasenbusch:2002ai} was at most $20\%$.

In order to generalize the leap frog integration scheme~(\ref{eq:leap-frog}) 
we assume, in the following, that we can bring the Hamiltonian to the form
\begin{equation}
  \label{eq:generalH}
  H = \frac{1}{2}\sum_{x,\mu}\Tr[\Pi(x,\mu)^2] + \sum_{k=0}^n S_k[U]\, ,
\end{equation}
with $n\geq 1$. For instance with $n=1$ $S_0$ might be identified with
the gauge action and $S_1$ with the pseudofermion action of
eq.~(\ref{hmc:hamiltonian}). 

Clearly, in order to keep the discretization errors small in an 
algorithm like leap frog, the time steps have to be small if the driving
forces are large. Hence multiple time scale integration is a valuable
tool, if the forces originating from the single parts in the
Hamiltonian (\ref{eq:generalH}) differ significantly in their absolute
values. Then the different parts in the Hamiltonian might be integrated
on time scales inversely proportional to the corresponding forces.

The leap frog integration scheme can be generalized to multiple time
scales as has been proposed in ref.~\cite{Sexton:1992nu} without loss of
reversibility and the area preserving property. The scheme with only
one time scale can be recursively extended by starting with the
definition
\begin{equation}
  T_0 = T_{\mathrm{S}_0}(\dtau_0/2)\ T_\mathrm{U}(\dtau_0)\ 
  T_{\mathrm{S}_0}(\dtau_0/2)\, ,
\end{equation}
with $T_\mathrm{U}$ defined as in eq.~(\ref{eq:hmc2}) and where
$T_{\mathrm{S}_k}(\dtau)$ is given by
\begin{equation}
  \label{eq:basicT}
  T_{\mathrm{S}_k}(\dtau_k)\quad : \quad P\quad\to\quad P - i \dtau_k F_k \, .
\end{equation}
As $\dtau_0$ will be the smallest time scale, we can recursively
define the basic update steps $T_k$, with time scales $\dtau_k$
as
\begin{equation}
  \label{eq:lfrecursive}
  T_k = T_{\mathrm{S}_k}(\dtau_k/2)\
  [T_{k-1}]^{N_{k-1}}\ T_{\mathrm{S}_k}(\dtau_k/2)\, ,
\end{equation}
with integers $N_k$ and $0<k\leq n$. One full trajectory $\tau$ is then composed
by $[T_n]^{N_n}$. The different time scales $\dtau_k$ in
eq.~(\ref{eq:lfrecursive}) must be chosen such that the total number
of steps on the $k$-th time scale $N_{\mathrm{MD}_k}$ times $\dtau_k$
is equal to the trajectory length $\tau$ for all $0\leq k\leq n$:
$N_{\mathrm{MD}_k}\dtau_k=\tau$. This is achieved by setting
\begin{equation}
  \label{eq:timescales}
  \dtau_k = \frac{\tau}{N_n\cdot N_{n-1}\cdot...\cdot N_k} =
  \frac{\tau}{N_{\mathrm{MD}_k}}\, ,\qquad 0\leq k\leq n\, ,
\end{equation}
where $N_{\mathrm{MD}_k}=N_n\cdot N_{n-1}\cdot...\cdot N_k$.

In ref.~\cite{Sexton:1992nu} also a partially improved integration scheme with multiple
time scales was introduced, which reduces the size of the discretization errors.

To summarize we decompose the pseudofermion action as
\be
S = \sum_{k=0}^n S_k[U]
\label{eq:fact2}
\ee
with
\be
S_0 = S_G[U], \qquad S_1 = (Q_1^{-1}\phi_k,Q_1^{-1}\phi_k) , 
\ee
\be
S_k = (Q_{k}^{-1}Q_{k-1}\phi_k,Q_k^{-1}Q_{k-1}\phi_k) \quad 1 \le k \le n
\ee
where
\be
Q_k = Q_{\rm W} + i\rho_k \qquad  \rho_k < \rho_{k-1}
\ee
and $Q_n$ is the operator we are interested in: $Q_{\rm W}$ if we want to perform
dynamical simulations with Wilson fermions or $Q_n = Q_{\rm W} + i \mu_{\rm q}$ if we want to perform
simulation with Wtm ($\mu_{\rm q} < \rho_k \forall k$).
The strategy is then to tune
$\rho_k$ and $n$ in eq.~(\ref{eq:fact2}) such that the more expensive the
computation of a certain $F_k$ is,  
the less it contributes to the total force. The different parts
of the action can then be integrated on different time scales
$\dtau_k$ chosen according to their force magnitude $F_k$, guided by
$\dtau_k F_k=\mathrm{const}$ for all $k$. 

\subsection{Results}

In ref.~\cite{Urbach:2005ji,Jansen:2005yp} it was shown that this idea proves to
be useful in practice. Good performances were found for the mtM-HMC 
({\it m}ultiple {\it t}ime scales {\it m}ass preconditioned-HMC)
compared to the DD-HMC of ref.~\cite{Luscher:2005rx} and to a plain HMC as
used in ref.~\cite{Orth:2005kq}. 
The algorithm was tested using the standard Wilson gauge action~(\ref{eq:Wilson_gauge}) 
and $N_f=2$ degenerate Wilson fermions (eq.~\ref{eq:WtmQCD} with $\mu_{\rm q} = 0$). 
The simulations were done on
$24^3\times32$ lattices with $\beta=5.6$ and estimated pseudoscalar masses of
$m_\mathrm{PS}=665\ \mathrm{MeV}$, $485\ \mathrm{MeV}$, $380\
\mathrm{MeV}$ and $300\ \mathrm{MeV}$ (runs $A$, $B$, $C$ and $D$). 
Details of the algorithm
parameters as well as results for several quantities such as the
plaquette expectation value or the vector mass $m_\mathrm{V}$ can be
found in ref.~\cite{Urbach:2005ji}.

The first important observation from this investigation is that for all
four aforementioned simulation points the preconditioning masses and
time scales can be tuned such that simulations are
stable. Examples for Monte Carlo histories of the plaquette
expectation value or $\Delta H$ can be found in
ref.~\cite{Urbach:2005ji}. 

In order to compare the performance of the mtM-HMC to other
HMC variants one could choose two different measures. The first is the
\emph{performance figure} 
$\nu = 10^{-3}(2N_n+3)\tau_\mathrm{int}(P)$ as introduced in
ref.~\cite{Luscher:2005rx}. $\tau_\mathrm{int}(P)$ is the integrated
autocorrelation time of the plaquette and $n$ is the number of
integration steps for the physical operator $Q_W^2$ necessary for one
trajectory. $\nu$ represents the number of inversions of the
operator $Q_W$ in thousands needed in order to obtain one independent
configuration. It is clearly algorithm and machine independent, but it
does not account for the preconditioning overhead, which is, at least
for the mtM-HMC, not completely negligible. 
The estimate was that this overhead was roughly a factor of two, 
and this has been recently confirmed in~\cite{Meyer:2006hx}.

\begin{table}[t!]
  \centering
  \begin{tabular*}{.9\textwidth}{@{\extracolsep{\fill}}lcccc}
    \hline\hline
    $\bigl.\Bigr.$& $\kappa$ & $\nu$\cite{Jansen:2005yp} (mtM-HMC) & 
$\nu$\cite{Luscher:2005rx,Luscher:2005mv} (DD-HMC)& $\nu$\cite{Orth:2005kq} (HMC)\\
    \hline
    $A$ & $0.15750$ & $0.09(3)$ & $0.69(29)$ & $1.8(8)$\\

    $B$ & $0.15800$ & $0.11(3)$ & $0.50(17)$ & $5.1(5)$\\

    $C$ & $0.15825$ & $0.23(9)$ & $0.62(23)$ & -\\
    
    $D$ & $0.15835$ & $\simeq 0.35$    & $0.74(18)$ & -\\
    \hline\hline
  \end{tabular*}
  \caption{Comparison of $\nu$ values from ref.~\cite{Jansen:2005yp},
    ref.~\cite{Luscher:2005rx} (with updates from \cite{Luscher:2005mv}
    and ref.~\cite{Orth:2005kq}). The $\nu$-value
for simulation point $D$ is only based on an extrapolation of
$\tau_\mathrm{int}(P)$ in $1/m_\mathrm{PS}^2$.} 
  \label{tab:nu}
\end{table}

The results for the $\nu$-values are summarized in table \ref{tab:nu}
and, while the $\nu$-values for the mtM-HMC and the DD-HMC 
are comparable, they are
significantly smaller than the values extracted for the plain HMC
algorithm used in ref.~\cite{Orth:2005kq}.

\newpage
\section*{Concluding remarks}
\label{sec:conclu}

I have presented an overview of the theoretical properties and numerical
results of Wilson twisted mass QCD. 
To warm up, I started by considering a classical theory with a twisted mass
term (tmQCD). In the continuum a twisted mass term can always be rotated away
by a non-anomalous change of fermion variables in the functional integral.

On the lattice the twisted mass cannot be rotated away: the standard Wilson
lattice action and the Wilson twisted mass (Wtm) lattice action are not related by a
change of variables, and as a consequence have different discretization errors.
The difference between the two discretizations is governed by the amount of
disaglinement, also called twist, in the chiral space between the Wilson term and the mass term.
After describing the theoretical properties of Wtm QCD I have shown how the choice
of working at full twist, i.e. maximally disaligning the Wilson term and the
mass term, is rewarding in many respects.
At full twist, the dimension five operators describing the leading
discretization errors in physical quantities break chiral symmetry in an
``orthogonal'' direction respect to the mass term. As a consequence their insertions
in correlation functions of multiplicatively renormalizable operators vanish.
Physical quantities are then automatically O($a$) improved if we stay at
full twist at finite lattice spacing, without the knowledge of any of the
improvement coefficients required by the implementation of the Symanzik
improvement program.
After discussing many technical issues concerning the choice of the critical
mass, I have rederived some of the results using the so called physical basis,
where the twist is applied to the Wilson term, keeping the mass term in the
standard chiral direction.

The disaglinement in chiral space between the mass term and Wilson term
causes, at finite lattice spacing, the breaking of isospin and parity
symmetry.
Both symmetry breakings appear as O($a^2$) cutoff effect. The consequences of the
breaking of parity symmetry at finite lattice spacing can be analyzed in the
correlation functions with standard techniques. Numerical results seem to
indicate that parity breaking cutoff effects are well under control.
Isospin breaking cutoff effects induce splittings among flavour multiplets.
In particular the neutral and the charged pions are not degenerate at finite
lattice spacing, even if they contain degenerate quarks.
The splitting is an O($a^2$) effects but numerically not negligible.
I have shown that one way to partially mitigate this problem is to use a
discretization for the valence quarks which do not break flavour symmetry like
the Osterwalder-Seiler action or a Ginsparg-Wilson action. Carefully matching
the renormalized quark masses for the valence and the sea quark action, it is
then possible to eliminate the isospin splitting at the valence level.
At the level of virtual quarks and virtual pions the problem cannot be
eliminated, and this is certainly an issue that has to be investigated both
theoretically and numerically in the future. 
Isospin splitting in other sectors, like in baryon multiplets, seems to be
numerically under control.

The O($a^2$) cutoff effects of a Wilson-like theory deform the diagram of the
chiral phase transition. The chiral phase diagram can be analyzed using a
generalization of chiral perturbation theory ($\chi$PT) at finite lattice
spacing to include the O($a$) and O($a^2$) cutoff effects. This generalization
called W(ilson)$\chi$PT predicts two possible scenarios for the
chiral phase diagram: the Aoki scenario, the Sharpe-Singleton scenario. There
are numerical evidences that close enough to the continuum limit, with a Wilson lattice gauge action and Wtm lattice
fermion action the Aoki scenario applies for the quenched model while the Sharpe-Singleton scenario
applies for dynamical quarks.
A consequence of the Sharpe-Singleton scenario is that even if the twisted
mass term gives a sharp infrared cutoff to the lattice theory on each gauge
background, there is a minimal value of the twisted mass which can be
simulated before encountering a second order phase transition point where the
neutral pion is massless. The position of this second order point goes to zero
in the continuum as O($a^2$) and depends on all the details of the lattice
action used. In particular I have shown how using a slightly modified gauge
action, like the tree-level Symanzik (tlSym) improved gauge action can
mitigate this problem allowing simulations at smaller quark masses given a
certain lattice spacing. 
In particular the tlSym gauge action and the Wtm fermion action allow
simulations at a lattice spacing of $a\simeq 0.1$ fm with $M_{\pi} \simeq 300$ MeV.

A further advantage of Wtm is the possibility to ease the renormalization
patterns of phenomenologically relevant physical quantities.
The pseudoscalar decay constant, contrary to what happens with Wilson and
clover improved fermions, can be computed without the knowledge of any
renormalization constant, because at full twist, the decay constant
is related to the vector current which is protected from renormalization,
because of the exact flavour symmetry of Wtm in the massless limit.
Recent results with $N_f=2$ dynamical simulations show that the chiral
behaviour in the pseudoscalar sector can be analyzed very precisely and, within
the current understanding of all the systematic effects, it is consistent with NLO $\chi$PT.
A second example is the provided by $B_K$ which parametrizes the
non-perturbative contribution to the indirect CP violation in the kaon sector.
Two different strategies have been presented which remove the mixing of the
relevant four-fermion operator, both based on the observation that $B_K$ can
be evaluated also from a parity violating four fermion operator which
renormalizes multiplicatively and can be extracted using twisted mass
regularizations.

I have concluded, in the last section, with a brief digression on recent 
algorithmic developments to simulate dynamical Wilson-like fermions.

Wtm can be used to describe also non-degenerate quarks. I have discussed two
different discretizations which describe non-degenerate quarks in the context
of tm QCD. First results with {\it up}, {\it down}, {\it
  strange} and {\it charm} dynamical quarks show that simulations of realistic
QCD with realistic physical parameters are within reach of the current
machines and algorithms.

Although presently not all aspects of Wtm are fully investigated, Wtm is
a powerful discretization of lattice QCD, and it
certainly belongs to the pool of well founded fermion actions that ought to be
used to control the continuum limit of physical quantities of interest.

I hope that the ongoing theoretical and numerical investigations related to
Wtm might help in the quest for solving QCD.

\newpage
\section*{Acknowledgements}
\label{sec:ack}

I am indebted to all those people who have contributed, in one way or another,
to the development and understanding of Wilson twisted mass QCD. Special
thanks go to Roberto Frezzotti, Karl Jansen and Giancarlo Rossi for many enlightening
discussions and suggestions in several subjects covered by this
report, and for a careful reading of parts of this manuscript.
I thank the mysterious referee for a careful reading and many interesting remarks.
Special thanks to all the members of the European twisted mass collaboration
(ETMC) for a most enjoyable collaboration and for triggering many of the 
issues investigated in this work.
In particular I thank P. Dimopoulos, R. Frezzotti, G. Herdoiza and C. Michael
for sending me numerical data. 
I acknowledge also discussions with B. Blossier, 
N. Garron, G. Herdoiza and C. Michael.
Further thanks go to the students in DESY-Zeuthen for their feedback
while writing this report, in particular I. Hailperin and J. Gonzalez for
checking some of the formul{\ae} presented here.
It is a pleasure to thank Chris Michael to go through the first version of the
entire report to correct and improve my English. 
Last but not least I want to thank my parents Vincenza Cardarelli and 
Maurice Shindler, and my grandfather Harry Shindler for a persistent and decisive 
encouragement.  

\newpage
\appendix
\section{Definitions and conventions}
\label{app:A}

\subsection{Index conventions}

Lorentz indices are taken from the middle Greek alphabet $\mu,\nu, \ldots $ and run from 0 to 3.
Latin indices $k$, $l$, $\ldots$ are used to label the components of the spatial vectors and run from 1 to 3.
For Dirac indices we use letters $\alpha$, $\beta$, $\ldots$ from the beginning of the Greek alphabet 
and they run from 1 to 4. Colour vectors in the fundamental representation of SU($N_{\rm c}$) carry indices 
$A$,$B$, $\ldots$ ranging from 1 to $N_{\rm c}$, while for vectors in the adjoint representation capital letters
$a$, $b$, $\ldots$ running from 1 to $N_{\rm c}^2-1$ are employed.
By abuse of notation for flavour vectors in the fundamental representation of SU($N_{\rm f}$) we use latin
indices $i$,$j$,$\ldots$ ranging from 1 to $N_{\rm f}$ while for vectors in the adjoint representation 
indices $a$,$b$,$\ldots$ running from 1 to $N_{\rm f}^2-1$ are used. 
It will be clear from the context to which case we are refering to.
Repeated indices are always summed over unless stated and scalar products 
are taken with euclidean metric.

\subsection{Dirac matrices}

We choose a chiral representation for the Dirac matrices, where

\begin{displaymath}
\gamma_\mu =
\left(\begin{array}{ccc}
0 & \phantom{0} & e_\mu \\
e_\mu^\dagger & \phantom{0} & 0
\end{array}\right).
\end{displaymath}
The $2 \times 2$ matrices $e_\mu$ are taken to be
\be
e_0 = -1, \qquad e_k = -i\sigma_k ,
\ee
with $\sigma_k$ the Pauli matrices
\begin{displaymath}
\sigma_1 = 
\left(\begin{array}{c c c}
0 & \phantom{0} & \phantom{-}1 \\
1 & \phantom{0} & \phantom{-}0
\end{array}\right), \qquad
\sigma_2 = 
\left(\begin{array}{c c c}
0 & \phantom{0} &-i \\
i & \phantom{0} & 0
\end{array}\right), \qquad
\sigma_3 = 
\left(\begin{array}{c c c}
1 & \phantom{0} & 0 \\
0 & \phantom{0} & -1
\end{array}\right).
\end{displaymath}
It is then easy to check that 
\be
\gamma_\mu = \gamma_\mu^\dagger , \qquad \{\gamma_\mu , \gamma_\nu \} = 2 \delta_{\mu \nu} . 
\ee
Furthermore we define 
\begin{displaymath}
\gamma_5 = \gamma_0 \gamma_1 \gamma_2 \gamma_3  \qquad  \Rightarrow  \qquad \gamma_5 =
\left(\begin{array}{c c c}
1 & \phantom{0} & 0 \\
0 & \phantom{0} & -1
\end{array}\right).
\end{displaymath}
In particular
\be
\gamma_5 = \gamma_5^\dagger , \qquad \gamma_5^2 = 1 ,
\ee
and the hermitian matrices
\be
\sigma_{\mu \nu} = \frac{i}{2}[\gamma_\mu , \gamma_\nu], \qquad \sigma_{\mu \nu} = \sigma_{\mu \nu}^\dagger , 
\ee
are explicitly given by
\begin{displaymath}
\sigma_{0 k} = 
\left(\begin{array}{c c c}
\sigma_k & \phantom{0} & 0 \\
0 & \phantom{0} & -\sigma_k
\end{array}\right) , \qquad
\sigma_{i j} = - \epsilon_{i j k} 
\left(\begin{array}{c c c}
\sigma_k & \phantom{0} & 0 \\
0 & \phantom{0} & \sigma_k
\end{array}\right) , 
\end{displaymath}
where $\epsilon_{i j k}$ is the totally antisymmetric tensor with $\epsilon_{1 2 3} = 1$.

\subsection{Gauge group}

A representation of the Lie algebra of SU($N_{\rm c}$) is given by complex $N_{\rm c} \times N_{\rm c}$ 
matrices $X_{A B}$ which satisfy
\be
X^\dagger = - X , \qquad {\rm tr} \{ X \} = X_{A A} = 0 . 
\ee
It is possible to choose a basis in this matrix space $T^a$, $a=1$,$2$,$\ldots N_{\rm c}^2-1$ 
that satisfies
\be
{\rm tr} \{ T^a T^b \} = -\frac{1}{2}\delta^{a b}.
\ee 
For $N_{\rm c}=3$ the standard basis is 
\be
T^a = \frac{\lambda^a}{2 i} , \qquad a=1, \ldots 8,
\ee
where $\lambda^a$ denote the Gell-Mann matrices.
With this convention the structure constants $f^{a b c}$, defined by
\be
[ T^a , T^b ] = f^{a b c} T^c ,
\ee
are real and totally antisymmetric.

\subsection{Lattice derivatives}

Ordinary lattice forward and backward derivatives are diagonal in colour space and are defined by
\be
\drv{\mu} f(x) = \frac{1}{a} [ f(x + a \hat{\mu}) - f(x) ] , 
\label{eq:partialmu}
\ee
\be
\drvstar{\mu} f(x) = \frac{1}{a} [ f(x) - f(x - a \hat{\mu}) ] , 
\ee
where $\hat{\mu}$ denotes the unit vector in direction $\mu$.
The gauge covariant derivatives acting on a quark fields are not trivial anymore in colour space 
and they are defined by
\be
\nab{\mu} \chi(x) = \frac{1}{a} [ U(x , \mu) \chi(x + a \hat{\mu}) - \chi(x) ] , 
\ee
\be
\nabstar{\mu} \chi(x) = \frac{1}{a} [ \chi(x) - U(x - a \hat{\mu} , \mu)^{-1} \chi(x - a \hat{\mu}) ] .
\ee
The left action of the lattice derivative operators is defined by
\be
\chibar(x)\lvec{\nab{\mu}}  = 
\frac{1}{a} [ \chibar(x + a \hat{\mu})U(x , \mu )^{-1}  - \chibar(x) ] , 
\ee
\be
\chibar(x)\lvec{\nabstar{\mu}}  = 
\frac{1}{a} [ \chibar(x) - \chibar(x - a \hat{\mu}) U(x - a \hat{\mu} , \mu)  ] .
\ee

\subsection{Continuum gauge fields}

The gauge field in a continuum gauge theory belongs to the algebra of the gauge group and may be written as
\be
G_\mu(x) = G_\mu^a(x)T^a
\ee
with real components $G_\mu^a(x)$. The field strength tensor
\be
F_{\mu \nu} = \partial_\mu G_\nu(x) - \partial_\nu G_\mu(x) + [G_\mu(x) , G_\nu(x) ] ,
\ee
can be also decomposed in the same way.
The right and left covariant derivatives are defined by
\be
D_\mu \chi(x) = (\partial_\mu + G_\mu) \chi(x)
\label{eq:Dmur}
\ee
\be
\chibar(x) \stackrel{\leftarrow}{D}_\mu  = \chibar(x)(\stackrel{\leftarrow}{\partial}_\mu - G_\mu). 
\label{eq:Dmul}
\ee
We stress that $\partial_\mu$ in formul\ae (\ref{eq:Dmur},\ref{eq:Dmul}) are continuum partial derivatives, 
and do not have to be confused with the lattice derivative (\ref{eq:partialmu}).

\section{Symmetries in the twisted basis}
\label{app:B}

We list here the form of the relevant symmetries for a generic angle $\omega$.
The $SU(2)$ axial and vector twisted transformations take the form
\be
SU_{\rm V}(2)_\omega \colon
\begin{cases}
   \chi(x)     \longrightarrow 
               \exp(-i\frac{\omega}{2}\gamma_5\tau^3)\exp(i\frac{\alpha_V^a}{2}\tau^a)
\exp(i\frac{\omega}{2}\gamma_5\tau^3)\chi(x),\\
   \chibar(x)  \longrightarrow 
               \chibar(x)\exp(i\frac{\omega}{2}\gamma_5\tau^3)\exp(-i\frac{\alpha_V^a}{2}\tau^a)
\exp(-i\frac{\omega}{2}\gamma_5\tau^3) .
\end{cases}
\label{eq:tv}
\ee
\be
SU_{\rm A}(2)_\omega \colon
\begin{cases}
   \chi(x)     \longrightarrow 
               \exp(-i\frac{\omega}{2}\gamma_5\tau^3)\exp(i\frac{\alpha_A^a}{2}\gamma_5\tau^a)
\exp(i\frac{\omega}{2}\gamma_5\tau^3)\chi(x),\\
   \chibar(x)  \longrightarrow 
               \chibar(x)\exp(i\frac{\omega}{2}\gamma_5\tau^3)\exp(i\frac{\alpha_A^a}{2}\gamma_5\tau^a)
\exp(-i\frac{\omega}{2}\gamma_5\tau^3) .
\end{cases}
\label{eq:ta}
\ee

The twisted discrete symmetries that involve axis reflections (parity and time reversal), 
using the gamma matrix representation given above, are 
\be
\mathcal{P}_\omega \colon 
\begin{cases}   U(x_0,{\bf x};0) \longrightarrow  U(x_0,-{\bf x};0), \quad  
   U(x_0,{\bf x};k)  \longrightarrow  U^{-1}(x_0,-{\bf x} - a \hat{k};k), \quad k = 1,2,3 \\
   \chi(x_0, \bx)    \longrightarrow  
               \gamma_0\exp(i\omega\gamma_5\tau^3)\chi(x_0,-\bx),\\
   \chibar(x_0,\bx)  \longrightarrow 
               \chibar(x_0,-\bx)\exp(i\omega\gamma_5\tau^3)\gamma_0,
\end{cases}
\label{eq:pomega}
\ee
\be
\mathcal{T}_\omega \colon
\begin{cases}
   U(x_0,{\bf x};0) \longrightarrow  U^{-1}(-x_0-a,{\bf x};0), \quad 
   U(x_0,{\bf x};k)  \longrightarrow  U(-x_0,{\bf x};k), \quad k = 1,2,3 \\
   \chi(x_0,\bx)     \longrightarrow  
               i\gamma_0\gamma_5\exp(i\omega\gamma_5\tau^3)\chi(-x_0,\bx),\\
   \chibar(x_0,\bx)  \longrightarrow 
               -i \chibar(-x_0,\bx)\exp(i\omega\gamma_5\tau^3)\gamma_5\gamma_0.
\end{cases}
\label{eq:tomega}
\ee
Charge conjugation takes a form that is invariant under the 
change of variables~(\ref{eq:axial})
\be
\mathcal{C} \colon
\begin{cases}
   U(x;\mu) \longrightarrow  U(x;\mu)^*, \\
   \chi(x)    \longrightarrow  
                C^{-1}\chibar(x)^T,\\
   \chibar(x) \longrightarrow 
               - \chi(x)^T C ,
\end{cases}
\label{eq:chargeconj}
\ee
where $C$ satisfies
\be
- \gamma_\mu^T = C \gamma_\mu  C^{-1} , \qquad \gamma_5 = C \gamma_5  C^{-1}. 
\ee
In the chosen gamma matrix representation a possible choice is $C=i\gamma_0 \gamma_2$
so that $C=C^\dagger=C^{-1}$.

\section{Transfer matrix}
\label{app:C}

We use here the original notation of ref.~\cite{Luscher:1976ms}.
The transfer matrix as an operator in Fock space and as an
integral kernel with respect to the gauge fields
has the structure
\begin{equation}
  {T}_0[U,U'] = \hat T_{\rm F}^\dagger(U)K_0[U,U']\hat T_{\rm F}^{}(U'),
\label{transfer}
\end{equation}
with pure gauge kernel $K_0$ and the fermionic part
\begin{equation}
   \hat T_{\rm F}(U)=\det(2\kappa B)^{1/4}
   \exp(\hat{\eta}^\dagger P_-C\hat\eta)
   \exp(-\hat{\eta}^\dagger\gamma_0 M\hat\eta).        
\end{equation}
Here, the operators $\hat{\eta}_i({\bf x})$ are canonical 
($i$ is a shorthand for colour, spin and flavour indices)  
\begin{equation}
   \{\hat\eta_i^{}({\bf x}),\hat\eta_j^\dagger({\bf y})\}
   =\delta_{ij}a^{-3}\delta_{{\bf x}{\bf y}},  
\end{equation}
and $B$ and  $C$ are matrix representations of the
difference operators
\be
B(\bx,\by)_{A\alpha;B\beta} = \delta_{\alpha\beta}\Big\{\delta_{AB}\delta(\bx,\by) - \kappa
\sum_{k=1}^{3}\Big[U(\bx;k)_{AB}\delta(\bx+a\hat{k},\by) + U(\by;k)^{-1}_{AB}\delta(\by+a\hat{k},\bx)\Big]\Big\}
\label{eq:B}
\ee
\be
C(\bx,\by)_{A\alpha;B\beta} = \sum_{k=1}^3 (\gamma_k)_{\alpha \beta} \frac{1}{2} 
\Big[U(\bx;k)_{AB}\delta(\bx+a\hat{k},\by) - U(\by;k)^{-1}_{AB}\delta(\by+a\hat{k},\bx)\Big]
\label{eq:C}
\ee

The matrix $M$ is defined as
\begin{equation}
   M = \frac12 \ln\left(\frac{1}{2\kappa} B\right),
\end{equation}
In order to prove the positivity of the transfer matrix it is enough to show that 
$\hat T_{\rm F}$ is bounded and invertible and that $K_0[U,U']$ is positive. $K_0[U,U']$
depends only on the gauge action and the proof of its positivity for the
Wilson gauge action can be found in~\cite{Luscher:1976ms}.
What remains to be proven is the positivity of $B$.
Let's indicate with $\hat{\mathcal{U}}_i$ the following unitary operator

\be
\hat{\mathcal{U}}_i(\psi)(x) = U_i(x)\psi(x+a\hat{i})
\label{eq:T}
\ee

then the operator whose kernel is B~(\ref{eq:B}) can be written as

\be
\hat{B} = 1 - \kappa\sum_{i=1}^{3}\Big[\hat{\mathcal{U}}_i+\hat{\mathcal{U}}_i^{\dagger}\Big] 
\label{eq:Bop}
\ee
We can then write the eigenvalues of this matrix in the following way
\be
\lambda(\hat{B}) = 1 - \kappa\sum_{i=1}^{3}\Big[2 {{\mathbb R}\rm{e}}\lambda(\hat{\mathcal{U}}_i)\Big]
\label{eq:lambda}
\ee
and given the unitarity of $\hat{\mathcal{U}}_i$ we can conclude
\be
1-6 \kappa \le \lambda(\hat{B}) \le  1+6 \kappa 
\label{eq:kappa}
\ee

From this we can conclude that if $|\kappa| < {1 \over 6}$ then the operator $B$
is positive definite.

We remark here that in principle physical positivity can be violated 
at finite lattice spacing, but one should have available a clean definition 
of the transfer matrix and the corresponding eigenvalues.
Moreover positivity should be lost only at the cutoff scale, in order to become
unimportant in the continuum limit.
In ref.~\cite{Luscher:1984is} it has been shown indeed that this is the case 
for improved gauge actions.

In ref.~\cite{Frezzotti:2001ea} it was shown that Wtm for degenerate quarks 
has also a well defined and positive tranfer matrix.
The twisted mass term can be added to the antihermitian $C$ matrix without changing the given proof
\bea
C_{\rm Wtm}(\bx,\by)_{A\alpha i;B\beta j} &=& 
\delta_{ij} \sum_{k=1}^3 (\gamma_k)_{\alpha \beta} \frac{1}{2} 
\Big[U(\bx;k)_{AB}\delta(\bx+a\hat{k},\by) - U(\by;k)^{-1}_{AB}\delta(\by+a\hat{k},\bx)\Big]  \nonumber \\
&+& ia\mu_{\rm q}(\gamma_5)_{\alpha \beta}(\tau^3)_{ij} \delta_{AB} \ ,
\label{eq:C}
\eea
except that $\mu_{\rm q}$ must be real.

But if we want to add a twisted term for non-degenerate quarks as given in eq.~(\ref{eq:WtmQCDnondeg})
one needs to add this to the hermitian matrix $B$.
To keep the matrix $B$ positive the constraint on the values of $\kappa$ has to be changed
In fact the matrix $B$ is given now by
\be
\delta_{\alpha\beta}\Big\{\delta_{AB}\delta(\bx,\by)\Big[1 + 2 \kappa a \epsilon_{\rm q} \tau^1 \Big]_{ij} 
- \delta_{ij} \kappa \sum_{k=1}^{3}\Big[U(\bx;k)_{AB}\delta(\bx+a\hat{k},\by) + 
U(\by;k)^{-1}_{AB}\delta(\by+a\hat{k},\bx)\Big]\Big\} \ .
\label{eq:B}
\ee
Repeating the same argument given in eqs.~(\ref{eq:Bop}-~\ref{eq:kappa}) we obtain the new constraint
\be
|\kappa| < \frac{1}{6 + 2a\epsilon_{\rm q}}, \qquad \epsilon_{\rm q} > 0.
\label{eq:kappa_new}
\ee

\section{O($a$) improvement}
\label{app:D}

In this appendix we give more technical details of the dimension 5 operators describing
the O($a$) effects of the Wtm lattice action.
In sect.~\ref{sec:basic} we have analyzed the form of the continuum action $S_0$ 
using the symmetries of the lattice theory.

We now construct $S_1$ in eq.~(\ref{eq:eff_action}), which contains dimension five operators.
To classify all the possible operators,
we have to use again the symmetries of the lattice action, 
and make use of partial integration in 
(\ref{eq:operators}) given the space integration over $y$.
As for $S_0$ the residual $U_{\rm V}(1)_3$ flavour symmetry forbids bilinears with $\tau^{1,2}$.
A number of potential terms can be excluded by the
$\widetilde{P} = P \times (\mu_{\rm q} \rightarrow -\mu_{\rm q})$ symmetry, because it implies that
parity violating dimension 5 fields have to be multiplied with a twisted mass term 
to an odd power. We can exclude then terms like $m_{\rm q}\mu_{\rm q}\chibar\chi$,
$m_{\rm q}^2 \chibar i\gamma_5\tau_3 \chi$,
$\mu_{\rm q}^2 \chibar i\gamma_5\tau_3 \chi$,
$\chibar D^2 i\gamma_5 \tau_3 \chi$ and
$\chibar i \sigma_{\mu\nu} F_{\mu\nu} i \gamma_5 \tau_3 \chi$.
The last of these, the ``twisted Pauli term'', requires a factor
of $\mu_{\rm q}$ and thus appears only in $S_2$.
Charge conjugation symmetry excludes a term like $i\mu_{\rm q} 
\{ \chibar \gamma_5 \tau^3 \gamma_\mu \rvec{D_{\mu}}\chi + 
\chibar \lvec{D_{\mu}}  \gamma_\mu \gamma_5 \tau^3 \chi \}$.

The list of the possible fields contributing to ${\mathcal L}_1(y)$ then 
is~\cite{Luscher:1996sc,Frezzotti:2001ea}
\be
{\mathcal O}_1 = i \chibar\sigma_{\mu\nu} F_{\mu\nu} \chi,
\label{eq:o1app}
\ee
\be
{\mathcal O}_2 = m_{\rm q} \tr \{ F_{\mu\nu} F_{\mu\nu} \},
\label{eq:o2app}
\ee
\be
{\mathcal O}_3 = m_{\rm q}^2\chibar \chi,
\label{eq:o3app}
\ee
\be
{\mathcal O}_4=m_{\rm q}\mu_{\rm q} i \chibar\gamma_5\tau^3\chi,
\label{eq:o4app}
\ee
\be
{\mathcal O}_5=\mu_{\rm q}^2\chibar\chi,
\label{eq:o5app}
\ee
\be
{\mathcal O}_6=m_{\rm q}\{\chibar\gamma_\mu \rvec{D_\mu}\chi - \chibar \lvec{D_\mu} \gamma_\mu \chi\},
\label{eq:o6app}
\ee
\be
{\mathcal O}_7=\{\chibar \rvec{D_\mu} \rvec{D_\mu}\chi + \chibar \lvec{D_\mu} \lvec{D_\mu} \chi\}.
\label{eq:o7app}
\ee

To apply the Symanzik improvement program one needs to improve also the local fields
in the correlation functions. 
To be concrete we give here the example of the currents $A_\mu^a$, $V_\mu^a$
and $P^a$ which appear in the Ward identities~(\ref{eq:PCAC},\ref{eq:PCVC}). 
The local operators contributing to the leading correction term in the effective axial current
$A_\mu^a$ are~\cite{Heatlie:1990kg,Luscher:1996sc,Frezzotti:2001ea}: 
\be
({\mathcal O}_8)_\mu^a = \chibar\gamma_5\frac{\tau^a}{2}\sigma_{\mu \nu}\rvec{D_\nu}\chi - 
\chibar\lvec{D_\nu}\sigma_{\mu \nu}\gamma_5\frac{\tau^a}{2}\chi,
\label{eq:O8app}
\ee
\be
({\mathcal O}_9)_\mu^a = \chibar\gamma_5\frac{\tau^a}{2}\rvec{D_\mu}\chi + 
\chibar\lvec{D_\mu}\gamma_5\frac{\tau^a}{2}\chi,
\ee
\be
({\mathcal O}_{10})_\mu^a = m_{\rm q} \chibar\gamma_\mu\gamma_5\frac{\tau^a}{2}\chi,
\ee
\be
({\mathcal O}_{11})_\mu^a =  \mu_{\rm q} \epsilon^{3ab} \chibar\gamma_\mu\frac{\tau^b}{2}\chi \ ,
\ee
for the pseudoscalar density $P^a$ we need
\be
({\mathcal O}_{12})^a = m_{\rm q} \chibar\gamma_5\frac{\tau^a}{2}\chi \ ,
\ee
and for the vector current $V_\mu^a$
\be
({\mathcal O}_{13})_\mu^a = \chibar\sigma_{\mu \nu} \frac{\tau^a}{2}\rvec{D_\nu}\chi +
\chibar\lvec{D_\nu}\sigma_{\mu \nu} \frac{\tau^a}{2}\chi,
\ee
\be
({\mathcal O}_{14})_\mu^a = m_{\rm q} \chibar\gamma_\mu\frac{\tau^a}{2}\chi,
\ee
\be
({\mathcal O}_{15})_\mu^a =  \mu_{\rm q} \epsilon^{3ab} \chibar\gamma_\mu\gamma_5\frac{\tau^b}{2}\chi \ .
\label{eq:O15}
\ee

In general we consider a generic connected lattice correlation function 
$\langle \Phi \rangle$~(\ref{eq:phi}) at $x_1 \neq x_2 \neq \cdots \neq x_n$.
In the effective theory up to order $a$, $\langle \Phi \rangle$ is given by eq.~(\ref{eq:sym_exp1}).
Without contact terms the continuum equation of motions can be used in order to constrain the set of operators
${\mathcal O}_i$. Since we consider correlation functions with $x_1 \neq x_2 \neq \cdots \neq x_n$,
the only contact terms arise when the argument $y$ of ${\mathcal L}_1(y)$ is equal to the argument $x_k$ of 
one of the fields in the correlation function. But as we have discussed 
in sect.~\ref{ssec:sym} this simply amounts to a redefinition of $\phi_1$, so the equations of motion 
of the continuum theory can be used to simplify the field basis for the effective action and operators.
But it is important to remember that the coefficients appearing in the linear combination
$\phi_1$ will depend on the choice of the basis and on the value of the 
coefficients of ${\mathcal L}_1$.

The list of operators given above for the O($a$) lattice correction 
can then be reduced using the field equations. Formal application of the field equations
gives
\be
{\mathcal O}_1 - {\mathcal O}_7 + 2 {\mathcal O}_5 + 2 {\mathcal O}_3 = 0
\label{eq:eom1}
\ee
\be
{\mathcal O}_6 + 2 {\mathcal O}_3 + 2 {\mathcal O}_4 = 0 , 
\label{eq:eom2}
\ee
which allows to eleminate ${\mathcal O}_6$ and 
${\mathcal O}_7$.

A technical remark has to be made here.
The naive relations~(\ref{eq:eom1},\ref{eq:eom2}) are obtained at tree-level of perturbation theory.
At higher order they should be replaced by linear combinations of all basis elements with coefficients 
that depend on the coupling. Nevertheless the simple existence of such relations allows us to eliminate
${\mathcal O}_6$ and ${\mathcal O}_7$.

The same procedure can be adopted to eliminate redundant terms for the operators 
describing the leading O($a$) corrections to the pseudoscalar density and the axial and vector current,
similarly to what we have done for the operators contributing to the effective action, 

For example for the axial current the operator $({\mathcal O}_8)_\mu^a$ can be eliminated using 
the equations of motion.

\section{Automatic O($a$) improvement}
\label{app:E}

In this appendix we will give the basic idea of the first proof of automatic
O($a$) improvement~\cite{Frezzotti:2003ni}. We do not give the whole proof
and we invite the reader to look at the original paper~\cite{Frezzotti:2003ni} 
for more details.

The form of the unimproved lattice action is $S = S_G + S_F$
and the Wtm quark action $S_F$ is defined in eq.~(\ref{eq:WtmQCD}). 
The form of the gauge action is not relevant because it has leading O($a^2$) discretization errors.
A general Symanzik expansion of a connected lattice correlation function~(\ref{eq:sym_exp1}) 
of multiplicatively renormalized fields contains coefficients (e.g. in the linear combinations
which define $\mathcal{L}_1$) that will depend on the bare parameters of the lattice action.

We consider first the plain Wilson action setting $\mu_{\rm q} = 0$ in eq.~(\ref{eq:WtmQCD}),
and we define here the discrete axial-vector transformation\footnote{This transformation 
is not anomalous since is the product of 2 non-anomalous vector and axial transformations 
in the same isospin direction.}
\be
\mathcal{R}_5 \colon
\begin{cases}
\chi(x_0,{\bf x}) \rightarrow \gamma_5  \chi(x_0,{\bf x}) \\
\chibar(x_0,{\bf x}) \rightarrow  -\chibar(x_0,{\bf x}) \gamma_5
\end{cases}
\ee
 The lattice action is invariant if after a 
$\mathcal{R}_5$ transformation on the fields we change also the sign of the Wilson term and the mass term.
In particular the lattice action~(\ref{eq:WtmQCD}), with $\mu_{\rm q} = 0$, 
is invariant under the following transformation
\be
\mathcal{R}_5^{\rm sp} \equiv \mathcal{R}_5 \times [r \rightarrow -r] \times [m_0 \rightarrow -m_0].
\ee
This happens because both Wilson and mass terms are odd under the $\mathcal{R}_5$ symmetry.
It is also interesting to note that both the Wilson and the mass term have odd dimensions, 
if we exclude the bare quark mass $m_0$ and the lattice spacing $a$ in the counting.
As a consequence of this, all the terms of O($a$) in the Symanzik expansion~(\ref{eq:sym_exp1}) 
will have opposite definite properties under $r \rightarrow -r$, such that averages of correlation
functions computed on the lattice with opposite values of $r$ have a faster approach to the continuum
limit, i.e. the O($a$) effects are 
cancelled\footnote{Ideas in this direction were already put forward in \cite{Jacobs:1983ph,Aoki:1984qi}.}.

If the theory is tuned to full twist, $\omega=\pi/2$, this averaging procedure is automatic.
In fact if we consider the action~(\ref{eq:tWilson}) the transformation
$r \rightarrow -r$ is equivalent to $\omega \rightarrow \omega + \pi$. 

\section{Nucleon operators in the twisted basis}
\label{app:F}

In this appendix we show how the nucleon operators change in the twisted
basis. This is a rather simple exercise and it is shown here just as an
example how to proceed in a general case.
The starting point is the lattice action in eq.~(\ref{eq:WtmQCD}). The
connection between correlators in standard QCD and tmQCD can be inferred from
classical consideration.
We consider the doublet 
\begin{displaymath}
\chi =
\left( \begin{array}{c}
u \\
d
\end{array} \right) .
\end{displaymath}
The axial rotation in eq.~(\ref{eq:axial}) for single flavour takes the form
\be
u(x) \rightarrow {\rm exp}\big(i\frac{\omega}{2}\gamma_5\big) u(x)
\ee
\be
d(x) \rightarrow {\rm exp}\big(-i\frac{\omega}{2}\gamma_5\big) d(x) 
\ee
and the corresponding for the antiquark fields.
The proton field 
\be
\mathscr{P} = u_A[u_B^TC^{-1}\gamma_5d_C]\epsilon_{ABC}, 
\label{eq:proton_app}
\ee
transforms in
\be
\mathscr{P} \rightarrow {\rm
  exp}\left(i\frac{\omega}{2}\gamma_5\right)\mathscr{P} .
\ee
In QCD the proton correlation function 
\be
G_{\mathscr{P}}(x_0) = \int {\rm d}^3x \langle \mathscr{P}(x_0,{\bf x}) 
\bar{\mathscr{P}}(0,{\bf 0}) \rangle_{\rm QCD} , 
\ee
has the general decomposition (for simplicity we consider only the leading exponential contributions)
\bea
G_{\mathscr{P}}(x_0) &=& P_+ \theta(x_0) A_{\mathscr{P}^+} e^{-M_{\mathscr{P}^+} x_0} +
P_- \theta(-x_0) A_{\mathscr{P}^+} e^{M_{\mathscr{P}^+} x_0} - \nonumber \\
&& P_+ \theta(-x_0) A_{\mathscr{P}^-} e^{M_{\mathscr{P}^-} x_0} -
P_- \theta(x_0) A_{\mathscr{P}^-} e^{-M_{\mathscr{P}^-} x_0} ,
\label{eq:prot_dec}
\eea
where
\be
P_{\pm} = \frac{1}{2}\left( 1 \pm \gamma_0 \right) .
\ee
The backward propagating contributions correspond to the antiparticles 
of the forward propagating states with opposite parity.
The desired state is obtained using the appropriate projection operator
$P_{\pm}$ and the appropriate direction of propagation.
Given the classical correspondence between QCD and tmQCD~(\ref{eq:continuum_rot})
the mass of the desired state in the proton channel can be extracted from the exponential behaviour
of the following correlator in tm QCD
\be
\Tr\left[P_{\pm}G_{\mathscr{P}}^{tm}(x_0)\right]
\ee
where
\be
G_{\mathscr{P}}^{tm}(x_0) = \int {\rm d}^3x \left\langle e^{\left(i\frac{\omega}{2}\gamma_5\right)} 
\mathscr{P}(x_0,{\bf x}) \bar{\mathscr{P}}(0,{\bf 0}) e^{\left(i\frac{\omega}{2}\gamma_5\right)}
\right\rangle_{\rm tmQCD}.
\label{eq:tm_proton_corr}
\ee
The correlation function in eq.~(\ref{eq:tm_proton_corr}) can now be computed on the lattice
with the Wtm lattice action~(\ref{eq:WtmQCD}).

\newpage

\bibliographystyle{h-elsevier}    
\bibliography{tmrep}      
\end{document}